\documentclass[article,10pt,aps,prx,twocolumn,superscriptaddress,english,longbibliography,floatfix]{revtex4-2}

\usepackage{bm,amssymb,amsmath,dsfont,mathrsfs,amstext,latexsym,physics,relsize}
\usepackage{graphicx}

\usepackage[usenames,dvipsnames]{xcolor}

\usepackage{hyperref}
\usepackage[all]{hypcap}
\usepackage{hyperref}
\usepackage[normalem]{ulem}
\usepackage[T1]{fontenc}
\usepackage{float}

\newcommand{\phdagger}[0]{{\phantom{\dagger}}}
\newcommand{\Te}[0]{\mathrm{T}e}
\newcommand{\aTe}[0]{\widetilde{\mathrm{T}}e}

\newcommand{\pbra}[1]{(#1|}
\newcommand{\pket}[1]{|#1)}

\newcommand{\figpanel}[2]{\hyperref[#1]{\ref{#1}#2}}

\begin{document}

\title{Fighting Exponentially Small Gaps by Counterdiabatic Driving}

\author{Andr\'as Grabarits\href{https://orcid.org/0000-0002-0633-7195}{\includegraphics[scale=0.05]{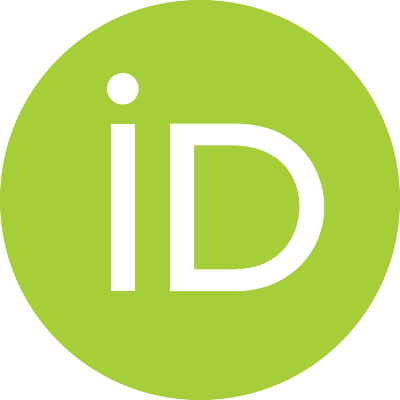}}}
\email{andras.grabarits@uni.lu}
\affiliation{Department  of  Physics  and  Materials  Science,  University  of  Luxembourg,  L-1511  Luxembourg, Luxembourg}

\author{Federico Balducci\href{https://orcid.org/0000-0002-4798-6386}{\includegraphics[scale=0.05]{orcidid.pdf}}}
\email{fbalducci@pks.mpg.de}
\affiliation{Department  of  Physics  and  Materials  Science,  University  of  Luxembourg,  L-1511  Luxembourg, Luxembourg}
\affiliation{Max Planck Institute for the Physics of Complex Systems, N\"othnitzer Str. 38, 01187 Dresden, Germany}

\author{Adolfo del Campo\href{https://orcid.org/0000-0003-2219-2851}{\includegraphics[scale=0.05]{orcidid.pdf}}}
\affiliation{Department  of  Physics  and  Materials  Science,  University  of  Luxembourg,  L-1511  Luxembourg, Luxembourg}
\affiliation{Donostia International Physics Center,  E-20018 San Sebasti\'an, Spain}

\date{\today}


\begin{abstract}
We investigate the efficiency of approximate counterdiabatic driving (CD) in accelerating adiabatic passage through exponentially small gaps. First, we analyze a minimal spin-glass bottleneck model that is analytically tractable and exhibits both an exponentially small gap at the transition point and a change in the ground state that involves a macroscopic rearrangement of spins. Using the variational Floquet-Krylov expansion to construct CD terms, we find that while the formation of excitations is significantly suppressed, achieving a fully adiabatic evolution remains challenging. Extending our investigation to realistic NP-hard spin-glass problems---specifically, the $3$-regular \textsc{Max Cut} and $3$-\textsc{XORSAT}---we find again that local CD expansions lead to negligible improvements in the final ground state fidelity. These results highlight the limited impact of local CD methods in overcoming the bottlenecks associated with first-order quantum phase transitions. To address this limitation, we propose an alternative method, termed quantum brachistochrone counterdiabatic driving (QBCD), which employs the approximate full CD connecting the ground state and the first excited state at a single parameter value close to the critical point. In the minimal spin-glass model, QBCD enables exponentially faster adiabatic evolution than the local strategies.  To alleviate the challenges of its experimental and classical implementation for realistic \textsc{NP}-hard problems, we exponentially reduce the non-locality of the QBCD Hamiltonian by sparsifying its matrix elements to the density of the local expansions. Despite this drastic simplification, sparsified QBCD maintains finite ground-state fidelity at driving times exponentially shorter than in local strategies and counterdiabatic optimized local driving (COLD), demonstrating that even an exponentially reduced fraction of non-locality near the critical point is sufficient to achieve significant speedup across spin-glass bottlenecks. At the same time, reduced non-locality facilitates classical implementations via advanced numerical methods, while on quantum hardware its truncation serves as a resource-efficient approximation scheme for higher-order CD realizations.

\end{abstract}

\maketitle

\section{Introduction} 

The breaking of adiabatic dynamics in many-body spin systems arises in a wide variety of scenarios ranging from condensed matter physics to quantum simulation and quantum computing. According to the adiabatic theorem, the driving of a quantum system in a time shorter than the inverse of the spectral gaps generally induces diabatic transitions. Nonadiabatic dynamics limit the preparation of novel ground-state phases of matter in quantum simulation. Excitations further account for errors in adiabatic quantum computation, quantum annealing, and quantum optimization algorithms. They are also responsible for quantum friction in finite-time thermodynamics, limiting the scaling of quantum heat engines and refrigerators, among other quantum devices. 

Enforcing adiabaticity in many-body quantum systems is thus broadly desirable. Yet, it faces the need to overcome small gaps in the spectrum of the driven system.
By way of example, hard instances of combinatorial problems might be difficult to solve by quantum annealers because of the presence of gaps on the driving path that are exponentially small in the system size~\cite{Young2008Size,*Young2010First,Amin2008Effect,Amin2009First,Altshuler2009Adiabatic,*Altshuler2010Anderson}. Such gaps render quantum adiabatic evolution unfeasible in practice since the adiabatic theorem requires the driving time $T$ to be larger than the inverse square of the minimal gap $\Delta_{\min}$~\cite{Marzlin2004Inconsistency,Jansen2007Bounds}. The physical origin of these gaps remains a topic of active debate.  Some authors have used perturbation theory to link them to first-order quantum phase transitions~\cite{Amin2008Effect,Amin2009First}, suggesting they arise from Anderson localization ~\cite{Altshuler2009Adiabatic,Altshuler2010Anderson,Laumann2015Quantum}. Others argue that the gaps have a non-perturbative origin~\cite{Knysh2010Relevance} or result from the clustering of solutions typical in spin-glass phases~\cite{Foini2010Solvable}. Despite the debate, large-scale numerical simulations have consistently demonstrated that certain models display exponentially small gaps~\cite{Young2008Size,*Young2010First,Jorg2008Simple,Jorg2010First}.

Facing the presence of small gaps and their dependence on the system size has led to reformulations of the adiabatic theorem in a many-body setting~\cite{Bachmann17,Bachmann2018}, and the notion of quasi-adiabatic continuation used to determine the existence and stability of phases of matter~\cite{Hastings04,*Hastings2007Quasi,*Hastings2010Locality,*Hastings2010Quasi,Hastings05,Hastings2015Quantization,DeRoeck15}. 

These efforts have been accompanied by a closely related quest for fast quantum control of many-body systems,  tackling the presence of small gaps on the adiabatic path. 
In the context of adiabatic quantum computing, tailoring the mixer or problem Hamiltonian has been proposed to remove the exponential gaps from the path~\cite{Farhi2011Quantum, Choi2011Avoid,*Choi2011Different,*Choi2011Different2, Dickson2011Elimination, Dickson2011Does, Choi2020Effects}. Alternative methods involve additional ``catalyst'' Hamiltonians, i.e.,\ operators that act only at intermediate times to amplify the gap~\cite{Suzuki_2007_originalFerroGapamplification, Seki_2012_AntiFerroFluct_Gapamplification, Seoane_2012_ClassicalMEthodforGapAmplfiication, Somma2013Spectral, CrossonFarhi2014Gap_amplication, Hormozi_NonStoqCatalyst_gapAmplificatio_2017, Susa2018_Inhomdriving_gap_amplification, Albash2019_NonStoqGap_amplification, Cao2021Catatlyst, Mehta_2SAT_NonSotq_Gapamplification}. 
Other strategies rely on time discretization combined with the use of classical optimization, leading to the quantum approximate optimization algorithm (QAOA)~\cite{Farhi2014Quantum, Zhou2020Quantum}, and the use of quantum optimal control to tailor non-linear parameter driving~\cite{Yang2017Optimizing, Lin2019Application,Brady2021Optimal,Cote2023Diabatic,Grabarits2024Non}.

The problem of speeding up the adiabatic evolution was also addressed in other contexts, such as the application of quantum control in chemistry and physics~\cite{Rice2000}. Seminal works~\cite{Demirplak2003Adiabatic,*Demirplak2005Assisted,*Demirplak2008Consistency,Berry2009Transitionless} showed that adiabatic evolution for a generic Hamiltonian can be achieved in any finite time, provided an auxiliary \emph{counterdiabatic driving} (CD) control term is implemented. The CD term is constructed in such a way as to cancel all diabatic transitions, realizing exactly the adiabatic evolution of the reference uncontrolled Hamiltonian. While its explicit form is familiar from early proofs of the adiabatic theorem~\cite{Kato1950Adiabatic} and can be obtained directly given the spectral decomposition of the instantaneous Hamiltonian, the CD Hamiltonian is generically non-local and involves multiple-body interaction terms ~\cite{delCampo2012Assisted,Takahashi2013Transitionless,Damski2014Counterdiabatic,Saberi2014Adiabatic}. However, it can also be obtained variationally via controlled expansions in increasingly non-local terms~\cite{Saberi2014Adiabatic,Kolodrubetz2017Geometry,Sels2017Minimizing,Claeys2019Floquet,Takahashi2024Shortcuts,Bhattacharjee2023Lanczos}, with exact adiabatic dynamics being recovered when all the terms in the expansion are retained. As a result, contrary to catalysts and ad-hoc deformations of the driving Hamiltonian, CD provides a coherent framework to address the problem of exponentially small gaps. CD has been shown to boost the performance of QAOA, motivating its use in quantum optimization ~\cite{Chandarana2022Digitized,Hegade2022Digitized,Wurtz2022Counterdiabaticity,hegade2023digitizedCDFact,Chandarana2023ProteinDigitizedCD,Malla2024,Chandarana2024}. It should be noted that the required  CD terms cannot generally be realized in analog quantum annealers and simulators. However, they can be implemented by digital~\cite{delCampo2012Assisted,Saberi2014Adiabatic,Chandarana2022Digitized,Hegade2022Digitized,Chandarana2023ProteinDigitizedCD,Hatomura23,Hatomura24,Malla2024,Chandarana2024} or hybrid digital-analog quantum simulation~\cite{Lamata18,Kumar24}. Within such approaches, the implementation is generally facilitated by the locality of the interactions.

Despite a surge of interest in harnessing CD for many-body interacting systems~\cite{delCampo2012Assisted,Takahashi2013Transitionless,Saberi2014Adiabatic,Damski2014Counterdiabatic,Campbell15,Mukherjee16,Sels2017Minimizing,Kolodrubetz2017Geometry,Claeys2019Floquet,Carolan22,Hartmann2022Polynomial,Chandarana2022Digitized,Hegade2022Digitized,Wurtz2022Counterdiabaticity,hegade2023digitizedCDFact,Chandarana2023ProteinDigitizedCD,Cepaite2023Counterdiabatic,Takahashi2024Shortcuts,Bhattacharjee2023Lanczos,McKeever24,Malla2024}, crucial aspects remain to be elucidated. 
Early studies pointed out the interplay between CD and the Kibble-Zurek (KZ) scaling at second-order quantum phase transitions, which describes the breakdown of adiabaticity while crossing critical point in terms of universal equilibrium critical exponents~\cite{Damski05,Zurek2005Dynamics,Dziarmaga05,Polkovnikov2005Universal,Damski06,delCampo2014Universality}. It was found that non-local CD terms are needed to suppress the KZ scaling and restore adiabaticity~\cite{delCampo2012Assisted,delCampo2014Universality,Damski2014Counterdiabatic}, which is at odds with the quest for approximate CD terms with locality tailored for implementations. 
The breaking of adiabaticity described by KZ mechanism in first-order phase transitions has been theoretically and experimentally studied~\cite{Qiu20,Suzuki24}, while the use of bias fields in digitized quantum optimization assisted by CD has only recently been put forward~\cite{Cadavid2024bias}. 
The performance of CD in a problem with an exponentially small gap and frustration is yet to be elucidated. Some works~\cite{Passarelli2020Counterdiabatic,Prielinger2021Two} have addressed the CD-assisted annealing of ferromagnetic $p$-spin models, showing that CD improves the final fidelity while still remaining exponentially small in the system size. The effect of frustration, however, being at the basis of the exponential difficulty of solving \textsc{NP}-hard problems, was not considered. Furthermore, while in certain $p$-spin models, simple antiferromagnetic catalysts can yield a polynomial scaling of the gap~\cite{Albash2019_NonStoqGap_amplification}, this cannot happen if the system is frustrated~\cite{Mehta_2SAT_NonSotq_Gapamplification}. 

Several vital questions thus arise: How efficiently can approximate counterdiabatic (CD) driving improve adiabatic passage through exponentially small gaps between frustrated energy levels? Can alternative strategies be developed to overcome these regimes more effectively? Most importantly, what additional, minimal information beyond local terms is required to achieve exponential speedups in adiabaticity? And how far can the required non-locality and the evaluation range of parameters be restricted?

In this work, we investigate whether local, approximate counterdiabatic (CD) driving can effectively suppress excitations and accelerate adiabatic dynamics in many-body quantum systems exhibiting exponentially small gaps and frustration. To this end, we evaluate the performance using three experimentally relevant metrics: the density of non-adiabatic excitations, the ground-state gap, and the final ground-state fidelity\cite{Gong2016,Cui16,Cui20,Keesling2019,Gardas2018Defects,Weinberg2020Scaling,Bando2020Probing,King2022Coherent,King2024}. We find that while local approximate CD can significantly reduce excitations away from the adiabatic limit, its impact diminishes near the adiabatic timescale. To address this limitation, we introduce quantum brachistochrone counterdiabatic driving (QBCD), which yields a remarkable improvement by relying solely on approximate knowledge of the ground and first excited states at a single point in parameter space. We further validate these findings on two realistic \textsc{NP}-hard problems, demonstrating the superiority of an advanced, sparsified QBCD method---featuring exponentially reduced non-locality---over both the local expansion schemes and the counterdiabatic local optimized driving (COLD) approach.

\subsection{Summary of results} 

To address the questions above, we study a recently introduced minimal model of a spin-glass bottleneck~\cite{Roberts2020Noise}. The model is an Ising-like spin chain that displays both a small gap and frustration: not only does the gap to the first excited state scale exponentially with the system size $L$, but also $O(L)$ spins need to be flipped at the avoided crossing to track adiabatically the ground state. These features are due to a simple localization mechanism reminiscent of the formation of bound states for delta-function potentials on the real line. The model can be mapped onto a chain of free fermions, which allows a complete analytical understanding of the small gap formation~\cite{Roberts2020Noise} and the simulation of large systems to access the asymptotic scaling. Although this model exhibits most of the essential features of realistic spin glass problems, we further extend the investigation to two realistic \textsc{NP}-hard problems, the \textsc{Max Cut} and the $3$-\textsc{XORSAT} problems~\cite{Young2010First}, in which spin glass behavior and an exponentially small gap emerges for numerically accessible system sizes, $L\gtrsim10$.

First of all, we present a thorough analysis of the model properties in Sec.~\ref{sec:model}. While most properties were already derived in the original reference~\cite{Roberts2020Noise}, we elucidate the role of a chiral symmetry that was previously mistaken for a regular symmetry and provide a detailed derivation of the free-fermionic representation of the model. Then, after briefly reviewing counterdiabatic driving (CD) in Sec.~\ref{sec:CD_review}, we compute the free-fermionic representation of approximate CD for the model under consideration in Sec.~\ref{sec:approximate_CD}.

In Sec.~\ref{sec:results}, we present our main results for the minimal spin glass model. We show that local, approximate CD terms obtained without prior knowledge of the spectrum provide reasonable speed-up only for a given regime of driving times. We do so with two complementary approaches. First, we consider the effect of CD terms on the minimal gap along the adiabatic path, i.e., \ we study the performance of CD as a gap amplification technique. Since the CD Hamiltonian depends on the total evolution time, defining a gap requires care. A  solution is presented in Sec.~\ref{sec:CD_gap_amplification},  showing that if the problem Hamiltonian displays a minimal gap $\Delta_{\min} \sim e^{-\alpha L}$, then adding CD terms leads to $\Delta_{\min{},\mathrm{CD}} \sim e^{-\alpha_\mathrm{CD} L}$ with $\alpha_\mathrm{CD}$ being smaller than $\alpha$ but remaining finite. Therefore, while CD accelerates the driving protocol exponentially, the scaling remains exponentially hard. Second, in Sec.~\ref{sec:performance_CD}, we confirm that CD does not help the system overcome the smallest gap by directly simulating the time evolution of the system. We consider the number of kinks at the end of the schedule, displaying relevant suppression only for the short-time evolutions. Meanwhile, significant improvement for slower schedules can only be achieved for a sufficiently large number of expansion terms due to the macroscopic rearrangement of the ground states around the critical point. In the language of free fermions, local CD maps to local fermion hopping terms, while a non-local population transfer is needed to track adiabatically the ground state.

In Sec.~\ref{sec:CD_LR}, we introduce a novel approach to the bottleneck problem by addressing directly the closing of the gap. The QBCD Hamiltonian is constructed using approximate knowledge of the eigenstates at a single parameter value near the transition. In the minimal spin glass model, the QCDB can be understood as driving the localized edge states associated with the minimal gap.  We show that this approach achieves exponentially shorter adiabatic time scales by exploiting only an exponentially small fraction of the full counterdiabatic Hamiltonian. Combined with its system-size-independent energetic cost, these features support near-term experimental implementation.

We then test the generality of our findings in two realistic spin glass models representing \textsc{NP}-hard optimization problems, the \textsc{Max Cut} and the $3$-\textsc{XORSAT}. First, we demonstrate the same limited efficiency of the first- and second-order local expansion via the ground state gap and the final fidelity. Next, we introduce a yet further enhanced version of the QBCD Hamiltonian by sparsifying it to the density of the local expansion. Remarkably, this exponentially reduced non-locality still suffices to achieve a comparable adiabatic speedup as the full QBCD. These features significantly ease the challenges of implementation, making the approach promising for both experimental platforms and large-scale classical simulations. Finally, we test the COLD approach in the \textsc{Max Cut} problem and find that, even when implemented along optimized annealing paths, its advantage over local expansion quickly diminishes with increasing $L$, whereas the sparsified QBCD consistently outperforms it. In addition, the numerical effort required by COLD grows rapidly, restricting its applicability to very small system sizes ($L \lesssim 10$), at variance with our approach.

A closing discussion in relation to previous findings follows in Sec.~\ref{sec:discussion}.

\section{Minimal model of spin-glass bottleneck}
\label{sec:model}

\begin{figure}[t]
    \centering
    \includegraphics[width=\columnwidth]{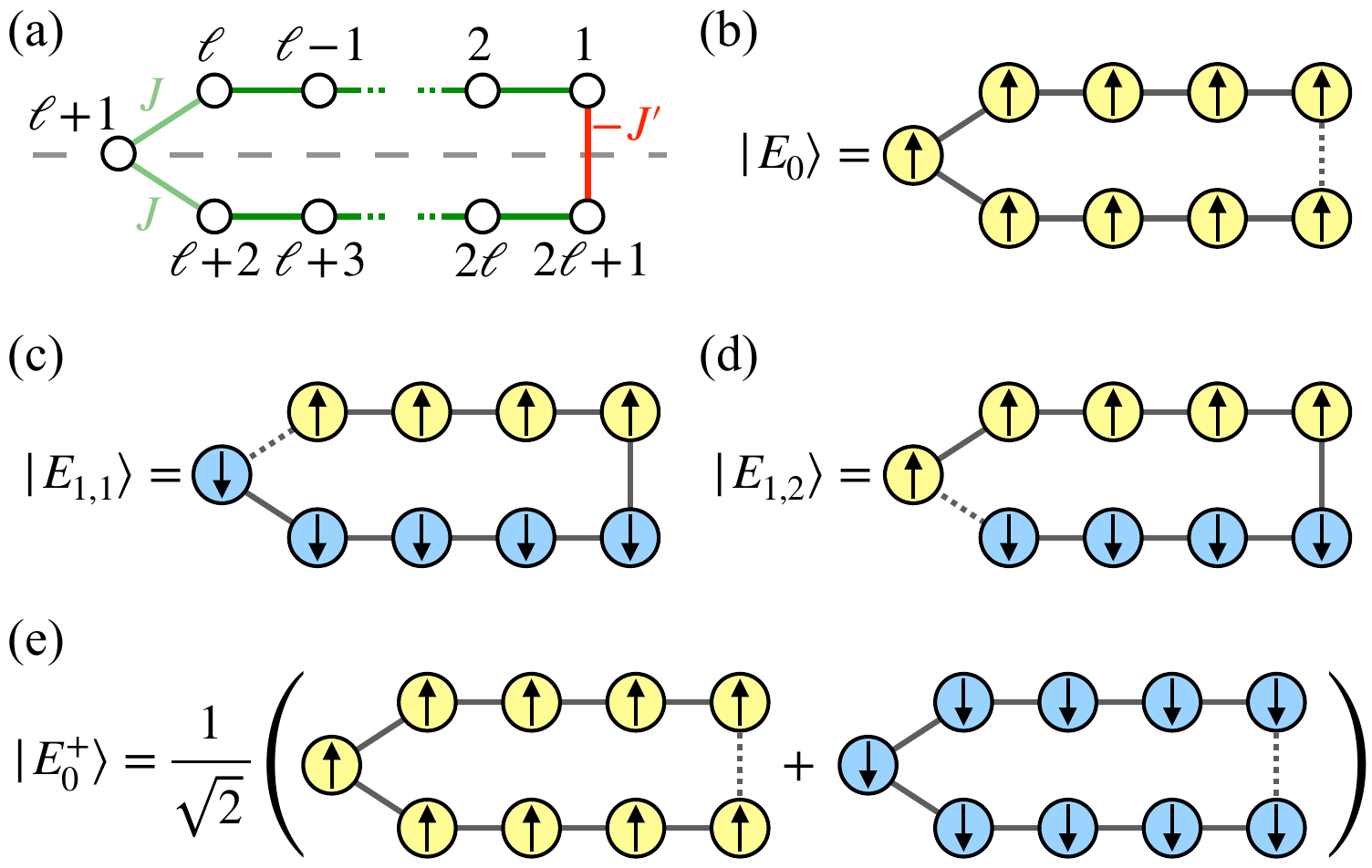}
    \caption{(a) Ising chain of Eqs.~\eqref{eq:H}--\eqref{eq:Jj}. The chain is $L = 2\ell+1$ sites long, with uniform ferromagnetic bonds (green lines), except for two neighboring, weakly ferromagnetic bonds (light green lines) and a single antiferromagnetic bond (red line). The dashed gray line indicates the presence of the $\mathbb{Z}_2$ reflection symmetries, Eqs.~\eqref{eq:reflection_parity} and \eqref{eq:chiral_reflection_parity}. (b) The ground state at $\lambda=1$ (fixing $L=9$) has only the antiferromagnetic bond frustrated (dotted line). The other degenerate ground state with opposite fermion parity is not shown. (c)--(d) Two degenerate first excited states at $\lambda=1$ (with $L=9$), characterized by only one weak ferromagnetic bond frustrated (dotted line). Two other degenerate first excited states are obtained by flipping all the spins. (e) The restriction to the even fermion parity sector makes the ground state nondegenerate and the first excited manifold two-dimensional. The positive fermion parity states are obtained by means of symmetric combinations, as shown here for the ground state.}
    \label{fig:chain}
\end{figure}

Consider the quantum many-body Hamiltonian~\cite{Roberts2020Noise}
\begin{equation}
    \label{eq:H}
    H[\lambda] = -(1-\lambda) \sum_{j=1}^L \sigma_j^x - \lambda \sum_{j=1}^L J_j \sigma_j^z \sigma_{j+1}^z,
\end{equation}
where $\lambda \in [0,1]$ is the driving parameter, and the chain has an odd number of sites $L=2\ell+1$, with a periodic boundary. We use square brackets to denote the dependence on the parameter $\lambda$, while we use parentheses for the other functional dependencies, e.g., of $\lambda(t)$ on time. The couplings are set to
\begin{equation}
    \label{eq:Jj}
    J_j =
    \begin{cases}
        J   &j = \ell,\ell+1 \\
        -J' &j = 2\ell+1 \\
        1   &\text{otherwise,}
    \end{cases} 
\end{equation}
where it is assumed
\begin{equation}
    \label{eq:J_range}
    0 < J ' < J < 1, \qquad J^2 < J'.
\end{equation}
From the pictorial representation in Fig.~\figpanel{fig:chain}{a}, it can be seen that Eq.~\eqref{eq:H} describes an Ising chain with uniform ferromagnetic couplings, except for three weaker couplings placed in diametrically opposite positions: two weak ferromagnetic ones ($J$) next to each other, and one weaker antiferromagnetic one ($-J'$) on the other side. The condition $J^2<J'$ in Eq.~\eqref{eq:J_range} is more technical, and is explained in App.~\ref{app:sec:avoided_crossing}.

The Hamiltonian in Eq.~\eqref{eq:H} posseses both $\mathbb{Z}_2$ fermion parity and $\mathbb{Z}_2$ reflection parity symmetry:
\begin{gather}
    \label{eq:fermion_parity}
    \Pi_\mathrm{F} = \prod_{j=1}^L \sigma_j^x,\quad
    \big[ \Pi_\mathrm{F}, H[\lambda] \big] = 0; \\
    \label{eq:reflection_parity}
    \Pi_\mathrm{R} \sigma_j^\alpha \Pi_\mathrm{R} = \sigma_{L-j+1}^\alpha,\quad\left[\Pi_\mathrm{R},H[\lambda]\right]=0,\quad \alpha = x,y,z,
\end{gather}
The second one is highlighted with a grey dashed line in Fig.~\ref{fig:chain}a. In the following, we focus on the even fermion parity sector, dynamically accessible from the ground state at $\lambda = 0$.

\subsection{Simple limits}

To analyze the model, we consider the lowest energy states at the beginning ($\lambda=0$) and end ($\lambda=1$) of the driving schedule. At $\lambda=0$, the ground state is the product state $\ket{E_0[\lambda=0]} = \prod_j (\ket{\uparrow}_j + \ket{\downarrow}_j)/\sqrt{2}$, with even fermion parity, $\Pi_\mathrm{F} \ket{E_0[\lambda=0]} = \ket{E_0[\lambda=0]}$. The first-excited states are obtained by flipping a single spin.  

At $\lambda=1$, the presence of a single antiferromagnetic bond in the loop leads to frustration. For $0 < J' < J < 1$, see Eq.~\eqref{eq:J_range}, the ground-state energy is $E_0 = -(L-3) -2J + J'$, and the twofold degeneracy is lifted when restricting to the even-parity sector, yielding the symmetric state $\ket{E_0^+}$. The first excited states correspond to configurations with a single frustrated bond at $j=2\ell+1$ and energies $E_1 = -(L-3)-J'$, with symmetric combinations in the even parity sector.

The ground state undergoes an exponentially small avoided crossing with the first excited state at an intermediate value of $\lambda$, denoted by $\lambda_c$. Its explicit expression is discussed in App.~\ref{app:sec:avoided_crossing}. Across this critical value, the ground state changes from a dressed version of the states in Fig.~\figpanel{fig:chain}{c--d} ($\lambda < \lambda_c$) to a dressed version of the state in Fig.~\figpanel{fig:chain}{b} ($\lambda > \lambda_c$), involving a macroscopic flipping of half the chain. This behavior resembles transitions found in spin-glass models.

In the remainder of this section, we provide a thorough description of the avoided crossing, correcting minor omissions in Ref.~\cite{Roberts2020Noise} while preserving the key results.

\subsection{Fermionic representation and reflection parity symmetry}
\label{sec:fermions}

By means of a Jordan-Wigner transformation, the spin chain Eq.~\eqref{eq:H} can be mapped to non-interacting free fermions, as shown in App.~\ref{app:sec:H_Dirac}). Upon introducing a Majorana fermion representation and by a further transformation using fermionic operators accounting for reflection symmetry (see App.~\ref{app:sec:Majorana} and App.~\ref{app:sec:Gamma}, respectively), one arrives at a one-dimensional hopping model with on-site potentials at the ends of the chain,
\begin{equation}
    \label{eq:H_Gamma}
    \begin{aligned}
        H[\lambda] = &2\lambda \sum_{j=1}^{\ell} J_j \left( \Gamma_{2j}^+ \Gamma_{2j+1}^- + \Gamma_{2j+1}^+ \Gamma_{2j}^- \right) \\
        &-2 (1-\lambda) \sum_{j=1}^{\ell} \left( \Gamma_{2j-1}^+ \Gamma_{2j}^- + \Gamma_{2j}^+ \Gamma_{2j-1}^- \right) \\
        &- \lambda J_L \left( 2 \Gamma_1^+ \Gamma_1^- - 1 \right)
        - (1-\lambda) \left( 2 \Gamma_L^+ \Gamma_L^- - 1 \right), 
    \end{aligned}
\end{equation}
where the $\Gamma$ operators satisfy fermionic anti-commutation relations and possess reflection parity symmetry:
\begin{subequations}
\begin{gather}
    \label{eq:Gamma_commutation}
    \{ \Gamma_k^+, \Gamma_{k'}^-\} = \delta_{k,k'}, \qquad
    \{ \Gamma_k^+, \Gamma_{k'}^+\} = \{ \Gamma_k^-, \Gamma_{k'}^-\} = 0, \\
    \label{eq:Gamma_reflection}
    \widetilde{\Pi}_\mathrm{R} \Gamma_k^\pm \widetilde{\Pi}_\mathrm{R} = \pm i \Gamma_k^\mp, \qquad
    (\Gamma^\pm_k)^\dagger = \Gamma^\mp_k.
\end{gather}
\end{subequations}
These identities make use of the chiral reflection parity operator $\widetilde{\Pi}_\mathrm{R}$, which is discussed in detail in App.~\ref{app:sec:reflection_parity}.

Introducing the vectors
\begin{equation}
    \vec{\Gamma}^+ = 
    \begin{pmatrix}
        \Gamma^+_1 &
        \Gamma^+_2 &
        \dots      &
        \Gamma^+_L
    \end{pmatrix}, \qquad
    \vec{\Gamma}^- = \big( \vec{\Gamma}^+ \big)^\dagger,
\end{equation}
and the effective single-particle Hamiltonian
\begin{equation}
    \label{eq:H_hopping}
	\mathcal{H}[\lambda] = 
    \begin{pmatrix}
        -\lambda J_L &\lambda-1   &            &            &            &                 &  \\
        \lambda-1    &0           &\lambda J_1 &            &            &                 &  \\
                     &\lambda J_1 &0           &\lambda-1   &            &                 &  \\
                     &            &\lambda-1   &0           &\lambda J_2 &                 &  \\
                     &            &            &\lambda J_2 &\ddots      &\ddots           &  \\
                     &            &            &            &\ddots      &0                &\lambda J_{\ell}\\
                     &            &            &            &            &\lambda J_{\ell} &\lambda-1
    \end{pmatrix},
\end{equation}
the matrix representation of the Hamiltonian~\eqref{eq:H_Gamma} becomes
\begin{equation}
    \label{eq:H_Gamma_matrix}
	H[\lambda] = 2\, \vec{\Gamma}^+ \mathcal{H}[\lambda] \vec{\Gamma}^- + \lambda J_L + (1-\lambda).
\end{equation}
Thus, diagonalizing $H[\lambda]$ reduces to diagonalizing the simpler one-particle hopping matrix $\mathcal{H}[\lambda]$. Here, calligraphic letters denote operators on this effective space, and we use $\pket{\psi}$ for the corresponding single-particle modes.

\subsection{Description of the avoided crossing}

Although the matrix $\mathcal{H}[\lambda]$ in  Eq.~\eqref{eq:H_hopping} cannot be diagonalized generally, the entire spectrum can be studied by a Fourier ansatz, due to its translational invariance in the bulk. Thus, the eigenstates must have a plane (or evanescent) wave structure in the bulk, and the on-site potentials acting on the ends of the chains only affect the boundary conditions, as in the case of a piecewise constant (or delta-function) potential.

\begin{figure}
    \centering
    \includegraphics[width=\columnwidth]{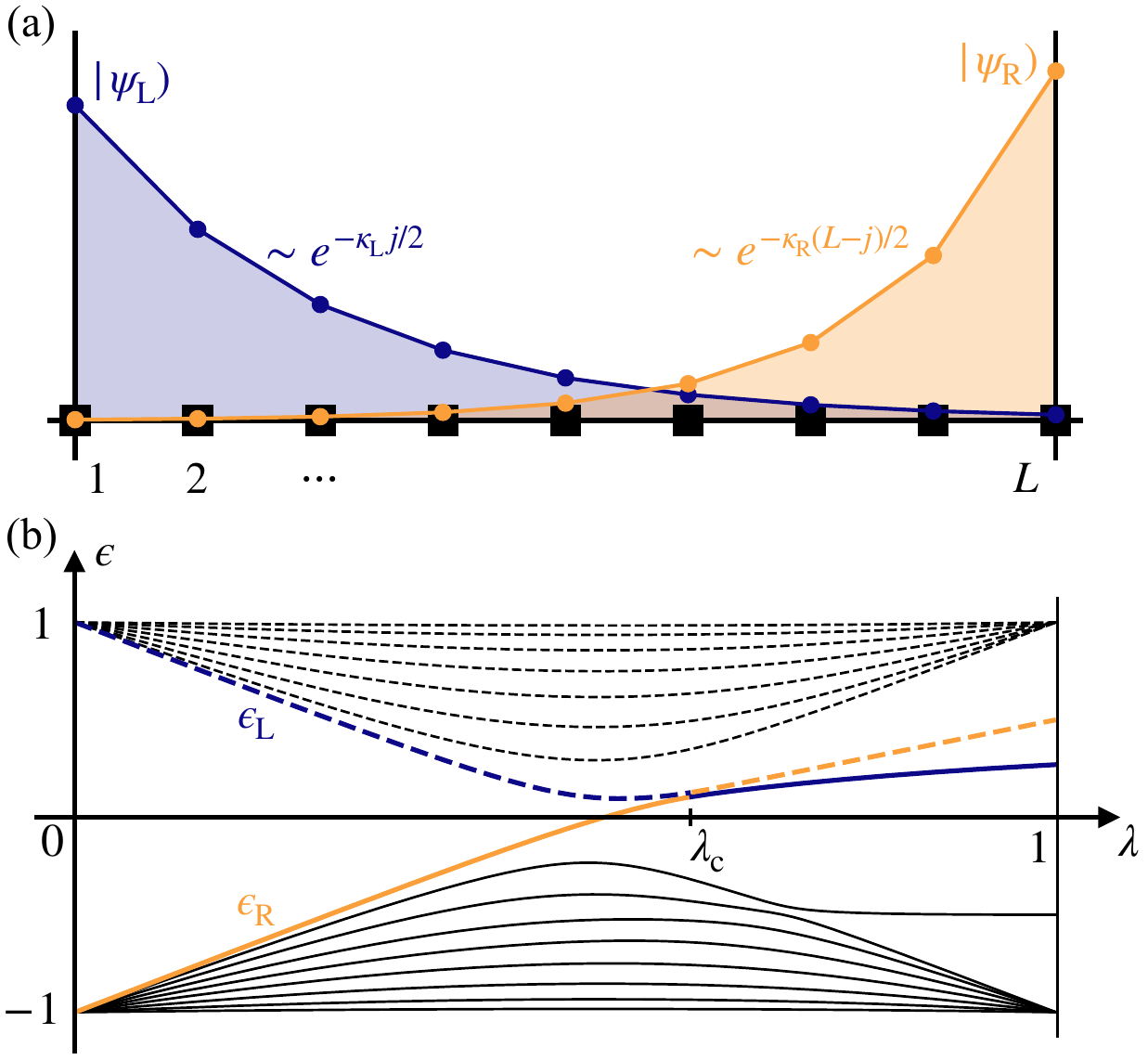}
    \caption{(a) Sketch of the boundary-localized eigenmodes $\pket{\psi_\mathrm{L}}$,$\pket{\psi_\mathrm{R}}$ of the effective hopping Hamiltonian~\eqref{eq:H_hopping}. (b) Spectrum of the hopping Hamiltonian~\eqref{eq:H_hopping} for $\ell=8$, $J=0.5$, $J'=0.27$. The many-body ground state of the $\Gamma$ Hamiltonian in Eq.~\eqref{eq:H_Gamma} is formed by populating all the negative-energy modes at $\lambda=0$ (solid lines) and leaving the positive-energy modes empty (dashed lines). The two localized modes $\epsilon_\mathrm{L}$ (blue) and $\epsilon_\mathrm{R}$ (orange) undergo an avoided crossing at $\lambda=\lambda_\mathrm{c}$: while $\epsilon_\mathrm{R}$ contributes to the ground state for $\lambda<\lambda_\mathrm{c}$ (continuous portion of the orange line), it is replaced by $\epsilon_\mathrm{L}$ for $\lambda>\lambda_\mathrm{c}$ (the continuous portion of the blue line). Since the spatial overlap between $\pket{\psi_\mathrm{L}}$ and $\pket{\psi_\mathrm{R}}$ is exponentially small in $L$, also the gap $\Delta_\mathrm{min}$ remains exponentially small according to Eq.~\eqref{eq:Delta_min}. }
    \label{fig:psiLR}
\end{figure}

As shown in App.~\ref{app:sec:diagonalization}, the spectrum of the effective hopping model, Eq.~\eqref{eq:H_hopping}, is essentially composed of plane wave eigenmodes. Under the condition $J^2 < J'$, however, a couple of boundary modes are present (see also Eq.~\eqref{eq:ineq_Js}): $\pket{\psi_\mathrm{L}}$, localized at $j=0$, and $\pket{\psi_\mathrm{R}}$, at $j=L$. They are characterized by the exponential decay,
\begin{equation}
\psi_{\mathrm{L},j}\sim e^{-\kappa_\mathrm{L} j/2}, \qquad
\psi_{\mathrm{R},j}\sim e^{-\kappa_\mathrm{R} (L-j)/2};
\end{equation}
see Fig.~\ref{fig:psiLR}a. Their eigenvalues undergo an avoided crossing at $\lambda=\lambda_c$ with an exponentially small gap
\begin{equation}
    \label{eq:Delta_min}
    \Delta_\mathrm{min} \propto e^{-\alpha L}, \qquad
\alpha=\frac{1}{2}\ln \frac{J'(1-J^2)}{J^2-(J')^2}>0,
\end{equation}
where $\lambda_c$ depends on $J$ and $J'$; see App.~\ref{app:sec:avoided_crossing}. 
This gap limits adiabatic driving, since the ground state switches from $\pket{\psi_\mathrm{R}}$ to $\pket{\psi_\mathrm{L}}$ across $\lambda_c$ (Fig.~\ref{fig:psiLR}b), requiring the transfer of a particle across the chain, i.e., flipping an extensive number of spins.

\section{Counterdiabatic driving}
\label{sec:CD}

\subsection{Brief review of counterdiabatic driving}
\label{sec:CD_review}

The adiabatic theorem in quantum mechanics states that the time-evolved state of a system driven by a slowly varying Hamiltonian $H[\lambda(t)]$ closely follows the corresponding instantaneous eigenstate. A sufficient condition for adiabaticity is
\begin{equation}
\dot\lambda  \frac{|\bra{m[\lambda(t)]}\partial_\lambda H[\lambda(t)]\ket{n[\lambda(t)]}|}{(E_n[\lambda(t)]-E_m[\lambda(t)])^2} \ll 1,
\end{equation}
where $\ket{n[\lambda(t)]}$ and $E_n[\lambda(t)]$ denote the instantaneous eigenstates and eigenvalues. If this condition is met, nonadiabatic transitions between eigenstates are strongly suppressed~\cite{Kato1950Adiabatic}; see also Refs.~\cite{Marzlin2004Inconsistency,Jansen2007Bounds} for refinements and rigorous bounds.
Yet, a perfect adiabatic tracking can be obtained at any \emph{finite} driving rate if the dynamics is generated by the modified counterdiabatic driving (CD) Hamiltonian~\cite{Demirplak2003Adiabatic,*Demirplak2005Assisted,*Demirplak2008Consistency,Berry2009Transitionless}
\begin{equation}
    \label{eq:H_CD}
    H_\mathrm{CD}(t) = H[\lambda(t)] + \partial_t\lambda \cdot H_1[\lambda(t)],
\end{equation}
with
\begin{align}
    \label{eq:H1_spectral}
    H_1[\lambda] &= i \sum_n \big( \dyad{\partial_\lambda n}{n} - \braket{n}{\partial_\lambda n} \dyad{n}{n} \big) \\
    &= i \sum_n \sum_{m\neq n} \frac{\ket{m} \bra{m} \partial_\lambda H \ket{n} \bra{n}}{E_n-E_m},
\end{align}
where the last equation assumes that the spectrum of $H[\lambda(t)]$ is nondegenerate \cite{Berry2009Transitionless}. The first term on the right-hand side of $H_1[\lambda]$ is the generator of parallel transport, familiar from proofs of the adiabatic theorem. The second term, which is diagonal in the instantaneous eigenbasis, provides the correct Berry phase associated with a perfectly adiabatic trajectory~\cite{Berry1984Quantal}.

The challenge in utilizing the CD Hamiltonian, such as the one in Eq.~\eqref{eq:H1_spectral}, comes from the fact that the terms involved are highly non-local and multiple-body, being $\dyad{\partial_\lambda n}{n}$ similar in structure to a projector~\cite{delCampo2012Assisted}. 
For this reason, a growing body of works has investigated the possibility of truncating the CD Hamiltonian to few-body local terms,  while retaining the suppression of diabatic transitions~\cite{delCampo2012Assisted,Takahashi2013Transitionless,Saberi2014Adiabatic,Damski2014Counterdiabatic,Sels2017Minimizing,Claeys2019Floquet,Hartmann2022Polynomial,Cepaite2023Counterdiabatic}. A convenient framework for this purpose relies on the Floquet-Krylov expansion of the CD term~\cite{Claeys2019Floquet,Takahashi2024Shortcuts,Bhattacharjee2023Lanczos}. Using the equivalent representation
\begin{equation}
    \label{eq:H1_integral}
    H_1[\lambda] = - \int_0^{+\infty} dx \, e^{-\epsilon |x|} \,  e^{i x H[\lambda]} \partial_\lambda H[\lambda] e^{-i x H[\lambda]},
\end{equation}
with $\epsilon=0^+$, one can expand the integrand in powers of $x$ and find, upon integration,
\begin{equation}
    \label{eq:H1_series}
    H_1[\lambda] = i \sum_{n=1}^\infty \alpha_n \underbrace{[H,\cdots,[H}_{2n-1 \text{ times}},\partial_\lambda H] \cdots].
\end{equation}
Here, $\alpha_n = (-1)^n \epsilon^{-2n}$ is formally diverging, but allowing the $\alpha_n$'s to be free variational parameters, one obtains a controlled expansion of the CD term with increasingly non-local operators~\cite{Claeys2019Floquet}. To fix the parameters $\alpha_n$, one may employ a QAOA-inspired procedure, in which they are updated recursively by using as a cost function the energy at the end of the driving process~\cite{Yao2021,Chandarana2022Digitized,Hegade2022Digitized,Wurtz2022Counterdiabaticity}, or an a priori method like the minimization of the action~\cite{Sels2017Minimizing}
\begin{equation}
	\label{eq:action}
    \mathcal{S}(H_1) := \Tr \big( G^\dagger(H_1) G(H_1) \big),
\end{equation}
where
\begin{equation}  
	G(H_1) := \partial_t H[\lambda(t)] -i [H[\lambda(t)], H_1],
\end{equation}
with respect to the parameters $\alpha_n$. We choose to follow the second route, since it allows greater analytical control and does not rely on heavy numerical optimization.

\subsection{Approximate counterdiabatic driving for the spin-glass bottleneck model}
\label{sec:approximate_CD}

\begin{figure}
    \centering
    \includegraphics[width=\columnwidth]{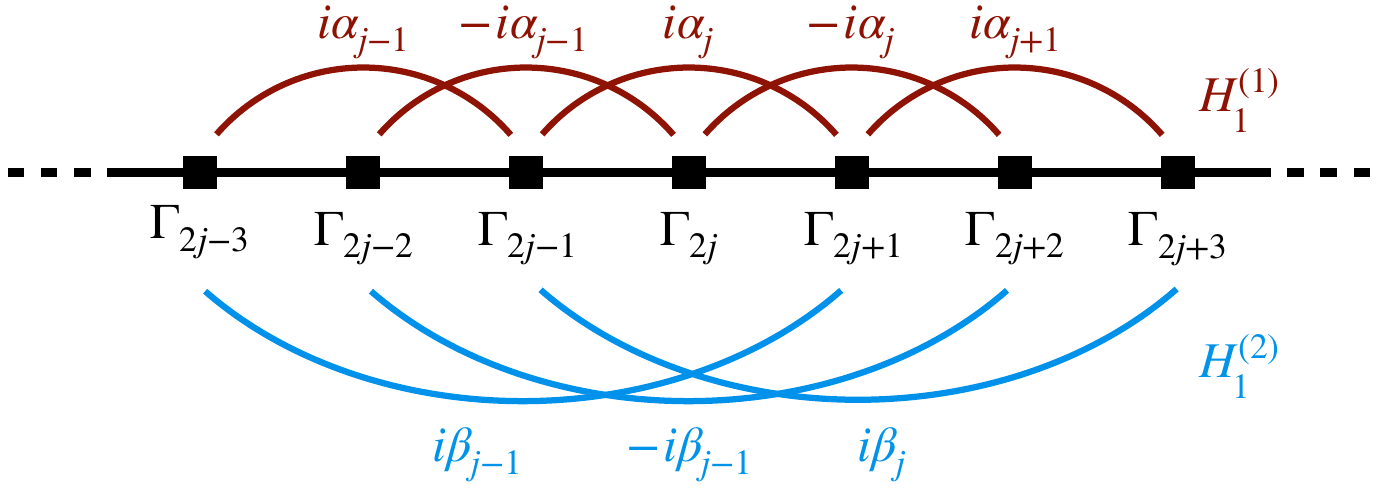}
    \caption{The local approximate counterdiabatic Hamiltonians $H_1^{(1)}$, Eq.~\eqref{eq:H1_1st_order}, and $H_1^{(2)}$, Eq.~\eqref{eq:H1_2nd_order}, take a simple form in the $\Gamma$ free fermionic representation, Eq.~\eqref{eq:H_Gamma}: they are a longer-range hopping terms, with fine-tuned coefficients $i\alpha_j,i\beta_j$. Higher-order variational ansatzes yield a similar structure with longer-range couplings.}
    \label{fig:chain_CD}
\end{figure}

Following Refs.~\cite{Sels2017Minimizing,Claeys2019Floquet}, consider the lowest-order approximant to the CD term in the Floquet-Krylov expansion. At the first order, the commutator $[H, \partial_\lambda H]$, with $H$ in Eq.~\eqref{eq:H}, contains terms $\sigma_j^y \sigma_{j+1}^z + \sigma_j^z \sigma_{j+1}^y$, motivating the variational ansatz
\begin{equation}
    \label{eq:H1_1st_order}
    H_1^{(1)}[\lambda] = \sum_{j=1}^L \alpha_j \left(\sigma_j^y \sigma_{j+1}^z + \sigma_j^z \sigma_{j+1}^y \right).
\end{equation}
 The coefficients $\alpha_j$ can be found by minimizing the action~\eqref{eq:action}, as given by Eq.~\eqref{eq:variational_eq_1st} in App.~\ref{app:sec:H1_1st_order}.
 Going to the second order, from the computation of the nested commutator $[H,[H,[H, \partial_\lambda H]]]$, one determines the variational ansatz $H_1^{(1)}[\lambda]+H_1^{(2)}[\lambda]$, where
\begin{equation}
    \label{eq:H1_2nd_order}
    H_1^{(2)}[\lambda] = \sum_{j=1}^L \beta_j \left(\sigma_j^y \sigma_{j+1}^x \sigma_{j+2}^z + \sigma_j^z \sigma_{j+1}^x \sigma_{j+2}^y \right).
\end{equation}
Again, the coefficients $\{\alpha_j, \beta_j\}$ can be determined by solving a $2L\times2L$ linear system of equations (Eq.~\eqref{eq:variational_eq_2nd}). Generalizing,  one needs to solve a $nL \times nL$ linear system for the $n$-th order term of the Floquet-Krylov expansion. Thus, for finite $n$, the CD term can be determined efficiently with the help of a classical computer.

At all orders of the Floquet-Krylov expansion of the CD Hamiltonian, Eq.~\eqref{eq:H1_series}, the approximate CD term for the spin-glass bottleneck model admits a free-fermionic representation. The full derivation is reported in App.~\ref{app:sec:H1_free_fermions}, and it leads to
\begin{align}\label{eq:H_CDapprox}
    H_1^{(1)}[\lambda] &= -2i \sum_{j=1}^L \tilde{\alpha}_j \left[ c^\dagger_j c^\dagger_{j+1} + c_j c_{j+1} \right] , \\
	H_1^{(2)}[\lambda] &= -2i \sum_{j=1}^L \tilde{\beta}_j \left[ c^\dagger_j c^\dagger_{j+2} + c_j c_{j+2} \right].
\end{align}
Notice that the CD terms take the form of longer-range hoppings. A similar longer-range-hopping form is acquired in the $\Gamma_j^\pm$ basis as well, see Eqs.~\eqref{app:eq:H1_1_Gamma} and \eqref{app:eq:H1_2_Gamma}. Thus, CD helps the driving by preventing localization with two-fermion longer-range hopping terms; see also Sec.~\ref{sec:discussion} for further discussion.

\section{Results}
\label{sec:results}

Here, we present our main results regarding the behavior of CD in the quantum driving of the spin-glass bottleneck model. While the Hamiltonian in  Eq.~\eqref{eq:H} could be studied fully analytically, the introduction of the approximate CD terms forces one to resort to numerical methods: the variational equations~\eqref{eq:variational_eq_1st} and \eqref{eq:variational_eq_2nd} do not seem to have an explicit analytical solution in the general case. Nevertheless, the simple structure of the model, together with the mappings to $c$ and $\Gamma$ fermions performed above, allows for a thorough understanding of the effect of CD on the driving process.

\subsection{Counterdiabatic driving and gap amplification}
\label{sec:CD_gap_amplification}

According to the adiabatic theorem, adiabaticity can be preserved upon reducing the driving time, provided that the gap is increased by the same factor. However, the construction of the CD term outlined in Sec.~\ref{sec:CD_review} indicates that CD is more than a mere gap amplification method: the operators involved are constructed in such a way to transport a state along its adiabatic path, and they do so by enforcing parallel transport while accounting for the corresponding quantum phase. By contrast, catalysts and gap-amplifying terms~ just enhance the dephasing due to highly oscillatory terms $e^{i(E_m-E_n)t}$ for $m \neq n$, making the adiabatic approximation hold to a better degree~\cite{Suzuki_2007_originalFerroGapamplification, Seki_2012_AntiFerroFluct_Gapamplification, Seoane_2012_ClassicalMEthodforGapAmplfiication, Somma2013Spectral, CrossonFarhi2014Gap_amplication, Hormozi_NonStoqCatalyst_gapAmplificatio_2017, Susa2018_Inhomdriving_gap_amplification, Albash2019_NonStoqGap_amplification, Cao2021Catatlyst, Mehta_2SAT_NonSotq_Gapamplification}.

Notice that using a schedule $\lambda(t)$ with vanishing derivative at the endpoints, the eigenstates of $H_\mathrm{CD}[\lambda(t)]$, Eq.~\eqref{eq:H_CD}, coincide with those of $H[\lambda(t)]$ at the beginning ($\lambda(0)=0$) and the end ($\lambda(T)=1$) of the protocol. Therefore, when using approximate CD, one can apply the adiabatic theorem to the dynamics generated by $H_\mathrm{CD}[\lambda(t)]$ and try to infer the success of the driving process from the minimal gap $\Delta_{\mathrm{min,CD}}$ along the path (by construction, the exact CD term yield success with probability 1 and this study would not make sense there). Noticing that the approximate $H_\mathrm{CD}$ depends on the final time $T$,  one cannot blindly use the adiabaticity condition to estimate the time $T_\mathrm{ad}$ needed for adiabaticity to hold. Nevertheless, it can be used to determine self-consistently $T_\mathrm{ad}$, if the dependence $\Delta_{\mathrm{min,CD}} = \Delta_{\mathrm{min,CD}}(T)$ is known.

\begin{figure}
    \centering
    \includegraphics[width=\columnwidth]{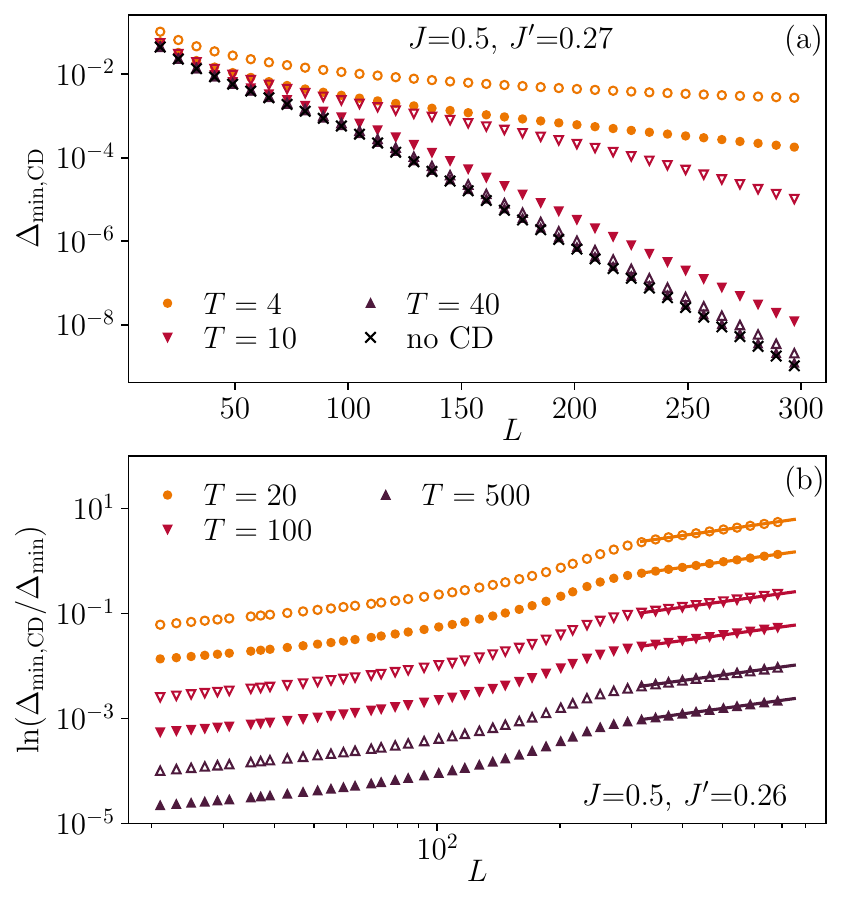}
    \caption{Gap amplification from counterdiabatic driving (CD). (a) For any driving rate, the minimal gap $\Delta_\mathrm{min,CD}$ along the CD-assisted adiabatic path remains exponentially small in the system size, both for the 1st order (Eq.~\eqref{eq:H1_1st_order}, filled markers) and 2nd order (Eq.~\eqref{eq:H1_2nd_order}, empty markers) variational ansatzes. The bare minimal gap $\Delta_\mathrm{min}$ is shown for comparison with black crosses. (b) The CD-assisted gap $\Delta_\mathrm{min,CD}$, despite being exponentially small in the system size, is still exponentially larger than the bare gap $\Delta_\mathrm{min}$: all the log-log fits (lines) are compatible with a growth $\Delta_\mathrm{min,CD} \sim \Delta_\mathrm{min} e^{c(T)L}$, with $c(T) = \alpha-\alpha_\mathrm{CD}(T)$.}
    \label{fig:gap_L}
\end{figure}

In Fig.~\ref{fig:gap_L}, the numerical results for the gap $\Delta_{\mathrm{min,CD}}(T)$ of the CD Hamiltonian are shown. The cubic ramp $\lambda(t) = 3(t/T)^2-2(t/T)^3$, that satisfies $\dot{\lambda}(0) = \dot{\lambda}(T) = 0$, is used. From Fig.~\figpanel{fig:gap_L}{a}, one can see that, for large system sizes and long driving times, the minimum of the gap remains exponentially small in the system size: $\Delta_\mathrm{min,CD} \propto e^{-\alpha_\mathrm{CD}L}$. From Fig.~\figpanel{fig:gap_L}{b}, however, one can notice that $\Delta_\mathrm{min,CD}$ is still exponentially larger than $\Delta_\mathrm{min}$, as a function of the system size $L$ and for every fixed time $T$. In other words, $\alpha_\mathrm{CD} < \alpha$. Also, there is no appreciable difference in the values of $\alpha_\mathrm{CD}$ extracted from either the first- or second-order expansions; only the prefactor of the exponential is larger for the second-order one. 

\begin{figure}
    \centering
    \includegraphics[width=\columnwidth]{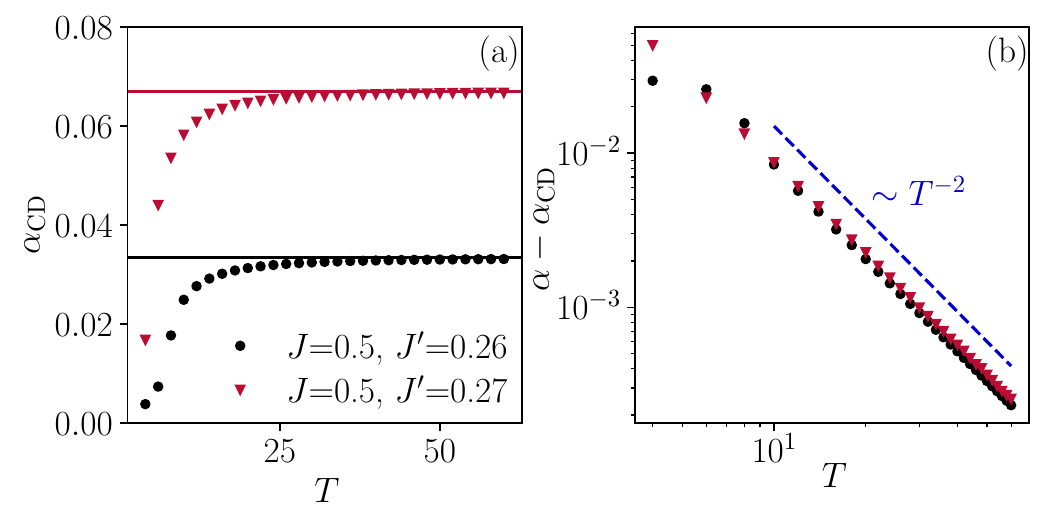}
    \caption{The approximate CD path displays a minimal gap $\Delta_\mathrm{min,CD} \propto e^{-\alpha_\mathrm{CD}(T)L}$. (a) Rate $\alpha_\mathrm{CD}$ as a function of the driving time $T$: the value of $\alpha_\mathrm{CD}$ (dots) saturates at the analytical value $\alpha$  in the absence of CD in Eq.~\eqref{eq:Delta_min} (solid lines), already for moderately short times. (b) The approach to the asymptotic value is well fitted by a power law: $\alpha-\alpha_\mathrm{CD}(T) \sim T^{-2}$.}
    \label{fig:gap_T}
\end{figure}

From plots like that in Fig.~\figpanel{fig:gap_L}{a}, the exponential decay rate $\alpha_\mathrm{CD}$ of $\Delta_\mathrm{min,CD}$ is extracted for many values of $T$, and the values are reported in Fig.~\ref{fig:gap_T}. The fits are performed on the first-order expansion gaps; quantitatively similar results follow from using the second-order expansion gaps. One can see that already the first-order ansatz is sufficient to provide an exponential speed-up for moderate driving times, $T\lesssim25$, as $\alpha_\mathrm{CD}$ is substantially smaller than $\alpha$. Increasing the driving time, however, $\alpha_\mathrm{CD}$ approaches $\alpha$: higher order CD terms would be required to maintain the same gap amplification.

The conclusion above can be put on quantitative grounds by estimating the adiabatic timescale for approximate CD with the inverse of the minimal gap, $T_\mathrm{ad,CD}\sim\Delta^{-2}_\mathrm{min,CD}(T)$. Using $\Delta_\mathrm{min,CD}(T) \simeq \Delta_0 e^{-\alpha_\mathrm{CD}(T)L}$, with $\alpha_\mathrm{CD}(T) \simeq \alpha - \delta/T^2$ that is shown to hold up to high precision in Fig.~\figpanel{fig:gap_T}{b} ($\Delta_0$ and $\delta$ are two fitting coefficients), one obtains
\begin{equation}
    T_\mathrm{ad,CD}\approx T_\mathrm{ad}\,e^{-2\delta L/T^2_\mathrm{ad,CD}} \approx T_\mathrm{ad}\left(1-\frac{2\delta L}{T^2_\mathrm{ad}}\right),
\end{equation}
where in the last step we expanded up to first order the exponential and approximated $T_\mathrm{ad,CD}$ with $T_\mathrm{ad}$. One can see that the solution $T_\mathrm{ad,CD}$ is significantly reduced for moderate driving times, and it converges to the bare value, $T_\mathrm{ad}$, upon increasing $T$. Let us remark that both this empirical power law and the coefficients are expected to change in favour of smaller adiabatic times for larger order expansions.

\subsection{Performance of counterdiabatic driving}
\label{sec:performance_CD}

Above, the adiabatic theorem was used to predict the performance of approximate CD. It was found that the gap remains exponentially small in the system size, with the relative advantage of CD decreasing for longer protocols. To quantify more precisely how much CD improves the adiabatic evolution, we next consider the average number of bonds violated compared to the final ground state.

As before, a cubic ramp $\lambda(t) = 3(t/T)^2-2(t/T)^3$, satisfying $\dot{\lambda}(0) = \dot{\lambda}(T) = 0$, is employed. Upon initializing the system  in the paramagnetic ground state of $H[\lambda=0]$, we consider the evolution with the approximate CD Hamiltonians $H[\lambda(t)] + \dot{\lambda}(t) H^{(1)}[\lambda(t)]$ (first-order ansatz) or $H[\lambda(t)] + \dot{\lambda}(t) [ H^{(1)}[\lambda(t)] + H^{(2)}[\lambda(t)]]$ (second-order ansatz). Because of the free-fermionic nature of the model under consideration, we evolve the single-particle operators $(c_j^\phdagger,c_j^\dagger)$ or $\Gamma_k^\pm$ by means of Bogoliubov-de Gennes equations, and then compute the desired expectation values at the end of the protocol with the appropriate contractions; see App.~\ref{app:sec:numerics} for all the details. This technique allows one to reach much larger system sizes, which are needed in the present case to extract the asymptotic scalings. 

\begin{figure}
    \centering
    \includegraphics[width=\columnwidth]{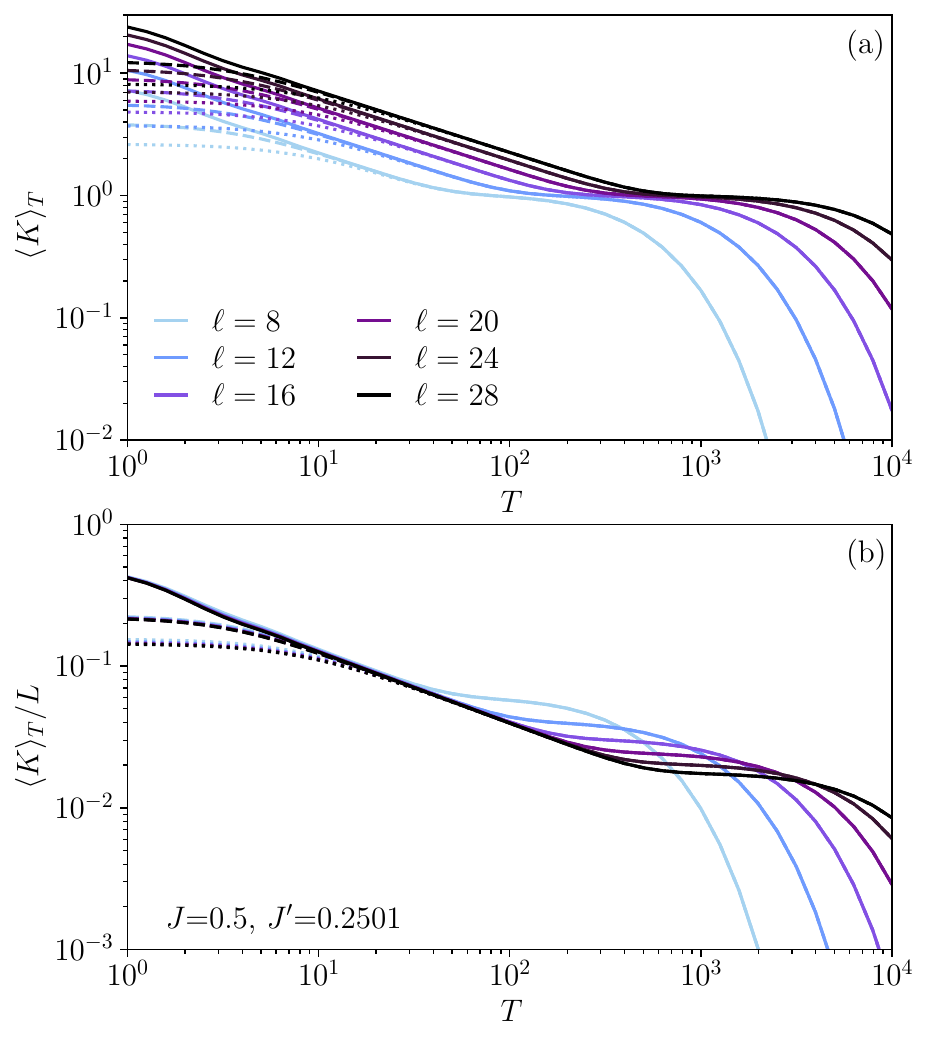}
    \caption{Approximate CD performance measured by the expectation value of the kink number on the final state: $\ev{K}_T \equiv \ev{K}{\psi(T)}$, Eq.~\eqref{eq:K}. (a) The kink number decreases until it reaches the plateau unit value, corresponding to a single frustrated bond. A driving time exponentially large in the system size is needed to make $\ev{K}_T<1$ and enter the adiabatic regime. The effect of approximate CD is to reduce the kink number at smaller driving times (dashed lines: 1st order ansatz; dotted lines: 2nd order ansatz) with respect to\ the bare driving with only the Hamiltonian (solid lines). (b) Rescaling the kink number with the system size, a good collapse is obtained at short times, signaling that approximate CD is improving $\ev{K}_T$ by an extensive amount. However, the smallest exponential gap remains as hard to be crossed as for the bare dynamics without CD terms.}
    \label{fig:kinks}
\end{figure}

The results of our numerical simulations to assess the performance of the driving are presented in Figs.~\ref{fig:kinks}.  First, we compute the kink number in Fig.~\ref{fig:kinks}, i.e., the number of frustrated bonds at the end of the process. It is given by the expectation value of the operator on the final state
\begin{equation}
    \label{eq:K}
    K := \frac{1}{2} \sum_{j=1}^L \left[ 1 - \mathrm{sign}(J_j)  \sigma_j^z \sigma_{j+1}^z \right].
\end{equation}
This is an extensive observable that can be computed with free-fermion techniques; the details are shown in App.~\ref{app:sec:numerics_observables}. As shown in Fig.~\figpanel{fig:kinks}{a}, the bottleneck of adiabaticity is represented by one remaining frustrated bond, corresponding to the transition $\pket{\psi_\mathrm{R}} \to \pket{\psi_\mathrm{L}}$. The time required to anneal this last bond increases with the size of the system, as indicated by the increasing length of the plateau of $\ev{K}_T \equiv \ev{K}{\psi(T)}$. As illustrated in Fig.~\figpanel{fig:kinks}{b}, the effect of CD is more pronounced at fast driving and results in an excitation suppression that scales extensively with the system size. The improvement increases with the order of the variational ansatz, yet addressing the exponentially small gap remains beyond the reach of low-order schemes and would necessitate higher-order corrections.

The findings of Fig.~\ref{fig:kinks} are consistent with the Kibble-Zurek (KZ) prediction for the density of defects formed in a quantum phase transition~\cite{Damski05,Zurek2005Dynamics,Dziarmaga05,Polkovnikov2005Universal,Damski06}. The collapse in Fig.~\figpanel{fig:kinks}{b} confirms the expected scaling $\sim T^{-\nu/(1+z\nu)} = T^{-1/2}$, with $\nu=1$ and $z=1$ for the transverse-field Ising model. The three weak links affect the dynamics only at long times, where the KZ scaling gives way to a plateau for the defect density. For fast driving protocols, deviations induced by approximate CD terms reproduce the results of Ref.~\cite{delCampo2012Assisted}. Overall, CD improves the performance in the fast-driving regime but remains ineffective in mitigating excitations across the exponentially small gap, as discussed in Sec.~\ref{sec:discussion}.

\section{Quantum Brachistochrone Counterdiabatic Driving (QBCD):  exponential speedup}
\label{sec:CD_LR}

In this section, we show how the efficiency of the driving can be improved beyond low-order CD expansions by canceling ground–first-excited transitions near the critical point $\lambda=\lambda_c$.
\subsection{Comparison to the local approach in the spin glass toy model}\label{subsec: MinSG}
While earlier ansatze required no spectral information, here we assume approximate knowledge of the spectrum at a single-parameter point. This resembles reverse quantum annealing, where an approximate ground state improves adiabatic performance~\cite{Ohkuwa2018Reverse,King2018Observation,Yamashiro2019Dynamics}. Unlike many early many-body CD approaches based on truncating nonlocal terms~\cite{delCampo2012Assisted,Saberi2014Adiabatic}, which can be as costly as full diagonalization, methods like reverse annealing~\cite{Ohkuwa2018Reverse,King2018Observation,Yamashiro2019Dynamics} or quantum Monte Carlo methods~\cite{Young2008Size,Young2010First,Boixo2014Evidence} can provide the needed approximate spectral data at moderate cost.

The key idea is to target the bottleneck by using approximate left and right localized edge states, responsible for the exponentially small gap, to construct the CD term at the critical point. Multiplying it by the schedule derivative preserves the spectrum at $\lambda=0,1$ (see App.\ref{app:LRwavefunctions}). In the one-particle effective model, Eq.\eqref{eq:H_hopping}, and Eq.\eqref{eq:H_Gamma_matrix}, the QBCD Hamiltonian is
\begin{equation}
\label{eq:H_LR}
\mathcal{H}^{\mathrm{QBCD}} = i \frac{\pbra{\psi_\mathrm{R}} \partial_\lambda \mathcal{H}[\lambda_c] \pket{\psi_\mathrm{L}}}{\Delta_{\min}} \big[\pket{\psi_\mathrm{R}} \pbra{\psi_\mathrm{L}} - \pket{\psi_\mathrm{L}} \pbra{\psi_\mathrm{R}} \big],
\end{equation}
with $\lambda_c$ the minimal-gap point, $\Delta_{\min}$ the gap, and $\ket{\psi_\mathrm{L,R}}$ the edge states. This expression, equivalent to $\sigma^y$ in the two-level subspace, is reminiscent of the solution to the quantum brachistochrone problem~\cite{Carlini06,Bender09,Takahashi13}, motivating the term QBCD.

\begin{figure}
    \centering
    \includegraphics[width=\columnwidth]{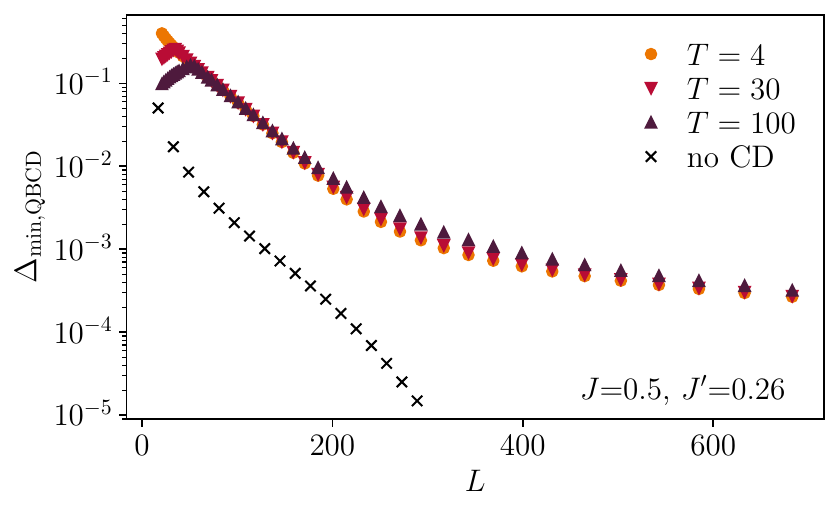}
    \caption{Gap amplification from the QBCD term, Eq.~\eqref{eq:H_LR}. Comparing with Fig.~\ref{fig:gap_L}, one can see that $H^{\mathrm{QBCD}}$ is able to significantly enlarge the minimal gap on the adiabatic path for all system sizes and driving times.}
    \label{fig:gap_LR}
\end{figure}

\begin{figure}
    \centering
    \includegraphics[width=\columnwidth]{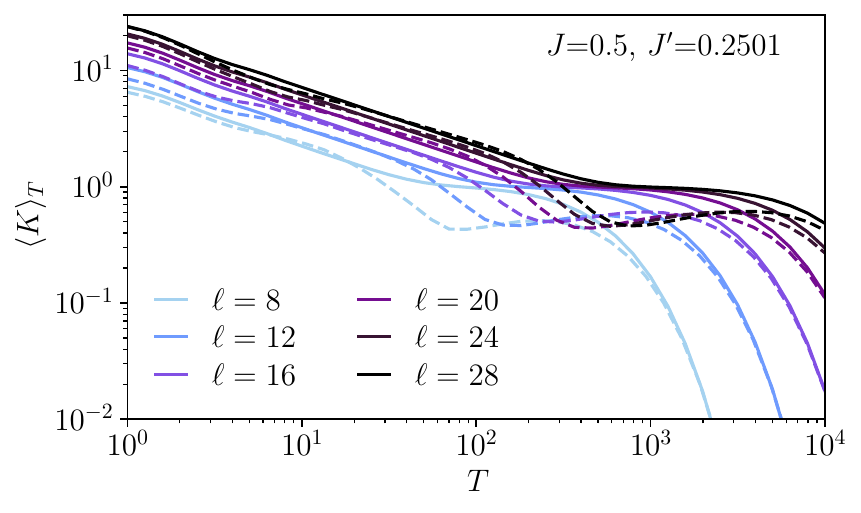}
    \caption{Performance of the driving assisted by the QBCD term, Eq.~\eqref{eq:H_LR}. The expectation value of the kink number on the final state plateaus for intermediate driving times, but the actual value of $\ev{K}_T$ is reduced when employing $H^{\mathrm{QBCD}}_1$ (dashed lines) with respect to the bare evolution (solid lines).}
    \label{fig:kinks_LR}
\end{figure}

The QBCD Hamiltonian, Eq.~\eqref{eq:H_LR}, cancels exactly the transition between the ground and first excited states only at the gap-closing point; however, it also provides an efficient CD term in the neighborhood thereof. This feature can be inferred from the results shown in Figs.~\ref{fig:gap_LR} and \ref{fig:kinks_LR}: both the gap $\Delta_{{\min},\mathrm{QBCD}}$ and the ground-state fidelity are significantly increased with respect to the bare Hamiltonian. Here, the fidelity is captured by the kink number, since a value $K < 1$ can arise only from the ground state. In particular, from Fig.~\figpanel{fig:kinks_LR}{a} one can see that for values of $T$ for which the bare Hamiltonian exhibits a kink number plateauing very close to 1, the QBCD term is able to lower it to $\ev{K}_T \approx 0.5$. This implies that the ground state is found at the end of the protocol with a finite probability of $\approx 0.5$, which is significantly greater than the one found with the bare protocol.

The effects of QBCD remain remarkably robust against deviations from the exact critical point $\lambda_c$, an essential feature in realistic settings where noise and fluctuations hinder precise parameter identification and control. Furthermore, this robustness reduces computational costs in classical simulations and provides a resource-efficient route to approximate local CD protocols on quantum hardware, particularly when suitable truncations are applied~\cite{Romero2025,romero2025proteinfolding}.

\begin{figure}
    \centering
    \includegraphics[width=\columnwidth]{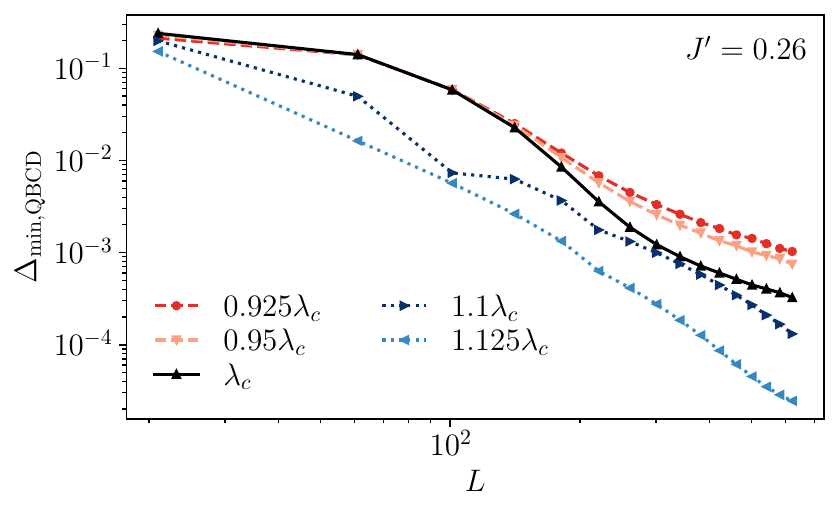}
    \caption{Robustness of the minimal gap scaling against deviations between the applied QBCD Hamiltonian and that taken exactly at $\lambda_c$. The red dashed lines correspond to the approximate analytical results, while the blue dotted lines follow from a purely numerical approach. The black solid line corresponds to the analytical approach at $\lambda_c=\lambda^*_c$. The curves clearly indicate a polynomial decay rather than an exponential one.}
    \label{fig:gap_error}
\end{figure}

To demonstrate this, we vary the analytically obtained critical point $\lambda_c^*$ (App.~\ref{app:LRwavefunctions}) within $[0.85\lambda_c^*,1.15\lambda_c^*]$ and study the behavior of the minimal gap. We also perform a purely numerical search for the ground state gap and eigenstates using only $40$ sampling points and vary $\lambda_c$ within the same interval in this approach. This requires far less computational effort than simulating the full QBCD protocol, yet captures the ingredients needed for its construction. As shown in Fig.~\ref{fig:gap_error}, the gap at the critical point preserves its polynomial decay with the system size $L$ up to $\sim10\%$ deviations from $\lambda_c$. Interestingly, for analytically approximated QBCD, the minimal gap can even increase slightly under small shifts, while for larger deviations (e.g., $\lambda_c=
1.125\lambda_c^*$), the numerical construction still outperforms local expansions by two orders of magnitude for system sizes up to $L=600$. Thus, constructing QBCD from approximate spectral data is sufficient to realize an exponential speedup of adiabatic time scales across an exponentially small gap.

Finally, let us investigate the efficiency of the variational and QBCD approaches presented from the perspective of energy resources.  Although reducing external resources is essential for experimental realizations~\cite{Abah_2019_Energetic_costs, Torrontegui_2017_EnergyConsumeSTA}, the energetic costs determine the speed at which the assisted process can stay adiabatic~\cite{Demirplak2008Consistency,Campbell2017,Funo2017}. The Hilbert-Schmidt norm of the Hamiltonians, defined as $\| H_1 \|^2=\mathrm{Tr}[H^\dagger_1\,H_1]$, provides a natural measure to quantify the cost of CD~\cite{Demirplak2008Consistency,delCampo2012Assisted,Takahashi2024Shortcuts,Bhattacharjee2023Lanczos}. It can be shown (see App.~\ref{app:E_cost}) that the energy costs of the low-order expansions grow as
\begin{equation}
    \left\|H_1^{(1,2)}\right\|^2\sim L
\end{equation}
for the first- and second-order variational expansions, respectively. The QBCD exhibits more favorable features (App.~\ref{app:E_cost}), as its norm is bounded in the leading order with $L$,
\begin{equation}
    \left\|\mathcal{H}^{\mathrm{QBCD}}\right\|^2\sim O(1)\,.
\end{equation}
This fact is a consequence of the exponentially localized nature of the two single-particle states involved in the avoided crossing and their appearance both in the gap and the transition matrix element.

\begin{figure}
    \centering
    \includegraphics[width=\columnwidth]{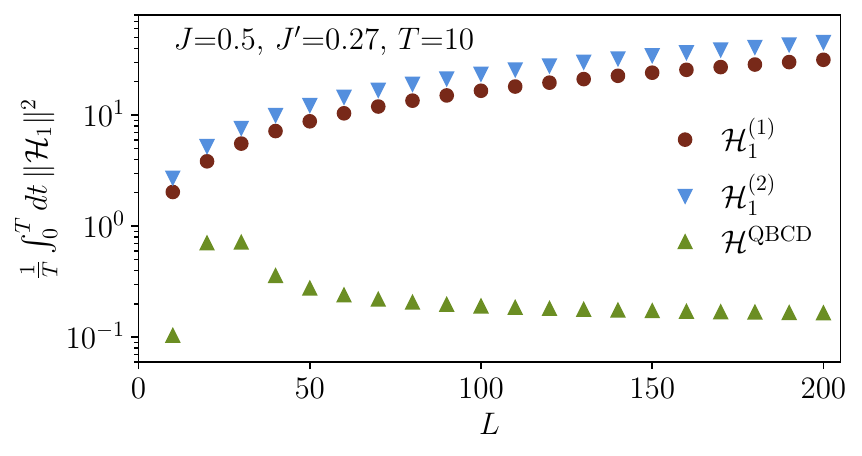}
    \caption{Energetic cost of implementing the counterdiabatic terms, as measured by the average Hilbert-Schmidt norm of the Hamiltonian along the driving path. While the first- and second-order variational terms have a cost that grows with the system size, the QBCD approach of Eq.~\eqref{eq:H_LR} converges to an $L$-independent value.}
    \label{fig:Ecost}
\end{figure}

To also incorporate the time dependence in the computation, in Fig.~\ref{fig:Ecost}, the quantities $\frac{1}{T}\int_0^T\mathrm dt\,\|H_1\|^2$ are presented. It is clearly demonstrated that $\|H_1^{(1)}\|$ and $\|H_1^{(2)}\|$ increase with the system size, while $\| \mathcal{H}^{\mathrm{QBCD}}\|$ remains independent of $L$ up to the leading order. Thus, not only does the QBCD approach achieve significantly larger fidelities to the target state, but it does so in a resource-efficient way.

We stress that the model’s specific features and free-fermion representation were used primarily to access large system sizes, compare numerical constructions with analytical approximations, and evaluate energetic costs. As we show in the following subsection, QBCD retains its efficiency in more realistic spin-glass models.

\subsection{QBCD in the $3$-regular \textsc{Max Cut} problem}
\label{sec:QBCDMax Cut}

Motivated by the efficiency of the QBCD approach in the minimal spin-glass bottleneck, we now consider a more realistic \textsc{NP}-hard problem, the $3$-regular \textsc{Max Cut}~\cite{FarhiMAXCUT_XORSAT2012,Albash2018Adiabatic}, to assess QBCD in a practical setting.
 The $3$-regular \textsc{Max Cut} problem describes the optimization task of dividing a $3$-regular random graph into two subgraphs such that the number of connecting edges is maximal. The solution for a given graph is equivalent to finding the ground state of the following Hamiltonian,
\begin{eqnarray}\label{eq:H_P_MAXCUT}
    &&H_P=\sum_{c=1}^{M}\frac{1+\sigma^z_{c_1}\sigma^z_{c_2}}{2},
\end{eqnarray}
where $c$ denotes the edge connecting the two spins that represent the vertices of the graph. Edge configurations are distributed uniformly so that all spins appear in exactly three interaction terms without repetition, $c_1\neq c_2$, implying a total $M=3L/2$ number of quadratic terms. In Ref.~\cite{FarhiMAXCUT_XORSAT2012}, it was shown that the ground state energy gap of \textsc{Max Cut} exhibits an exponential decay even for intermediate system sizes, $L=8,\dots,18$ for problem instances with two-fold ground state degeneracy. Correspondingly, we also restrict the exposition to these instances and consider the time evolution initiated from the fully paramagnetic ground state governed by the Hamiltonian
\begin{eqnarray}
    H(t)=-\lambda\sum_{c=1}^{M}\frac{1+\sigma^z_{c_1}\sigma^z_{c_2}}{2}-(1-\lambda)\sum_{j=1}^L\,\sigma^x_j,
\end{eqnarray}
where the time dependence of the linear protocol was omitted for brevity, $\lambda(t)=t/T,\quad t\in[0,T]$.

First, we investigate the efficiency of the local variational expansion. Similarly to Sect.~\ref{sec:approximate_CD}, we consider two- and three-body terms in which all interactions are restricted to spins connected by the graph edges. Allowing also for all local couplings to be optimized, the first- and second-order CD corrections read 
\begin{eqnarray}
    &&H^{(1)}_1=\sum_{c=1}^M\alpha_{c}(\sigma^z_{c_1}\sigma^y_{c_2}+\sigma^z_{c_1}\sigma^y_{c_2}),\\
    &&H^{(2)}_1=\sum_{c,c^\prime=1}^M\delta_{c_2,c^\prime_1}\beta_{c,c^\prime}(\sigma^z_{c_1}\sigma^x_{c_2}\sigma^y_{c^\prime_2}+\sigma^y_{c_1}\sigma^x_{c_2}\sigma^z_{c^\prime_2}).\label{eq:MAXCUT_second_order}
\end{eqnarray}
In the second-order expansion, $H^{(2)}_1$, three-body terms run over connected spins in the interaction graph of the \textsc{Max Cut} problem, indicated by the constraint $\delta_{c_2,c^\prime_1}$ and the adapted order of indices, $c_1<c_2=c^\prime_1<c^\prime_2$.
The corresponding minimization equation for the first- and second-order expansions can be found in App.~\ref{app:MAX_CUT}.

\begin{figure}
    \centering
    \includegraphics[width=\columnwidth]{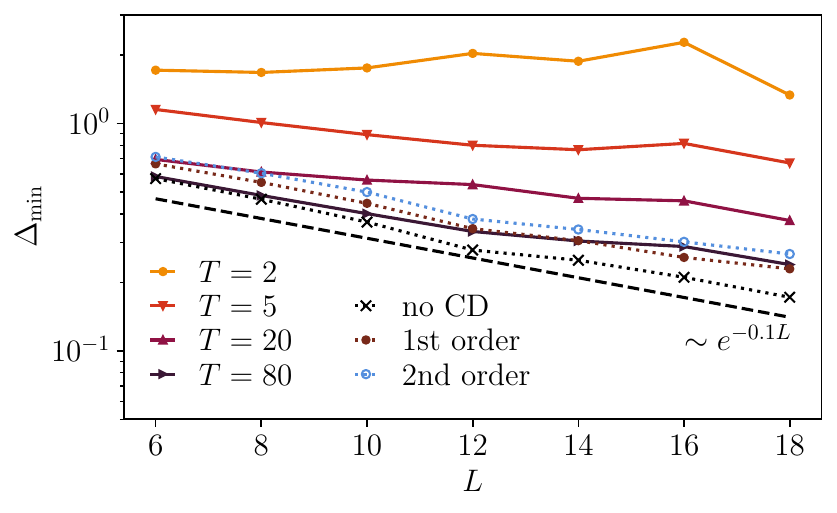}
    \caption{Ground state energy gaps for the \textsc{Max Cut} problem without CD, with first-order and second-order local expansions (dotted lines), and with the sparsified QBCD (solid lines). Local methods yield only a constant-factor improvement over the exponentially small gap of the bare protocol, as shown for short driving times ($T = 2$). The sparsified QBCD leads to exponential gap amplification across all timescales ($T = 2, 5, 20, 80$).}
    \label{fig:Gap_Max Cut}
\end{figure}

\begin{figure}
    \centering
    \includegraphics[width=\columnwidth]{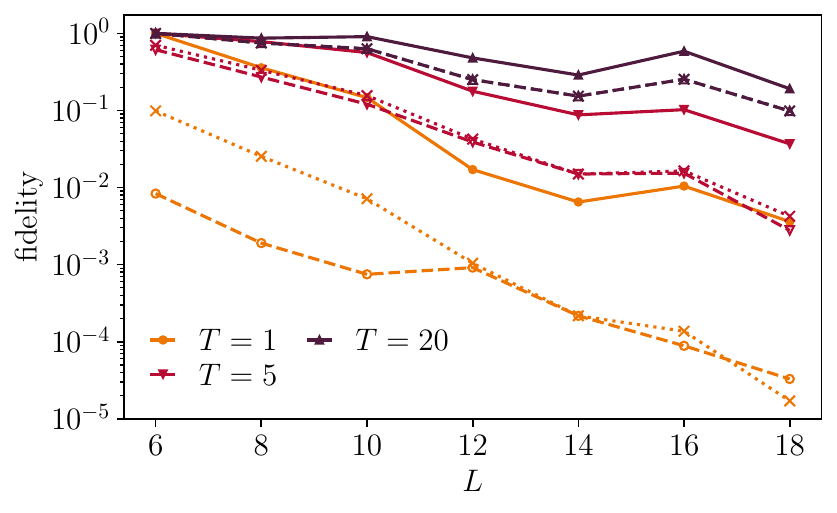}
    \caption{Ground-state fidelity for the \textsc{Max Cut} problem. Comparison of the second-order local CD (dashed lines) and no CD (dotted lines) approaches, with the sparsified QBCD method (solid lines). Local strategies offer negligible improvement even in slow processes, while the sparsified QBCD achieves finite fidelity already at short driving times, where the other methods yield exponentially small values.}
    \label{fig:fidelity_Max Cut}
\end{figure}

To assess adiabaticity for system sizes $L=6,\dots,18$, we consider both the ground-state gap and fidelity. As shown in Fig.\ref{fig:Gap_Max Cut}, the variational approach can amplify the gap only by a constant factor, even for short driving times, $T=2$ (with adiabaticity completely broken down due to excitations to higher energy modes), which is slightly more pronounced for the second-order expansion compared to the slow decay in the bare process; see App.\ref{app:Gap_difference}. The final ground-state fidelities show even weaker gains: for slow protocols, there is minimal improvement, while for faster processes, fidelities can drop below those of the bare evolution (Fig.\ref{fig:fidelity_Max Cut}). In contrast to this modest gap amplification and limited fidelity gains observed, QBCD retains its efficiency and robustness in these more realistic spin-glass models, similarly to the minimal spin-glass bottleneck. In particular, the QBCD Hamiltonian is constructed using the ground state gap, $\Delta$, and the many-body ground ($\lvert\mathrm{GS}\rangle$) and first excited states ( $\lvert\mathrm{ES}\rangle$)  near the critical point
\begin{eqnarray}\label{eq:MAXCUT_QBCD}
    &&H_\mathrm{QBCD}=\\
    &&-i\frac{\langle\mathrm{GS}\lvert\partial_\lambda H\rvert\mathrm{ES}\rangle}{\Delta}\left(\lvert\mathrm{ES}\rangle\langle\mathrm{GS}\rvert-\lvert\mathrm{GS}\rangle\langle\mathrm{ES}\rvert\right).\nonumber
\end{eqnarray}
The approximate knowledge of these quantities is obtained via the same purely numerical procedure employed for the minimal spin glass model, based on a search over only $40$ parameter values.

To improve its practicality, we sparsify the QBCD Hamiltonian by retaining only the $L(L+1)/2$ largest matrix elements, matching the density of the quadratic local expansion. Although the exact QBCD construction involves only two states, selected out of the $\sim 4^L$ possible energy level pairs of the full CD driving, the sparsification further reduces its non-locality exponentially, enabling a practically realizable approximate implementation through controlled truncation. This truncated scheme substantially alleviates the resource demands of current local CD protocols and can facilitate the efficient realization of higher-order CD expansions on quantum devices. Moreover, the computational cost of constructing this sparsified QBCD Hamiltonian grows only polynomially with the system size, making it accessible for classical large-scale algorithms as well~\cite{tindall2025,mauron2025}.

The sparsified QBCD is tested via the ground-state gap and final fidelity. As shown in Fig.~\ref{fig:Gap_Max Cut}, the gap is exponentially amplified, decaying only approximately polynomially even at intermediate driving times ($T=5,20$) and approaching the bare behavior near fully adiabatic timescales. Correspondingly, the final fidelity remains finite at short driving times ($T=1,5$), whereas the variational approach yields exponentially small fidelity except near adiabaticity, highlighting that sparsified QBCD simultaneously amplifies the gap and preserves high fidelity even far from the adiabatic regime. Thus, our results demonstrate that sparsified QBCD pinpoints an exponentially small non-local fraction whose inclusion alone suffices to achieve exponential speedups. Although extrapolation to significantly larger system sizes remains challenging, the underlying key microscopic mechanisms indicate that comparable performance may also extend to a broader range of instances.

We also compare the performance of the sparsified QBCD and the local expansion with the COLD approach~\cite{Cepaite2023Counterdiabatic}. Within this scheme, one introduces an additional local operator, $\mathcal O_\mathrm{opt}(\boldsymbol{\gamma},t)$, to the problem Hamiltonian with optimizable additional parameters,
\begin{eqnarray}
    &&H_\mathrm{COLD}(t,\boldsymbol{\gamma})=\\
    &&-\lambda\sum_{c=1}^{M}\frac{1+\sigma^z_{c_1}\sigma^z_{c_2}}{2}-(1-\lambda)\sum_{j=1}^L\,\sigma^x_j+ \mathcal O_\mathrm{opt}(t,\boldsymbol{\gamma}).\nonumber
\end{eqnarray}
Following the strategy of Ref.~\cite{LCD_Morawetz_2024}, we choose $\mathcal O_\mathrm{opt}(t,\boldsymbol{\gamma})$ as the second-order nested commutator
$\mathcal O_\mathrm{opt}(t,\boldsymbol{\gamma})=\gamma^{(1)}\sin(2\pi\lambda)\left[H_0\left[H_1,H_0\right]\right]+\gamma^{(2)}\sin(2\pi\lambda)\left[H_1\left[H_1,H_0\right]\right]$, providing a more efficient annealing path by extending the total Hamiltonian with the even nested commutators. The time- and parameter-dependent coefficients, $\gamma^{(1,2)}$, are subjected to further optimization within the local CD expansion framework, where the final fidelity of the evolved state serves as the objective function. Since the double commutator structure will contain terms like $\left[H_0\left[H_1,H_0\right]\right]\sim ZXZ + X$ and $\left[H_1\left[H_1,H_0\right]\right]\sim YY + ZZ$, the general form of the local CD expansion up to cubic terms matches that of Eq.~\eqref{eq:MAXCUT_second_order}
\begin{eqnarray}    H^{(2)}_\mathrm{COLD}=&&\sum_c\alpha_c(\sigma^z_{c_1}\sigma^y_{c_2}+\sigma^y_{c_1}\sigma^z_{c_2})\\
&&+\sum_{c,c^\prime=1}^M\delta_{c_2,c^\prime_1}\beta_{c,c^\prime}(\sigma^z_{c_1}\sigma^x_{c_2}\sigma^y_{c^\prime_2}+\sigma^y_{c_1}\sigma^x_{c_2}\sigma^z_{c^\prime_2}).\nonumber
\end{eqnarray}
The minimization principle is written in terms of this local CD term, together with the optimizable operator,
\begin{eqnarray}
G(H^{(2)}_\mathrm{COLD}) =&& \partial_tH(t)+\partial_t\mathcal O_\mathrm{opt}(t,\boldsymbol{\gamma}) \\
&&-i [ H(t),H^{(2)}_\mathrm{COLD}]-i[ \mathcal O_\mathrm{opt}(t,\boldsymbol{\gamma}),H^{(2)}_\mathrm{COLD}].\nonumber
\end{eqnarray}
The corresponding trace minimization and the resulting system of equations for the sets of $\alpha_c\left(\lambda,t,\boldsymbol{\gamma}\right)$ and $\beta_{c,c^\prime}\left(\lambda,t,\boldsymbol{\gamma}\right)$ are detailed in App.~\ref{app:COLD}.
To perform the numerical optimization, we employed MATLAB R2023's built-in genetic algorithm to search for the optimal set of Fourier coefficients $\gamma^{(1,2)}$ that maximizes the final fidelity. The optimization was repeated $100$ times to ensure the convergence of the results. As a first clear sign of the inferiority of the COLD approach compared to the QBCD, the procedure proved to be computationally demanding, restricting our analysis to systems with up to $L=10$ spins---only almost half the size studied in the QBCD framework, which already exceeded the QBCD running time by several orders of magnitude for the largest system sizes. Despite these constraints, our setting already surpasses that of Ref.~\cite{Cepaite2023Counterdiabatic,LCD_Morawetz_2024}, by using twice the number of spins, allowing for a more systematic exploration of its performance in spin glasses.
These results consistently show that even modest extensions of local CD methods rapidly incur severe computational bottlenecks while providing only marginal benefits in maintaining adiabaticity near the critical point, reinforcing the conclusion that QBCD offers a far more practical and scalable route for larger systems.

\begin{figure}
    \centering
    \includegraphics[width=\columnwidth]{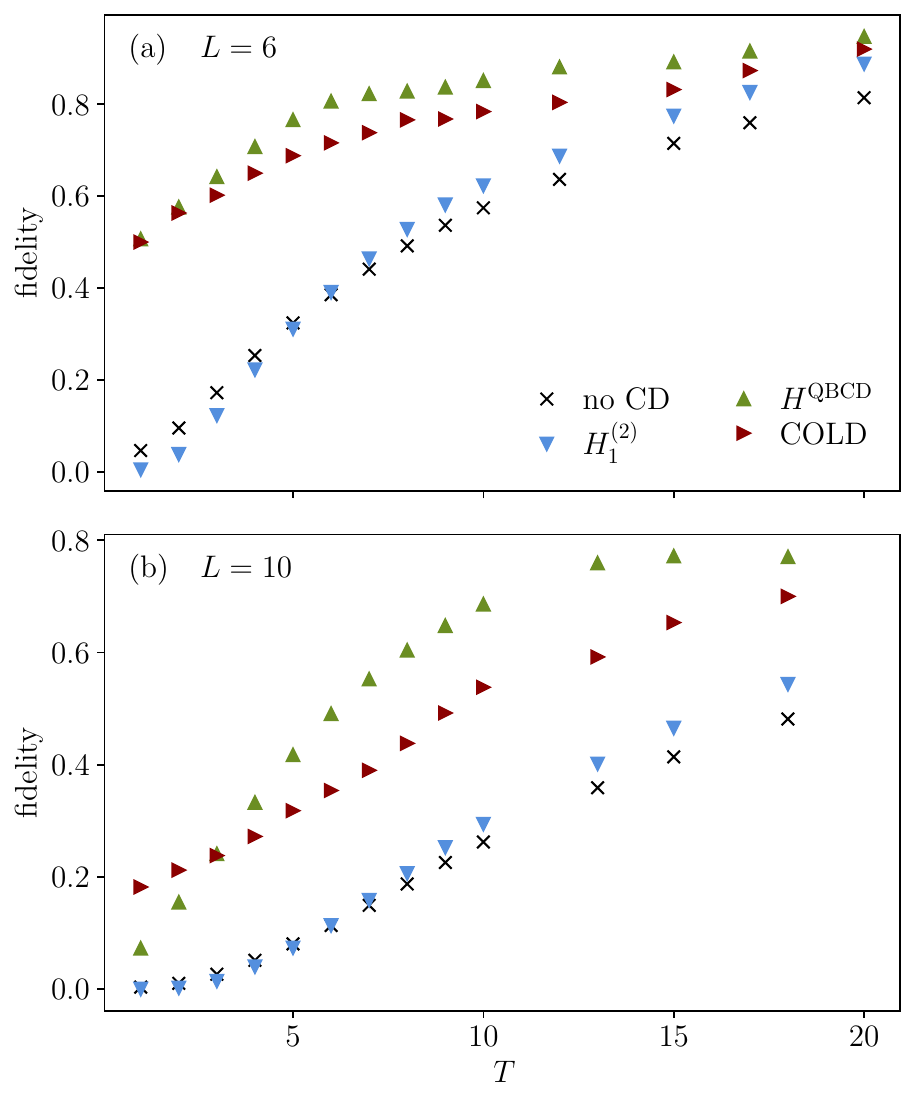}
    \caption{Fidelity as a function of the driving time for the sparsified QBCD, COLD approach, second-order local expansion, and for the bare process, for various system sizes. (a) Even for $L=6$ QBCD provides a sizable improvement at intermediate times. (b) For $L=10$ even close to adiabaticity, the QBCD clearly outperforms the COLD, while the latter gets closer to the second-order expansion and the bare process results.}
    \label{fig:COLD}
\end{figure}

Furthermore, we benchmarked the COLD method against the sparsified QBCD protocol and the second-order local CD expansion, Eq.~\eqref{eq:MAXCUT_second_order}, using the final ground-state fidelity for $L=6,10$ spins averaged over 10 random graph instances. As shown in Fig.~\ref{fig:COLD}, COLD benefits from its variational optimization at intermediate and long annealing times. For small systems, COLD can get close to the sparsified QBCD method, but this advantage diminishes with increasing system size due to stronger spin-glass effects and additional diabatic transitions. QBCD, in contrast, maintains near-unit fidelity in the modest driving regime ($T\lesssim 15$) by capturing global features and suppressing non-local corrections, consistently outperforming local CD approaches.

Finally, we compare the energetic requirements of the local CD expansion with those of the sparsified QBCD, by studying the system size dependence of the Hilbert-Schmidt norm, similar to the minimal spin-glass model [Sec.~\ref{subsec: MinSG}]. As demonstrated in Fig.~\ref{fig:Ecost_MAXCUT_XORSAT}, for the \textsc{Max Cut} it implies the following scalings:
\begin{equation}\label{eq:Ecost_MAXCUT}
    \lvert\lvert H^{(1)}_1\rvert\rvert^2\sim L \qquad \text{and} \qquad
    \lvert\lvert H^{(2)}_1\rvert\rvert^2\sim L^2.
\end{equation}
By contrast, the sparsified QBCD displays a much slower growth, following approximately a linear slope for $L\lesssim20$, owing to the exponentially reduced nonlocality and the finite deviations from the exact critical point (see Fig.~\ref{fig:Ecost_MAXCUT_XORSAT} as well). While the divergence of the energy denominator is significantly weakened by the finite deviations from the critical point in the approximate numerical construction, the numerator itself becomes exponentially small. This suppression stems from the sparse structure of $\partial_\lambda H$, the eigenstates, and their overall normalization factor scaling as $\sim2^{-L}$. Although the scaling is only approximately linear, the underlying microscopic mechanisms indicate that such moderate growth is a comprehensible scenario for larger systems. Taken together, they demonstrate that sparsified QBCD not only avoids the critical-point divergence but also achieves a more favorable energetic scaling while still retaining its efficiency in overcoming spin-glass bottlenecks.This further demonstrates how QBCD provides a new avenue for approximate CD, facilitating implementations that place substantially lower demands on external resources through systematic truncation.

\subsection{Demonstration on the \texorpdfstring{$3$-regular $3$-\textsc{XORSAT}}{3-regular 3-XORSAT}}

Finally, we test the QBCD method on the $3$-regular $3$-\textsc{XORSAT}, an optimization problem whose solutions correspond to the ground state of a spin-glass Hamiltonian with three-body interactions. In particular, the $3$-\textsc{XORSAT} problem can be formulated in terms of linear equations modulo $2$ with each of them containing three variables with the constraint that each variable appears in exactly three equations without repetition,
\begin{eqnarray}
    \underline{\underline{A}}\cdot\underline{x}=\underline{y}\qquad(\text{mod}\,2).
\end{eqnarray}
Here $\underline{\underline{A}}$ is an $ L\times L$ matrix with each row containing only three non-zero elements corresponding to the prescription of the problem. As the possible outcomes can only be $0$ or $1$, the assigned variables can take the same values as well, $x_i=0,1$. The corresponding Hamiltonian is given by
\begin{eqnarray}\label{eq:H_P_XORSAT}
    H_P=\sum_{c=1}^M\frac{1+J_c\sigma^z_{c_1}\sigma^z_{c_2}\sigma^z_{c_3}}{2},
\end{eqnarray}
with alternating randomly as $J_c=\pm1$ and with $M=L$. Following the strategy of Refs.~\cite{FarhiMAXCUT_XORSAT2012, Jorg2010First} we only consider such instances where the ground state is non-degenerate, allowing for a better characterization of the energy gap at the critical point. In these instances, one can drop the alternating coupling constant, $J_c=1$.
 The corresponding quantum annealing Hamiltonian is written as
\begin{eqnarray}
    H(t)=-\lambda \sum_c\frac{1-\sigma^z_{c_1}\sigma^z_{c_2}\sigma^z_{c_3}}{2}-(1-\lambda)\sum_{j=1}^L\sigma^x_j,
\end{eqnarray}
exhibiting an exponentially small gap around $\lambda_c=1/2$, where exponentially long driving times are required to reach near adiabatic dynamics. To this end, we probe the performance of both the sparsified QBCD Hamiltonian and the first-order locally optimized variational CD ansatz by the ground state fidelity and the minimal energy gap at the critical point. The first-order CD expansion is given by
\begin{eqnarray}\label{eq:H_1_XSAT}
H^{(1)}_1=\sum_c\alpha_c\left(\sigma^y_{c_1}\sigma^z_{c_2}\sigma^z_{c_3}+\sigma^z_{c_1}\sigma^y_{c_2}\sigma^z_{c_3}+\sigma^z_{c_1}\sigma^z_{c_2}\sigma^y_{c_3}\right),\nonumber\\
\end{eqnarray}
for which a similar system of equations can be derived based on the variational principle as in the case of the \textsc{Max-Cut} problem (see App.~\ref{app:XORSAT}).

Within the QBCD approach, we follow the same strategy similar to the previous subsection, Eq.~\eqref{eq:MAXCUT_QBCD} and determine the minimal gap, $\Delta$ and the corresponding eigenvectors, $\lvert\mathrm{GS}\rangle$ and $\lvert\mathrm{ES}\rangle$ with a mesh of solely $40$ points. Finally, the full QBCD was also sparsified to the same density as that of the leading order local CD expansion, Eq.~\eqref{eq:H_1_XSAT}.

\begin{figure}
    \centering
    \includegraphics[width=\columnwidth]{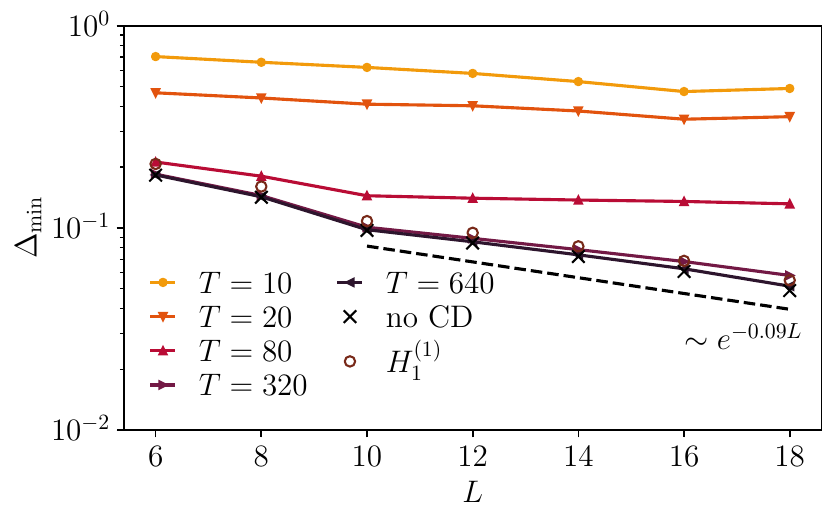}
    \caption{Ground state energy gap for the $3$-\textsc{XORSAT} model. Local CD expansion shows only negligible amplification even at short driving times (orange squares) $T=5$, compared to the bare process (red line). The sparsified QBCD exhibits an exponential amplification of the gap, as long as adiabaticity is not reached in the bare process.}
    \label{fig:Gap_XORSAT}
\end{figure}

\begin{figure}
    \centering
    \includegraphics[width=\columnwidth]{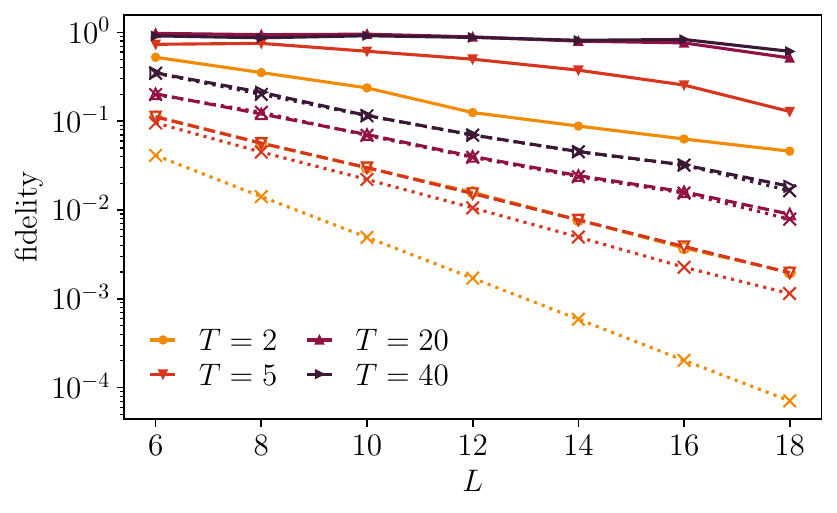}
    \caption{Ground-state fidelity for the $3$-\textsc{XORSAT} problem. The first-order local CD method yields a slower exponential decay at short driving times, but its advantage diminishes as the system approaches the adiabatic regime, where no significant improvement is observed.
In contrast, the sparsified QBCD exponentially enhances the fidelity across all timescales, yielding finite values already at $T = 2$ and approaching near-adiabatic fidelity for longer processes ($T = 5, 20, 40$). Solid lines refer to QBCD, dashed lines to 1st order CD, dotted lines to no CD.}
    \label{fig:fidelity_XORSAT}
\end{figure}

We again use the ground-state gap and final fidelity to assess adiabaticity. As shown in Fig.~\ref{fig:Gap_XORSAT}, even for short annealing times ($T=5$), the local expansion amplifies the exponential gap decay only by a constant factor. In contrast, the sparsified QBCD achieves exponentially larger gaps with approximately polynomial decay, even for longer times ($T=10,20,80$), with its effect disappearing only once adiabaticity is reached in the bare process. Further details are shown about the relative gap amplification in App.~\ref{app:Gap_difference}. Similar trends hold for the final fidelity (Fig.~\ref{fig:fidelity_XORSAT}): at short times ($T=2$), QBCD significantly outperforms both the local expansion and the bare process, even when the latter runs much longer ($T=40$). In this regime, QBCD fidelities remain near unity, while the local expansion and bare dynamics still show exponential decay (see Fig.~\ref{fig:Gap_XORSAT}).
Overall, the results for the \textsc{XORSAT} model further reinforce the conclusions drawn from the case of the \textsc{Max Cut}, highlighting both the efficiency and the prospects of the sparsified QBCD to boost currently realizable CD strategies.

\begin{figure}
    \centering
    \includegraphics[width=\columnwidth]{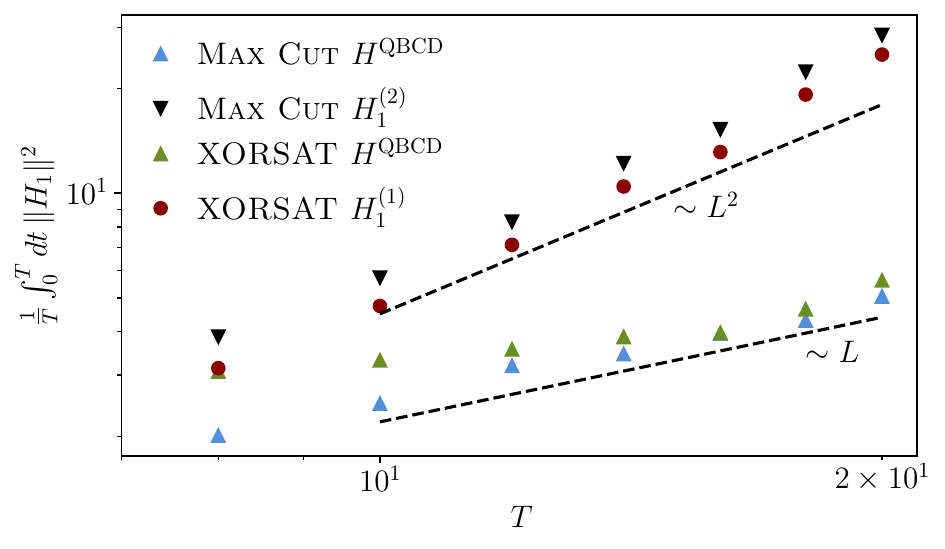}
    \caption{Energy costs of the second and first order local CD expansions of the $3$-\textsc{XORSAT} and \textsc{Max Cut} problems quantified by the Hilbert-Schmidt norms growing approximately as $\sim L^2$, in agreement with Eq.~\eqref{eq:Ecost_MAXCUT} and Eq.~\eqref{eq:Ecost_XORSAT}. On the other hand, the sparsified QBCD exhibits a much smaller norm growing approximately linearly for the accessible system sizes.}
    \label{fig:Ecost_MAXCUT_XORSAT}
\end{figure}

Closing this subsection, we perform the analogous comparison of the local CD expansion and the sparsified QBCD for the \textsc{XORSAT} model in terms of energetic requirements. In this case, the Hilbert–Schmidt norm of the second-order CD term scales as
\begin{eqnarray}\label{eq:Ecost_XORSAT}
\lvert\lvert H^{(2)}_1\rvert\rvert^2 \sim L^2,
\end{eqnarray}
as shown in Fig.~\ref{fig:Ecost_MAXCUT_XORSAT}. The sparsified QBCD exhibits a similar trend to that observed for the \textsc{Max Cut}, with a slightly faster yet still subleading growth. This parallel energetic scaling further illustrates the \textsc{Max Cut} results, showing that sparsified QBCD significantly alleviates energetic demands for implementation

\section{Discussion}
\label{sec:discussion}

In this study, we explored how quantum adiabatic dynamics can be enhanced to overcome spin-glass bottlenecks using various approximate counterdiabatic (CD) driving schemes. Specifically, we examined the variational local expansion and its improved version known as counterdiabatic optimized local driving (COLD). We also introduced a new scheme---quantum brachistochrone counterdiabatic driving (QBCD)---which focuses directly on the bottleneck region while relying only on the approximate knowledge of an exponentially small fraction of the full CD at a single parameter value. First, we tested the local expansion approaches on an exactly solvable minimal spin glass toy model described by an Ising chain admitting a free-fermionic description~\cite{Roberts2020Noise}. This model provides a valuable testbed with an exponentially small gap and two macroscopically distinct degenerate ground states. Its solvability gives access to larger system sizes and the extraction of asymptotic scalings, while the key ingredients and the level of efficiency observed here are further demonstrated in spin-glass models. To this end, we employed the variational principle to construct two- and three-body CD terms from the Floquet-Krylov expansion, which can efficiently be simulated on a classical computer~\cite{Sels2017Minimizing,Claeys2019Floquet} and has also recently been realized on quantum hardware~\cite{Romero2025,romero2025proteinfolding,visuri2025digitized}. A key finding is that, while these CD ansatzes can exponentially amplify the size of the minimal gap, it remains exponentially small in the system size $L$, and the final ground state probability is of the same order of magnitude for times shorter than $\sim e^{\alpha L}$, with $\alpha$ given in Eq.~\eqref{eq:Delta_min}.

Our analysis highlights key features of CD of frustrated spin systems. First, from the gap characterization in Sec.~\ref{sec:CD_gap_amplification}, it can be inferred that CD can amplify the gap exponentially in the diabatic regime, and thus substantially reduce the diabatic transitions. However, reaching the regime of driving times where diabatic transitions come solely from the exponentially small ground state gap, the first two CD terms provide negligible improvement (see Fig.~\ref{fig:kinks}) and higher order expansions are required. On the other hand, slowing down the driving schedule also reduces the overall strength of the CD term, given the $\partial_t \lambda$ prefactor in Eq.~\eqref{eq:H_CD}. These features prevent the construction of an effective local CD term that ensures adiabatic dynamics in the presence of exponentially small gaps between frustrated energy levels.

In this context, CD techniques, including the variational formulation~\cite{Sels2017Minimizing,Claeys2019Floquet,Takahashi2024Shortcuts}, aim to keep the evolved state close to the instantaneous ground state. In our simple model, a single transition spans the entire effective chain, Eq.~\eqref{eq:H_hopping}, meaning the state must be adjusted non-locally to follow the changing ground state.

As an alternative and novel approach, we introduced the QBCD Hamiltonian in Sec.~\ref{sec:CD_LR}, which shares the structure of the Hamiltonian for the time-optimal evolution in the quantum brachistochrone problem for the corresponding localized edge states. The QBCD Hamiltonian employs only an exponentially small fraction of the full CD at a single parameter value, focusing exclusively on the ground and first excited states around the critical point. With the help of it, the success rate of the protocol increases from a vanishing probability to almost $0.5$, as indicated by the average final kink number. While implementing nonlocal, multi-body interactions on quantum hardware remains challenging, QBCD shows that even an exponentially small fraction of the full CD Hamiltonian can suffice for exponential speed-up: approximate knowledge at a single parameter point is enough, and the scheme remains robust near the critical point, drastically alleviating implementation complexity. The constant energetic cost of QBCD, in contrast to the diverging ones in the local expansions, further underscores its practical advantages, paving the way for experimentally feasible and resource-efficient CD expansion schemes in both quantum and classical simulations.

In the second part of the paper, we examined how our findings extend to two realistic spin-glass models associated with \textsc{NP}-hard problems, namely the $3$-regular \textsc{Max Cut} and the $3$-\textsc{XORSAT} problems. These models exhibit an exponentially small gap already at system sizes $L\gtrsim 10$, accessible by classical numerical simulations. Specifically, the $3$-\textsc{XORSAT} problem extends the validity of our results to cubic spin-glass Hamiltonians. Due to the limitations in accessible system sizes, we used the final ground-state fidelity---rather than the excitation density---as a measure of adiabaticity, alongside the minimum gap. Our results show that local CD expansions have a similar efficiency as in the minimal spin-glass model. They yield only modest improvements outside the adiabatic regime: the gap is amplified primarily for fast quenches where, however, excitations to higher energy states break down adiabaticity completely.
On the other hand, gap amplification terminates for slightly shorter driving time scales than those where the bare process fidelity starts to depart from zero. At the same time, for intermediate and slow annealing times, the fidelity under local CD driving preserves the same exponential decay with system size as in the bare dynamics, and thus no significant gains are obtained in this regime.

For the QBCD method, we introduced an enhanced, sparsified version that matches the density of the local expansions. Our findings show that exponential speed-up across an exponentially small gap can be achieved with only an exponentially small nonlocal contribution from the already substantially reduced full CD term at a single parameter value, with persistent efficiency under finite deviations from the critical point. In this sparsified scheme, only a polynomially-sized subset of eigenstate components needs to be retained, and the ground-state gap can be efficiently approximated using advanced large-scale numerical techniques such as quantum Monte Carlo methods~\cite{Young2008Size,Young2010First,Boixo2014Evidence} or reverse annealing~\cite{Ohkuwa2018Reverse,King2018Observation,Yamashiro2019Dynamics}, substantially lowering computational resources and facilitating large-scale classical implementations~\cite{tindall2025,mauron2025}. From the viewpoint of quantum-device implementation, the controlled truncation of the residual non-locality guides the construction of approximate local CD terms with drastically reduced resource-requirements, enabling higher-order CD expansions on near-term quantum hardware~\cite{romero2025proteinfolding,Romero2025,visuri2025digitized}.

Finally, we also compared the efficiency of the COLD approach with the local expansion and the sparsified QBCD in the \textsc{Max Cut} problem. Although it provided a noticeable improvement over the local expansion, its performance remained consistently inferior to that of the sparsified QBCD. Even for small systems ($L=6$), spin-glass effects already reduce the efficiency of COLD below that of QBCD, and its numerical implementation proved challenging even at modest sizes ($L \lesssim 10$). As $L$ was increased and spin glass effects became stronger, the performance of COLD converged to that of the second-order local expansion. This highlights both its limited impact and the substantially higher computational costs compared to QBCD.

This combination of results suggests that the challenge of traversing exponentially small gaps in these spin-glass problems is primarily of entropic rather than energetic origin \cite{Bellitti2021Entropic}. In other words, overcoming the bottleneck does not require access to strong control fields but rather a minimal amount of information about where and how to steer the quantum state. This perspective further highlights the practical value of methods like QBCD, where even a small fraction of non-local terms is enough to achieve efficient evolution.

Our further results on the Hilbert–Schmidt norms in the spin-glass models demonstrate that sparsified QBCD is drastically less resource demanding than the corresponding local CD expansions. In both the \textsc{Max Cut} and \textsc{XORSAT} cases, the unfavorable polynomial growth of local CD expansions is replaced by a far more benign scaling under the sparsified QBCD. This improved behavior arises from the reduced non-locality and the sparsity of both $\partial_\lambda H$ and the eigenstates, which together suppress the divergences that make the norm of the exact QBCD diverge at critical points. Taken together, these results further support the notion that controlled truncation in sparsified QBCD can facilitate more resource-efficient implementations of approximate CD, providing a natural pathway to higher-order expansions.
  
Accordingly, while exploring alternative strategies beyond a priori action minimization remains an interesting direction, our results---and the preceding discussion---suggest that concentrating on progressively improving approximate implementations of QBCD may ultimately be more effective for overcoming bottlenecks. Approaches based on the integration of CD ansatzes with optimal control methods---such as QAOA \cite{Hegade2022Digitized,Chandarana2022Digitized,Wurtz2022Counterdiabaticity,hegade2023digitizedCDFact}---optimize control fields iteratively using only the final state reached by the dynamics \cite{Cepaite2023Counterdiabatic}. However, these methods often suffer from limited analytical transparency and can be hampered by complex control landscapes and barren plateaus \cite{McClean2018Barren,Day2019Glassy,Larocca2024Review,Beato2024Theory}. In contrast, the sparsified QBCD already demonstrates significant improvements and points towards a more efficient approximate implementation using experimentally accessible resources. Building on these advantages, future work could combine QBCD with complementary strategies, such as reverse annealing \cite{Ohkuwa2018Reverse,King2018Observation,Yamashiro2019Dynamics}, bias fields \cite{Rams2019,Cadavid2024bias}, Lyapunov control \cite{Malla2024,Chandarana2024}, quantum Monte Carlo methods~\cite{Young2008Size,Young2010First,Boixo2014Evidence} or targeted control engineering near critical points.

\acknowledgments
It is a pleasure to thank Pranav Chandarana and Koushik Paul for comments on the manuscript
and Oskar A. Pro\'sniak and Vittorio Vitale for discussions. This project was supported by the Luxembourg National Research Fund (FNR Grant Nos.\ 17132054 and 16434093). It has also received funding from the QuantERA II Joint Programme and co-funding from the European Union’s Horizon 2020 research and innovation programme. The numerical simulations presented in this work were partly carried out using the HPC facilities of the University of Luxembourg.

\onecolumngrid
\appendix

\section{Free-fermion representation of the Hamiltonian}
\label{app:sec:H_free_fermions}

\subsection{Jordan-Wigner transformation and Dirac fermions}
\label{app:sec:H_Dirac}

Start from the Hamiltonian~\eqref{eq:H}. By applying a set of one-qubit gates
\begin{equation}
    \label{eq:U_rotation_xz}
    U=\frac{1}{\sqrt{2}}
    \begin{pmatrix}
		1	&-1 \\
		1	&1
	\end{pmatrix},
\end{equation}
by virtue of the identities
\begin{equation}
	U \sigma^x U^\dagger = -\tilde{\sigma}^z, \qquad
	U \sigma^y U^\dagger = \tilde{\sigma}^y, \qquad
	U \sigma^z U^\dagger = \tilde{\sigma}^x, 
\end{equation}
one can rewrite equivalently
\begin{equation}
    \label{eq:H_rotated}
	H[\lambda] = \sum_{j=1}^L \left[ -\lambda J_j \tilde{\sigma}^x_j \tilde{\sigma}^x_{j+1} + (1-\lambda) \tilde{\sigma}_j^z \right].
\end{equation}
Tildes are used to distinguish the two sets of Pauli matrices for later convenience. This form of the Hamiltonian is amenable to a Jordan-Wigner transformation, mapping spins $\{\tilde{\sigma}_j\}$ to Dirac fermions $\{c_j^\phdagger,c_j^\dagger\}$ by means of the following identities,
\begin{equation}
    \label{eq:Jordan-Wigner}
	\tilde{\sigma}_j^+ = e^{-i\pi \Sigma_j} c_j^\dagger, \qquad
	\tilde{\sigma}_j^- = e^{i\pi \Sigma_j} c_j , \qquad
    \tilde{\sigma}_j^z = 2 c_j^\dagger c_j^\phdagger - 1,
\end{equation} 
where the string operator $\Sigma_j$ in the exponent reads
\begin{equation}
	\Sigma_j := \sum_{k=1}^{j-1} c_k^\dagger c_k^\phdagger.
\end{equation}
Another possible way of writing the transformation is
\begin{equation}
    \label{eq:Jordan-Wigner2}
	c_j^\dagger = \prod_{k=1}^{j-1} \big(-\tilde{\sigma}_k^z\big) \, \tilde{\sigma}_j^+, \qquad
	 c_j =\prod_{k=1}^{j-1} \big(-\tilde{\sigma}_k^z\big) \, \tilde{\sigma}_j^-.
\end{equation}
Then, the fermionic vacuum $\ket{0_c}$ corresponds to spins positively aligned along $x$-axis in the original basis ($\ket{0_c} = \ket{\rightarrow}$), and the fully occupied fermionic state corresponds to the spins anti-aligned along $x$-axis  ($\ket{1_c} = \ket{\leftarrow}$).

Applying the Jordan-Wigner transformation to Eq.~\eqref{eq:H_rotated}, one finds
\begin{align}
	H[\lambda] &= \sum_{j=1}^L \left[ -\lambda J_j \left( e^{-i\pi \Sigma_j} c_j^\dagger + e^{i\pi \Sigma_j} c_j\right) \left( e^{-i\pi \Sigma_{j+1}} c_{j+1}^\dagger + e^{i\pi \Sigma_{j+1}} c_{j+1}\right) + (1-\lambda) \left( 2 c_j^\dagger c_j^\phdagger - 1 \right) \right] \\
	&= \sum_{j=1}^L \left[ - \lambda J_j \left(c_j^\dagger e^{-i\pi c_j^\dagger c_j^\phdagger} c_{j+1}^\dagger + c_j^\dagger e^{i\pi c_j^\dagger c_j^\phdagger} c_{j+1} + c_j e^{-i\pi c_j^\dagger c_j^\phdagger} c_{j+1}^\dagger + c_j e^{i\pi c_j^\dagger c_j^\phdagger} c_{j+1} \right) + 2 (1-\lambda) c_j^\dagger c_j^\phdagger \right] - (1-\lambda)L \\
    &= \sum_{j=1}^L \left[ - \lambda J_j \left(c_j^\dagger c_{j+1}^\dagger + c_j^\dagger c_{j+1}^\phdagger - c_j^\phdagger c_{j+1}^\dagger - c_j c_{j+1} \right) + 2 (1-\lambda) c_j^\dagger c_j^\phdagger \right] - (1-\lambda)L. 
\end{align}
The Hamiltonian is manifestly a quadratic form in the creation/annihilation operators. In order to fix the boundary conditions, one needs to look back at the original Hamiltonian~\eqref{eq:H}. In terms of the spins, the boundary conditions are periodic: $\sigma_{j+L} \equiv \sigma_j$. However, for the fermionic operators, it holds $c_{j+L} = + c_j$ or $c_{j+L} = - c_j$ depending on whether there is an odd or even number of fermions, respectively. Since the initial state (ground state at $\lambda=0$) is the fermionic vacuum $\ket{0_c}$, the correct sector to take is the even one.

In order to simplify things, we will not use the relation $c_{j+L} = - c_j$, but equivalently use $c_{j+L} = c_j$ and flip the sign of the last coupling $J_L \to -J_L$, while still working within the even parity sector. In order to keep track of this change, we use tildes:
\begin{equation}
    \label{eq:J_tilde}
    \tilde{J}_j :=
    \begin{cases}
        J_j    &\text{if } j \neq L \\
        -J_L   &\text{if } j = L. 
    \end{cases}
\end{equation}
Therefore, the Hamiltonian reads
\begin{align}
    \label{app:eq:H_Dirac}
    H[\lambda] &= \sum_{j=1}^L \left[ - \lambda \tilde{J}_j \left(c_j^\dagger c_{j+1}^\dagger + c_j^\dagger c_{j+1}^\phdagger - c_j^\phdagger c_{j+1}^\dagger - c_j c_{j+1} \right) + 2 (1-\lambda) c_j^\dagger c_j^\phdagger \right] - (1-\lambda)L \\
    &= \sum_{j=1}^L \left[ - \lambda \tilde{J}_j \left(c_j^\dagger - c_j^\phdagger \right) \left( c_{j+1}^\dagger + c_{j+1}^\phdagger \right) + 2 (1-\lambda) c_j^\dagger c_j^\phdagger \right] - (1-\lambda)L .
\end{align}
Notice that if any operator (different from the Hamiltonian) has terms coupling $c_L$ with $c_{L+1}$, also there the sign must be flipped. We will use tildes to remember this fact.

\subsection{Majorana fermions}
\label{app:sec:Majorana}

\begin{figure}
    \centering
    \includegraphics[width=0.5\textwidth]{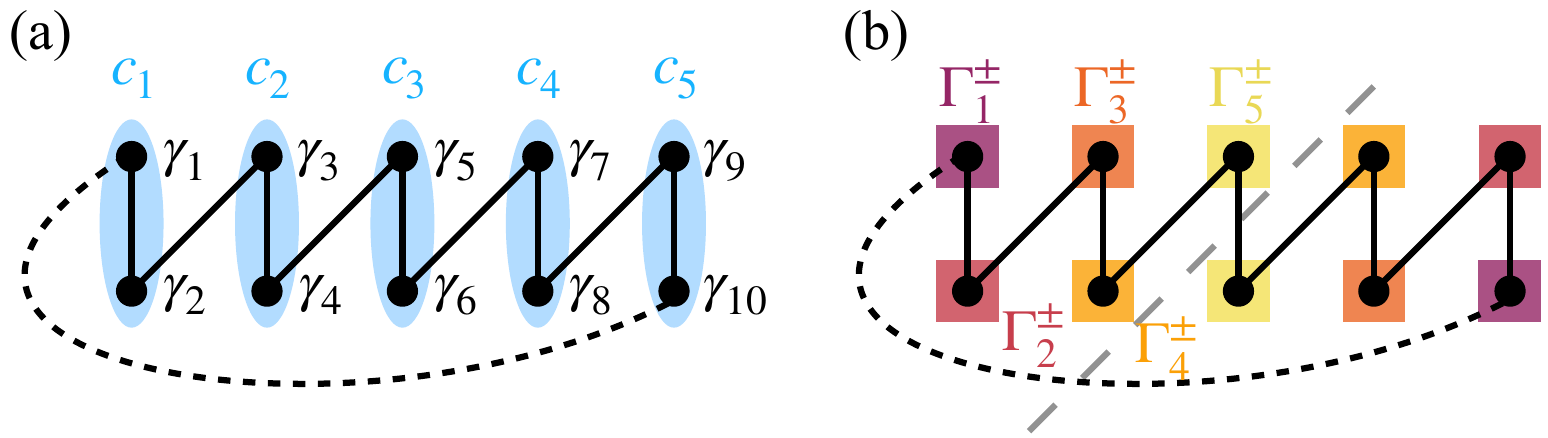}
    \caption{(a) Majorana representation of the chain for $L=5$. Each Dirac site is doubled into two Majorana sites, according to Eq.~\eqref{app:eq:Majorana_fermions}, and thus the length of the Majorana chain becomes $2L$ (i.e.,  $10$ in the picture). Notice that the sign of the coupling $J_L$ between the first and last Majorana fermions is flipped to take into account the even fermionic parity; the bond is dotted in the figure to highlight the fact. (b) Symmetrized Dirac representation of the chain. Each Majorana operator $\gamma_k$ is coupled with its symmetric partner $\gamma_{2L-k+1}$, according to Eqs.~\eqref{eq:def_Gamma}: in the picture, this is shown as a color-code matching of the sites. The chain goes back to a length of $L$ because the sites to the right of the symmetry cut (dashed gray line) become redundant.}
    \label{fig:chain_fermions}
\end{figure}

From the Dirac fermion representation $\{c_i^\phdagger,c_j^\dagger\}_{j=1}^L$, one can pass to an equivalent Majorana fermionic representation $\{\gamma_{2j-1}, \gamma_{2j}\}_{j=1}^L$ by effectively doubling the length of the chain. The defining relations are
\begin{equation}
    \label{app:eq:Majorana_fermions}
    \gamma_{2j-1} := \frac{1}{\sqrt{2}} \left( c_j^\phdagger+c_j^\dagger\right), \qquad
    \gamma_{2j} := \frac{i}{\sqrt{2}} \left( c_j^\phdagger-c_j^\dagger \right), 
\end{equation}
and the inverse transformation is
\begin{equation}
    \label{app:eq:Majorana_fermions_inv}
    c_j^\dagger = \frac{\gamma_{2j-1} + i \gamma_{2j}}{\sqrt{2}}, \qquad 
    c_j^\phdagger = \frac{\gamma_{2j-1} - i \gamma_{2j}}{\sqrt{2}},
\end{equation}
leading to anticommutation relations
\begin{equation}
    \left\{ \gamma_k, \gamma_{k'} \right\} = \delta_{k,k'}.
\end{equation}
The Majorana representation is shown pictorially in Fig.~\ref{fig:chain_fermions}(a). 
One can rewrite the Hamiltonian as
\begin{equation}
    \label{app:eq:H_Majorana}
    H[\lambda] = -2i \sum_{j=1}^L \big[ \lambda \tilde{J}_j \gamma_{2j} \gamma_{2j+1} + (1-\lambda) \gamma_{2j-1} \gamma_{2j} \big] .
\end{equation}
The boundary conditions inherited by the Dirac fermions entail $\gamma_{k+2L} = - \gamma_k$ (even parity sector) but, as stated above, we are employing $\gamma_{k+2L} = \gamma_k$ and the flipped couplings $\tilde{J}_j$.

\subsection{Reflection parity symmetry}
\label{app:sec:reflection_parity}

The Hamiltonian~\eqref{eq:H}, and thus its Majorana representation~\eqref{app:eq:H_Majorana}, is odd under the chiral reflection parity symmetry $\widetilde{\Pi}_\mathrm{R}$, defined by
\begin{equation}
    \label{eq:chiral_reflection_parity}
    \widetilde{\Pi}_\mathrm{R} \gamma_k \widetilde{\Pi}_\mathrm{R} = \gamma_{2L-k+1}.
\end{equation}
Indeed, the couplings satisfy $J_j = J_{L-j}$, and one can write
\begin{align}
    \widetilde{\Pi}_\mathrm{R} H \widetilde{\Pi}_\mathrm{R} &= -2i \sum_{j=1}^L \big[ \lambda \tilde{J}_j \gamma_{2L-2j+1} \gamma_{2L-2j} + (1-\lambda) \gamma_{2L-2j+2} \gamma_{2L-2j+1} \big] \\
    &= -2i \sum_{j=1}^L \big[ \lambda \tilde{J}_{L-j} \gamma_{2L-2j+1} \gamma_{2L-2j} + (1-\lambda) \gamma_{2L-2j+2} \gamma_{2L-2j+1} \big] \\
    &= -2i \sum_{l=0}^{L-1} \big[\lambda \tilde{J}_l \gamma_{2l+1} \gamma_{2l} + (1-\lambda) \gamma_{2l+2} \gamma_{2l+1} \big] 
    = -H.
\end{align}
In the first line, the operator $\widetilde{\Pi}_\mathrm{R}$ was applied on the $\gamma$'s; in the second line, $\tilde{J}_j$ was exchanged with $\tilde{J}_{L-j}$; in the third line, it was set $l:=L-j$. $H$ is odd under the reflection parity $\widetilde{\Pi}_\mathrm{R}$, thus $\widetilde{\Pi}_\mathrm{R}$ is a \emph{chiral symmetry} for $H$~\cite{Asboth2016Short}. 

Before moving on, it is interesting to express the chiral reflection parity symmetry also in terms of the Dirac operators:
\begin{subequations}
\begin{align}
    \widetilde{\Pi}_\mathrm{R} c_j \widetilde{\Pi}_\mathrm{R} &= \widetilde{\Pi}_\mathrm{R} \frac{\gamma_{2j-1} - i \gamma_{2j}}{\sqrt{2}} \widetilde{\Pi}_\mathrm{R} = \frac{\gamma_{2L-2j+2} - i \gamma_{2L-2j+1}}{\sqrt{2}} = - i c^\dagger_{L-j+1}, \\
    \widetilde{\Pi}_\mathrm{R} c_j^\dagger \widetilde{\Pi}_\mathrm{R} &= \widetilde{\Pi}_\mathrm{R} \frac{\gamma_{2j-1} + i \gamma_{2j}}{\sqrt{2}} \widetilde{\Pi}_\mathrm{R} = \frac{\gamma_{2L-2j+2} + i \gamma_{2L-2j+1}}{\sqrt{2}} = i c_{L-j+1},
\end{align}
\end{subequations}
which shows that $\widetilde{\Pi}_\mathrm{R}$ also exchanges particles and holes. In terms of the spin operators, it holds instead
\begin{equation}
    \widetilde{\Pi}_\mathrm{R} \sigma_j^x \widetilde{\Pi}_\mathrm{R} = -\widetilde{\Pi}_\mathrm{R} \tilde{\sigma}_j^z \widetilde{\Pi}_\mathrm{R} = - \widetilde{\Pi}_\mathrm{R} c_j^\dagger c_j^\phdagger \widetilde{\Pi}_\mathrm{R} = - i c_{L-j+1} (-i) c_{L-j+1}^\dagger = c_{L-j+1}^\dagger c_{L-j+1} = \tilde{\sigma}_{L-j+1}^z = -\sigma_{L-j+1}^x,
\end{equation}
while $\widetilde{\Pi}_\mathrm{R} \sigma_j^z \widetilde{\Pi}_\mathrm{R}$ involves also the transformation of the Jordan-Wigner string. Thus, the chiral reflection parity symmetry is non-local in the spin representation of the Hamiltonian.

The non-locality of $\widetilde{\Pi}_\mathrm{R}$ in the spin representation is ``dual'' to the non-locality of $\Pi_\mathrm{R}$ in the fermionic representation. Indeed, the presence of the Jordan-Wigner string entails the transformation rules 
\begin{subequations}
\begin{align}
    \Pi_\mathrm{R} c_j \Pi_\mathrm{R} &= \Pi_\mathrm{R} \prod_{k=1}^{j-1} \big( -\sigma_k^z \big) \sigma_j^- \Pi_\mathrm{R} = \prod_{k=1}^{j-1} \big( -\sigma_{L-k+1}^z \big) \sigma_{L-j+1}^- = (-1)^{N_\mathrm{F}} \prod_{k=1}^{L-j+1} \big( -\sigma_k^z \big) \sigma_{L-j+1}^- \nonumber \\
    &= (-1)^{N_\mathrm{F}} \left( 1-2 c_{L-j+1}^\dagger c_{L-j+1}^\phdagger \right) c_{L-j+1} = (-1)^{N_\mathrm{F}} c_{L-j+1}, \\
    \Pi_\mathrm{R} c_j^\dagger \Pi_\mathrm{R} &= \Pi_\mathrm{R} \prod_{k=1}^{j-1} \big( -\sigma_k^z \big) \sigma_j^+ \Pi_\mathrm{R} = \prod_{k=1}^{j-1} \big( -\sigma_{L-k+1}^z \big) \sigma_{L-j+1}^+ = (-1)^{N_\mathrm{F}} \prod_{k=1}^{L-j+1} \big( -\sigma_k^z \big) \sigma_{L-j+1}^+ \nonumber \\
    &= (-1)^{N_\mathrm{F}} \left( 1-2 c_{L-j+1}^\dagger c_{L-j+1}^\phdagger \right) c_{L-j+1}^\dagger = - (-1)^{N_\mathrm{F}} c_{L-j+1}^\dagger,
\end{align}
\end{subequations}
where $N_\mathrm{F} := \sum_{j=1}^L c_j^\dagger c_j^\phdagger$ is the total fermion number operator. The prefactor $(-1)^{N_\mathrm{F}}$ is a non-local operator that makes it more difficult to define creation and annihilation operators symmetrized with respect to\ $\Pi_\mathrm{R}$. This is the reason why it is convenient to use the chiral symmetry $\widetilde{\Pi}_\mathrm{R}$ instead of the unitary symmetry $\Pi_\mathrm{R}$ when working with the Dirac fermions. A similar situation takes place in the Majorana representation. A comprehensive, modern view on the subtleties of symmetries in Ising-like chains can be found in Ref.~\cite{Seiberg2024Majorana}.

\subsection{Reflection-symmetrized representation}
\label{app:sec:Gamma}

Let us introduce symmetrized operators 
\begin{subequations}
    \label{eq:def_Gamma}
\begin{alignat}{3}
    \Gamma^\pm_k &= \frac{\pm i \gamma_k + \gamma_{2L-k+1}}{\sqrt{2}}, \qquad &&k=1,3,\dots,L, \\
    \Gamma^\pm_k &= \frac{\gamma_k \pm i \gamma_{2L-k+1}}{\sqrt{2}}, \qquad &&k=2,4,\dots,L-1,
\end{alignat}
\end{subequations}
where the factors $\pm i$ were chosen so that 
\begin{subequations}
\begin{gather}
    \label{app:eq:Gamma_commutation}
    \{ \Gamma_k^+, \Gamma_{k'}^-\} = \delta_{k,k'}, \qquad
    \{ \Gamma_k^+, \Gamma_{k'}^+\} = \{ \Gamma_k^-, \Gamma_{k'}^-\} = 0, \\
    \label{app:eq:Gamma_reflection}
    \widetilde{\Pi}_\mathrm{R} \Gamma_k^\pm \widetilde{\Pi}_\mathrm{R} = \pm i \Gamma_k^\mp, \qquad
    (\Gamma^\pm_k)^\dagger = \Gamma^\mp_k.
\end{gather}
\end{subequations}
These correspond to Eqs.~\eqref{eq:Gamma_commutation} and \eqref{eq:Gamma_reflection} in the main text, and are shown pictorially in Fig.~\ref{fig:chain_fermions}(b). The inverse transformation is
\begin{subequations}    
\begin{alignat}{3}
    \gamma_k &= \frac{\Gamma_k^+ - \Gamma_k^-}{\sqrt{2}i}, \qquad &&k=1,3,\dots,L, \\ 
    \gamma_k &= \frac{\Gamma_k^+ + \Gamma_k^-}{\sqrt{2}}, \qquad &&k=2,4,\dots,L-1, \\
    \gamma_k &= \frac{\Gamma_{2L-k+1}^+ + \Gamma_{2L-k+1}^-}{\sqrt{2}}, \qquad &&k=L+1,L+3,\dots,2L, \\
    \gamma_k &= \frac{\Gamma_{2L-k+1}^+ - \Gamma_{2L-k+1}^-}{\sqrt{2}i}, \qquad &&k=L+2,L+3,\dots,2L-1. 
\end{alignat}
\end{subequations}

Let us split the Hamiltonian~\eqref{app:eq:H_Majorana} into the different contributions: for $-\lambda J$ or $(1-\lambda)$, and bulk or boundary. It holds for $j=1,2,\dots,\ell$ 
\begin{align}
    \tilde{J}_j \gamma_{2j} \gamma_{2j+1} + \tilde{J}_{L-j} \gamma_{2L-2j} \gamma_{2L-2j+1} 
    &= \tilde{J}_j \left[\frac{\Gamma_{2j}^+ + \Gamma_{2j}^-}{\sqrt{2}} \frac{\Gamma_{2j+1}^+ - \Gamma_{2j+1}^-}{\sqrt{2}i} + \frac{\Gamma_{2j+1}^+ + \Gamma_{2j+1}^-}{\sqrt{2}} \frac{\Gamma_{2j}^+ - \Gamma_{2j}^-}{\sqrt{2}i} \right] \nonumber \\
    &= -\frac{i\tilde{J}_j}{2} \left[ \left(\Gamma_{2j}^+ + \Gamma_{2j}^-\right) \left(\Gamma_{2j+1}^+ - \Gamma_{2j+1}^-\right) - \left(\Gamma_{2j}^+ - \Gamma_{2j}^- \right) \left( \Gamma_{2j+1}^+ + \Gamma_{2j+1}^- \right)  \right] \nonumber \\
    &= i \tilde{J}_j \left( \Gamma_{2j}^+ \Gamma_{2j+1}^- - \Gamma_{2j}^- \Gamma_{2j+1}^+ \right),
\end{align}
and
\begin{align}
    (1-\lambda) (\gamma_{2j-1} \gamma_{2j} + \gamma_{2L-2j-1} \gamma_{2L-2j})
    &= (1-\lambda) \left[ \frac{\Gamma_{2j-1}^+ - \Gamma_{2j-1}^-}{\sqrt{2}i} \frac{\Gamma_{2j}^+ + \Gamma_{2j}^-}{\sqrt{2}}  + \frac{\Gamma_{2j}^+ - \Gamma_{2j}^-}{\sqrt{2}i} \frac{\Gamma_{2j-1}^+ + \Gamma_{2j-1}^-}{\sqrt{2}}  \right] \nonumber \\
    &= -\frac{i(1-\lambda)}{2} \left[ \left(\Gamma_{2j-1}^+ - \Gamma_{2j-1}^-\right) \left(\Gamma_{2j}^+ + \Gamma_{2j}^-\right) - \left( \Gamma_{2j-1}^+ + \Gamma_{2j-1}^- \right) \left(\Gamma_{2j}^+ - \Gamma_{2j}^- \right) \right] \nonumber \\
    &= i (1-\lambda) \left( \Gamma_{2j-1}^- \Gamma_{2j}^+ - \Gamma_{2j-1}^+ \Gamma_{2j}^- \right),
\end{align}
while for the other sites
\begin{gather}
    \tilde{J}_L \gamma_{2L} \gamma_1
    = \tilde{J}_L \frac{\Gamma_1^+ + \Gamma_1^-}{\sqrt{2}} \frac{\Gamma_1^+ - \Gamma_1^-}{\sqrt{2}i} 
    = \frac{i\tilde{J}_L}{2} \left( \Gamma_1^+ \Gamma_1^- - \Gamma_1^- \Gamma_1^+ \right), \\
    (1-\lambda) \gamma_{L} \gamma_{L+1} 
    = (1-\lambda) \frac{\Gamma_L^+ - \Gamma_L^-}{\sqrt{2}i} \frac{\Gamma_L^+ + \Gamma_L^-}{\sqrt{2}} 
    = \frac{i}{2} (1-\lambda) \left( \Gamma_L^- \Gamma_L^+ - \Gamma_L^+ \Gamma_L^- \right) .
\end{gather}
The Hamiltonian acquires the form
\begin{multline}
    \label{app:eq:H_Gamma}
    H[\lambda] = 2\lambda \sum_{j=1}^{\ell} \tilde{J}_j \left( \Gamma_{2j}^+ \Gamma_{2j+1}^- - \Gamma_{2j}^- \Gamma_{2j+1}^+ \right)
    + 2 (1-\lambda) \sum_{j=1}^{\ell} \left( \Gamma_{2j-1}^- \Gamma_{2j}^+ - \Gamma_{2j-1}^+ \Gamma_{2j}^- \right) \\
    + \lambda \tilde{J}_L \left( \Gamma_1^+ \Gamma_1^- - \Gamma_1^- \Gamma_1^+ \right)
    + (1-\lambda) \left( \Gamma_L^- \Gamma_L^+ - \Gamma_L^+ \Gamma_L^- \right),
\end{multline}
which is the same of Eq.~\eqref{eq:H_Gamma} in the main text, once the $J$'s are restored from the $\tilde{J}$'s. One can verify that $\widetilde{\Pi}_\mathrm{R} H[\lambda] \widetilde{\Pi}_\mathrm{R} = -H[\lambda]$ directly from Eqs.~\eqref{eq:Gamma_commutation}--\eqref{eq:Gamma_reflection}.

\subsection{Transformation of states}

From the equations above, one can also find the relation between the vacuum state for the $c$'s and the one for the $\Gamma$'s. Defining $\ket{0_c}$ to be the state annihilated by all $c_j$'s, and $\ket{0_\Gamma}$ to be the one annihilated by all the $\Gamma_k^-$'s, it holds
\begin{equation}
    \ket{0_c} = \prod_{k=1}^{\ell} \left( 1 + \Gamma_{2k-1}^+ \Gamma_{2k}^+ \right) \ket{0_\Gamma},
\end{equation}
as one can check by directly applying each $c_j$ expressed in terms of the $\Gamma_k^\pm$'s.

\section{Diagonalization of the effective hopping model}
\label{app:sec:diagonalization}

The purpose of this Appendix is to provide the eigendecomposition of the system Hamiltonian~\eqref{eq:H} by diagonalizing the effective 1D hopping model Eq.~\eqref{eq:H_hopping}. It is convenient to rewrite the matrix as
\begin{equation}
    \label{app:eq:H_hopping}
	\mathcal{H} = \lambda
    \begin{pmatrix}
        -J_L &-B  &    &    &       &         &  \\
        -B   &0   &J_1 &    &       &         &  \\
             &J_1 &0   &-B  &       &         &  \\
             &    &-B  &0   &J_2    &         &  \\
             &    &    &J_2 &\ddots &\ddots   &  \\
             &    &    &    &\ddots &0        &J_{\ell}\\
             &    &    &    &       &J_{\ell} &-B
    \end{pmatrix},
\end{equation}
where $B:= (1-\lambda)/\lambda$.

As a first thing, notice that for the choice of $J_j$'s in Eq.~\eqref{eq:Jj}, the model is translation invariant in bulk (step 2), while only the sites $1$, $L-1$ and $L$ experience different couplings. This fact suggests that plane waves are eigenfunctions of the system, provided the boundary conditions are fixed.

Second, notice that for $B=0$ (i.e.,\ at the end of the anneal), there is an eigenstate localized at the left end $j=0$, namely $(1\ 0\ 0\dots)^T$. Similarly, when $B=\infty$ (i.e., \ at the beginning of the anneal), one eigenstate is localized at the right end, namely $(0\dots 0\ 1)^T$. In the following, these localized states are determined with a scattering approach.

\subsection{Plane waves in the bulk}

Let us start by constructing the plane wave eigenvectors of Eq.~\eqref{app:eq:H_hopping} in the bulk. The eigenvalue equation $\mathcal{H} \pket{\psi} = \lambda \epsilon \pket{\psi}$ reads, in components,
\begin{equation}
    \begin{cases}
        -B \psi_{2j} + \psi_{2j-2} = \epsilon \psi_{2j-1} \\
        \psi_{2j+1} - B \psi_{2j-1} = \epsilon \psi_{2j},
    \end{cases}
\end{equation}
where the couplings $J_j$ of Eq.~\eqref{eq:Jj} have been specified. Using the Fourier ansatz
\begin{equation}
    \label{eq:Fourier_ansatz}
     \begin{pmatrix}
         \psi_{2j-1} \\ \psi_{2j}
     \end{pmatrix}
     \sim e^{ikj}
     \begin{pmatrix}
         a_k    \\ b_k
     \end{pmatrix},
\end{equation}
the bulk eigenvalue equation acquires the form
\begin{equation}
    \begin{cases}
        -B b_k + e^{-ik} b_k = \epsilon_k a_k \\
        e^{ik} a_k - B a_k = \epsilon_k b_k.
    \end{cases}
\end{equation}
From the linear system above, one determines
\begin{equation}
    \label{eq:epsilon_k}
    (e^{-ik}-B)(e^{ik}-B) = \epsilon^2_k
\end{equation}
and
\begin{equation}
    \label{eq:a/b}
    \frac{a_k}{b_k} = \frac{\epsilon_k}{e^{ik}- B}.
\end{equation}
The first equation fixes the dispersion relation in the bulk, and the second the relative scale of $a_k$ to $b_k$. They need to be supplemented by the boundary conditions, which fix the admissible values of $k$, and the normalization condition, which fixes the overall scale of the eigenvector. In order to simplify things, it is instrumental to work already in the thermodynamic limit $L \to \infty$, since this way, the left and right boundary conditions are completely decoupled. The two will be considered separately in the following subsections.

\subsection{Left-localized state}

Here, consider a semi-infinite chain starting from $j=0$ and ending at $j=+\infty$. The only boundary condition one needs to take into account is provided by the eigenvalue equation at the first site:
\begin{equation}
    J' \psi_1 -B \psi_2 = \epsilon \psi_1,
\end{equation}
that can be rewritten as
\begin{equation}
    \label{eq:left_boundary}
    J' a_k - B b_k = \epsilon_k a_k \qquad \implies \qquad
    \frac{a_k}{b_k} = \frac{B}{J'-\epsilon_k}.
\end{equation}
When supplemented with Eqs.~\eqref{eq:epsilon_k}, \eqref{eq:a/b} and the normalization condition, one obtains four equations for the four unknowns $a_k$,$b_k$,$k$,$\epsilon_k$. Here, we consider only the localized edge state, for which $k \equiv i \kappa$, $\kappa>0$. Eliminating $a_k/b_k$ and $\epsilon_k$ from Eq.~\eqref{eq:left_boundary} via Eqs.~\eqref{eq:epsilon_k} and \eqref{eq:a/b}, one finds
\begin{equation}
    \frac{1-e^\kappa B}{1-e^{-\kappa}B} = (J')^2,
\end{equation}
which has a real positive solution
\begin{equation}
    \label{eq:kappa_L}
    \kappa = \ln \frac{1-(J')^2+\sqrt{4 B^2 (J')^2+[(J')^2-1]^2}}{2 B}.
\end{equation}

\subsection{Right-localized state}

Here, consider the half-infinite chain that starts at $j=-\infty$ and ends at $j=L$. Besides Eqs.~\eqref{eq:epsilon_k} and \eqref{eq:a/b} that describe the wave propagation in the bulk, the two boundary conditions at the right end read
\begin{equation}
    \label{eq:right_boundary}
    -B\psi_{L-2} + J_\ell \psi_L = \epsilon \psi_{L-1}, \qquad
    J_\ell \psi_{L-1} -B \psi_L = \epsilon \psi_L.
\end{equation}
Being the last site odd, the Fourier ansatz Eq.~\eqref{eq:Fourier_ansatz} needs to be supplemented by
\begin{equation}
    \psi_L \sim e^{ikL} c_k.
\end{equation}
The boundary conditions Eq.~\eqref{eq:right_boundary} read, in terms of the Fourier ansatz,
\begin{equation}
    -B a_k + J e^{ik} c_k = \epsilon_k b_k, \qquad
    J b_k e^{-ik} -B c_k = \epsilon_k c_k.
\end{equation}
From here, one can eliminate $c_k$, obtaining
\begin{equation}
    \label{eq:right_boundary2}
    \epsilon_k b_k + B a_k = \frac{J^2 b_k}{\epsilon_k+B}.
\end{equation}
Now, Eqs.~\eqref{eq:epsilon_k}, \eqref{eq:a/b}, \eqref{eq:right_boundary2} and the normalization condition are again four equations for the four unknowns $a_k$,$b_k$,$k$,$\epsilon_k$. The right-localized state will have an imaginary momentum $k \equiv - i \kappa$, $\kappa>0$, that can be determined by combining the cited equations. Since this would lead to cumbersome equations, it is convenient to utilize a workaround to determine the avoided crossing directly, as in Ref.~\cite{Roberts2020Noise}.

\subsection{Avoided crossing}
\label{app:sec:avoided_crossing}

The avoided crossing takes place at the value of $B$ such that $\epsilon_k$ is the same for the left ($\mathrm{L}$) and right ($\mathrm{R}$) localized states. In principle, one should impose $(\epsilon_k)_\mathrm{L} \equiv (\epsilon_k)_\mathrm{R}$ by using the expression \eqref{eq:epsilon_k} with the values of $\kappa_\mathrm{L}$ and $\kappa_\mathrm{R}$ determined above. However, it is convenient to notice that, because of Eq.~\eqref{eq:epsilon_k}, it must hold $\kappa_\mathrm{L} = \kappa_\mathrm{R}$, and from Eq.~\eqref{eq:a/b} follows $(a_k/b_k)_\mathrm{L} = (b_k/a_k)_\mathrm{R}$ (being $k_\mathrm{R} = -k_\mathrm{L}$ for the choice of $i$'s). Thus, one determines $(a_k/b_k)_\mathrm{L}$ from Eq.~\eqref{eq:left_boundary}, and equates it to $(b_k/a_k)_\mathrm{R}$ from Eq.~\eqref{eq:right_boundary2}, finding
\begin{equation}
    \label{eq:combined_boundaries}
    \epsilon + B = \frac{J^2}{J'}.
\end{equation}
Plugging above $\epsilon$ from Eq.~\eqref{eq:epsilon_k}, with $\kappa$ taken from Eq.~\eqref{eq:kappa_L}, one determines the value of $B$ at which the crossing takes place:
\begin{equation}
    B_\mathrm{c} = \frac{(1-J^2)[J^2-(J')^2]}{J'[1-2J^2+(J')^2]}.
\end{equation}
The corresponding critical parameter value is given by $\lambda_c=1/(1+B_c)$ changing in between $0.5$ and $0.6$ for the used values of $J'$ and $J$ in the main text.
From Eq.~\eqref{eq:combined_boundaries} follows
\begin{equation}
    \epsilon_\mathrm{c} = \frac{(J')^2-J^4}{J'[1-2J^2+(J')^2]},
\end{equation}
and in turn from Eq.~\eqref{eq:epsilon_k}
\begin{equation}
    \kappa_\mathrm{c} = \ln \frac{J'(1-J^2)}{J^2-(J')^2}.
\end{equation}
From the equation above, the condition $\kappa>0$ implies for the couplings
\begin{equation}
    \label{eq:ineq_Js}
    J^2 < J',
\end{equation}
which explains the requirement stated in Eq.~\eqref{eq:J_range}.

The last step consists of determining the magnitude of the gap at the avoided level crossing. Since $\pket{\psi_\mathrm{L}}$ and $\pket{\psi_\mathrm{R}}$ determined above are \emph{not} eigenfunctions of the chain of finite length, one can argue that
\begin{equation}
    \label{app:eq:Delta_min}
    \Delta_{\min} \simeq \pbra{\psi_\mathrm{L}} \mathcal{H} \pket{\psi_\mathrm{R}}
    \sim \sum_{j=1}^L e^{-\kappa_c j/2} e^{-\kappa_c (L-j)/2}
    \sim e^{-\kappa_c L/ 2},
\end{equation}
which corresponds to Eq.~\eqref{eq:Delta_min} in the main text. Indeed, the hybridization of $\pket{\psi_\mathrm{L}}$ and $\pket{\psi_\mathrm{R}}$ is approximately given by the off-diagonal matrix element $\pbra{\psi_\mathrm{L}} \mathcal{H} \pket{\psi_\mathrm{R}}$ in the reduced 2$\times$2 subspace spanned by $\pket{\psi_\mathrm{L}}$ and $\pket{\psi_\mathrm{R}}$ themselves. All the factors dropped in Eq.~\eqref{app:eq:Delta_min} are subleading with respect to\ the exponential term.

\section{Free-fermion representation of the counterdiabatic Hamiltonian}
\label{app:sec:H1_free_fermions}

\subsection{Chiral reflection parity of the CD Hamiltonian}
\label{app:sec:H1_parity}

As a first thing, let us investigate how the reflection parity symmetry acts on the CD Hamiltonian, Eq.~\eqref{eq:H1_integral}. Using the fact that $\widetilde{\Pi}_\mathrm{R} H \widetilde{\Pi}_\mathrm{R} = - H$, as proven in Sec.~\ref{app:sec:reflection_parity}, it follows
\begin{align}
    \widetilde{\Pi}_\mathrm{R} H_1 \widetilde{\Pi}_\mathrm{R} &= -\frac{1}{2} \int_{-\infty}^{+\infty} dx \, f(x) \, \widetilde{\Pi}_\mathrm{R} e^{i x H} \partial_\lambda H e^{-i x H} \widetilde{\Pi}_\mathrm{R} \\
    &= \frac{1}{2} \int_{-\infty}^{+\infty} dx \, f(x) \,  e^{-i x H} \partial_\lambda H e^{i x H} \\
    &= - \frac{1}{2} \int_{-\infty}^{+\infty} dy \, f(y) \,  e^{i y H} \partial_\lambda H e^{-i y H} = + H_1,
\end{align}
having set $y := -x$. Thus, $H_1$ is even under the chiral reflection parity, Eq.~\eqref{eq:chiral_reflection_parity}, while $H$ is odd. The same conclusion can be reached by noticing that in Eq.~\eqref{eq:H1_series}, there is an even number of $H$'s at every order of the expansion.

At the same time, $H_1$ is even under the standard reflection parity, Eq.~\eqref{eq:reflection_parity}, as can be checked directly from the spin representation.

\subsection{First-order ansatz}
\label{app:sec:H1_1st_order}

Let us move to the Hamiltonian~\eqref{eq:H1_1st_order}. By applying the set of one-qubit gates defined in Eq.~\eqref{eq:U_rotation_xz}, one can rewrite 
\begin{align}
    H_1^{(1)}[\lambda] &= \sum_{j=1}^L \alpha_j \left(\tilde{\sigma}_j^y \tilde{\sigma}_{j+1}^x + \tilde{\sigma}_j^x \tilde{\sigma}_{j+1}^y \right) \\
    &= 2i \sum_{j=1}^L \alpha_j \left( \tilde{\sigma}_j^- \tilde{\sigma}_{j+1}^- - \tilde{\sigma}_j^+ \tilde{\sigma}_{j+1}^+ \right).
\end{align}
This form of the CD Hamiltonian is expressed in terms of Dirac fermions, Eq.~\eqref{eq:Jordan-Wigner}, as
\begin{align}
	H_1^{(1)}[\lambda] &= -2i \sum_{j=1}^L \tilde{\alpha}_j \left( e^{-i\pi \Sigma_j} c_j^\dagger e^{-i\pi \Sigma_{j+1}} c_{j+1}^\dagger \right) + \mathrm{h.c.} \\
    &= -2i \sum_{j=1}^L \tilde{\alpha}_j \left( c_j^\dagger e^{-i\pi c^\dagger_j c^\phdagger_{j}} c_{j+1}^\dagger \right) + \mathrm{h.c.} \\
    \label{eq:H1_1st_order_Dirac}
    &= -2i \sum_{j=1}^L \tilde{\alpha}_j \left[ c^\dagger_j c^\dagger_{j+1} + c_j c_{j+1} \right] ,
\end{align}
where there is a tilde on the $\alpha_j$'s because of the fermion parity symmetry, see above Eq.~\eqref{eq:J_tilde}.

Next, one has to rewrite the Hamiltonian above in terms of Majorana fermions, Eq.~\eqref{app:eq:Majorana_fermions_inv}:
\begin{align}
	H_1^{(1)}[\lambda] &= -2i \sum_{j=1}^L \tilde{\alpha}_j \left[ \frac{1}{2} (\gamma_{2j-1} +i\gamma_{2j}) (\gamma_{2j+1} +i \gamma_{2j+2}) + \frac{1}{2} (\gamma_{2j-1} -i \gamma_{2j}) (\gamma_{2j+1} -i \gamma_{2j+2}) \right] \\
    &= 2i \sum_{j=1}^L \tilde{\alpha}_j \left[ \gamma_{2j} \gamma_{2j+2} - \gamma_{2j-1} \gamma_{2j+1} \right].
\end{align}
It is convenient to check the reflection parity of the expression above for $H_1$:
\begin{align}
    \widetilde{\Pi}_\mathrm{R} H_1^{(1)}[\lambda] \widetilde{\Pi}_\mathrm{R} &= 2i \sum_{j=1}^L \tilde{\alpha}_j \left[ \gamma_{2L-2j+1} \gamma_{2L-2j-1} - \gamma_{2L-2j+2} \gamma_{2L-2j} \right] \\
    &= 2i \sum_{j=1}^L \tilde{\alpha}_{L-j} \left[ \gamma_{2j+1} \gamma_{2j-1} - \gamma_{2j+2} \gamma_{2j} \right] \\
    &= 2i \sum_{j=1}^L \tilde{\alpha}_{L-j} \left[ \gamma_{2j} \gamma_{2j+2} - \gamma_{2j-1} \gamma_{2j+1} \right],
\end{align}
where in the first line Eq.~\eqref{eq:chiral_reflection_parity} was used, and in the second it was changed $j \to L-j$. In order to have $\widetilde{\Pi}_\mathrm{R} H_1 \widetilde{\Pi}_\mathrm{R} = H_1$, it must hold $\alpha_j \equiv \alpha_{L-j}$, which indeed follows from the variational equations~\eqref{eq:variational_eq_1st}. 

At this point, one can pass to the symmetrized operators $\Gamma^\pm$. Using
\begin{subequations}
\begin{alignat}{3}
    -\tilde{\alpha}_j \gamma_{2j-1} \gamma_{2j+1} + \tilde{\alpha}_{L-j} \gamma_{2L-2j} \gamma_{2L-2j+2} &= -\tilde{\alpha}_j \left( \Gamma_{2j-1}^- \Gamma_{2j+1}^+ + \Gamma_{2j-1}^+ \Gamma_{2j+1}^- \right), \qquad
    && j=1,2,\dots,\ell, \\
    \tilde{\alpha}_j \gamma_{2j} \gamma_{2j+2} - \tilde{\alpha}_{L-j} \gamma_{2L-2j-1} \gamma_{2L-2j+1} &= \tilde{\alpha}_j \left( \Gamma_{2j}^+ \Gamma_{2j+2}^- + \Gamma_{2j}^- \Gamma_{2j+2}^+ \right), 
    && j=1,2,\dots,\ell-1,\\
    \tilde{\alpha}_\ell \gamma_{2\ell} \gamma_{2\ell+2} - \tilde{\alpha}_{\ell+1}\gamma_{2\ell+1} \gamma_{2\ell+3} &= \tilde{\alpha}_\ell \left( \Gamma_{2\ell}^- \Gamma_{2\ell+1}^+ + \Gamma_{2\ell}^+ \Gamma_{2\ell+1}^- \right), &&\\
    \tilde{\alpha}_L \left( -\gamma_{2L-1} \gamma_1 + \gamma_{2L} \gamma_2 \right) &= \tilde{\alpha}_L \left( \Gamma_1^+ \Gamma_2^- + \Gamma_1^- \Gamma_2^+ \right), &&
\end{alignat}
\end{subequations}
it follows
\begin{multline}
    \label{app:eq:H1_1_Gamma}
    H_1^{(1)}[\lambda] = 2i \sum_{j=1}^{\ell} \tilde{\alpha}_j \left( \Gamma_{2j+1}^+ \Gamma_{2j-1}^- - \Gamma_{2j-1}^+ \Gamma_{2j+1}^-\right) 
    +2i \sum_{j=1}^{\ell-1} \tilde{\alpha}_j \left( \Gamma_{2j}^+ \Gamma_{2j+2}^- - \Gamma_{2j+2}^+ \Gamma_{2j}^- \right) \\
    +2 i\tilde{\alpha}_\ell \left( \Gamma_{2\ell}^+ \Gamma_{2\ell+1}^- - \Gamma_{2\ell+1}^+ \Gamma_{2\ell}^- \right)
    +2 i\tilde{\alpha}_L \left( \Gamma_1^+ \Gamma_2^- - \Gamma_2^+ \Gamma_1^- \right).
\end{multline}
The matrix representation reads (for $\ell = 3$, thus $L = 7$)
\begin{equation}
	H_1^{(1)}[\lambda] = 2\, \vec{\Gamma}^+
    \begin{pmatrix}
        0         &-i\alpha_7 &-i\alpha_1 &           &           &          &  \\
        i\alpha_7 &0          &0          &i\alpha_1  &           &          &  \\
        i\alpha_1 &0          &0          &0          &-i\alpha_2 &          &  \\
                  &-i\alpha_1 &0          &0          &0          &i\alpha_2 &  \\
                  &           &i\alpha_2  &0          &0          &0         &-i\alpha_3  \\
                  &           &           &-i\alpha_2 &0          &0         &i\alpha_3   \\
                  &           &           &           &i\alpha_3  &-i\alpha_3& 0\\
    \end{pmatrix}
    \vec{\Gamma}^- .
\end{equation}
The variational coefficients $\alpha_j$ are given by the solution of the system of equations obtained by the action minimization of Eq.~\eqref{eq:action},
    \begin{eqnarray}
    \label{eq:variational_eq_1st}
    \left[8 (1-\lambda)^2 + \lambda^2 \left(J_{j-1}^2 + 2 J_j^2 + J_{j+1}^2\right)\right] \alpha_j
    + 2 \lambda^2 J_j \left(J_{j-1} \alpha_{j-1} + J_{j+1} \alpha_{j+1} \right) = -J_j,
\end{eqnarray}
with periodic boundary conditions implied, which can be solved via linear algebra techniques with a cost polynomial in $L$.

\subsection{Second-order ansatz}
\label{app:sec:H1_2nd_order}

When considered together, the first- and second-order approximate CD Hamiltonians $H_1^{(1)} + H_1^{(2)}$, one can check that the variational equations, obtained from the minimization of the action Eq.~\eqref{eq:action}, are
\begin{gather}
    \left[8 (1-\lambda)^2 + \lambda^2 \left(J_{j-1}^2 + 2 J_j^2 + J_{j+1}^2\right)\right] \alpha_j + 2 \lambda^2 J_j \left(J_{j-1} \alpha_{j-1} + J_{j+1} \alpha_{j+1} \right) + 4\lambda(\lambda-1) (\beta_{j-1}J_{j-1} + \beta_j J_{j+1})= -J_j, \nonumber \\
    \label{eq:variational_eq_2nd}
    \left[8 (1-\lambda)^2 + \lambda^2 \left(J_{j-1}^2 + J_j^2 + J_{j+1}^2 + J_{j+2}^2\right)\right] \beta_j + 2 \lambda^2 \left(J_{j-1} J_{j+1} \beta_{j-1} + J_j J_{j+2} \beta_{j+1} \right) + 4\lambda(\lambda-1) (\alpha_{j+1} J_j + \alpha_j J_{j+1})= 0,
\end{gather}
Thanks to the relation $J_j = J_{L-j}$, these equations are symmetric under the exchange
\begin{alignat}{4}
    \alpha_j & \longleftrightarrow \, &&\alpha_{L-j}, \qquad &&j=1,2,\dots,L \nonumber \\
    \beta_j &\longleftrightarrow &&\beta_{L-j-1}, \qquad &&j=1,2,\dots,L-2 \\
    \beta_{L-1} &\longleftrightarrow &&\beta_{L}.  \qquad &&  \nonumber
\end{alignat}

One can check that the second-order ansatz for the CD Hamiltonian, Eq.~\eqref{eq:H1_2nd_order}, is expressed in terms of Dirac fermions as
\begin{equation}
	H_1^{(2)}[\lambda] = -2i \sum_{j=1}^L \tilde{\beta}_j \left[ c^\dagger_j c^\dagger_{j+2} + c_j c_{j+2} \right].
\end{equation}
Indeed, the same computation of App.~\ref{app:sec:H1_1st_order} applies, and the intermediate operator $\sigma_{j+1}^x$ just cancels the Jordan-Wigner string. The tilde on the $\beta_j$'s is again to remind that the sign in front of any operator that connects the last sites of the chain to the firsts needs to be flipped. Contrary to the cases above, where only the last coupling $J_L$ or $\alpha_L$ needed to be flipped, here also $\beta_{L-1}$ needs to.
 
Proceeding in the same fashion of App.~\ref{app:sec:H1_1st_order}, one finds
\begin{equation}
    H_1^{(2)}[\lambda] = 2i \sum_{j=1}^L \tilde{\beta}_j \left[ \gamma_{2j} \gamma_{2j+4} - \gamma_{2j-1} \gamma_{2j+3} \right]
\end{equation}
and
\begin{multline}
    \label{app:eq:H1_2_Gamma}
    H_1^{(2)}[\lambda] = 2i \sum_{j=1}^{\ell-1} \tilde{\beta}_j \left( \Gamma_{2j+3}^+ \Gamma_{2j-1}^- - \Gamma_{2j-1}^+ \Gamma_{2j+3}^-\right) 
    +2i \sum_{j=1}^{\ell-2} \tilde{\beta}_j \left( \Gamma_{2j}^+ \Gamma_{2j+4}^- - \Gamma_{2j+4}^+ \Gamma_{2j}^- \right) \\
    +2 i\tilde{\beta}_{\ell-1} \left( \Gamma_{2\ell-2}^+ \Gamma_{2\ell+1}^- - \Gamma_{2\ell+1}^+ \Gamma_{2\ell-2}^- \right) 
    +2 i\tilde{\beta}_\ell \left( \Gamma_{2\ell}^+ \Gamma_{2\ell-1}^- - \Gamma_{2\ell-1}^+ \Gamma_{2\ell}^- \right) \\
    +2 i\tilde{\beta}_{L-1} \left( \Gamma_1^+ \Gamma_4^- - \Gamma_4^+ \Gamma_1^- \right) 
    +2 i\tilde{\beta}_L \left( \Gamma_3^+ \Gamma_2^- - \Gamma_2^+ \Gamma_3^- \right).
\end{multline}
In matrix form (for $\ell = 3$, thus $L = 7$):
\begin{equation}
	H_1^{(2)}[\lambda] = 2\, \vec{\Gamma}^+
    \begin{pmatrix}
        0        &0         &0        &-i\beta_6 &-i\beta_1 &0         &0 \\
        0        &0         &i\beta_7 &0         &0         &i\beta_1  &0 \\
        0        &-i\beta_7 &0        &0         &0         &0         &-i\beta_2  \\
        i\beta_6 &0         &0        &0         &0         &0         &i\beta_2  \\
        i\beta_1 &0         &0        &0         &0         &-i\beta_3 &0 \\
        0        &-i\beta_1 &0        &0         &i\beta_3  &0         &0 \\
        0        &0         &i\beta_2 &-i\beta_2 &0         &0         &0 \\
    \end{pmatrix}
    \vec{\Gamma}^- .
\end{equation}

\subsection{Quantum Brachistochrone Counterdiabatic Driving (QBCD) in the minimal spin glass model}
\label{app:LRwavefunctions}

We report the details of how to compute $ \mathcal{H}^{\mathrm{QBCD}}$, Eq.~\eqref{eq:H_LR}. The left- and right-localized edge states of the effective hopping model, $\pket{\psi_\mathrm{L}}$ and $\pket{\psi_\mathrm{R}}$ respectively, were derived in App.~\ref{app:sec:diagonalization} for an infinitely long chain:
\begin{equation}
    \pket{\psi_\mathrm{L}} =
    \begin{pmatrix}
        a_\mathrm{L} & b_\mathrm{L} & e^{-\kappa_\mathrm{L}}a_\mathrm{L} & e^{-\kappa_\mathrm{L}} b_\mathrm{L} & e^{-2\kappa_\mathrm{L}} a_\mathrm{L} & \cdots    
    \end{pmatrix}^T, \qquad
    \pket{\psi_\mathrm{R}} =
    \begin{pmatrix}
        \cdots &e^{-2\kappa_\mathrm{R}} a_\mathrm{R} & e^{-\kappa_\mathrm{R}}b_\mathrm{R} & e^{-\kappa_\mathrm{R}}a_\mathrm{R} & b_\mathrm{R} &  a_\mathrm{R} & c_\mathrm{R}    
    \end{pmatrix}^T.
\end{equation}
Here, however, they are needed for a chain of finite length $L$. First of all, we set $a_\mathrm{L} = a_\mathrm{R}$, $b_\mathrm{L}=b_\mathrm{R}$ and $\kappa_\mathrm{L}=\kappa_\mathrm{R}$, which from App.~\ref{app:sec:avoided_crossing} is known to be valid at the avoided crossing. As explained in the main text, we want to obtain a CD term from the knowledge of the avoided crossing alone. Let us also parametrize the ratio of $a$ and $b$ by $e^{-\mu} \equiv a/b$. Plugging $\pket{\psi_\mathrm{L}}$ into the eigenvalue equation of $\mathcal{H}$, Eq.~\eqref{app:eq:H_hopping}, one can find the relation between $\mu$, $\kappa$ and the eigenvalue $\epsilon$:
\begin{equation}
    e^{\kappa}=J' \,e^{-\mu}, \qquad
    \epsilon = J' - Be^{\mu}.
\end{equation}
The right-end eigenvalue equation instead leads to
\begin{equation}
    B + \epsilon=\frac{J^2}{J'}.
\end{equation}
From here, one can derive the results
\begin{equation}
    B = \frac{(1-J^2)[J^2-(J')^2]}{J'[(J')^2+1-2J^2]}, \qquad
    \epsilon =  \frac{(J')^2-J^4}{J'[(J')^2+1-2J^2]}, \qquad
    \mu = \ln \frac{J^2-\left(J'\right)^2}{(1-J^2)}, \qquad  
    k=\ln \frac{J'(1-J^2)}{[J^2-(J')^2]},
\end{equation}
which are consistent with what found in App.~\ref{app:sec:avoided_crossing}. Additionally, the last and first elements of the left and right localized states, respectively, are found to be 
\begin{equation}
    c = \frac{J B e^{-\ell k}}{J'} b\,.
\end{equation}
At this point, the two localized edge modes read
\begin{equation}
    \pket{\psi_\mathrm{L}} \simeq \frac{1}{\sqrt{Z_\mathrm{L}}}
    \begin{pmatrix}
        &e^{-\mu}\\
        &1\\
        &e^{-\kappa-\mu}\\
        &e^{-\kappa}\\
        &e^{-2\kappa-\mu}\\
        &e^{-2\kappa}\\
        &\vdots\\   
        &e^{-(\ell-1) \kappa-\mu}\\
        &e^{-(\ell-1) \kappa}\\
        &J B e^{-\ell \kappa}/J'
    \end{pmatrix}, \qquad
    \pket{\psi_\mathrm{R}} \simeq \frac{1}{\sqrt{Z_\mathrm{R}}}
    \begin{pmatrix}
        &e^{-(\ell-1) \kappa}\\
        &e^{-(\ell-1) \kappa-\mu}\\
        &\vdots\\        
        &e^{-2\kappa}\\
        &e^{-2\kappa-\mu}\\
        &e^{-\kappa}\\
        &e^{-\kappa-\mu}\\
        &1\\
        &e^{-\mu}\\
        &J B e^{-\mu}/J'
    \end{pmatrix},
\end{equation}
where the normalization factors $Z_\mathrm{L,R}$ need not be specified for the moment. 

The derivative with respect to $\lambda$ of the effective Hamiltonian is
\begin{equation}
	\partial_\lambda \mathcal{H} =
    \begin{pmatrix}
        J' &1  &    &    &       &         &  \\
        1  &0   &1 &    &       &         &  \\
             &1 &0   &1  &       &         &  \\
             &    &1  &0   &1    &         &  \\
             &    &    &1 &\ddots &\ddots   &  \\
             &    &    &    &\ddots &0        &J\\
             &    &    &    &       &J &1
    \end{pmatrix}.
\end{equation}    
Correspondingly, the approximate energy gap takes the form
\begin{multline}
    \label{eq:Delta_min_LR}
    \Delta_{\min} \approx \pbra{\psi_\mathrm{R}} \mathcal{H} \pket{\psi_\mathrm{L}}
    = J'\lambda\,e^{-(\ell-1)\kappa-\mu} + \frac{J^2(\lambda-1)^3}{\lambda^2(J')^2}\,e^{-\ell \kappa-\mu} + (\lambda-1)\ell  e^{-(\ell-1)\kappa}(1+e^{-2\mu})\\
    +(\ell-1)\lambda\left(e^{-\ell \kappa-2\mu}+e^{-(\ell-2)\kappa}\right)+J\lambda \left(\frac{J}{J'}\frac{\lambda-1}{\lambda}\right)^2e^{-\ell \kappa-\mu}\,(1+e^\kappa),
\end{multline}
while the matrix element of the parametric derivative of the Hamiltonian is
\begin{multline}
    \label{eq:dH_matrixelement}
    \pbra{\psi_\mathrm{R}} \partial_\lambda\mathcal{H}[\lambda_c] \pket{\psi_\mathrm{L}} = J'\,e^{-(\ell-1)\kappa-\mu}+\frac{J^2(\lambda-1)^2}{\lambda^2(J')^2}\,e^{-\ell \kappa-\mu} +\ell e^{-(\ell-1)\kappa}(1+e^{-2\mu}) \\
    +(\ell-1)\left(e^{-\ell \kappa-2\mu}+e^{-(\ell-2)\kappa}\right)+J \left(\frac{J}{J'}\frac{\lambda-1}{\lambda}\right)^2e^{-\ell \kappa-\mu}\,(1+e^\kappa).
\end{multline}
With all this information, the QBCD term $ \mathcal{H}^{\mathrm{QBCD}}$, Eq.~\eqref{eq:H_LR}, can be constructed. Notice that the normalization factors $Z_\mathrm{L,R}$ cancel out and need not be computed explicitly.

\section{Energy cost of the counterdiabatic Hamiltonians}
\label{app:E_cost}

In this section, we compute the trace norms of the CD Hamiltonians. Starting with the QBCD ansatz, Eq.~\eqref{eq:H_LR}, the trace norm of the square results in two projectors, leading to 
\begin{equation}
\begin{split}
    &\left\|\mathcal{H}^{\mathrm{QBCD}}\right\|^2=\frac{|\pbra{\psi_\mathrm{R}} \partial_\lambda \mathcal{H} \pket{\psi_\mathrm{L}}|^2}{\Delta_{\min}^2}
    \mathrm{Tr}\big[ \pket{\psi_\mathrm{R}} \pbra{\psi_\mathrm{L}}
    + \pket{\psi_\mathrm{L}}\pbra{\psi_\mathrm{R}} \big] = O(1).
    \end{split}
\end{equation}
Indeed, the ratio of the matrix element and the gap is size-independent to leading order, as follows from Eqs.~\eqref{eq:Delta_min_LR} and \eqref{eq:dH_matrixelement}. Note that the non-locality and the exponentially small gap do not affect the overall energy cost.

Concerning the approximate first- and second-order variational CD terms, since the coefficients $\alpha_i,\beta_i\sim\mathrm O(J)$ individually, one gets the typical scale for the norm
\begin{align}
    \left\| H^{(1)}_1\right\|^2 &=\sum_{n,m=1}^L\,\Big| H^{(1)}_1\Big|^2_{n,m} = 4\sum_{n=1}^\ell |\alpha|^2_n + 2 |\alpha_L|^2\sim L,\\
    \left\| H^{(2)}_1\right\|^2 &=\sum_{n,m=1}^L\,\Big| H^{(2)}_1 \Big|^2_{n,m} = 4\sum_{n=1}^\ell |\beta|^2_n + 2 |\beta_L|^2\sim L,
\end{align}
that is diverging in the thermodynamic limit.

\section{Details of numerical simulations}
\label{app:sec:numerics}

\subsection{Particle-number-nonconserving free fermions}
\label{app:sec:numerics_non_conserving}

Suppose one wants to simulate numerically the quantum driving protocol by working at the level of Dirac fermions rather than Majoranas. Both $H[\lambda]$ and $H[\lambda]+H_1[\lambda]$ belong to the more general family of fermionic quadratic Hamiltonians
\begin{equation}
    H = \sum_{i,j=1}^L c_i \mathcal{H}_{ij}^{--} c_j + \sum_{i,j=1}^L c_i \mathcal{H}_{ij}^{-+} c_j^\dagger + \sum_{i,j=1}^L c_i^\dagger \mathcal{H}_{ij}^{+-} c_j + \sum_{i,j=1}^L c_i^\dagger \mathcal{H}_{ij}^{++} c_j^\dagger .
\end{equation}
For now, it is assumed that $H$ (and thus the matrices $\mathcal{H}$) are time-independent. The time-dependent case is recovered easily, as shown later on. 

Introducing the vector
\begin{equation}
    \Psi = 
    \begin{pmatrix}
        c_1  &c_2  &\dots  &c_L  &c^\dagger_1  &c^\dagger_N  &\dots  &c^\dagger_L 
    \end{pmatrix}^T,
\end{equation}
one can rewrite succinctly
\begin{equation}
    H = \Psi^\dagger \mathcal{H} \Psi,
\end{equation}
where the matrix $\mathcal{H}$ reads
\begin{equation}
	\mathcal{H} = 
	\left(
    \begin{array}{c|c}
        \mathcal{H}^{+-}   &\mathcal{H}^{++} \\ \hline
        \mathcal{H}^{--}   &\mathcal{H}^{-+}
    \end{array} \right).
\end{equation}
By Hermiticity of $H$, one can choose the four submatrices above to respect
\begin{equation}
    \label{eq:symmetry_H_general_FF}
    \mathcal{H}^{+-} = - \left(\mathcal{H}^{-+}\right)^*, \qquad
    \mathcal{H}^{++} = -\left(\mathcal{H}^{--}\right)^*,
\end{equation}
and
\begin{equation}
    \label{eq:symmetry_H_general_FF_2}
    \mathcal{H}^{+-} = \left(\mathcal{H}^{+-} \right)^\dagger, \qquad 
    \mathcal{H}^{++} = -\left(\mathcal{H}^{++} \right)^T.
\end{equation}

To diagonalize $H$, it is sufficient to diagonalize $\mathcal{H}$ via the introduction of Bogoliubov operators $\xi$. Define $\mathcal{U}$ to be the matrix that diagonalizes $\mathcal{H}$, i.e.\
\begin{equation}
    \mathcal{H} = \mathcal{U} \, \mathcal{H}_\mathrm{diag} \, \mathcal{U}^\dagger.
\end{equation}
Because of the symmetry in Eq.~\eqref{eq:symmetry_H_general_FF}, it holds~\cite{Lieb1967Two,Prosniak2019Size,mbeng2020quantum,Surace2022Fermionic}
\begin{equation}
    \mathcal{U}^\dagger := 
	\left(
    \begin{array}{c|c}
        u^*   &v \\ \hline
        v^*   &u
    \end{array} \right),
\end{equation}
with $u,v$ being two $L \times L$ matrices, and
\begin{equation}
    \mathcal{H}_\mathrm{diag} = \left(
    \begin{array}{c|c}
        \epsilon   &0 \\ \hline
        0          &-\epsilon
    \end{array} \right),
\end{equation}
where $\epsilon$ is the diagonal matrix of eigenvalues (that must appear in positive/negative couples, as entailed by the symmetries in Eqs.~\eqref{eq:symmetry_H_general_FF}--\eqref{eq:symmetry_H_general_FF_2}). Finally, introducing
\begin{equation}
    \Phi :=
    \begin{pmatrix}
        \xi \\ \xi^\dagger
    \end{pmatrix}
    := \mathcal{U}^\dagger \Psi \, ,
\end{equation}
one gets to the diagonal form
\begin{equation}
    H = \Phi^\dagger \mathcal{H}_\mathrm{diag} \, \Phi = \sum_{j=1}^L \epsilon_j \left(\xi^\dagger_j \xi^\phdagger_j - \xi^\phdagger_j \xi^\dagger_j \right) = \sum_{j=1}^L \epsilon_j \left(2 \xi^\dagger_j \xi^\phdagger_j - 1 \right).
\end{equation}
The Heisenberg time evolution of the operators $c,c^\dagger$ follows:
\begin{equation}
    \label{eq:Psi(t)_time_indep}
    \Psi(t) = e^{iHt} \Psi e^{-iHt} = e^{iHt} \mathcal{U} \Phi e^{-iHt}
    = \mathcal{U} \left(
    \begin{array}{c|c}
        e^{-2i\epsilon t}   &0 \\ \hline
        0          &e^{2i\epsilon t}
    \end{array} \right) \Phi
    = \mathcal{U} \left(
    \begin{array}{c|c}
        e^{-2i\epsilon t}   &0 \\ \hline
        0          &e^{2i\epsilon t}
    \end{array} \right) \mathcal{U}^\dagger \Psi
    = e^{-2i \mathcal{H} t} \Psi.
\end{equation}
Thus, in order to time evolve the operators $c,c^\dagger$, one just needs to compute the exponential of the $2L \times 2L$ matrix $\mathcal{H}$.

Let us move to the time-dependent case $H=H(t)$. Recall that states evolve with the time-ordered exponential: $\ket{\psi(t)} = \Te^{-i\int_0^t H(t')dt'} \ket{\psi(0)}$. Thus, passing to the Schr\"odinger picture, operators evolve as
\begin{equation}
    O(t) = \aTe^{i\int_0^t H(t')dt'} O(0) \Te^{-i\int_0^t H(t')dt'},
\end{equation}
where $\widetilde{\mathrm{T}}$ denotes the \emph{anti}-time ordering. Notice that, upon trotterization in time steps of $dt$, it is the \emph{last} time to act first on the operator:
\begin{equation}
    O(t) \simeq e^{i H(0)dt} \cdots e^{i H(t-dt)dt} O(0) e^{-iH(t-dt)dt} \cdots e^{-iH(0)dt}  .
\end{equation}
For our $\Psi$ operator, the first Trotter step to apply is thus
\begin{equation}
    e^{iH(t-dt)dt} \Psi_\alpha e^{-iH(t-dt)dt} = \big( e^{-2i\mathcal{H}(t-dt) dt} \big)_{\alpha\beta} \Psi_\beta,
\end{equation}
where the indices of the matrices are written down explicitly for convenience. The next time step reads
\begin{align}
    e^{iH(t-2dt)dt} e^{iH(t-dt)dt} \Psi_\alpha e^{-iH(t-dt)dt} e^{iH(t-2dt)dt}
    &= \big( e^{-2i\mathcal{H}(t-dt) dt} \big)_{\alpha\beta} \, e^{iH(t-2dt)dt} \Psi_\beta e^{-iH(t-2dt)dt} \\
    &= \big( e^{-2i\mathcal{H}(t-dt) dt} \big)_{\alpha\beta} \big( e^{-2i\mathcal{H}(t-2dt) dt} \big)_{\beta\gamma} \Psi_\gamma.
\end{align}
Iterating, one gets to
\begin{equation}
    \Psi(t) = \Te^{-2i \int_0^t \mathcal{H}(t') dt'} \Psi.
\end{equation}
This form could have been guessed directly from Eq.~\eqref{eq:Psi(t)_time_indep}, but all the steps were performed for clarity, given the subtlety of the anti-time ordering.

\subsection{Specification of the problem under consideration} 

The formulas above apply to the Dirac Hamiltonian, Eq.~\eqref{app:eq:H_Dirac}, upon setting
\begin{subequations}
\begin{gather}
    \mathcal{H}^{+-}[\lambda] =
    \begin{pmatrix}
        1-\lambda      &-\lambda J_1/2 &               &                   &\lambda J_L/2      \\
        -\lambda J_1/2 &1-\lambda      &-\lambda J_2/2 &                   &                   \\
                       &-\lambda J_2/2 &\ddots         &\ddots             &                   \\
                       &               &\ddots         &1-\lambda          &-\lambda J_{L-1}/2 \\
       \lambda J_L/2   &               &               &-\lambda J_{L-1}/2 &1-\lambda          \\
   \end{pmatrix}, \\
   \mathcal{H}^{++}[\lambda] =
    \begin{pmatrix}
        0             &-\lambda J_1/2 &               &                   &-\lambda J_L/2     \\
        \lambda J_1/2 &0              &-\lambda J_2/2 &                   &                   \\
                      &\lambda J_2/2  &\ddots         &\ddots             &                   \\
                      &               &\ddots         &0                  &-\lambda J_{L-1}/2 \\
       \lambda J_L/2  &               &               &\lambda J_{L-1}/2  &0                  \\
   \end{pmatrix}.
\end{gather}
\end{subequations}
One can check that the conjugation properties, Eq.~\eqref{eq:symmetry_H_general_FF_2}, are satisfied. The first-order CD Hamiltonian, Eq.~\eqref{eq:H1_1st_order_Dirac}, reads instead
\begin{equation}
    \mathcal{H}^{+-}[\lambda] = 0, \qquad
    \mathcal{H}^{++}[\lambda] =
    \begin{pmatrix}
        0         &-i\alpha_1 &           &              &-i\alpha_L  \\
        i\alpha_1 &0          &-i\alpha_2 &              &            \\
                  &i\alpha_1  &\ddots     &\ddots        &              \\
                  &           &\ddots     &0             &-i\alpha_{L-1} \\
        i\alpha_L &           &           &i\alpha_{L-1} &0             \\
   \end{pmatrix},
\end{equation}
while the second-order one reads
\begin{equation}
    \mathcal{H}^{+-}[\lambda] = 0, \qquad
    \mathcal{H}^{++}[\lambda] =
    \begin{pmatrix}
        0            &0        &-i\beta_1 &            &             &-i\beta_{L-1} &  \\
        0            &0        &0         &-i\beta_2   &             &              &-i\beta_L \\
        i\beta_1     &0        &0         &0           &\ddots       &              & \\
                     &i\beta_2 &0         &\ddots      &\ddots       &\ddots        & \\
                     &         &\ddots    &\ddots      &\ddots       &\ddots        &-i\beta_{L-2} \\
        i\beta_{L-1} &         &          &\ddots      &\ddots       &0             &0 \\
                     &i\beta_L &          &            &i\beta_{L-2} &0             &0\\
   \end{pmatrix}.
\end{equation}
We are using the same letter $\mathcal{H}$ for all the different Hamiltonians, bare or counterdiabatic, in order not to overload the notation.

The QBCD Hamiltonian, Eq.~\eqref{eq:H_LR}, instead, reads in the $\Gamma$ fermion representation
\begin{equation}
    H^{\mathrm{QBCD}}_1=\sum_{i,j=1}^{L} \Gamma^+_i \mathcal{H}^{\mathrm{QBCD}}_{ij} \Gamma^-_j.
\end{equation}
In order to find the $c$-fermion representation of $H^{\mathrm{QBCD}}_1$, it is convenient to transform back the $\Gamma^\pm$ operators to the Dirac representation:
\begin{equation}
    \begin{cases}
        \displaystyle \Gamma^+_k=\frac{i\,\gamma_k+\gamma_{2L-k+1}}{\sqrt 2}\,, &k=1,3,\dots,L\\
        \displaystyle \Gamma^+_k=\frac{\gamma_k+i\,\gamma_{2L-k+1}}{\sqrt 2}\,, &k=2,4,\dots,L-1
    \end{cases}
    \quad \implies \quad
    \begin{cases}
        \displaystyle \Gamma^+_{2j-1}=i\frac{c_j+c^\dagger_j+c_{L-j+1}-c^\dagger_{L-j+1}}{2}, &j=1,\dots \ell+1\\
        \displaystyle \Gamma^+_{2j}=i\,\frac{c_j-c^\dagger_j+c_{L-j+1}+c^\dagger_{L-j+1}}{2}, &j=1,\dots \ell.
    \end{cases}
\end{equation}
From here, the corresponding Dirac Hamiltonian reads
\begin{equation}
    \begin{aligned}
        H^{\mathrm{QBCD}}_1=\frac{1}{4}\sum_{i,j}
        &\left[\mathcal H^{\mathrm{QBCD}}_{2i-1,2j-1}\,\left(c_i+c^\dagger_i+c_{L-i+1}-c^\dagger_{L-i+1}\right)\left(c^\dagger_j+c_j+c^\dagger_{L-j+1}-c_{L-j+1}\right) \right.\\
        &+ \mathcal{H}^{\mathrm{QBCD}}_{2i,2j}\,\left(c_i-c^\dagger_i+c_{L-i+1} +c^\dagger_{L-i+1}\right)\left(c^\dagger_j-c_j+c^\dagger_{L-j+1}+c_{L-j+1}\right)\\
        &+ \mathcal{H}^{\mathrm{QBCD}}_{2i-1,2j}\left(c_i+c^\dagger_i+c_{L-i+1}-c^\dagger_{L-i+1}\right)\left(c^\dagger_j-c_j+c^\dagger_{L-j+1}+c_{L-j+1}\right)\\
        &\left. + \mathcal{H}^{\mathrm{QBCD}}_{2i,2j-1}\,\left(c_i-c^\dagger_i+c_{L-i+1} +c^\dagger_{L-i+1}\right)\left(c^\dagger_j+c_j+c^\dagger_{L-j+1}-c_{L-j+1}\right)\right],
    \end{aligned}
\end{equation}
where the summations are running until $\ell+1$ for odd indices and until $\ell$ for the even ones. Above, $\mathcal{H}_{ij}$ are the matrix elements of the single-particle QBCD Hamiltonian. Decomposing further the expression above in the particle number-conserving and non-conserving blocks, one finds four different matrices $\mathcal{H}^{++}$, $\mathcal{H}^{+-}$, $\mathcal{H}^{-+}$, and $\mathcal{H}^{--}$. For the sake of brevity, we use a non-symmetrized form of these submatrices, which does not obey Eqs.~\eqref{eq:symmetry_H_general_FF}--\eqref{eq:symmetry_H_general_FF_2}:
\begin{subequations}
\begin{align}
    \mathcal H^{++}_{ij} &=\frac{1}{4}\left(\mathcal H^{\mathrm{QBCD}}_{2i-1,2j-1}-\mathcal H^{\mathrm{QBCD}}_{2i,2j}+\mathcal H^{\mathrm{QBCD}}_{2i-1,2j}-\mathcal H^{\mathrm{QBCD}}_{2i,2j-1}\right),\\
    \mathcal H^{++}_{L-i+1,L-j+1} &=\frac{1}{4}\left(-\mathcal H^{\mathrm{QBCD}}_{2i-1,2j-1}+\mathcal H^{\mathrm{QBCD}}_{2i,2j}-\mathcal H^{\mathrm{QBCD}}_{2i-1,2j}+\mathcal H^{\mathrm{QBCD}}_{2i,2j-1}\right), \\
    \mathcal H^{++}_{L-i+1,j} &=\frac{1}{4}\left(-\mathcal H^{\mathrm{QBCD}}_{2i-1,2j-1}+\mathcal H^{\mathrm{QBCD}}_{2i,2j}-\mathcal H^{\mathrm{QBCD}}_{2i-1,2j}+\mathcal H^{\mathrm{QBCD}}_{2i,2j-1}\right), \\
    \mathcal H^{++}_{i,L-j+1} &=\frac{1}{4}\left(\mathcal H^{\mathrm{QBCD}}_{2i-1,2j-1}-\mathcal H^{\mathrm{QBCD}}_{2i,2j}+\mathcal H^{\mathrm{QBCD}}_{2i-1,2j}-\mathcal H^{soor\mathrm{QBCD}}_{2i,2j-1}\right),
\end{align}
\end{subequations}
for $\mathcal{H}^{++}$,
\begin{subequations}
\begin{align}
    \mathcal H^{+-}_{ij} &=\frac{1}{4}\left(\mathcal H^{\mathrm{QBCD}}_{2i-1,2j-1}+\mathcal H^{\mathrm{QBCD}}_{2i,2j}-\mathcal H^{\mathrm{QBCD}}_{2i-1,2j}-\mathcal H^{\mathrm{QBCD}}_{2i,2j-1}\right), \\
    \mathcal H^{+-}_{L-i+1,L-j+1} &=\frac{1}{4}\left(\mathcal H^{\mathrm{QBCD}}_{2i-1,2j-1}+\mathcal H^{\mathrm{QBCD}}_{2i,2j}-\mathcal H^{\mathrm{QBCD}}_{2i-1,2j}-\mathcal H^{\mathrm{QBCD}}_{2i,2j-1}\right), \\
    \mathcal H^{+-}_{L-i+1,j} &=\frac{1}{4}\left(-\mathcal H^{\mathrm{QBCD}}_{2i-1,2j-1}-\mathcal H^{\mathrm{QBCD}}_{2i,2j}+\mathcal H^{\mathrm{QBCD}}_{2i-1,2j}+\mathcal H^{\mathrm{QBCD}}_{2i,2j-1}\right), \\
    \mathcal H^{+-}_{i,L-j+1} &=\frac{1}{4}\left(-\mathcal H^{\mathrm{QBCD}}_{2i-1,2j-1}-\mathcal H^{\mathrm{QBCD}}_{2i,2j}+\mathcal H^{\mathrm{QBCD}}_{2i-1,2j}+\mathcal H^{\mathrm{QBCD}}_{2i,2j-1}\right),
\end{align}
\end{subequations}
for $\mathcal{H}^{+-}$,
\begin{subequations}
\begin{align}
    \mathcal H^{-+}_{ij} &=\frac{1}{4}\left(\mathcal H^{\mathrm{QBCD}}_{2i-1,2j-1}+\mathcal H^{\mathrm{QBCD}}_{2i,2j}+\mathcal H^{\mathrm{QBCD}}_{2i-1,2j}+\mathcal H^{\mathrm{QBCD}}_{2i,2j-1}\right), \\
    \mathcal H^{-+}_{L-i+1,L-j+1} &=\frac{1}{4}\left(\mathcal H^{\mathrm{QBCD}}_{2i-1,2j-1}+\mathcal H^{\mathrm{QBCD}}_{2i,2j}+\mathcal H^{\mathrm{QBCD}}_{2i-1,2j}+\mathcal H^{\mathrm{QBCD}}_{2i,2j-1}\right), \\
    \mathcal H^{-+}_{L-i+1,j} &=\frac{1}{4}\left(\mathcal H^{\mathrm{QBCD}}_{2i-1,2j-1}+\mathcal H^{\mathrm{QBCD}}_{2i,2j}+\mathcal H^{\mathrm{QBCD}}_{2i-1,2j}+\mathcal H^{\mathrm{QBCD}}_{2i,2j-1}\right), \\
    \mathcal H^{-+}_{i,L-j+1} &=\frac{1}{4}\left(\mathcal H^{\mathrm{QBCD}}_{2i-1,2j-1}+\mathcal H^{\mathrm{QBCD}}_{2i,2j}+\mathcal H^{\mathrm{QBCD}}_{2i-1,2j}+\mathcal H^{\mathrm{QBCD}}_{2i,2j-1}\right),
\end{align}
\end{subequations}
for $\mathcal{H}^{-+}$, and
\begin{subequations}
\begin{align}
    \mathcal H^{--}_{ij} &=\frac{1}{4}\left(\mathcal H^{\mathrm{QBCD}}_{2i-1,2j-1}-\mathcal H^{\mathrm{QBCD}}_{2i,2j}-\mathcal H^{\mathrm{QBCD}}_{2i-1,2j}+\mathcal H^{\mathrm{QBCD}}_{2i,2j-1}\right), \\
    \mathcal H^{--}_{L-i+1,L-j+1} &=\frac{1}{4}\left(-\mathcal H^{\mathrm{QBCD}}_{2i-1,2j-1}+\mathcal H^{\mathrm{QBCD}}_{2i,2j}+\mathcal H^{\mathrm{QBCD}}_{2i-1,2j}-\mathcal H^{\mathrm{QBCD}}_{2i,2j-1}\right),\\
    \mathcal H^{--}_{L-i+1,j} &=\frac{1}{4}\left(\mathcal H^{\mathrm{QBCD}}_{2i-1,2j-1}-\mathcal H^{\mathrm{QBCD}}_{2i,2j}-\mathcal H^{\mathrm{QBCD}}_{2i-1,2j}+\mathcal H^{\mathrm{QBCD}}_{2i,2j-1}\right), \\
    \mathcal H^{--}_{i,L-j+1}&=\frac{1}{4}\left(-\mathcal H^{\mathrm{QBCD}}_{2i-1,2j-1}+\mathcal H^{\mathrm{QBCD}}_{2i,2j}+\mathcal H^{\mathrm{QBCD}}_{2i-1,2j}-\mathcal H^{\mathrm{QBCD}}_{2i,2j-1}\right),
\end{align}
\end{subequations}
for $\mathcal{H}^{--}$. Notice that the block Hamiltonians above separate into four further blocks, depending on whether the Dirac fermions are to the left or right of site $\ell$.

\subsection{Expectation values}
\label{app:sec:numerics_observables}

Once the operators $c(t), c^\dagger(t)$ are obtained, expectation values can be computed (note instead that accessing the fidelity is more complicated). For instance, the instantaneous energy expectation reads
\begin{align}
    \ev{H[\lambda](t)}{\mathrm{GS}} &= -\lambda \sum_{j=1}^L J_j \ev{\sigma_j^z(t) \sigma_{j+1}^z(t)}{\rightarrow} - (1-\lambda) \sum_{j=1}^L \ev{\sigma_j^x(t)}{\rightarrow} \\
    &= -\lambda \sum_{j=1}^L \tilde{J}_j \ev{ \left(c_j^\dagger(t) c_{j+1}^\dagger(t) + c_j^\dagger(t) c_{j+1}(t) - c_j(t) c_{j+1}^\dagger(t) - c_j(t) c_{j+1}(t) \right) }{0_c} \nonumber \\
    &\hspace{2cm}+2(1-\lambda) \sum_{j=1}^L \ev{c_j^\dagger(t) c_j(t)}{0_c} - (1-\lambda)L,
\end{align}
where $H[\lambda=1](t)$ is the Hamiltonian corresponding to the parameter $\lambda$, evolved in the Heisenberg picture for a time $t$. Let us compute each term separately. Calling $\mathcal{V} := \Te^{-2i\int_0^t \mathcal{H}(t') dt'}$, it holds
\begin{equation}
    \sum_{j=1}^L \tilde{J}_j \ev{ c_j^\dagger(t) c_{j+1}^\dagger(t) }{0_c} = 
    \sum_{j=1}^L \sum_{\alpha,\beta=1}^{2L} \tilde{J}_j \ev{ \mathcal{V}_{j^\dagger,\alpha} \mathcal{V}_{(j+1)^\dagger,\beta} \Psi_\alpha \Psi_\beta}{0_c}
    = \sum_{i,j=1}^L \tilde{J}_j \mathcal{V}_{j^\dagger,i} \mathcal{V}_{(j+1)^\dagger,i^\dagger},
\end{equation}
where $\Psi_{i}$ refers to $c_i$, and $\Psi_{i^\dagger}$ to $c_i^\dagger$ (and similarly for the corresponding rows and columns of $\mathcal{V}$). A similar computation leads to
\begin{align}
    \sum_{j=1}^L \tilde{J}_j \ev{ c_j^\dagger(t) c_{j+1}(t) }{0_c} &= \sum_{i,j=1}^L \tilde{J}_j \mathcal{V}_{j^\dagger,i} \mathcal{V}_{j+1,i^\dagger}, \\
    \sum_{j=1}^L \tilde{J}_j \ev{ c_j(t) c_{j+1}^\dagger(t) }{0_c} &= \sum_{i,j=1}^L \tilde{J}_j \mathcal{V}_{j,i} \mathcal{V}_{(j+1)^\dagger,i^\dagger}, \\
    \sum_{j=1}^L \tilde{J}_j \ev{ c_j(t) c_{j+1}(t) }{0_c} &= \sum_{i,j=1}^L \tilde{J}_j \mathcal{V}_{j,i} \mathcal{V}_{j+1,i^\dagger}, \\
    \sum_{j=1}^L \ev{ c_j^\dagger(t) c_j(t) }{0_c} &= \sum_{i,j=1}^L \mathcal{V}_{j^\dagger,i} \mathcal{V}_{j,i^\dagger}.
\end{align}
Putting everything together, one finds
\begin{multline}
    \ev{H[\lambda](t)}{0_c} = -\lambda \sum_{i,j=1}^L \tilde{J}_j \left[ \mathcal{V}_{j^\dagger,i} \mathcal{V}_{(j+1)^\dagger,i^\dagger} + \mathcal{V}_{j^\dagger,i} \mathcal{V}_{j+1,i^\dagger} - \mathcal{V}_{j,i} \mathcal{V}_{(j+1)^\dagger,i^\dagger} - \mathcal{V}_{j,i} \mathcal{V}_{j+1,i^\dagger} \right] \\
    +2(1-\lambda) \sum_{i,j=1}^L \mathcal{V}_{j^\dagger,i} \mathcal{V}_{j,i^\dagger} - (1-\lambda) L .
\end{multline}
Similarly, the kink number reads
\begin{equation}
    \ev{K(t)}{0_c} = \frac{L}{2} - \frac{1}{2}\sum_{i,j=1}^L \tilde{1}_j \left[ \mathcal{V}_{j^\dagger,i} \mathcal{V}_{(j+1)^\dagger,i^\dagger} + \mathcal{V}_{j^\dagger,i} \mathcal{V}_{j+1,i^\dagger} - \mathcal{V}_{j,i} \mathcal{V}_{(j+1)^\dagger,i^\dagger} - \mathcal{V}_{j,i} \mathcal{V}_{j+1,i^\dagger} \right],
\end{equation}
where
\begin{equation}
    \tilde{1}_j = 
    \begin{cases}
        1   &j=1,2,\dots,L-1 \\
        -1  &j=L
    \end{cases}
\end{equation}
takes into account the restriction to the even parity sector.

\section{Derivation of the minimization equations for the \textsc{Max Cut}}
\label{app:MAX_CUT}

In this section, we derive the equations for the coefficients for the first- and second-order CD expansions for the \textsc{Max Cut} problem. The enlisted Hamiltonians at a given $\lambda$ value read as
\begin{eqnarray}
    &&H_P=-\lambda\sum_c\frac{1-\sigma^z_{c_1}\sigma^z_{c_2}}{2}-(1-\lambda)\sum_j\sigma^x_j,\\
    &&H^{(1)}_1=\sum_c\alpha_c\left(\sigma^z_{c_1}\sigma^y_{c_2}+\sigma^y_{c_1}\sigma^z_{c_2}\right),\\
    &&H^{(2)}_1=\sum_{c,c^\prime}\beta_{c,c^\prime}\delta_{c_2,c^\prime_1}\left(\sigma^z_{c_1}\sigma^x_{c_2}\sigma^y_{c^\prime_2}+\sigma^y_{c_1}\sigma^x_{c_2}\sigma^z_{c^\prime_2}\right).
\end{eqnarray}
In the first-order local expansion, the action integrand reads
\begin{eqnarray}
    G(H^{(1)}_1)=&&-\sum_{c=1}^M\frac{1+\sigma^z_{c_1}\sigma^z_{c_2}}{2}+\sum_{c=1}^M\frac{1+2\lambda\alpha_c}{2}[\sigma^x_{c_1}+\sigma^x_{c_2}]\\
    &&+\lambda\sum_{c,c^\prime}\alpha_c\left[(1-\delta_{c_1,c^\prime_1})\delta_{c_2,c^\prime_2}\sigma^z_{c_1}\sigma^z_{c^\prime_1}\sigma^x_{c_2}+\delta_{c_1,c^\prime_1}(1-\delta_{c_2,c^\prime_2})\sigma^x_{c_1}\sigma^z_{c_2}\sigma^z_{c^\prime_2}\right]\nonumber\\
    &&+4(1-\lambda)\sum_{c=1}^M\alpha_{c}(\sigma^y_{c_1}\sigma^y_{c_2}-\sigma^z_{c_1}\sigma^z_{c_2})
    +\lambda\sum_{c, c^\prime}^M\alpha_c(\sigma^z_{c_1}\sigma^x_{c_2}\sigma^z_{c^\prime_2}\delta_{c_2,c^\prime_1}+\sigma^z_{c^\prime_1}\sigma^x_{c_1}\sigma^z_{c_2}\delta_{c_1,c^\prime_2})\nonumber\\
    &&=-\sum_{c=1}^M\frac{1+8(1-\lambda)\alpha_c}{2}\sigma^z_{c_1}\sigma^z_{c_2}+\sum_{c=1}^M\frac{1+2\lambda\alpha_c}{2}(\sigma^x_{c_1}+\sigma^x_{c_2})+4(1-\lambda)\sum_{c=1}^M\alpha_{c}\sigma^y_{c_1}\sigma^y_{c_2}
    \nonumber\\
    &&+\lambda\sum_{c,c^\prime}\alpha_c\left[(1-\delta_{c_1,c^\prime_1})\delta_{c_2,c^\prime_2}\sigma^z_{c_1}\sigma^z_{c^\prime_1}\sigma^x_{c_2}+\delta_{c_1,c^\prime_1}(1-\delta_{c_2,c^\prime_2})\sigma^x_{c_1}\sigma^z_{c_2}\sigma^z_{c^\prime_2}+\delta_{c_2,c^\prime_1}\sigma^z_{c_1}\sigma^x_{c_2}\sigma^z_{c^\prime_2}+\delta_{c_1,c^\prime_2}\sigma^z_{c^\prime_1}\sigma^x_{c_1}\sigma^z_{c_2}\right],\nonumber  
\end{eqnarray}
where all terms have been ordered so that the strictly greater inequality is always understood between the subindices, 
 $c_1<c_2$ and $c^\prime_1<c^\prime_2$.
The corresponding trace norm reads
\begin{eqnarray}
    &&\mathrm{Tr}\left[G^\dagger\,G\right]=\sum_c\frac{[1+8(1-\lambda)\alpha_c]^2}{4}\nonumber\\
    &&+\frac{1}{2}\sum_{c,c^\prime}(2\lambda^2\alpha_c\alpha_{c^\prime}+\lambda(\alpha_c+\alpha_{c^\prime}))\left[\delta_{c_1,c^\prime_1}(1-\delta_{c_2,c^\prime_2})+(1-\delta_{c_1,c^\prime_1})\delta_{c_2,c^\prime_2}+2\delta_{c_1,c^\prime_1}\delta_{c_2,c^\prime_2}+\delta_{c_1,c^\prime_2}+\delta_{c_2,c^\prime_1}\right]\nonumber\\
    &&+\lambda^2\sum_{c}(\alpha_c+\alpha_{c^\prime})^2\delta_{c_1,c^\prime_2}+\lambda^2\sum_c\alpha^2_c\left[\delta_{c_1,c^\prime_1}(1-\delta_{c_2,c^\prime_2})+(1-\delta_{c_1,c^\prime_1})\delta_{c_2,c^\prime_2}\right]+16(1-\lambda)^2\sum_c\alpha^2_c.
\end{eqnarray}

Taking now the derivative with respect to a given $\alpha_c$ the coefficient of $\lambda^2\alpha_c$ will equal the number of ways the given spins of $\alpha_c$ appear in other interactions, that is $6$ altogether for the two spins in $\alpha_c$. At the same time, all $\alpha_{c^\prime}$ will be coupled to all other interactions that share at least one spin with it, plus the additional correlation term of $\delta_{c_1,c^\prime_2}+\delta_{c_2,c^\prime_1}$. This yields
\begin{eqnarray}
    \frac{\delta \mathrm{Tr}(G^\dagger G)}{\delta\alpha_c}&&=\left[64(1-\lambda)^2+12\lambda^2\right]\alpha_c+2\lambda^2\sum_{c^\prime}\alpha_{c^\prime}\left[\delta_{c_1,c^\prime_1}(1-\delta_{c_2,c^\prime_2})+(1-\delta_{c_1,c^\prime_1})\delta_{c_2,c^\prime_2}+2\delta_{c_1,c^\prime_2}+2\delta_{c_2,c^\prime_1}\right]\nonumber\\
    &&=-\lambda\sum_{c^\prime}\left(\delta_{c_1,c^\prime_1}+\delta_{c_2,c^\prime_2}+\delta_{c_1,c^\prime_2}+\delta_{c_2,c^\prime_1}\right)-4(1-\lambda)=-2(\lambda-2).
\end{eqnarray}

For the second order, the action integrated reads
\begin{eqnarray}
    &&G(H^{(2)}_1)=-\sum_{c=1}^M\frac{1+\sigma^z_{c_1}\sigma^z_{c_2}}{2}+\sum_{j=1}^L\sigma^x_j+4(1-\lambda)\sum_{c, c^\prime}^M\delta_{c_2,c^\prime_1}\beta_{c,c^\prime}\left(\sigma^y_{c_1}\sigma^x_{c_2}\sigma^y_{c^\prime_2}-\sigma^z_{c_1}\sigma^x_{c_2}\sigma^z_{c^\prime_2}\right)\nonumber\\
    &&+\lambda\sum_{c,c^\prime,c^{\prime\prime}}(\delta_{c_1,c^\prime_2}\beta_{c^\prime,c}\delta_{c^\prime_1,c^{\prime\prime}_2}\sigma^z_{c^{\prime\prime}_1}\sigma^x_{c^{\prime\prime}_2}\sigma^x_{c_1}\sigma^z_{c_2}+\delta_{c_2,c^\prime_1}\beta_{c,c^\prime}\delta_{c^{\prime\prime}_1,c^{\prime}_2}\sigma^z_{c_1}\sigma^x_{c_2}\sigma^x_{c^{\prime\prime}_1}\sigma^z_{c^{\prime\prime}_2})
    -\lambda\sum_{c,c^\prime}(\delta_{c_2,c^\prime_1}\beta_{c,c^\prime}+\delta_{c^\prime_2,c_1}\beta_{c^\prime,c})\sigma^y_{c_1}\sigma^y_{c_2}\nonumber\\
    &&-\lambda\sum_{c,c^\prime,c^{\prime\prime}}(\beta_{c,c^\prime}\delta_{c_2,c^\prime_1}+\beta_{c^{\prime\prime},c^\prime}\delta_{c^{\prime\prime}_2,c^\prime_1})(1-\delta_{c_1,c^{\prime\prime}_1})\delta_{c_2,c^{\prime\prime}_2}\sigma^z_{c^{\prime\prime}_1}\sigma^z_{c_1}\sigma^y_{c_2}\sigma^y_{c^\prime_2}+(\beta_{c,c^\prime}\delta_{c_2,c^\prime_1}\delta_{c_1,c^{\prime}_1}+\beta_{c,c^{\prime\prime}}\delta_{c_2,c^{\prime\prime}_1}\delta_{c_1,c^{\prime\prime}_1})(1-\delta_{c^\prime_2,c^{\prime\prime}_2})\sigma^y_{c_1}\sigma^y_{c_2}\sigma^z_{c^\prime_2}\sigma^z_{c^{\prime\prime}_2}\nonumber\\
&&+\lambda\sum_{c,c^\prime,c^{\prime\prime}}\beta_{c,c^\prime}\delta_{c_2,c^\prime_1}\left[
    \sigma^z_{c_1}(1-\delta_{c^{\prime\prime}_1,c_2})\delta_{c^{\prime\prime}_2,c^\prime_2}(\sigma^z_{c^{\prime\prime}_1}\sigma^x_{c_2}+\sigma^x_{c_2}\sigma^z_{c^{\prime\prime}_1})\sigma^x_{c^\prime_2}-
    \sigma^z_{c_1}\delta_{c^{\prime\prime}_1,c_2}(1-\delta_{c^{\prime\prime}_2,c^\prime_2})\sigma^y_{c_2}(\sigma^z_{c^{\prime\prime}_2}\sigma^y_{c^\prime_2}+\sigma^y_{c^\prime_2}\sigma^z_{c^{\prime\prime}_2})\right]\nonumber\\
    &&-\lambda\sum_{c,c^\prime,c^{\prime\prime}}\beta_{c,c^\prime}\delta_{c_2,c^\prime_1}\left[
    (1-\delta_{c^{\prime\prime}_1,c_1})\delta_{c^{\prime\prime}_2,c_2}(\sigma^z_{c^{\prime\prime}_1}\sigma^y_{c_1}+\sigma^y_{c_1}\sigma^z_{c^{\prime\prime}_1})\sigma^y_{c_2}\sigma^z_{c^\prime_2}-
    \delta_{c^{\prime\prime}_1,c_1}(1-\delta_{c^{\prime\prime}_2,c_2})\sigma^x_{c_1}(\sigma^z_{c^{\prime\prime}_2}\sigma^x_{c_2}+\sigma^x_{c_2}\sigma^z_{c^{\prime\prime}_2})\sigma^z_{c^\prime_2}\right],\nonumber\\
\end{eqnarray}
where the ordering in the indices, $c_1<c_2<c^\prime_2$ of the connected interaction edges is fixed, and those denoted by $c^{\prime\prime}$ refer to the ones of the second-order CD term. 
Here, the first term originates simply from the fact that in every three-body terms, the corresponding $\sigma^x_j$ will lead to either $\sigma^z_j$ or $\sigma^y_j$.
The factor of three in the second line comes from the freedom of choosing one of the $c^{\prime\prime}_{i=1,2}$ spins in the construction.
The third and fourth lines correspond to the cases when the $\sigma^z\sigma^z$ terms can potentially have non-zero commutator with the $\sigma^y\sigma^x$ and $\sigma^x\sigma^y$ terms.

The total second-order term thus reads as
\begin{eqnarray}
    &&G(H^{(2)}_1+H^{(1)}_1)=-\sum_{c=1}^M\frac{1+8(1-\lambda)\alpha_c}{2}\sigma^z_{c_1}\sigma^z_{c_2}+\sum_{c=1}^M\frac{1+2\lambda\alpha_c}{2}(\sigma^x_{c_1}+\sigma^x_{c_2})\\
    &&+\lambda\sum_{c,c^\prime}\alpha_c\left[(1-\delta_{c_1,c^\prime_1})\delta_{c_2,c^\prime_2}\sigma^z_{c_1}\sigma^z_{c^\prime_1}\sigma^x_{c_2}+\delta_{c_1,c^\prime_1}(1-\delta_{c_2,c^\prime_2})\sigma^x_{c_1}\sigma^z_{c_2}\sigma^z_{c^\prime_2}\right]\nonumber\\
    &&+4(1-\lambda)\sum_{c, c^\prime}^M\delta_{c_2,c^\prime_1}\beta_{c,c^\prime}\sigma^y_{c_1}\sigma^x_{c_2}\sigma^y_{c^\prime_2}-\sum_{c,c^\prime}\delta_{c_2,c^\prime_1}\left[4(1-\lambda)\beta_{c,c^\prime}-\lambda(\alpha_c+\alpha_{c^\prime})\right]\sigma^z_{c_1}\sigma^x_{c_2}\sigma^z_{c^\prime_2}\nonumber\\
    &&+\lambda\sum_{c,c^\prime,c^{\prime\prime}}(\delta_{c_1,c^\prime_2}\beta_{c^\prime,c}\delta_{c^\prime_1,c^{\prime\prime}_2}\sigma^z_{c^{\prime\prime}_1}\sigma^x_{c^{\prime\prime}_2}\sigma^x_{c_1}\sigma^z_{c_2}+\delta_{c_2,c^\prime_1}\beta_{c,c^\prime}\delta_{c^{\prime\prime}_1,c^{\prime}_2}\sigma^z_{c_1}\sigma^x_{c_2}\sigma^x_{c^{\prime\prime}_1}\sigma^z_{c^{\prime\prime}_2})
    -\sum_{c}\left(\lambda\sum_{c^\prime}\delta_{c_2,c^\prime_1}\beta_{c,c^\prime}+\delta_{c^\prime_2,c_1}\beta_{c^\prime,c}-4(1-\lambda)\alpha_c\right)\sigma^y_{c_1}\sigma^y_{c_2}\nonumber\\
    &&-\lambda\sum_{c,c^\prime,c^{\prime\prime}}(\beta_{c,c^\prime}\delta_{c_2,c^\prime_1}+\beta_{c^{\prime\prime},c^\prime}\delta_{c^{\prime\prime}_2,c^\prime_1})(1-\delta_{c_1,c^{\prime\prime}_1})\delta_{c_2,c^{\prime\prime}_2}\sigma^z_{c^{\prime\prime}_1}\sigma^z_{c_1}\sigma^y_{c_2}\sigma^y_{c^\prime_2}+(\beta_{c,c^{\prime}}\delta_{c_2,c^\prime_1}\delta_{c_2,c^{\prime\prime}_1}+\beta_{c,c^{\prime\prime}}\delta_{c_2,c^{\prime\prime}_1}\delta_{c_2,c^{\prime}_1})(1-\delta_{c^\prime_2,c^{\prime\prime}_2})\sigma^y_{c_1}\sigma^y_{c_2}\sigma^z_{c^\prime_2}\sigma^z_{c^{\prime\prime}_2}\nonumber\\
&&+\lambda\sum_{c,c^\prime,c^{\prime\prime}}\beta_{c,c^\prime}\delta_{c_2,c^\prime_1}\left[
    \sigma^z_{c_1}(1-\delta_{c^{\prime\prime}_1,c_2})\delta_{c^{\prime\prime}_2,c^\prime_2}(\sigma^z_{c^{\prime\prime}_1}\sigma^x_{c_2}+\sigma^x_{c_2}\sigma^z_{c^{\prime\prime}_1})\sigma^x_{c^\prime_2}-
    \sigma^z_{c_1}\delta_{c^{\prime\prime}_1,c_2}(1-\delta_{c^{\prime\prime}_2,c^\prime_2})\sigma^y_{c_2}(\sigma^z_{c^{\prime\prime}_2}\sigma^y_{c^\prime_2}+\sigma^y_{c^\prime_2}\sigma^z_{c^{\prime\prime}_2})\right]\nonumber\\
    &&-\lambda\sum_{c,c^\prime,c^{\prime\prime}}\beta_{c,c^\prime}\delta_{c_2,c^\prime_1}\left[
    (1-\delta_{c^{\prime\prime}_1,c_1})\delta_{c^{\prime\prime}_2,c_2}(\sigma^z_{c^{\prime\prime}_1}\sigma^y_{c_1}+\sigma^y_{c_1}\sigma^z_{c^{\prime\prime}_1})\sigma^y_{c_2}\sigma^z_{c^\prime_2}-
    \delta_{c^{\prime\prime}_1,c_1}(1-\delta_{c^{\prime\prime}_2,c_2})\sigma^x_{c_1}(\sigma^z_{c^{\prime\prime}_2}\sigma^x_{c_2}+\sigma^x_{c_2}\sigma^z_{c^{\prime\prime}_2})\sigma^z_{c^\prime_2}\right].\nonumber
\end{eqnarray}

Now, the trace of the square reads
\begin{eqnarray}
    &&\mathrm{Tr}\left[G^\dagger G\right]=\sum_c\frac{[1+8(1-\lambda)\alpha_c]^2}{4}\\
    &&+\frac{1}{2}\sum_{c,c^\prime}(2\lambda^2\alpha_c\alpha_{c^\prime}+\lambda(\alpha_c+\alpha_{c^\prime}))\left[\delta_{c_1,c^\prime_1}(1-\delta_{c_2,c^\prime_2})+(1-\delta_{c_1,c^\prime_1})\delta_{c_2,c^\prime_2}+2\delta_{c_1,c^\prime_1}\delta_{c_2,c^\prime_2}+\delta_{c_1,c^\prime_2}+\delta_{c_2,c^\prime_1}\right]\nonumber\\
    &&+\lambda^2\sum_{c,c^\prime}\alpha^2_c\left[\delta_{c_1,c^\prime_1}(1-\delta_{c_2,c^\prime_2})+(1-\delta_{c_1,c^\prime_1})\delta_{c_2,c^\prime_2}\right]\nonumber\\
    &&+16(1-\lambda)^2\sum_{c,c^\prime}\delta_{c_2,c^\prime_1}\beta^2_{c,c^\prime}+\sum_{c,c^\prime}\delta_{c_2,c^\prime_1}\left[4(1-\lambda)\beta_{c,c^\prime}-\lambda(\alpha_c+\alpha_{c^\prime})\right]^2+\lambda^2\sum_{c,c^\prime,c^{\prime\prime}}\left(\delta_{c^\prime_2,c_1}\beta^2_{c^\prime,c}\delta_{c^\prime_1,c^{\prime\prime}_2}+\delta_{c_2,c^\prime_1}\beta^2_{c,c^\prime}\delta_{c^{\prime\prime}_1,c^{\prime}_2}\right)\nonumber\\
    &&+\sum_{c}\left(\sum_{c^\prime}\lambda\delta_{c_2,c^\prime_1}\beta_{c,c^\prime}+\lambda\delta_{c^\prime_2,c_1}\beta_{c^\prime,c}-4(1-\lambda)\alpha_c\right)^2\nonumber\\
    &&+\lambda^2\sum_{c,c^\prime,c^{\prime\prime}}(\beta_{c,c^\prime}\delta_{c_2,c^\prime_1}+\beta_{c^{\prime\prime},c^\prime}\delta_{c^{\prime\prime}_2,c^\prime_1})^2(1-\delta_{c^{\prime\prime}_1,c_1})\delta_{c_2,c^{\prime\prime}_2}+(\beta_{c,c^\prime}\delta_{c_2,c^\prime_1}\delta_{c_2,c^{\prime\prime}_1}+\beta_{c,c^{\prime\prime}}\delta_{c_2,c^{\prime\prime}_1}\delta_{c_2,c^{\prime}_1})^2(1-\delta_{c^\prime_2,c^{\prime\prime}_2})\nonumber\\
    &&+2\lambda^2\sum_{c,c^\prime,c^{\prime\prime}}\beta^2_{c,c^\prime}\delta_{c_2,c^\prime_1}\left[(1-\delta_{c^{\prime\prime}_1,c_2})\delta_{c^{\prime\prime}_2,c^\prime_2}+\delta_{c^{\prime\prime}_1,c_2}(1-\delta_{c^{\prime\prime}_2,c^\prime_2})+
    (1-\delta_{c^{\prime\prime}_1,c_1})\delta_{c^{\prime\prime}_2,c_2}+
    \delta_{c^{\prime\prime}_1,c_1}(1-\delta_{c^{\prime\prime}_2,c_2})\right].\nonumber
\end{eqnarray}
For computing the derivatives, we again use combinatorial arguments. As all terms in $G\left(H^{(1)}_1+H^{(2)}_1\right)$ originating from different ordering of the connected spins are orthogonal to each other the constant prefactor in front of $\lambda^2\beta_{c,c^\prime}$ is given by four times the number of ways how a given spin can be involved in other interactions, i.e., by $12$. This can be seen easily by the fact, that one gets a non-zero operator for the commutator whenever the first two operators of $\sigma^z_{c^\prime_1}\sigma^x_{c^\prime_2}\sigma^y_{c^\prime_3}$ match any of the three in $\sigma^z_{c_1}\sigma^z_{c_2}\sigma^z_3$. As both of the resulting commutators give orthogonal operators, eventually one obtains two times the number of times the given $c_1,\,c_2,\,c_3$ spins appear in the interactions. The off-diagonal terms connecting different $\beta_{c,c^{\prime\prime}}$-s are obtained similarly as in the first-order case, by enlisting all possible connections to other interaction pairs of $c,c^{\prime\prime}$ or $c^\prime c^{\prime\prime}$. One finds
\begin{eqnarray}
    &&\frac{\delta \mathrm{Tr}(G^\dagger G)}{\delta\alpha_c}=0\nonumber\\
    &&\Rightarrow\left[64(1-\lambda)^2+12\lambda^2\right]\alpha_c +2\lambda^2\sum_{c^\prime}\alpha_{c^\prime}\left[\delta_{c_1,c^\prime_1}(1-\delta_{c_2,c^\prime_2})+(1-\delta_{c_1,c^\prime_1})\delta_{c_2,c^\prime_2}+2\delta_{c_1,c^\prime_2}+2\delta_{c_2,c^\prime_1}\right]\\
    &&=16\lambda(1-\lambda)\sum_{c^\prime}(\beta_{c,c^\prime}\delta_{c^\prime_1,c_2}+\beta_{c^\prime,c}\delta_{c^\prime_2,c_1})-2(\lambda-2)\nonumber,\\
    &&\frac{\delta \mathrm{Tr}(G^\dagger G)}{\delta\beta_{c,c^\prime}}=0\\
    &&\Rightarrow \left[64(1-\lambda)^2+24\lambda^2\right]\beta_{c,c^\prime}\delta_{c_2,c^\prime_1}\nonumber\\
    &&+2\lambda^2\sum_{c^{\prime\prime}}(\delta_{c^\prime_2,c^{\prime\prime}_1}\beta_{c^\prime,c^{\prime\prime}}+\delta_{c^{\prime\prime}_2,c_1}\beta_{c^{\prime\prime},c})+2\delta_{c^{\prime\prime}_2,c^\prime_1}\beta_{c^{\prime\prime},c^\prime}+\delta_{c_1,c^{\prime\prime}_1}(1-\delta_{c_2,c^{\prime\prime}_2})+2\delta_{c_2,c^{\prime\prime}_1}\beta_{c,c^{\prime\prime}}(1-\delta_{c^\prime_2,c^{\prime\prime}_2})\nonumber\\
    &&=16\lambda(1-\lambda)\delta_{c_2,c^\prime_1}(\alpha_c+\alpha_{c^\prime}).\nonumber
\end{eqnarray}

\section{Gap amplification within the local CD expansion}
\label{app:Gap_difference}

\begin{figure*}
    \centering
    \includegraphics[width=.5\columnwidth]{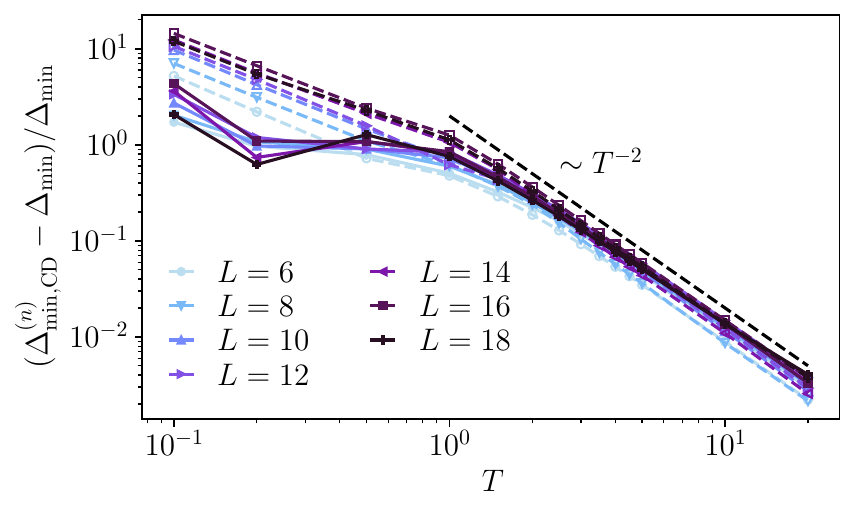}
    \caption{Relative difference between ground-state gaps without CD and with the local CD approaches. The improvement achieved by first-order (solid lines, filled dots) and second-order (dashed lines, empty dots) local expansions exhibits an approximate quadratic decay with system size. Notably, the amplification of the gap vanishes at driving times just preceding the onset of finite fidelity.}
    \label{fig:gapFirstorder_Max Cut}
\end{figure*}

\begin{figure*}
    \centering
    \includegraphics[width=.5\columnwidth]{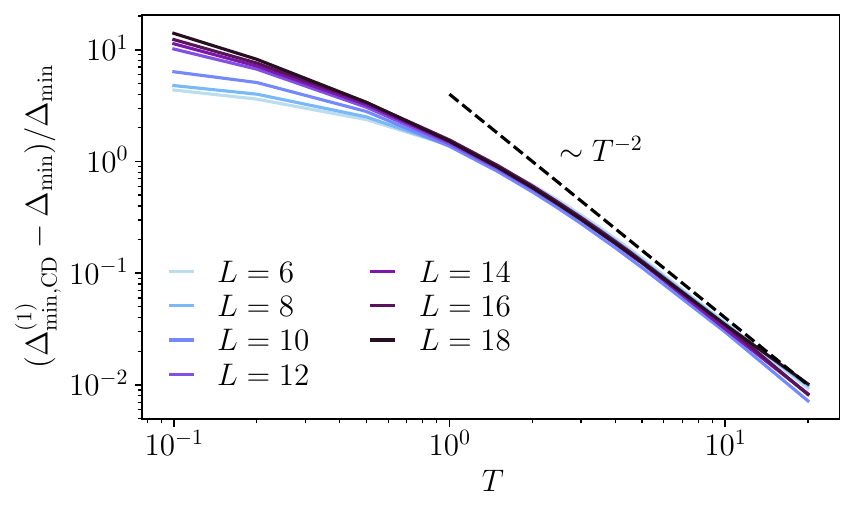}
    \caption{Relative difference between the energy gaps without CD and with the first-order ansatz decaying approximately as the minimal spin glass bottleneck independently of the system size. No essential gap amplification is present around the driving times, where even without CD, the fidelity departs from zero.}
    \label{fig:gapFirstorder_XORSAT}
\end{figure*}

In this section, we provide further details about the gap amplification of the local CD expansion in the \textsc{Max Cut} and $3$-\textsc{XORSAT} models.
The relative difference energy gaps in the bare process without CD and in that assisted by local CD admits a remarkably universal form of  $(\Delta^{(1,2)}_{\mathrm{min,CD}}-\Delta_\mathrm{min})/\Delta_\mathrm{min}\sim T^{-2}$ for both models, as shown in Fig.~\ref{fig:gapFirstorder_Max Cut} and in Fig.~\ref{fig:gapFirstorder_XORSAT}.

\section{Derivation of the variational equations with COLD for the \textsc{Max Cut}}
\label{app:COLD}

In this section, we derive the equations for the optimized COLD approach with the second-order commutator as optimizable operators. Due to the structure of these extra operators, the general form of the local CD up to three-body terms takes the same form as that of the second-order local expansion. First, we write out explicitly the optimizable operator,
\begin{eqnarray}
\mathcal O_\mathrm{opt} 
&=& \beta^{(1)} \left[ H_0, [H_1, H_0] \right] + \beta^{(2)} \left[ H_1, [H_1, H_0] \right] \nonumber\\[1mm]
&=& \beta^{(1)} \, \sin(2\pi \lambda) \, \Bigg[
3 \sum_{j=1}^L \sigma^x_j
+ \sum_{c,c'} \delta_{c_1,c'_1} (1-\delta_{c_2,c'_2}) \, \sigma^x_{c_1} \sigma^z_{c_2} \sigma^z_{c'_2} \nonumber\\
&& \quad + \sum_{c,c'} \delta_{c_2,c'_2} (1-\delta_{c_1,c'_1}) \, \sigma^z_{c_1} \sigma^z_{c'_1} \sigma^x_{c_2} 
+ 2 \sum_{c,c'} \delta_{c_2,c'_1} \, \sigma^z_{c_1} \sigma^x_{c_2} \sigma^z_{c'_2} 
\Bigg] \nonumber\\[1mm]
&& + 2 \beta^{(2)} \, \sin(2\pi \lambda) \, \sum_{c} \Big( \sigma^z_{c_1} \sigma^z_{c_2} - \sigma^y_{c_1} \sigma^y_{c_2} \Big),
\end{eqnarray}
where $H_1=\sum_{j=1}^L\sigma^x_j$ and $H_P=\sum_{c=1}^{M}\frac{1+\sigma^z_{c_1}\sigma^z_{c_2}}{2}$ is the problem Hamiltonian, Eq.~\eqref{eq:H_P_MAXCUT}, for the \textsc{Max Cut}, respectively.
From this, the full Hamiltonian reads
\begin{eqnarray}
H^{(2)}_\mathrm{COLD} &=& 
\frac{\lambda + 4 \beta^{(2)} \sin(2\pi \lambda)}{2} \sum_c \sigma^z_{c_1} \sigma^z_{c_2} 
- \big((1-\lambda) - 3 \beta^{(1)} \sin(2\pi \lambda)\big) \sum_j \sigma^x_j 
- 2 \beta^{(2)} \sin(2\pi \lambda) \sum_c \sigma^y_{c_1} \sigma^y_{c_2} 
\nonumber\\
&& + 2 \beta^{(1)} \sin(2\pi \lambda) \sum_{c,c'} \delta_{c_2,c'_1} \, 
\sigma^z_{c_1} \sigma^x_{c_2} \sigma^z_{c'_2} 
\nonumber\\
&& + \beta^{(1)} \sin(2\pi \lambda) \sum_{c,c'} \Big[
\delta_{c_1,c'_1} (1-\delta_{c_2,c'_2}) \, \sigma^x_{c_1} \sigma^z_{c_2} \sigma^z_{c'_2} 
+ \delta_{c_2,c'_2} (1-\delta_{c_1,c'_1}) \, \sigma^z_{c_1} \sigma^z_{c'_1} \sigma^x_{c_2} 
+ 2 \delta_{c_2,c'_1} \, \sigma^z_{c_1} \sigma^x_{c_2} \sigma^z_{c'_2} 
\Big] 
\nonumber\\
&& + \sum_c \alpha_c \big( \sigma^z_{c_1} \sigma^y_{c_2} + \sigma^y_{c_1} \sigma^z_{c_2} \big) 
+ \sum_{c,c'} \beta_{c,c'} \delta_{c_2,c'_1} \big( \sigma^z_{c_1} \sigma^x_{c_2} \sigma^y_{c'_2} + \sigma^y_{c_1} \sigma^x_{c_2} \sigma^z_{c'_2} \big).
\end{eqnarray}

The corresponding $G$ operator measuring the deviations from the exact CD, 
\begin{eqnarray}
G=H_1-H_P+\partial_\lambda\mathcal O_\mathrm{opt}-i\left[-\lambda H_p-(1-\lambda)H_1+\mathcal O_\mathrm{opt},H^{(1)}_1+H^{(2)}_1\right],    
\end{eqnarray}
can be written in the full explicit form as
\begin{eqnarray}
G &=& \sum_{j=1}^{L} \sigma^x_j 
    - \frac{1}{2}\sum_{c} 1
    + \left(-\frac{1}{2}+4\pi\cos(2\pi\lambda)\beta^{(2)}\right)\sum_{c}\sigma^z_{c_1}\sigma^z_{c_2}
    -4\pi\cos(2\pi\lambda)\beta^{(2)}\sum_{c}\sigma^y_{c_1}\sigma^y_{c_2}
    \nonumber\\
&& + 2\pi\cos(2\pi\lambda)\,\beta^{(1)}\Bigg[
     3\sum_{j=1}^{L}\sigma^x_j
     + \sum_{c,c'}\delta_{c_1,c'_1}\big(1-\delta_{c_2,c'_2}\big)\,\sigma^x_{c_1}\sigma^z_{c_2}\sigma^z_{c'_2}
     + \sum_{c,c'}\delta_{c_2,c'_2}\big(1-\delta_{c_1,c'_1}\big)\,\sigma^z_{c_1}\sigma^z_{c'_1}\sigma^x_{c_2}
     + 2\sum_{c,c'}\delta_{c_2,c'_1}\,\sigma^z_{c_1}\sigma^x_{c_2}\sigma^z_{c'_2}
     \Bigg]
    \nonumber\\
&& -\lambda\sum_{c,c'}\alpha_c\,\delta_{c_2,c'_1}\,
     \big(\sigma^z_{c_1}\sigma^x_{c_2}\sigma^y_{c'_2}+\sigma^y_{c_1}\sigma^x_{c_2}\sigma^z_{c'_2}\big) -\lambda\sum_{c,c'}\alpha_c\,\delta_{c_1,c'_1}\big(1-\delta_{c_2,c'_2}\big)\,
     \big(\sigma^x_{c_1}\sigma^z_{c_2}\sigma^y_{c'_2}+\sigma^y_{c_1}\sigma^z_{c_2}\sigma^x_{c'_2}\big)
    \nonumber\\
&& -\lambda\sum_{c,c'}\alpha_c\,\delta_{c_2,c'_2}\big(1-\delta_{c_1,c'_1}\big)\,
     \big(\sigma^z_{c_1}\sigma^x_{c_2}\sigma^y_{c'_1}+\sigma^y_{c_1}\sigma^x_{c_2}\sigma^z_{c'_1}\big) -2\lambda\sum_{c,c',c''}\beta_{c,c'}\,\delta_{c'_2,c''_1}\,
     \big(\sigma^z_{c_1}\sigma^x_{c_2}\sigma^y_{c''_2}+\sigma^y_{c_1}\sigma^x_{c_2}\sigma^z_{c''_2}\big)
    \nonumber\\
&& -2(1-\lambda)\sum_{c,c'}\beta_{c,c'}\,\delta_{c_2,c'_1}\,
     \big(\sigma^x_{c_2}\sigma^y_{c_1}\sigma^z_{c'_2}+\sigma^x_{c_1}\sigma^y_{c_2}\sigma^z_{c'_2}\big)
    \nonumber\\
&&-2 \, \beta^{(1)} \, \sin(2\pi \lambda) \Bigg[
\sum_{j=1}^{L} \sum_c \alpha_c \Big( \delta_{j,c_1} (\sigma^y_{j} \sigma^y_{c_2} - \sigma^z_{j} \sigma^z_{c_2}) 
+ \delta_{j,c_2} (\sigma^y_{j} \sigma^y_{c_1} - \sigma^z_{j} \sigma^z_{c_1}) \Big) \nonumber\\
&& + \sum_{c,c'} \sum_d \alpha_d \Big( 
\delta_{c_1,d_1} (1-\delta_{c_2,c'_2}) \, \sigma^y_{c_1} \sigma^z_{c_2} \sigma^z_{c'_2} \sigma^y_{d_2} 
- \delta_{c_1,d_2} (1-\delta_{c_2,c'_2}) \, \sigma^y_{d_1} \sigma^y_{c_1} \sigma^z_{c_2} \sigma^z_{c'_2} \nonumber\\
&& + \delta_{c_2,d_2} (1-\delta_{c_1,c'_1}) \, \sigma^z_{c_1} \sigma^z_{c'_1} \sigma^y_{c_2} \sigma^y_{d_1} 
- \delta_{c_2,d_1} (1-\delta_{c_1,c'_1}) \, \sigma^y_{d_2} \sigma^z_{c_1} \sigma^z_{c'_1} \sigma^y_{c_2} \nonumber\\
&& + 2 \delta_{c_2,c'_1} \delta_{c_1,d_1} \, \sigma^y_{c_1} \sigma^x_{c_2} \sigma^z_{c'_2} \sigma^y_{d_2} 
- 2 \delta_{c_2,c'_1} \delta_{c_1,d_2} \, \sigma^y_{d_1} \sigma^z_{c_1} \sigma^x_{c_2} \sigma^z_{c'_2} \Big) \nonumber\\
&& + \sum_{c,c'} \sum_{d,d'} \beta_{d,d'} \, \delta_{d_2,d'_1} \Big(
\delta_{c_1,d_1} (1-\delta_{c_2,c'_2}) \, \sigma^y_{c_1} \sigma^z_{c_2} \sigma^z_{c'_2} \sigma^x_{d_2} \sigma^y_{d'_2} 
- \delta_{c_1,d_2} (1-\delta_{c_2,c'_2}) \, \sigma^y_{d_1} \sigma^y_{c_1} \sigma^z_{c_2} \sigma^z_{c'_2} \sigma^x_{d_2} \sigma^y_{d'_2} \nonumber\\
&& + \delta_{c_2,d_2} (1-\delta_{c_1,c'_1}) \, \sigma^z_{c_1} \sigma^z_{c'_1} \sigma^y_{c_2} \sigma^x_{d_1} \sigma^y_{d'_2} 
- \delta_{c_2,d_1} (1-\delta_{c_1,c'_1}) \, \sigma^y_{d_2} \sigma^z_{c_1} \sigma^z_{c'_1} \sigma^y_{c_2} \sigma^x_{d_1} \sigma^y_{d'_2} \nonumber\\
&& + 2 \delta_{c_2,c'_1} \delta_{c_1,d_1} \, \sigma^y_{c_1} \sigma^x_{c_2} \sigma^z_{c'_2} \sigma^x_{d_2} \sigma^y_{d'_2} 
- 2 \delta_{c_2,c'_1} \delta_{c_1,d_2} \, \sigma^y_{d_1} \sigma^z_{c_1} \sigma^x_{c_2} \sigma^z_{c'_2} \sigma^x_{d_2} \sigma^y_{d'_2} \Big)
\Bigg] \nonumber\\
&& - 4  \, \beta^{(2)} \, \sin(2\pi \lambda) \Bigg[
\sum_c \sum_d \alpha_d \Big( 
\delta_{c_1,d_1} (\sigma^x_{c_1} \sigma^z_{c_2} \sigma^y_{d_2} - \sigma^x_{c_1} \sigma^y_{c_2} \sigma^z_{d_2}) 
+ \delta_{c_2,d_2} (\sigma^z_{c_1} \sigma^x_{c_2} \sigma^y_{d_1} - \sigma^y_{c_1} \sigma^x_{c_2} \sigma^z_{d_1}) \Big) \nonumber\\
&& + \sum_{c,c'} \sum_{d,d'} \beta_{d,d'} \, \delta_{d_2,d'_1} \Big(
\delta_{c_1,d_1} \, \sigma^x_{c_1} \sigma^x_{c_2} \sigma^z_{c'_2} \sigma^y_{d'_2} 
- \delta_{c_1,d_2} \, \sigma^y_{d_1} \sigma^x_{c_2} \sigma^x_{c_1} \sigma^z_{c'_2} \sigma^z_{d'_2} \nonumber\\
&& + \delta_{c_2,d_2} \, \sigma^z_{c_1} \sigma^x_{c_2} \sigma^x_{c'_1} \sigma^y_{d_1} \sigma^y_{d'_2} 
- \delta_{c_2,d_1} \, \sigma^y_{d_2} \sigma^z_{c_1} \sigma^x_{c_2} \sigma^x_{c'_1} \sigma^y_{d'_2} \Big)
\Bigg].\nonumber\\
\label{eq:G_simplified}
\end{eqnarray}

The corresponding action is given by the same terms as in the second-order case, supplemented by the proper additional terms. Denoting again the terms from the second-order expansion by $(1-\lambda)^2S_X$ and $\lambda^2S_{ZZ}$, one obtains
\begin{eqnarray}
&&\mathrm{Tr}\left[G^\dagger G\right]=\sum_c\frac{[1+8(1-\lambda)\alpha_c]^2}{4}\\
    &&+\frac{1}{2}\sum_{c,c^\prime}(2\lambda^2\alpha_c\alpha_{c^\prime}+\lambda(\alpha_c+\alpha_{c^\prime}))\left[\delta_{c_1,c^\prime_1}(1-\delta_{c_2,c^\prime_2})+(1-\delta_{c_1,c^\prime_1})\delta_{c_2,c^\prime_2}+2\delta_{c_1,c^\prime_1}\delta_{c_2,c^\prime_2}+\delta_{c_1,c^\prime_2}+\delta_{c_2,c^\prime_1}\right]\nonumber\\
    &&+\lambda^2\sum_{c,c^\prime}\alpha^2_c\left[\delta_{c_1,c^\prime_1}(1-\delta_{c_2,c^\prime_2})+(1-\delta_{c_1,c^\prime_1})\delta_{c_2,c^\prime_2}\right]\nonumber\\
    &&+16(1-\lambda)^2\sum_{c,c^\prime}\delta_{c_2,c^\prime_1}\beta^2_{c,c^\prime}+\sum_{c,c^\prime}\delta_{c_2,c^\prime_1}\left[4(1-\lambda)\beta_{c,c^\prime}-\lambda(\alpha_c+\alpha_{c^\prime})\right]^2+\lambda^2\sum_{c,c^\prime,c^{\prime\prime}}\left(\delta_{c^\prime_2,c_1}\beta^2_{c^\prime,c}\delta_{c^\prime_1,c^{\prime\prime}_2}+\delta_{c_2,c^\prime_1}\beta^2_{c,c^\prime}\delta_{c^{\prime\prime}_1,c^{\prime}_2}\right)\nonumber\\
    &&+\sum_{c}\left(\sum_{c^\prime}\lambda\delta_{c_2,c^\prime_1}\beta_{c,c^\prime}+\lambda\delta_{c^\prime_2,c_1}\beta_{c^\prime,c}-4(1-\lambda)\alpha_c\right)^2\nonumber\\
    &&+\lambda^2\sum_{c,c^\prime,c^{\prime\prime}}(\beta_{c,c^\prime}\delta_{c_2,c^\prime_1}+\beta_{c^{\prime\prime},c^\prime}\delta_{c^{\prime\prime}_2,c^\prime_1})^2(1-\delta_{c^{\prime\prime}_1,c_1})\delta_{c_2,c^{\prime\prime}_2}+(\beta_{c,c^\prime}\delta_{c_2,c^\prime_1}\delta_{c_2,c^{\prime\prime}_1}+\beta_{c,c^{\prime\prime}}\delta_{c_2,c^{\prime\prime}_1}\delta_{c_2,c^{\prime}_1})^2(1-\delta_{c^\prime_2,c^{\prime\prime}_2})\nonumber\\
    &&+ \sum_{c}\left(\left[-\tfrac{1}{2}+4\pi\cos(2\pi\lambda)\beta^{(2)}\right]^2+\left[-4\pi\cos(2\pi\lambda)\beta^{(2)}\right]^2\right)\nonumber\\
&&+ \big(6\beta^{(1)}\sin(2\pi\lambda)\big)^2 \sum_{c,d}\alpha_c\alpha_d\big(\delta_{c_1,d_1}+\delta_{c_1,d_2}+\delta_{c_2,d_1}+\delta_{c_2,d_2}\big) \\[2mm]
&&+ \big(\beta^{(1)}\sin(2\pi\lambda)\big)^2 \sum_{c,c',d,d'} \alpha_{c'}\alpha_{d'} \Big[
\delta_{c_1,c'_1}(1-\delta_{c_2,c'_2})\delta_{d_1,d'_1}(1-\delta_{d_2,d'_2})(\delta_{c_1,d_1}\delta_{c_2,d_2}\delta_{c'_2,d'_2}+\delta_{c_2,d_1}\delta_{c_1,d_2}\delta_{c'_2,d'_2}+\delta_{c'_2,d_1}\delta_{c_2,d_2}\delta_{c_1,d'_2}) \nonumber\\
&& + \delta_{c_2,c'_2}(1-\delta_{c_1,c'_1})\delta_{d_2,d'_2}(1-\delta_{d_1,d'_1})(\delta_{c_1,d_1}\delta_{c'_1,d'_1}\delta_{c_2,d_2}+\delta_{c_1,d_2}\delta_{c'_1,d'_1}\delta_{c_2,d_1}+\delta_{c'_1,d_2}\delta_{c_1,d_1}\delta_{c_2,d'_1})
\Big] \nonumber\\
&&+ \big(\beta^{(1)}\sin(2\pi\lambda)\big)^2 \sum_{c,c',d,d'} \beta_{c,c'}\beta_{d,d'} \delta_{c_2,c'_1}\delta_{d_2,d'_1}\big(\delta_{c_1,d_1}\delta_{c_2,d_2}\delta_{c'_2,d'_2}+\delta_{c_1,d_2}\delta_{c_2,d_1}\delta_{c'_2,d'_2}\big) \nonumber\\
&&+ \big(2\beta^{(2)}\sin(2\pi\lambda)\big)^2 \sum_{c,d} \alpha_c\alpha_d \big(\delta_{c_1,d_1}\delta_{c_2,d_2}+\delta_{c_1,d_2}\delta_{c_2,d_1}\big) \nonumber\\
&&+ \big(2\beta^{(2)}\sin(2\pi\lambda)\big)^2 \sum_{c,c',d,d'} \beta_{c,c'}\beta_{d,d'} \delta_{c_2,c'_1}\delta_{d_2,d'_1} \big(\delta_{c_1,d_1}\delta_{c_2,d_2}+\delta_{c_1,d_2}\delta_{c_2,d_1}\big) \nonumber\\
&&+ 2\big(\beta^{(1)}\sin(2\pi\lambda)\big)\big(2\beta^{(2)}\sin(2\pi\lambda)\big) \sum_{c,c',d,d'} \beta_{c,c'}\beta_{d,d'} \delta_{c_2,c'_1}\delta_{d_2,d'_1} \big(\delta_{c_1,d_1}\delta_{c_2,d_2}+\delta_{c_1,d_2}\delta_{c_2,d_1}\big)\nonumber\\
&&+ \sum_{c,c',d,d'} \beta_{c,c'} \beta_{d,d'} \, 2 \, \beta^{(1)} \sin(2\pi\lambda) \, \delta_{c_2,c'_1} \delta_{d_2,d'_1} \Big[
\delta_{c_1,d_1}(1-\delta_{c_2,c'_2})(1-\delta_{d_2,d'_2}) + \delta_{c_2,d_2}(1-\delta_{c_1,c'_1})(1-\delta_{d_1,d'_1}) \nonumber \\
&& + 2 \delta_{c_2,c'_1} \delta_{c_1,d_1} \delta_{d_1,d'_2} - 2 \delta_{c_2,c'_1} \delta_{c_1,d_2} \delta_{d_1,d'_2} \Big] \nonumber\\
&&+ \sum_{c,d,d'} \alpha_c \beta_{d,d'} \, 2 \, \beta^{(1)} \sin(2\pi\lambda) \Big[
\delta_{c_1,d_1} (1-\delta_{c_2,d'_2}) + \delta_{c_2,d_2} (1-\delta_{c_1,d'_1}) + \delta_{c_1,d_2} + \delta_{c_2,d_1} \Big] + \sum_c \alpha_c \, 6 \, \beta^{(1)} \sin(2\pi\lambda) \nonumber\\
&&+ \sum_{c,c'} \beta_{c,c'} \, 2 \, \beta^{(1)} \sin(2\pi\lambda) \, \delta_{c_2,c'_1} + \sum_{c,d} \alpha_c \alpha_d \, 2 \, \beta^{(2)} \sin(2\pi\lambda) \Big[
\delta_{c_1,d_1} + \delta_{c_2,d_2} - \delta_{c_1,d_2} - \delta_{c_2,d_1} \Big] \nonumber\\
&&+ \sum_{c,c',d,d'} \beta_{c,c'} \beta_{d,d'} \, 2 \, \beta^{(2)} \sin(2\pi\lambda) \, \delta_{c_2,c'_1} \delta_{d_2,d'_1} \Big[
\delta_{c_1,d_1} + \delta_{c_2,d_2} - \delta_{c_1,d_2} - \delta_{c_2,d_1} \Big] \nonumber\\
&&+ \sum_{c,d,d'} \alpha_c \beta_{d,d'} \, 2 \, \beta^{(2)} \sin(2\pi\lambda) \Big[
\delta_{c_1,d_1} + \delta_{c_2,d_2} - \delta_{c_1,d_2} - \delta_{c_2,d_1} \Big].
\end{eqnarray}

From here, the variational equations for $\alpha_c$ read
\begin{eqnarray}
&&\frac{\partial}{\partial \alpha_c} \mathrm{Tr}(G^\dagger G)=0\nonumber\\ 
&&\Rightarrow 2 (6 \beta^{(1)} \sin(2\pi \lambda))^2 \sum_d \alpha_d \big(\delta_{c_1,d_1} + \delta_{c_1,d_2} + \delta_{c_2,d_1} + \delta_{c_2,d_2}\big) \nonumber\\
&& + 2 (2 \beta^{(2)} \sin(2\pi\lambda))^2 \sum_d \alpha_d \big(\delta_{c_1,d_1}\delta_{c_2,d_2} + \delta_{c_1,d_2}\delta_{c_2,d_1}\big) \nonumber\\
&& + (\beta^{(1)} \sin(2\pi\lambda))^2 \sum_{c',d,d'} \Big[
\delta_{c,c'} \alpha_{d'} \Big( \delta_{c_1,d_1}\delta_{c_2,d_2}\delta_{c'_2,d'_2}(1-\delta_{c_2,c'_2})(1-\delta_{d_2,d'_2}) + \delta_{c_2,d_2}\delta_{c_1,d_1}\delta_{c'_1,d'_1}(1-\delta_{c_1,c'_1})(1-\delta_{d_1,d'_1}) \Big) \nonumber\\
&& + \delta_{c,d'} \alpha_{c'} \Big( \delta_{c_1,d_1}\delta_{c_2,d_2}\delta_{c'_2,d'_2}(1-\delta_{c_2,c'_2})(1-\delta_{d_2,d'_2}) + \delta_{c_2,d_2}\delta_{c_1,d_1}\delta_{c'_1,d'_1}(1-\delta_{c_1,c'_1})(1-\delta_{d_1,d'_1}) \Big)
\Big] \nonumber\\
&& + 2 \beta^{(1)} \sin(2\pi\lambda) \sum_{d,d'} \beta_{d,d'} \Big[ 
\delta_{c_1,d_1} (1-\delta_{c_2,d'_2}) + \delta_{c_2,d_2} (1-\delta_{c_1,d'_1}) + \delta_{c_1,d_2} + \delta_{c_2,d_1} \Big] \nonumber\\
&&   + 2 \beta^{(2)} \sin(2\pi\lambda) \sum_d \alpha_d \Big[ \delta_{c_1,d_1} + \delta_{c_2,d_2} - \delta_{c_1,d_2} - \delta_{c_2,d_1} \Big]  + 4 \beta^{(2)} \sin(2\pi\lambda) \sum_{d,d'} \beta_{d,d'} \Big[ \delta_{c_1,d_1} + \delta_{c_2,d_2} - \delta_{c_1,d_2} - \delta_{c_2,d_1} \Big]  \nonumber\\
&&+\left[64(1-\lambda)^2+12\lambda^2\right]\alpha_c +2\lambda^2\sum_{c^\prime}\alpha_{c^\prime}\left[\delta_{c_1,c^\prime_1}(1-\delta_{c_2,c^\prime_2})+(1-\delta_{c_1,c^\prime_1})\delta_{c_2,c^\prime_2}+2\delta_{c_1,c^\prime_2}+2\delta_{c_2,c^\prime_1}\right]\\
    &&-16\lambda(1-\lambda)\sum_{c^\prime}(\beta_{c,c^\prime}\delta_{c^\prime_1,c_2}+\beta_{c^\prime,c}\delta_{c^\prime_2,c_1})=-2(\lambda-2)-6 \beta^{(1)} \sin(2\pi\lambda).
\end{eqnarray}
On the other hand, for $\beta_{c,c'}$ one finds
\begin{eqnarray}
&&\frac{\partial}{\partial \beta_{u,v}} \mathrm{Tr}(G^\dagger G) =0\nonumber\\
&&\Rightarrow 2 (\beta^{(1)} \sin(2\pi\lambda))^2 \sum_{d,d'} \beta_{d,d'} \delta_{u_2,v_1}\delta_{d_2,d'_1} (\delta_{u_1,d_1}\delta_{v_2,d_2} + \delta_{u_1,d_2}\delta_{v_2,d_1}) \nonumber\\
&& + 2 (2 \beta^{(2)} \sin(2\pi\lambda))^2 \sum_{d,d'} \beta_{d,d'} \delta_{u_2,v_1}\delta_{d_2,d'_1} (\delta_{u_1,d_1}\delta_{v_2,d_2} + \delta_{u_1,d_2}\delta_{v_2,d_1}) \nonumber\\
&& + 4 (\beta^{(1)} \sin(2\pi\lambda)) (2 \beta^{(2)} \sin(2\pi\lambda)) \sum_{d,d'} \beta_{d,d'} \delta_{u_2,v_1}\delta_{d_2,d'_1} (\delta_{u_1,d_1}\delta_{v_2,d_2} + \delta_{u_1,d_2}\delta_{v_2,d_1}) \nonumber\\
&& + 2 \beta^{(1)} \sin(2\pi\lambda) \sum_{c,d,d'} \alpha_c \delta_{u,v=d,d'} \Big[ \delta_{c_1,d_1} (1-\delta_{c_2,d'_2}) + \delta_{c_2,d_2} (1-\delta_{c_1,d'_1}) + \delta_{c_1,d_2} + \delta_{c_2,d_1} \Big] \nonumber\\
&& + 2 \beta^{(2)} \sin(2\pi\lambda) \sum_{c,d,d'} \alpha_c \delta_{u,v=d,d'} \Big[ \delta_{c_1,d_1} + \delta_{c_2,d_2} - \delta_{c_1,d_2} - \delta_{c_2,d_1} \Big] +\left[64(1-\lambda)^2+24\lambda^2\right]\beta_{c,c^\prime}\delta_{c_2,c^\prime_1}\nonumber\\
    &&+2\lambda^2\sum_{c^{\prime\prime}}(\delta_{c^\prime_2,c^{\prime\prime}_1}\beta_{c^\prime,c^{\prime\prime}}+\delta_{c^{\prime\prime}_2,c_1}\beta_{c^{\prime\prime},c})+2\delta_{c^{\prime\prime}_2,c^\prime_1}\beta_{c^{\prime\prime},c^\prime}+\delta_{c_1,c^{\prime\prime}_1}(1-\delta_{c_2,c^{\prime\prime}_2})+2\delta_{c_2,c^{\prime\prime}_1}\beta_{c,c^{\prime\prime}}(1-\delta_{c^\prime_2,c^{\prime\prime}_2})\nonumber\\
    &&-16\lambda(1-\lambda)\delta_{c_2,c^\prime_1}
    (\alpha_c+\alpha_{c^\prime})\nonumber\\
    &&=- 2 \beta^{(1)} \sin(2\pi\lambda) \sum_{c,c',d,d'} \delta_{u,v=c,c'} \delta_{c_2,c'_1}\delta_{d_2,d'_1} \Big[
\delta_{c_1,d_1}(1-\delta_{c_2,c'_2})(1-\delta_{d_2,d'_2}) + \delta_{c_2,d_2}(1-\delta_{c_1,c'_1})(1-\delta_{d_1,d'_1}) \nonumber\\
&& + 2 \delta_{c_2,c'_1} \delta_{c_1,d_1} \delta_{d_1,d'_2} - 2 \delta_{c_2,c'_1} \delta_{c_1,d_2} \delta_{d_1,d'_2} \Big]  -2 \beta^{(1)} \sin(2\pi\lambda) \delta_{u_2,v_1}\nonumber\\
&& - 2 \beta^{(2)} \sin(2\pi\lambda) \sum_{c,c',d,d'} \delta_{u,v=c,c'} \delta_{c_2,c'_1}\delta_{d_2,d'_1} \Big[ \delta_{c_1,d_1} + \delta_{c_2,d_2} - \delta_{c_1,d_2} - \delta_{c_2,d_1} \Big].
\end{eqnarray}

\section{First-order CD ansatz for the \texorpdfstring{$3$-regular $3$-\textsc{XORSAT}}{3-regular 3-XORSAT}}
\label{app:XORSAT}

In this section, we show the steps for the derivation of the minimization equation for the $\alpha_c$ coefficients in the \textsc{XORSAT} problem. The first-order CD term already contains three spin interactions written as
\begin{eqnarray}    H^{(1)}_1=\sum_c\,\alpha_c\left(\sigma^y_{c_1}\sigma^z_{c_2}\sigma^z_{c_3}+\sigma^z_{c_1}\sigma^y_{c_2}\sigma^z_{c_3}+\sigma^z_{c_1}\sigma^z_{c_2}\sigma^y_{c_3}\right).
\end{eqnarray}
with the ordering $c_1<c_2<c_3$. The argument of the action thus takes the form of
\begin{eqnarray}
    &&G(H^{(1)}_1)=-\sum_c\frac{1+12(1-\lambda)}{2}\sigma^z_{c_1}\sigma^z_{c_2}\sigma^z_{c_3}+\sum_{c}\frac{1+3\lambda\alpha_c}{3}(\sigma^x_{c_1}+\sigma^x_{c_2}+\sigma^x_{c_3})\\
&&+\lambda\sum_{c,c^\prime}\alpha_c\left(\delta_{c_1,c^\prime_2}\delta_{c_2,c^\prime_3}\sigma^z_{c^\prime_1}(\sigma^x_{c_1}+\sigma^x_{c_2})\sigma^z_{c_3}+\delta_{c_2,c^\prime_1}\delta_{c_3,c^\prime_2}\sigma^z_{c_1}(\sigma^x_{c_2}+\sigma^x_{c_3})\sigma^z_{c^\prime_3}\right)\nonumber\\
&&+\lambda\sum_{c,c^\prime}\alpha_c\left(\delta_{c_1,c^\prime_1}\delta_{c_2,c^\prime_2}(1-\delta_{c_3,c^\prime_3})(\sigma^x_{c_1}+\sigma^x_{c_2})\sigma^z_{c_3}\sigma^z_{c^\prime_3}+\delta_{c_1,c^\prime_1}(1-\delta_{c_2,c^\prime_2})\delta_{c_3,c^\prime_3}(\sigma^x_{c_1}\sigma^z_{c_2}\sigma^z_{c^\prime_2}+\sigma^z_{c_2}\sigma^z_{c^\prime_2}\sigma^x_{c_3})\right)\nonumber\\
&&+\lambda\sum_c\alpha_c\left((1-\delta_{c_1,c^\prime_1})\delta_{c_2,c^\prime_2}\delta_{c_3,c^\prime_3}\sigma^z_{c_1}\sigma^z_{c^\prime_1}(\sigma^x_{c_2}+\sigma^x_{c_3})\right)\nonumber\\
&&+\lambda\sum_{c,c^\prime}\alpha_c\left(\delta_{c_1,c^\prime_3}\sigma^z_{c^\prime_1}\sigma^z_{c^\prime_2}\sigma^x_{c_1}\sigma^z_{c_2}\sigma^z_{c_3}+\delta_{c_3,c^\prime_1}\sigma^z_{c_1}\sigma^z_{c_2}\sigma^x_{c_3}\sigma^z_{c^\prime_2}\sigma^z_{c^\prime_3}+(1-\delta_{c_1,c^\prime_1})\delta_{c_2,c^\prime_2}(1-\delta_{c_3,c^\prime_3})\sigma^z_{c_1}\sigma^z_{c^\prime_1}\sigma^x_{c_2}\sigma^z_{c_3}\sigma^z_{c^\prime_3}\right)\nonumber\\
&&+\lambda\sum_{c,c^\prime}\alpha_c\left(\delta_{c_1,c^\prime_2}(1-\delta_{c_2,c^\prime_3})\sigma^z_{c^\prime_1}\sigma^x_{c_1}\sigma^z_{c_2}\sigma^z_{c^\prime_3}\sigma^z_{c_3}+\delta_{c_2,c^\prime_1}(1-\delta_{c_3,c^\prime_2})\sigma^z_{c_1}\sigma^x_{c_2}\sigma^z_{c^\prime_2}\sigma^z_{c_3}\sigma^z_{c^\prime_3}\right)\nonumber\\
&&+\lambda\sum_{c,c^\prime}\alpha_c\left((1-\delta_{c_1,c^\prime_2})\delta_{c_2,c^\prime_3}\sigma^z_{c^\prime_1}\sigma^z_{c_1}\sigma^z_{c^\prime_2}\sigma^x_{c_2}\sigma^z_{c_3}+(1-\delta_{c_2,c^\prime_1})\delta_{c_3,c^\prime_2}\sigma^z_{c_1}\sigma^z_{c^\prime_1}\sigma^z_{c_2}\sigma^x_{c_3}\sigma^z_{c^\prime_3}\right)\nonumber\\
&&+\lambda\sum_{c,c^\prime}\alpha_c\left(\delta_{c_1,c^\prime_1}(1-\delta_{c_2,c^\prime_2})(1-\delta_{c_3,c^\prime_3})\sigma^x_{c_1}\sigma^z_{c_2}\sigma^z_{c^\prime_2}\sigma^z_{c_3}\sigma^z_{c^\prime_3}+(1-\delta_{c_1,c^\prime_1})(1-\delta_{c_2,c^\prime_2})\delta_{c_3,c^\prime_3}\sigma^z_{c_1}\sigma^z_{c^\prime_1}\sigma^z_{c_2}\sigma^z_{c^\prime_2}\sigma^x_{c_3}\right)\nonumber\\
&&+2(1-\lambda)\sum_c\alpha_c\left(\sigma^y_{c_1}\sigma^x_{c_2}\sigma^z_{c_3}+\sigma^y_{c_1}\sigma^z_{c_2}\sigma^x_{c_3}+\sigma^x_{c_1}\sigma^y_{c_2}\sigma^z_{c_3}+\sigma^z_{c_1}\sigma^y_{c_2}\sigma^x_{c_3}+\sigma^z_{c_1}\sigma^x_{c_2}\sigma^y_{c_3}+\sigma^x_{c_1}\sigma^z_{c_2}\sigma^y_{c_3}\right).\nonumber
\end{eqnarray}
Note that the last five-body term in the $5$-th line is orthogonal to the first two due to the underlying graph structure. This is the consequence of the rule that every spin can only appear in three interactions, while matching the indices of $(1-\delta_{c_1,c^\prime_1})\delta_{c_2,c^\prime_2}(1-\delta_{c_3,c^\prime_3})\sigma^z_{c_1}\sigma^z_{c^\prime_1}\sigma^x_{c_2}\sigma^z_{c_3}\sigma^z_{c^\prime_3}$ and of $\delta_{c_1,c^\prime_3}\sigma^z_{c^\prime_1}\sigma^z_{c^\prime_2}\sigma^x_{c_1}\sigma^z_{c_2}\sigma^z_{c_3}$ in the trace norm requires the matching of the third spin by four different interactions. Similar consideration imply that also the last term in the third line, $\delta_{c_1,c^\prime_1}(1-\delta_{c_2,c^\prime_2})\delta_{c_3,c^\prime_3}(\sigma^x_{c_1}\sigma^z_{c_2}\sigma^z_{c^\prime_2}+\sigma^z_{c_2}\sigma^z_{c^\prime_2}\sigma^x_{c_3})$ is orthogonal to the first two, $\delta_{c_1,c^\prime_1}\delta_{c_2,c^\prime_2}(1-\delta_{c_3,c^\prime_3})(\sigma^x_{c_1}+\sigma^x_{c_2})\sigma^z_{c_3}\sigma^z_{c^\prime_3}$, as matching the proper indices in the trace norm would require that the $c_1$ spin appears in four different interactions.
As a result, the trace of the square reads
\begin{eqnarray}
    &&\mathrm{Tr}\left[G^\dagger G\right]=\sum_{c}\frac{(1+12(1-\lambda)\alpha_c)^2}{4}\\
    &&+\frac{1}{3}\sum_{c,c^\prime}\left(3\lambda^2\alpha_c\alpha_{c^\prime}+\lambda(\alpha_c+\alpha_{c^\prime})\right)\left[\delta_{c_1,c^\prime_1}(1-\delta_{c_2,c^\prime_2})(1-\delta_{c_3,c^\prime_3})+\delta_{c_1,c^\prime_1}\delta_{c_2,c^\prime_2}(1-\delta_{c_3,c^\prime_3})+\delta_{c_1,c^\prime_1}(1-\delta_{c_2,c^\prime_2})\delta_{c_3,c^\prime_3}\right]\nonumber\\
    &&+\frac{1}{3}\sum_{c,c^\prime}\left(3\lambda^2\alpha_c\alpha_{c^\prime}+\lambda(\alpha_c+\alpha_{c^\prime})\right)\left[(1-\delta_{c_1,c^\prime_1})\delta_{c_2,c^\prime_2}(1-\delta_{c_3,c^\prime_3})+
    (1-\delta_{c_1,c^\prime_1})\delta_{c_2,c^\prime_2}\delta_{c_3,c^\prime_3}+(1-\delta_{c_1,c^\prime_1})(1-\delta_{c_2,c^\prime_2})\delta_{c_3,c^\prime_3}\right]\nonumber\\
    &&+\frac{1}{3}\sum_{c,c^\prime}\left(3\lambda^2\alpha_c\alpha_{c^\prime}+\lambda(\alpha_c+\alpha_{c^\prime})\right)\left[\delta_{c_1c^\prime_2}+\delta_{c_1c^\prime_3}+\delta_{c_2c^\prime_3}+\delta_{c_2c^\prime_1}+\delta_{c_3c^\prime_1}+\delta_{c_3c^\prime_2}+3\delta_{c_1,c^\prime_1}\delta_{c_2,c^\prime_2}\delta_{c_3,c^\prime_3}\right]\nonumber\\
    &&+2\lambda^2\sum_{c,c^\prime}(\alpha_c+\alpha_{c^\prime})^2\delta_{c_1,c^\prime_2}\delta_{c_2,c^\prime_3}+2\lambda^2\sum_{c,c^\prime}\alpha^2_c\left(\delta_{c_1,c^\prime_1}\delta_{c_2,c^\prime_2}(1-\delta_{c_3,c^\prime_3})+\delta_{c_1,c^\prime_1}(1-\delta_{c_2,c^\prime_2})\delta_{c_3,c^\prime_3}+(1-\delta_{c_1,c^\prime_1})\delta_{c_2,c^\prime_2}\delta_{c_3,c^\prime_3}\right)\nonumber\\
    &&+\lambda^2\sum_{c,c^\prime}\left(\alpha_c+\alpha_{c^\prime}\right)^2\delta_{c_1,c^\prime_3}+\lambda^2\sum_{c,c^\prime}(\alpha_c+\alpha_{c^\prime})^2\delta_{c_1,c^\prime_2}(1-\delta_{c_2,c^\prime_3})+\lambda^2\sum_{c,c^\prime}(\alpha_c+\alpha_{c^\prime})^2\delta_{c_2,c^\prime_3}(1-\delta_{c_1,c^\prime_2})\nonumber\\
    &&+\lambda^2\sum_{c,c^\prime}\alpha^2_c\left[\delta_{c_1,c^\prime_1}(1-\delta_{c_2,c^\prime_2})(1-\delta_{c_3,c^\prime_3})+(1-\delta_{c_1,c^\prime_1})(1-\delta_{c_2,c^\prime_2})\delta_{c_3,c^\prime_3}+(1-\delta_{c_1,c^\prime_1})\delta_{c_2,c^\prime_2}(1-\delta_{c_3,c^\prime_3})\right]\nonumber\\
    &&+12(1-\lambda)^2\sum_c\alpha^2_c,\nonumber
\end{eqnarray}
where we have enlisted in the first three lines all the terms that distinguish between the $c\neq c^\prime$ and $c=c^\prime$ cases.

Although these expressions might appear slightly complicated at first glance, the terms $\alpha_c\sum_{c^\prime}\delta_{\dots}$ simplify to a constant for all $c$ as it counts how many times the given edges of an interaction $c$ appear in other interactions. Consequently, it gives a factor of $9$. Additional differentiations also induce an additional factor of $2$. The coupled $\alpha_{c^\prime}$ follow the same rule, those $\alpha_c$ and $\alpha_{c^\prime}$ are connected that share at least one spin.
\begin{eqnarray}
    &&18\lambda^2\alpha_c+96(1-\lambda)^2+2\lambda^2\sum_{c^\prime}\alpha_{c^\prime}\left[ 3\delta_{c_1,c^\prime_2}+3\delta_{c_2,c^\prime_1}+2\delta_{c_1,c^\prime_3}+2\delta_{c_3,c^\prime_1}\right]+2\lambda^2\sum_{c^\prime\neq c}\alpha_{c^\prime}\left[\delta_{c_1,c^\prime_1}+\delta_{c_2,c^\prime_2}+\delta_{c_3,c^\prime_3}\right]\nonumber\\
    &&=-6(1-\lambda)-\frac{2}{3}\lambda\sum_{c^\prime\neq c}\left[\delta_{c_1,c^\prime_1}+\delta_{c_2,c^\prime_2}+\delta_{c_3,c^\prime_3}+\delta_{c_1c^\prime_2}+\delta_{c_1c^\prime_3}+\delta_{c_2c^\prime_3}+\delta_{c_2c^\prime_1}+\delta_{c_3c^\prime_1}+\delta_{c_3c^\prime_2}\right].
\end{eqnarray}
As a final remark, the last term counts how many interactions have at least one shared spin with $c$. This final result is reported in the main text via simplified notation.
The constraint $c\neq c^\prime$ provides a more compact notation and ensures that only non-equal $\alpha_{c^\prime}$-s are considered.

\twocolumngrid
\bibliography{references}

\begin{thebibliography}{132}%
\makeatletter
\providecommand \@ifxundefined [1]{%
 \@ifx{#1\undefined}
}%
\providecommand \@ifnum [1]{%
 \ifnum #1\expandafter \@firstoftwo
 \else \expandafter \@secondoftwo
 \fi
}%
\providecommand \@ifx [1]{%
 \ifx #1\expandafter \@firstoftwo
 \else \expandafter \@secondoftwo
 \fi
}%
\providecommand \natexlab [1]{#1}%
\providecommand \enquote  [1]{``#1''}%
\providecommand \bibnamefont  [1]{#1}%
\providecommand \bibfnamefont [1]{#1}%
\providecommand \citenamefont [1]{#1}%
\providecommand \href@noop [0]{\@secondoftwo}%
\providecommand \href [0]{\begingroup \@sanitize@url \@href}%
\providecommand \@href[1]{\@@startlink{#1}\@@href}%
\providecommand \@@href[1]{\endgroup#1\@@endlink}%
\providecommand \@sanitize@url [0]{\catcode `\\12\catcode `\$12\catcode
  `\&12\catcode `\#12\catcode `\^12\catcode `\_12\catcode `\%12\relax}%
\providecommand \@@startlink[1]{}%
\providecommand \@@endlink[0]{}%
\providecommand \url  [0]{\begingroup\@sanitize@url \@url }%
\providecommand \@url [1]{\endgroup\@href {#1}{\urlprefix }}%
\providecommand \urlprefix  [0]{URL }%
\providecommand \Eprint [0]{\href }%
\providecommand \doibase [0]{https://doi.org/}%
\providecommand \selectlanguage [0]{\@gobble}%
\providecommand \bibinfo  [0]{\@secondoftwo}%
\providecommand \bibfield  [0]{\@secondoftwo}%
\providecommand \translation [1]{[#1]}%
\providecommand \BibitemOpen [0]{}%
\providecommand \bibitemStop [0]{}%
\providecommand \bibitemNoStop [0]{.\EOS\space}%
\providecommand \EOS [0]{\spacefactor3000\relax}%
\providecommand \BibitemShut  [1]{\csname bibitem#1\endcsname}%
\let\auto@bib@innerbib\@empty
\bibitem [{\citenamefont {Young}\ \emph {et~al.}(2008)\citenamefont {Young},
  \citenamefont {Knysh},\ and\ \citenamefont {Smelyanskiy}}]{Young2008Size}%
  \BibitemOpen
  \bibfield  {author} {\bibinfo {author} {\bibfnamefont {A.~P.}\ \bibnamefont
  {Young}}, \bibinfo {author} {\bibfnamefont {S.}~\bibnamefont {Knysh}},\ and\
  \bibinfo {author} {\bibfnamefont {V.~N.}\ \bibnamefont {Smelyanskiy}},\
  }\bibfield  {title} {\bibinfo {title} {{Size Dependence of the Minimum
  Excitation Gap in the Quantum Adiabatic Algorithm}},\ }\href
  {https://doi.org/10.1103/PhysRevLett.101.170503} {\bibfield  {journal}
  {\bibinfo  {journal} {Phys. Rev. Lett.}\ }\textbf {\bibinfo {volume} {101}},\
  \bibinfo {pages} {170503} (\bibinfo {year} {2008})}\BibitemShut {NoStop}%
\bibitem [{\citenamefont {Young}\ \emph {et~al.}(2010)\citenamefont {Young},
  \citenamefont {Knysh},\ and\ \citenamefont {Smelyanskiy}}]{Young2010First}%
  \BibitemOpen
  \bibfield  {author} {\bibinfo {author} {\bibfnamefont {A.~P.}\ \bibnamefont
  {Young}}, \bibinfo {author} {\bibfnamefont {S.}~\bibnamefont {Knysh}},\ and\
  \bibinfo {author} {\bibfnamefont {V.~N.}\ \bibnamefont {Smelyanskiy}},\
  }\bibfield  {title} {\bibinfo {title} {{First-Order Phase Transition in the
  Quantum Adiabatic Algorithm}},\ }\href
  {https://doi.org/10.1103/PhysRevLett.104.020502} {\bibfield  {journal}
  {\bibinfo  {journal} {Phys. Rev. Lett.}\ }\textbf {\bibinfo {volume} {104}},\
  \bibinfo {pages} {020502} (\bibinfo {year} {2010})}\BibitemShut {NoStop}%
\bibitem [{\citenamefont {Amin}(2008)}]{Amin2008Effect}%
  \BibitemOpen
  \bibfield  {author} {\bibinfo {author} {\bibfnamefont {M.~H.~S.}\
  \bibnamefont {Amin}},\ }\bibfield  {title} {\bibinfo {title} {{Effect of
  Local Minima on Adiabatic Quantum Optimization}},\ }\href
  {https://doi.org/10.1103/PhysRevLett.100.130503} {\bibfield  {journal}
  {\bibinfo  {journal} {Phys. Rev. Lett.}\ }\textbf {\bibinfo {volume} {100}},\
  \bibinfo {pages} {130503} (\bibinfo {year} {2008})}\BibitemShut {NoStop}%
\bibitem [{\citenamefont {Amin}\ and\ \citenamefont
  {Choi}(2009)}]{Amin2009First}%
  \BibitemOpen
  \bibfield  {author} {\bibinfo {author} {\bibfnamefont {M.~H.~S.}\
  \bibnamefont {Amin}}\ and\ \bibinfo {author} {\bibfnamefont {V.}~\bibnamefont
  {Choi}},\ }\bibfield  {title} {\bibinfo {title} {{First-order quantum phase
  transition in adiabatic quantum computation}},\ }\href
  {https://doi.org/10.1103/PhysRevA.80.062326} {\bibfield  {journal} {\bibinfo
  {journal} {Phys. Rev. A}\ }\textbf {\bibinfo {volume} {80}},\ \bibinfo
  {pages} {062326} (\bibinfo {year} {2009})}\BibitemShut {NoStop}%
\bibitem [{\citenamefont {Altshuler}\ \emph {et~al.}(2009)\citenamefont
  {Altshuler}, \citenamefont {Krovi},\ and\ \citenamefont
  {Roland}}]{Altshuler2009Adiabatic}%
  \BibitemOpen
  \bibfield  {author} {\bibinfo {author} {\bibfnamefont {B.}~\bibnamefont
  {Altshuler}}, \bibinfo {author} {\bibfnamefont {H.}~\bibnamefont {Krovi}},\
  and\ \bibinfo {author} {\bibfnamefont {J.}~\bibnamefont {Roland}},\
  }\href@noop {} {\bibinfo {title} {{Adiabatic quantum optimization fails for
  random instances of NP-complete problems}}} (\bibinfo {year} {2009}),\
  \Eprint {https://arxiv.org/abs/0908.2782} {arXiv:0908.2782} \BibitemShut
  {NoStop}%
\bibitem [{\citenamefont {Altshuler}\ \emph {et~al.}(2010)\citenamefont
  {Altshuler}, \citenamefont {Krovi},\ and\ \citenamefont
  {Roland}}]{Altshuler2010Anderson}%
  \BibitemOpen
  \bibfield  {author} {\bibinfo {author} {\bibfnamefont {B.}~\bibnamefont
  {Altshuler}}, \bibinfo {author} {\bibfnamefont {H.}~\bibnamefont {Krovi}},\
  and\ \bibinfo {author} {\bibfnamefont {J.}~\bibnamefont {Roland}},\
  }\bibfield  {title} {\bibinfo {title} {{Anderson localization makes adiabatic
  quantum optimization fail}},\ }\href
  {https://doi.org/10.1073/pnas.1002116107} {\bibfield  {journal} {\bibinfo
  {journal} {Proc. Natl. Acad. Sci. USA}\ }\textbf {\bibinfo {volume} {107}},\
  \bibinfo {pages} {12446} (\bibinfo {year} {2010})}\BibitemShut {NoStop}%
\bibitem [{\citenamefont {Marzlin}\ and\ \citenamefont
  {Sanders}(2004)}]{Marzlin2004Inconsistency}%
  \BibitemOpen
  \bibfield  {author} {\bibinfo {author} {\bibfnamefont {K.-P.}\ \bibnamefont
  {Marzlin}}\ and\ \bibinfo {author} {\bibfnamefont {B.~C.}\ \bibnamefont
  {Sanders}},\ }\bibfield  {title} {\bibinfo {title} {{Inconsistency in the
  Application of the Adiabatic Theorem}},\ }\href
  {https://doi.org/10.1103/PhysRevLett.93.160408} {\bibfield  {journal}
  {\bibinfo  {journal} {Phys. Rev. Lett.}\ }\textbf {\bibinfo {volume} {93}},\
  \bibinfo {pages} {160408} (\bibinfo {year} {2004})}\BibitemShut {NoStop}%
\bibitem [{\citenamefont {Jansen}\ \emph {et~al.}(2007)\citenamefont {Jansen},
  \citenamefont {Ruskai},\ and\ \citenamefont {Seiler}}]{Jansen2007Bounds}%
  \BibitemOpen
  \bibfield  {author} {\bibinfo {author} {\bibfnamefont {S.}~\bibnamefont
  {Jansen}}, \bibinfo {author} {\bibfnamefont {M.-B.}\ \bibnamefont {Ruskai}},\
  and\ \bibinfo {author} {\bibfnamefont {R.}~\bibnamefont {Seiler}},\
  }\bibfield  {title} {\bibinfo {title} {{Bounds for the adiabatic
  approximation with applications to quantum computation}},\ }\href
  {https://doi.org/10.1063/1.2798382} {\bibfield  {journal} {\bibinfo
  {journal} {J. Math. Phys.}\ }\textbf {\bibinfo {volume} {48}},\ \bibinfo
  {pages} {102111} (\bibinfo {year} {2007})}\BibitemShut {NoStop}%
\bibitem [{\citenamefont {Laumann}\ \emph {et~al.}(2015)\citenamefont
  {Laumann}, \citenamefont {Moessner}, \citenamefont {Scardicchio},\ and\
  \citenamefont {Sondhi}}]{Laumann2015Quantum}%
  \BibitemOpen
  \bibfield  {author} {\bibinfo {author} {\bibfnamefont {C.~R.}\ \bibnamefont
  {Laumann}}, \bibinfo {author} {\bibfnamefont {R.}~\bibnamefont {Moessner}},
  \bibinfo {author} {\bibfnamefont {A.}~\bibnamefont {Scardicchio}},\ and\
  \bibinfo {author} {\bibfnamefont {S.~L.}\ \bibnamefont {Sondhi}},\ }\bibfield
   {title} {\bibinfo {title} {{Quantum annealing: The fastest route to quantum
  computation?}},\ }\href {https://doi.org/10.1140/epjst/e2015-02344-2}
  {\bibfield  {journal} {\bibinfo  {journal} {Eur. Phys. J. Special Topics}\
  }\textbf {\bibinfo {volume} {224}},\ \bibinfo {pages} {75} (\bibinfo {year}
  {2015})}\BibitemShut {NoStop}%
\bibitem [{\citenamefont {Knysh}\ and\ \citenamefont
  {Smelyanskiy}(2010)}]{Knysh2010Relevance}%
  \BibitemOpen
  \bibfield  {author} {\bibinfo {author} {\bibfnamefont {S.}~\bibnamefont
  {Knysh}}\ and\ \bibinfo {author} {\bibfnamefont {V.}~\bibnamefont
  {Smelyanskiy}},\ }\href@noop {} {\bibinfo {title} {{On the relevance of
  avoided crossings away from quantum critical point to the complexity of
  quantum adiabatic algorithm}}} (\bibinfo {year} {2010}),\ \Eprint
  {https://arxiv.org/abs/1005.3011} {arXiv:1005.3011} \BibitemShut {NoStop}%
\bibitem [{\citenamefont {Foini}\ \emph {et~al.}(2010)\citenamefont {Foini},
  \citenamefont {Semerjian},\ and\ \citenamefont
  {Zamponi}}]{Foini2010Solvable}%
  \BibitemOpen
  \bibfield  {author} {\bibinfo {author} {\bibfnamefont {L.}~\bibnamefont
  {Foini}}, \bibinfo {author} {\bibfnamefont {G.}~\bibnamefont {Semerjian}},\
  and\ \bibinfo {author} {\bibfnamefont {F.}~\bibnamefont {Zamponi}},\
  }\bibfield  {title} {\bibinfo {title} {{Solvable Model of Quantum Random
  Optimization Problems}},\ }\href
  {https://doi.org/10.1103/PhysRevLett.105.167204} {\bibfield  {journal}
  {\bibinfo  {journal} {Phys. Rev. Lett.}\ }\textbf {\bibinfo {volume} {105}},\
  \bibinfo {pages} {167204} (\bibinfo {year} {2010})}\BibitemShut {NoStop}%
\bibitem [{\citenamefont {J\"org}\ \emph {et~al.}(2008)\citenamefont {J\"org},
  \citenamefont {Krzakala}, \citenamefont {Kurchan},\ and\ \citenamefont
  {Maggs}}]{Jorg2008Simple}%
  \BibitemOpen
  \bibfield  {author} {\bibinfo {author} {\bibfnamefont {T.}~\bibnamefont
  {J\"org}}, \bibinfo {author} {\bibfnamefont {F.}~\bibnamefont {Krzakala}},
  \bibinfo {author} {\bibfnamefont {J.}~\bibnamefont {Kurchan}},\ and\ \bibinfo
  {author} {\bibfnamefont {A.~C.}\ \bibnamefont {Maggs}},\ }\bibfield  {title}
  {\bibinfo {title} {{Simple Glass Models and Their Quantum Annealing}},\
  }\href {https://doi.org/10.1103/PhysRevLett.101.147204} {\bibfield  {journal}
  {\bibinfo  {journal} {Phys. Rev. Lett.}\ }\textbf {\bibinfo {volume} {101}},\
  \bibinfo {pages} {147204} (\bibinfo {year} {2008})}\BibitemShut {NoStop}%
\bibitem [{\citenamefont {J\"org}\ \emph {et~al.}(2010)\citenamefont {J\"org},
  \citenamefont {Krzakala}, \citenamefont {Semerjian},\ and\ \citenamefont
  {Zamponi}}]{Jorg2010First}%
  \BibitemOpen
  \bibfield  {author} {\bibinfo {author} {\bibfnamefont {T.}~\bibnamefont
  {J\"org}}, \bibinfo {author} {\bibfnamefont {F.}~\bibnamefont {Krzakala}},
  \bibinfo {author} {\bibfnamefont {G.}~\bibnamefont {Semerjian}},\ and\
  \bibinfo {author} {\bibfnamefont {F.}~\bibnamefont {Zamponi}},\ }\bibfield
  {title} {\bibinfo {title} {{First-Order Transitions and the Performance of
  Quantum Algorithms in Random Optimization Problems}},\ }\href
  {https://doi.org/10.1103/PhysRevLett.104.207206} {\bibfield  {journal}
  {\bibinfo  {journal} {Phys. Rev. Lett.}\ }\textbf {\bibinfo {volume} {104}},\
  \bibinfo {pages} {207206} (\bibinfo {year} {2010})}\BibitemShut {NoStop}%
\bibitem [{\citenamefont {Bachmann}\ \emph {et~al.}(2017)\citenamefont
  {Bachmann}, \citenamefont {De~Roeck},\ and\ \citenamefont
  {Fraas}}]{Bachmann17}%
  \BibitemOpen
  \bibfield  {author} {\bibinfo {author} {\bibfnamefont {S.}~\bibnamefont
  {Bachmann}}, \bibinfo {author} {\bibfnamefont {W.}~\bibnamefont {De~Roeck}},\
  and\ \bibinfo {author} {\bibfnamefont {M.}~\bibnamefont {Fraas}},\ }\bibfield
   {title} {\bibinfo {title} {{Adiabatic Theorem for Quantum Spin Systems}},\
  }\href {https://doi.org/10.1103/PhysRevLett.119.060201} {\bibfield  {journal}
  {\bibinfo  {journal} {Phys. Rev. Lett.}\ }\textbf {\bibinfo {volume} {119}},\
  \bibinfo {pages} {060201} (\bibinfo {year} {2017})}\BibitemShut {NoStop}%
\bibitem [{\citenamefont {Bachmann}\ \emph {et~al.}(2018)\citenamefont
  {Bachmann}, \citenamefont {De~Roeck},\ and\ \citenamefont
  {Fraas}}]{Bachmann2018}%
  \BibitemOpen
  \bibfield  {author} {\bibinfo {author} {\bibfnamefont {S.}~\bibnamefont
  {Bachmann}}, \bibinfo {author} {\bibfnamefont {W.}~\bibnamefont {De~Roeck}},\
  and\ \bibinfo {author} {\bibfnamefont {M.}~\bibnamefont {Fraas}},\ }\bibfield
   {title} {\bibinfo {title} {{The Adiabatic Theorem and Linear Response Theory
  for Extended Quantum Systems}},\ }\href
  {https://doi.org/10.1007/s00220-018-3117-9} {\bibfield  {journal} {\bibinfo
  {journal} {Commun. Math. Phys.}\ }\textbf {\bibinfo {volume} {361}},\
  \bibinfo {pages} {997} (\bibinfo {year} {2018})}\BibitemShut {NoStop}%
\bibitem [{\citenamefont {Hastings}(2004)}]{Hastings04}%
  \BibitemOpen
  \bibfield  {author} {\bibinfo {author} {\bibfnamefont {M.~B.}\ \bibnamefont
  {Hastings}},\ }\bibfield  {title} {\bibinfo {title} {{Locality in Quantum and
  Markov Dynamics on Lattices and Networks}},\ }\href
  {https://doi.org/10.1103/PhysRevLett.93.140402} {\bibfield  {journal}
  {\bibinfo  {journal} {Phys. Rev. Lett.}\ }\textbf {\bibinfo {volume} {93}},\
  \bibinfo {pages} {140402} (\bibinfo {year} {2004})}\BibitemShut {NoStop}%
\bibitem [{\citenamefont {Hastings}(2007)}]{Hastings2007Quasi}%
  \BibitemOpen
  \bibfield  {author} {\bibinfo {author} {\bibfnamefont {M.~B.}\ \bibnamefont
  {Hastings}},\ }\bibfield  {title} {\bibinfo {title} {{Quasi-adiabatic
  continuation in gapped spin and fermion systems: Goldstone’s theorem and
  flux periodicity}},\ }\href
  {https://doi.org/10.1088/1742-5468/2007/05/P05010} {\bibfield  {journal}
  {\bibinfo  {journal} {J. Stat. Mech.: Theory Exp.}\ }\textbf {\bibinfo
  {volume} {2007}}\bibinfo  {number} { (05)},\ \bibinfo {pages}
  {P05010}}\BibitemShut {NoStop}%
\bibitem [{\citenamefont {Hastings}(2012)}]{Hastings2010Locality}%
  \BibitemOpen
\bibfield  {number} {  }\bibfield  {author} {\bibinfo {author} {\bibfnamefont
  {M.~B.}\ \bibnamefont {Hastings}},\ }\bibfield  {title} {\bibinfo {title}
  {{Locality in quantum systems}},\ }in\ \href
  {https://doi.org/10.1093/acprof:oso/9780199652495.003.0003} {\emph {\bibinfo
  {booktitle} {{Quantum Theory from Small to Large Scales: Lecture Notes of the
  Les Houches Summer School: Volume 95, August 2010}}}}\ (\bibinfo  {publisher}
  {Oxford University Press},\ \bibinfo {year} {2012})\BibitemShut {NoStop}%
\bibitem [{\citenamefont {Hastings}(2010)}]{Hastings2010Quasi}%
  \BibitemOpen
  \bibfield  {author} {\bibinfo {author} {\bibfnamefont {M.~B.}\ \bibnamefont
  {Hastings}},\ }\href@noop {} {\bibinfo {title} {{Quasi-adiabatic Continuation
  for Disordered Systems: Applications to Correlations, Lieb-Schultz-Mattis,
  and Hall Conductance}}} (\bibinfo {year} {2010}),\ \Eprint
  {https://arxiv.org/abs/1001.5280} {arXiv:1001.5280} \BibitemShut {NoStop}%
\bibitem [{\citenamefont {Hastings}\ and\ \citenamefont
  {Wen}(2005)}]{Hastings05}%
  \BibitemOpen
  \bibfield  {author} {\bibinfo {author} {\bibfnamefont {M.~B.}\ \bibnamefont
  {Hastings}}\ and\ \bibinfo {author} {\bibfnamefont {X.-G.}\ \bibnamefont
  {Wen}},\ }\bibfield  {title} {\bibinfo {title} {{Quasiadiabatic continuation
  of quantum states: The stability of topological ground-state degeneracy and
  emergent gauge invariance}},\ }\href
  {https://doi.org/10.1103/PhysRevB.72.045141} {\bibfield  {journal} {\bibinfo
  {journal} {Phys. Rev. B}\ }\textbf {\bibinfo {volume} {72}},\ \bibinfo
  {pages} {045141} (\bibinfo {year} {2005})}\BibitemShut {NoStop}%
\bibitem [{\citenamefont {Hastings}\ and\ \citenamefont
  {Michalakis}(2015)}]{Hastings2015Quantization}%
  \BibitemOpen
  \bibfield  {author} {\bibinfo {author} {\bibfnamefont {M.~B.}\ \bibnamefont
  {Hastings}}\ and\ \bibinfo {author} {\bibfnamefont {S.}~\bibnamefont
  {Michalakis}},\ }\bibfield  {title} {\bibinfo {title} {{Quantization of Hall
  Conductance for Interacting Electrons on a Torus}},\ }\href
  {https://doi.org/10.1007/s00220-014-2167-x} {\bibfield  {journal} {\bibinfo
  {journal} {Commun. Math. Phys.}\ }\textbf {\bibinfo {volume} {334}},\
  \bibinfo {pages} {433} (\bibinfo {year} {2015})}\BibitemShut {NoStop}%
\bibitem [{\citenamefont {De~Roeck}\ and\ \citenamefont
  {Schütz}(2015)}]{DeRoeck15}%
  \BibitemOpen
  \bibfield  {author} {\bibinfo {author} {\bibfnamefont {W.}~\bibnamefont
  {De~Roeck}}\ and\ \bibinfo {author} {\bibfnamefont {M.}~\bibnamefont
  {Schütz}},\ }\bibfield  {title} {\bibinfo {title} {{Local perturbations
  perturb—exponentially–locally}},\ }\href
  {https://doi.org/10.1063/1.4922507} {\bibfield  {journal} {\bibinfo
  {journal} {J. Math. Phys.}\ }\textbf {\bibinfo {volume} {56}},\ \bibinfo
  {pages} {061901} (\bibinfo {year} {2015})}\BibitemShut {NoStop}%
\bibitem [{\citenamefont {Farhi}\ \emph {et~al.}(2011)\citenamefont {Farhi},
  \citenamefont {Goldston}, \citenamefont {Gosset}, \citenamefont {Gutmann},
  \citenamefont {Meyer},\ and\ \citenamefont {Shor}}]{Farhi2011Quantum}%
  \BibitemOpen
  \bibfield  {author} {\bibinfo {author} {\bibfnamefont {E.}~\bibnamefont
  {Farhi}}, \bibinfo {author} {\bibfnamefont {J.}~\bibnamefont {Goldston}},
  \bibinfo {author} {\bibfnamefont {D.}~\bibnamefont {Gosset}}, \bibinfo
  {author} {\bibfnamefont {S.}~\bibnamefont {Gutmann}}, \bibinfo {author}
  {\bibfnamefont {H.~B.}\ \bibnamefont {Meyer}},\ and\ \bibinfo {author}
  {\bibfnamefont {P.}~\bibnamefont {Shor}},\ }\bibfield  {title} {\bibinfo
  {title} {{Quantum adiabatic algorithms, small gaps, and different paths}},\
  }\href {https://doi.org/10.5555/2011395.2011396} {\bibfield  {journal}
  {\bibinfo  {journal} {Quantum Info. Comput.}\ }\textbf {\bibinfo {volume}
  {11}},\ \bibinfo {pages} {181} (\bibinfo {year} {2011})}\BibitemShut
  {NoStop}%
\bibitem [{\citenamefont {Choi}(2011{\natexlab{a}})}]{Choi2011Avoid}%
  \BibitemOpen
  \bibfield  {author} {\bibinfo {author} {\bibfnamefont {V.}~\bibnamefont
  {Choi}},\ }\href@noop {} {\bibinfo {title} {{Avoid First Order Quantum Phase
  Transition by Changing Problem Hamiltonians}}} (\bibinfo {year}
  {2011}{\natexlab{a}}),\ \Eprint {https://arxiv.org/abs/1010.1220}
  {arXiv:1010.1220} \BibitemShut {NoStop}%
\bibitem [{\citenamefont {Choi}(2011{\natexlab{b}})}]{Choi2011Different}%
  \BibitemOpen
  \bibfield  {author} {\bibinfo {author} {\bibfnamefont {V.}~\bibnamefont
  {Choi}},\ }\href@noop {} {\bibinfo {title} {{Different Adiabatic Quantum
  Optimization Algorithms for the NP-Complete Exact Cover and 3SAT Problems}}}
  (\bibinfo {year} {2011}{\natexlab{b}}),\ \Eprint
  {https://arxiv.org/abs/1010.1221} {arXiv:1010.1221} \BibitemShut {NoStop}%
\bibitem [{\citenamefont {Choi}(2011{\natexlab{c}})}]{Choi2011Different2}%
  \BibitemOpen
  \bibfield  {author} {\bibinfo {author} {\bibfnamefont {V.}~\bibnamefont
  {Choi}},\ }\bibfield  {title} {\bibinfo {title} {{Different adiabatic quantum
  optimization algorithms for the NP-complete exact cover problem}},\ }\href
  {https://doi.org/10.1073/pnas.1018310108} {\bibfield  {journal} {\bibinfo
  {journal} {Proc. Natl. Acad. Sci. (USA)}\ }\textbf {\bibinfo {volume}
  {108}},\ \bibinfo {pages} {E19} (\bibinfo {year}
  {2011}{\natexlab{c}})}\BibitemShut {NoStop}%
\bibitem [{\citenamefont {Dickson}(2011)}]{Dickson2011Elimination}%
  \BibitemOpen
  \bibfield  {author} {\bibinfo {author} {\bibfnamefont {N.~G.}\ \bibnamefont
  {Dickson}},\ }\bibfield  {title} {\bibinfo {title} {{Elimination of
  perturbative crossings in adiabatic quantum optimization}},\ }\href
  {https://doi.org/10.1088/1367-2630/13/7/073011} {\bibfield  {journal}
  {\bibinfo  {journal} {New J. Phys.}\ }\textbf {\bibinfo {volume} {13}},\
  \bibinfo {pages} {073011} (\bibinfo {year} {2011})}\BibitemShut {NoStop}%
\bibitem [{\citenamefont {Dickson}\ and\ \citenamefont
  {Amin}(2011)}]{Dickson2011Does}%
  \BibitemOpen
  \bibfield  {author} {\bibinfo {author} {\bibfnamefont {N.~G.}\ \bibnamefont
  {Dickson}}\ and\ \bibinfo {author} {\bibfnamefont {M.~H.~S.}\ \bibnamefont
  {Amin}},\ }\bibfield  {title} {\bibinfo {title} {{Does Adiabatic Quantum
  Optimization Fail for NP-Complete Problems?}},\ }\href
  {https://doi.org/10.1103/PhysRevLett.106.050502} {\bibfield  {journal}
  {\bibinfo  {journal} {Phys. Rev. Lett.}\ }\textbf {\bibinfo {volume} {106}},\
  \bibinfo {pages} {050502} (\bibinfo {year} {2011})}\BibitemShut {NoStop}%
\bibitem [{\citenamefont {Choi}(2020)}]{Choi2020Effects}%
  \BibitemOpen
  \bibfield  {author} {\bibinfo {author} {\bibfnamefont {V.}~\bibnamefont
  {Choi}},\ }\bibfield  {title} {\bibinfo {title} {{The effects of the problem
  Hamiltonian parameters on the minimum spectral gap in adiabatic quantum
  optimization}},\ }\href {https://doi.org/10.1007/s11128-020-2582-1}
  {\bibfield  {journal} {\bibinfo  {journal} {Quantum Info. Processing}\
  }\textbf {\bibinfo {volume} {19}},\ \bibinfo {pages} {90} (\bibinfo {year}
  {2020})}\BibitemShut {NoStop}%
\bibitem [{\citenamefont {Suzuki}\ \emph {et~al.}(2007)\citenamefont {Suzuki},
  \citenamefont {Nishimori},\ and\ \citenamefont
  {Suzuki}}]{Suzuki_2007_originalFerroGapamplification}%
  \BibitemOpen
  \bibfield  {author} {\bibinfo {author} {\bibfnamefont {S.}~\bibnamefont
  {Suzuki}}, \bibinfo {author} {\bibfnamefont {H.}~\bibnamefont {Nishimori}},\
  and\ \bibinfo {author} {\bibfnamefont {M.}~\bibnamefont {Suzuki}},\
  }\bibfield  {title} {\bibinfo {title} {{Quantum annealing of the random-field
  Ising model by transverse ferromagnetic interactions}},\ }\href
  {https://doi.org/10.1103/PhysRevE.75.051112} {\bibfield  {journal} {\bibinfo
  {journal} {Phys. Rev. E}\ }\textbf {\bibinfo {volume} {75}},\ \bibinfo
  {pages} {051112} (\bibinfo {year} {2007})}\BibitemShut {NoStop}%
\bibitem [{\citenamefont {Seki}\ and\ \citenamefont
  {Nishimori}(2012)}]{Seki_2012_AntiFerroFluct_Gapamplification}%
  \BibitemOpen
  \bibfield  {author} {\bibinfo {author} {\bibfnamefont {Y.}~\bibnamefont
  {Seki}}\ and\ \bibinfo {author} {\bibfnamefont {H.}~\bibnamefont
  {Nishimori}},\ }\bibfield  {title} {\bibinfo {title} {{Quantum annealing with
  antiferromagnetic fluctuations}},\ }\href
  {https://doi.org/10.1103/PhysRevE.85.051112} {\bibfield  {journal} {\bibinfo
  {journal} {Phys. Rev. E}\ }\textbf {\bibinfo {volume} {85}},\ \bibinfo
  {pages} {051112} (\bibinfo {year} {2012})}\BibitemShut {NoStop}%
\bibitem [{\citenamefont {Seoane}\ and\ \citenamefont
  {Nishimori}(2012)}]{Seoane_2012_ClassicalMEthodforGapAmplfiication}%
  \BibitemOpen
  \bibfield  {author} {\bibinfo {author} {\bibfnamefont {B.}~\bibnamefont
  {Seoane}}\ and\ \bibinfo {author} {\bibfnamefont {H.}~\bibnamefont
  {Nishimori}},\ }\bibfield  {title} {\bibinfo {title} {{Many-body transverse
  interactions in the quantum annealing of the p-spin ferromagnet}},\ }\href
  {https://doi.org/10.1088/1751-8113/45/43/435301} {\bibfield  {journal}
  {\bibinfo  {journal} {J. Phys. A}\ }\textbf {\bibinfo {volume} {45}},\
  \bibinfo {pages} {435301} (\bibinfo {year} {2012})}\BibitemShut {NoStop}%
\bibitem [{\citenamefont {Somma}\ and\ \citenamefont
  {Boixo}(2013)}]{Somma2013Spectral}%
  \BibitemOpen
  \bibfield  {author} {\bibinfo {author} {\bibfnamefont {R.~D.}\ \bibnamefont
  {Somma}}\ and\ \bibinfo {author} {\bibfnamefont {S.}~\bibnamefont {Boixo}},\
  }\bibfield  {title} {\bibinfo {title} {{Spectral Gap Amplification}},\ }\href
  {https://doi.org/10.1137/120871997} {\bibfield  {journal} {\bibinfo
  {journal} {SIAM J. Comp.}\ }\textbf {\bibinfo {volume} {42}},\ \bibinfo
  {pages} {593} (\bibinfo {year} {2013})}\BibitemShut {NoStop}%
\bibitem [{\citenamefont {Crosson}\ \emph {et~al.}(2014)\citenamefont
  {Crosson}, \citenamefont {Farhi}, \citenamefont {Lin}, \citenamefont {Lin},\
  and\ \citenamefont {Shor}}]{CrossonFarhi2014Gap_amplication}%
  \BibitemOpen
  \bibfield  {author} {\bibinfo {author} {\bibfnamefont {E.}~\bibnamefont
  {Crosson}}, \bibinfo {author} {\bibfnamefont {E.}~\bibnamefont {Farhi}},
  \bibinfo {author} {\bibfnamefont {C.~Y.-Y.}\ \bibnamefont {Lin}}, \bibinfo
  {author} {\bibfnamefont {H.-H.}\ \bibnamefont {Lin}},\ and\ \bibinfo {author}
  {\bibfnamefont {P.}~\bibnamefont {Shor}},\ }\href@noop {} {\bibinfo {title}
  {{Different Strategies for Optimization Using the Quantum Adiabatic
  Algorithm}}} (\bibinfo {year} {2014}),\ \Eprint
  {https://arxiv.org/abs/1401.7320} {arXiv:1401.7320 [quant-ph]} \BibitemShut
  {NoStop}%
\bibitem [{\citenamefont {Hormozi}\ \emph {et~al.}(2017)\citenamefont
  {Hormozi}, \citenamefont {Brown}, \citenamefont {Carleo},\ and\ \citenamefont
  {Troyer}}]{Hormozi_NonStoqCatalyst_gapAmplificatio_2017}%
  \BibitemOpen
  \bibfield  {author} {\bibinfo {author} {\bibfnamefont {L.}~\bibnamefont
  {Hormozi}}, \bibinfo {author} {\bibfnamefont {E.~W.}\ \bibnamefont {Brown}},
  \bibinfo {author} {\bibfnamefont {G.}~\bibnamefont {Carleo}},\ and\ \bibinfo
  {author} {\bibfnamefont {M.}~\bibnamefont {Troyer}},\ }\bibfield  {title}
  {\bibinfo {title} {{Nonstoquastic Hamiltonians and quantum annealing of an
  Ising spin glass}},\ }\href {https://doi.org/10.1103/PhysRevB.95.184416}
  {\bibfield  {journal} {\bibinfo  {journal} {Phys. Rev. B}\ }\textbf {\bibinfo
  {volume} {95}},\ \bibinfo {pages} {184416} (\bibinfo {year}
  {2017})}\BibitemShut {NoStop}%
\bibitem [{\citenamefont {Susa}\ \emph {et~al.}(2018)\citenamefont {Susa},
  \citenamefont {Yamashiro}, \citenamefont {Yamamoto},\ and\ \citenamefont
  {Nishimori}}]{Susa2018_Inhomdriving_gap_amplification}%
  \BibitemOpen
  \bibfield  {author} {\bibinfo {author} {\bibfnamefont {Y.}~\bibnamefont
  {Susa}}, \bibinfo {author} {\bibfnamefont {Y.}~\bibnamefont {Yamashiro}},
  \bibinfo {author} {\bibfnamefont {M.}~\bibnamefont {Yamamoto}},\ and\
  \bibinfo {author} {\bibfnamefont {H.}~\bibnamefont {Nishimori}},\ }\bibfield
  {title} {\bibinfo {title} {{Exponential Speedup of Quantum Annealing by
  Inhomogeneous Driving of the Transverse Field}},\ }\href
  {https://doi.org/10.7566/JPSJ.87.023002} {\bibfield  {journal} {\bibinfo
  {journal} {J. Phys. Soc. Japan}\ }\textbf {\bibinfo {volume} {87}},\ \bibinfo
  {pages} {023002} (\bibinfo {year} {2018})}\BibitemShut {NoStop}%
\bibitem [{\citenamefont {Albash}(2019)}]{Albash2019_NonStoqGap_amplification}%
  \BibitemOpen
  \bibfield  {author} {\bibinfo {author} {\bibfnamefont {T.}~\bibnamefont
  {Albash}},\ }\bibfield  {title} {\bibinfo {title} {{Role of nonstoquastic
  catalysts in quantum adiabatic optimization}},\ }\href
  {https://doi.org/10.1103/PhysRevA.99.042334} {\bibfield  {journal} {\bibinfo
  {journal} {Phys. Rev. A}\ }\textbf {\bibinfo {volume} {99}},\ \bibinfo
  {pages} {042334} (\bibinfo {year} {2019})}\BibitemShut {NoStop}%
\bibitem [{\citenamefont {Cao}\ \emph {et~al.}(2021)\citenamefont {Cao},
  \citenamefont {Xue}, \citenamefont {Shannon},\ and\ \citenamefont
  {Joynt}}]{Cao2021Catatlyst}%
  \BibitemOpen
  \bibfield  {author} {\bibinfo {author} {\bibfnamefont {C.}~\bibnamefont
  {Cao}}, \bibinfo {author} {\bibfnamefont {J.}~\bibnamefont {Xue}}, \bibinfo
  {author} {\bibfnamefont {N.}~\bibnamefont {Shannon}},\ and\ \bibinfo {author}
  {\bibfnamefont {R.}~\bibnamefont {Joynt}},\ }\bibfield  {title} {\bibinfo
  {title} {{Speedup of the quantum adiabatic algorithm using delocalization
  catalysis}},\ }\href {https://doi.org/10.1103/PhysRevResearch.3.013092}
  {\bibfield  {journal} {\bibinfo  {journal} {Phys. Rev. Res.}\ }\textbf
  {\bibinfo {volume} {3}},\ \bibinfo {pages} {013092} (\bibinfo {year}
  {2021})}\BibitemShut {NoStop}%
\bibitem [{\citenamefont {Mehta}\ \emph {et~al.}(2021)\citenamefont {Mehta},
  \citenamefont {Jin}, \citenamefont {De~Raedt},\ and\ \citenamefont
  {Michielsen}}]{Mehta_2SAT_NonSotq_Gapamplification}%
  \BibitemOpen
  \bibfield  {author} {\bibinfo {author} {\bibfnamefont {V.}~\bibnamefont
  {Mehta}}, \bibinfo {author} {\bibfnamefont {F.}~\bibnamefont {Jin}}, \bibinfo
  {author} {\bibfnamefont {H.}~\bibnamefont {De~Raedt}},\ and\ \bibinfo
  {author} {\bibfnamefont {K.}~\bibnamefont {Michielsen}},\ }\bibfield  {title}
  {\bibinfo {title} {{Quantum annealing with trigger Hamiltonians: Application
  to 2-satisfiability and nonstoquastic problems}},\ }\href
  {https://doi.org/10.1103/PhysRevA.104.032421} {\bibfield  {journal} {\bibinfo
   {journal} {Phys. Rev. A}\ }\textbf {\bibinfo {volume} {104}},\ \bibinfo
  {pages} {032421} (\bibinfo {year} {2021})}\BibitemShut {NoStop}%
\bibitem [{\citenamefont {Farhi}\ \emph {et~al.}(2014)\citenamefont {Farhi},
  \citenamefont {Goldstone},\ and\ \citenamefont {Gutmann}}]{Farhi2014Quantum}%
  \BibitemOpen
  \bibfield  {author} {\bibinfo {author} {\bibfnamefont {E.}~\bibnamefont
  {Farhi}}, \bibinfo {author} {\bibfnamefont {J.}~\bibnamefont {Goldstone}},\
  and\ \bibinfo {author} {\bibfnamefont {S.}~\bibnamefont {Gutmann}},\
  }\href@noop {} {\bibinfo {title} {{A Quantum Approximate Optimization
  Algorithm}}} (\bibinfo {year} {2014}),\ \Eprint
  {https://arxiv.org/abs/1411.4028} {arXiv:1411.4028} \BibitemShut {NoStop}%
\bibitem [{\citenamefont {Zhou}\ \emph {et~al.}(2020)\citenamefont {Zhou},
  \citenamefont {Wang}, \citenamefont {Choi}, \citenamefont {Pichler},\ and\
  \citenamefont {Lukin}}]{Zhou2020Quantum}%
  \BibitemOpen
  \bibfield  {author} {\bibinfo {author} {\bibfnamefont {L.}~\bibnamefont
  {Zhou}}, \bibinfo {author} {\bibfnamefont {S.-T.}\ \bibnamefont {Wang}},
  \bibinfo {author} {\bibfnamefont {S.}~\bibnamefont {Choi}}, \bibinfo {author}
  {\bibfnamefont {H.}~\bibnamefont {Pichler}},\ and\ \bibinfo {author}
  {\bibfnamefont {M.~D.}\ \bibnamefont {Lukin}},\ }\bibfield  {title} {\bibinfo
  {title} {{Quantum Approximate Optimization Algorithm: Performance, Mechanism,
  and Implementation on Near-Term Devices}},\ }\href
  {https://doi.org/10.1103/PhysRevX.10.021067} {\bibfield  {journal} {\bibinfo
  {journal} {Phys. Rev. X}\ }\textbf {\bibinfo {volume} {10}},\ \bibinfo
  {pages} {021067} (\bibinfo {year} {2020})}\BibitemShut {NoStop}%
\bibitem [{\citenamefont {Yang}\ \emph {et~al.}(2017)\citenamefont {Yang},
  \citenamefont {Rahmani}, \citenamefont {Shabani}, \citenamefont {Neven},\
  and\ \citenamefont {Chamon}}]{Yang2017Optimizing}%
  \BibitemOpen
  \bibfield  {author} {\bibinfo {author} {\bibfnamefont {Z.-C.}\ \bibnamefont
  {Yang}}, \bibinfo {author} {\bibfnamefont {A.}~\bibnamefont {Rahmani}},
  \bibinfo {author} {\bibfnamefont {A.}~\bibnamefont {Shabani}}, \bibinfo
  {author} {\bibfnamefont {H.}~\bibnamefont {Neven}},\ and\ \bibinfo {author}
  {\bibfnamefont {C.}~\bibnamefont {Chamon}},\ }\bibfield  {title} {\bibinfo
  {title} {{Optimizing Variational Quantum Algorithms Using Pontryagin's
  Minimum Principle}},\ }\href {https://doi.org/10.1103/PhysRevX.7.021027}
  {\bibfield  {journal} {\bibinfo  {journal} {Phys. Rev. X}\ }\textbf {\bibinfo
  {volume} {7}},\ \bibinfo {pages} {021027} (\bibinfo {year}
  {2017})}\BibitemShut {NoStop}%
\bibitem [{\citenamefont {Lin}\ \emph {et~al.}(2019)\citenamefont {Lin},
  \citenamefont {Wang}, \citenamefont {Kolesov},\ and\ \citenamefont
  {Kalabi\'{c}}}]{Lin2019Application}%
  \BibitemOpen
  \bibfield  {author} {\bibinfo {author} {\bibfnamefont {C.}~\bibnamefont
  {Lin}}, \bibinfo {author} {\bibfnamefont {Y.}~\bibnamefont {Wang}}, \bibinfo
  {author} {\bibfnamefont {G.}~\bibnamefont {Kolesov}},\ and\ \bibinfo {author}
  {\bibfnamefont {U.}~\bibnamefont {Kalabi\'{c}}},\ }\bibfield  {title}
  {\bibinfo {title} {{Application of Pontryagin's minimum principle to Grover's
  quantum search problem}},\ }\href
  {https://doi.org/10.1103/PhysRevA.100.022327} {\bibfield  {journal} {\bibinfo
   {journal} {Phys. Rev. A}\ }\textbf {\bibinfo {volume} {100}},\ \bibinfo
  {pages} {022327} (\bibinfo {year} {2019})}\BibitemShut {NoStop}%
\bibitem [{\citenamefont {Brady}\ \emph {et~al.}(2021)\citenamefont {Brady},
  \citenamefont {Baldwin}, \citenamefont {Bapat}, \citenamefont {Kharkov},\
  and\ \citenamefont {Gorshkov}}]{Brady2021Optimal}%
  \BibitemOpen
  \bibfield  {author} {\bibinfo {author} {\bibfnamefont {L.~T.}\ \bibnamefont
  {Brady}}, \bibinfo {author} {\bibfnamefont {C.~L.}\ \bibnamefont {Baldwin}},
  \bibinfo {author} {\bibfnamefont {A.}~\bibnamefont {Bapat}}, \bibinfo
  {author} {\bibfnamefont {Y.}~\bibnamefont {Kharkov}},\ and\ \bibinfo {author}
  {\bibfnamefont {A.~V.}\ \bibnamefont {Gorshkov}},\ }\bibfield  {title}
  {\bibinfo {title} {{Optimal Protocols in Quantum Annealing and Quantum
  Approximate Optimization Algorithm Problems}},\ }\href
  {https://doi.org/10.1103/PhysRevLett.126.070505} {\bibfield  {journal}
  {\bibinfo  {journal} {Phys. Rev. Lett.}\ }\textbf {\bibinfo {volume} {126}},\
  \bibinfo {pages} {070505} (\bibinfo {year} {2021})}\BibitemShut {NoStop}%
\bibitem [{\citenamefont {C\^{o}t\'e}\ \emph {et~al.}(2023)\citenamefont
  {C\^{o}t\'e}, \citenamefont {Sauvage}, \citenamefont {Larocca}, \citenamefont
  {Jonsson}, \citenamefont {Cincio},\ and\ \citenamefont
  {Albash}}]{Cote2023Diabatic}%
  \BibitemOpen
  \bibfield  {author} {\bibinfo {author} {\bibfnamefont {J.}~\bibnamefont
  {C\^{o}t\'e}}, \bibinfo {author} {\bibfnamefont {F.}~\bibnamefont {Sauvage}},
  \bibinfo {author} {\bibfnamefont {M.}~\bibnamefont {Larocca}}, \bibinfo
  {author} {\bibfnamefont {M.}~\bibnamefont {Jonsson}}, \bibinfo {author}
  {\bibfnamefont {L.}~\bibnamefont {Cincio}},\ and\ \bibinfo {author}
  {\bibfnamefont {T.}~\bibnamefont {Albash}},\ }\bibfield  {title} {\bibinfo
  {title} {{Diabatic quantum annealing for the frustrated ring model}},\ }\href
  {https://doi.org/10.1088/2058-9565/acfbaa} {\bibfield  {journal} {\bibinfo
  {journal} {Quantum Sci. Tech.}\ }\textbf {\bibinfo {volume} {8}},\ \bibinfo
  {pages} {045033} (\bibinfo {year} {2023})}\BibitemShut {NoStop}%
\bibitem [{\citenamefont {Grabarits}\ \emph {et~al.}(2024)\citenamefont
  {Grabarits}, \citenamefont {Balducci}, \citenamefont {Sanders},\ and\
  \citenamefont {del Campo}}]{Grabarits2024Non}%
  \BibitemOpen
  \bibfield  {author} {\bibinfo {author} {\bibfnamefont {A.}~\bibnamefont
  {Grabarits}}, \bibinfo {author} {\bibfnamefont {F.}~\bibnamefont {Balducci}},
  \bibinfo {author} {\bibfnamefont {B.~C.}\ \bibnamefont {Sanders}},\ and\
  \bibinfo {author} {\bibfnamefont {A.}~\bibnamefont {del Campo}},\ }\href@noop
  {} {\bibinfo {title} {{Non-Adiabatic Quantum Optimization for Crossing
  Quantum Phase Transitions}}} (\bibinfo {year} {2024}),\ \Eprint
  {https://arxiv.org/abs/2407.09596} {arXiv:2407.09596} \BibitemShut {NoStop}%
\bibitem [{\citenamefont {Rice}\ and\ \citenamefont {Zhao}(2000)}]{Rice2000}%
  \BibitemOpen
  \bibfield  {author} {\bibinfo {author} {\bibfnamefont {S.}~\bibnamefont
  {Rice}}\ and\ \bibinfo {author} {\bibfnamefont {M.}~\bibnamefont {Zhao}},\
  }\href@noop {} {\emph {\bibinfo {title} {{Optical Control of Molecular
  Dynamics}}}},\ Baker Lecture Series\ (\bibinfo  {publisher} {Wiley},\
  \bibinfo {year} {2000})\BibitemShut {NoStop}%
\bibitem [{\citenamefont {Demirplak}\ and\ \citenamefont
  {Rice}(2003)}]{Demirplak2003Adiabatic}%
  \BibitemOpen
  \bibfield  {author} {\bibinfo {author} {\bibfnamefont {M.}~\bibnamefont
  {Demirplak}}\ and\ \bibinfo {author} {\bibfnamefont {S.~A.}\ \bibnamefont
  {Rice}},\ }\bibfield  {title} {\bibinfo {title} {{Adiabatic Population
  Transfer with Control Fields}},\ }\href {https://doi.org/10.1021/jp030708a}
  {\bibfield  {journal} {\bibinfo  {journal} {J. Phys. Chem. A}\ }\textbf
  {\bibinfo {volume} {107}},\ \bibinfo {pages} {9937} (\bibinfo {year}
  {2003})}\BibitemShut {NoStop}%
\bibitem [{\citenamefont {Demirplak}\ and\ \citenamefont
  {Rice}(2005)}]{Demirplak2005Assisted}%
  \BibitemOpen
  \bibfield  {author} {\bibinfo {author} {\bibfnamefont {M.}~\bibnamefont
  {Demirplak}}\ and\ \bibinfo {author} {\bibfnamefont {S.~A.}\ \bibnamefont
  {Rice}},\ }\bibfield  {title} {\bibinfo {title} {{Assisted Adiabatic Passage
  Revisited}},\ }\href {https://doi.org/10.1021/jp040647w} {\bibfield
  {journal} {\bibinfo  {journal} {J. Phys. Chem. B}\ }\textbf {\bibinfo
  {volume} {109}},\ \bibinfo {pages} {6838} (\bibinfo {year}
  {2005})}\BibitemShut {NoStop}%
\bibitem [{\citenamefont {Demirplak}\ and\ \citenamefont
  {Rice}(2008)}]{Demirplak2008Consistency}%
  \BibitemOpen
  \bibfield  {author} {\bibinfo {author} {\bibfnamefont {M.}~\bibnamefont
  {Demirplak}}\ and\ \bibinfo {author} {\bibfnamefont {S.~A.}\ \bibnamefont
  {Rice}},\ }\bibfield  {title} {\bibinfo {title} {{On the consistency,
  extremal, and global properties of counterdiabatic fields}},\ }\href
  {https://doi.org/10.1063/1.2992152} {\bibfield  {journal} {\bibinfo
  {journal} {J. Chem. Phys.}\ }\textbf {\bibinfo {volume} {129}},\ \bibinfo
  {pages} {154111} (\bibinfo {year} {2008})}\BibitemShut {NoStop}%
\bibitem [{\citenamefont {Berry}(2009)}]{Berry2009Transitionless}%
  \BibitemOpen
  \bibfield  {author} {\bibinfo {author} {\bibfnamefont {M.~V.}\ \bibnamefont
  {Berry}},\ }\bibfield  {title} {\bibinfo {title} {{Transitionless quantum
  driving}},\ }\href {https://doi.org/10.1088/1751-8113/42/36/365303}
  {\bibfield  {journal} {\bibinfo  {journal} {J. Phys. A}\ }\textbf {\bibinfo
  {volume} {42}},\ \bibinfo {pages} {365303} (\bibinfo {year}
  {2009})}\BibitemShut {NoStop}%
\bibitem [{\citenamefont {Kato}(1950)}]{Kato1950Adiabatic}%
  \BibitemOpen
  \bibfield  {author} {\bibinfo {author} {\bibfnamefont {T.}~\bibnamefont
  {Kato}},\ }\bibfield  {title} {\bibinfo {title} {{On the Adiabatic Theorem of
  Quantum Mechanics}},\ }\href {https://doi.org/10.1143/JPSJ.5.435} {\bibfield
  {journal} {\bibinfo  {journal} {J. Phys. Soc. Japan}\ }\textbf {\bibinfo
  {volume} {5}},\ \bibinfo {pages} {435} (\bibinfo {year} {1950})}\BibitemShut
  {NoStop}%
\bibitem [{\citenamefont {del Campo}\ \emph {et~al.}(2012)\citenamefont {del
  Campo}, \citenamefont {Rams},\ and\ \citenamefont
  {Zurek}}]{delCampo2012Assisted}%
  \BibitemOpen
  \bibfield  {author} {\bibinfo {author} {\bibfnamefont {A.}~\bibnamefont {del
  Campo}}, \bibinfo {author} {\bibfnamefont {M.~M.}\ \bibnamefont {Rams}},\
  and\ \bibinfo {author} {\bibfnamefont {W.~H.}\ \bibnamefont {Zurek}},\
  }\bibfield  {title} {\bibinfo {title} {{Assisted Finite-Rate Adiabatic
  Passage Across a Quantum Critical Point: Exact Solution for the Quantum Ising
  Model}},\ }\href {https://doi.org/10.1103/PhysRevLett.109.115703} {\bibfield
  {journal} {\bibinfo  {journal} {Phys. Rev. Lett.}\ }\textbf {\bibinfo
  {volume} {109}},\ \bibinfo {pages} {115703} (\bibinfo {year}
  {2012})}\BibitemShut {NoStop}%
\bibitem [{\citenamefont
  {Takahashi}(2013{\natexlab{a}})}]{Takahashi2013Transitionless}%
  \BibitemOpen
  \bibfield  {author} {\bibinfo {author} {\bibfnamefont {K.}~\bibnamefont
  {Takahashi}},\ }\bibfield  {title} {\bibinfo {title} {{Transitionless quantum
  driving for spin systems}},\ }\href
  {https://doi.org/10.1103/PhysRevE.87.062117} {\bibfield  {journal} {\bibinfo
  {journal} {Phys. Rev. E}\ }\textbf {\bibinfo {volume} {87}},\ \bibinfo
  {pages} {062117} (\bibinfo {year} {2013}{\natexlab{a}})}\BibitemShut
  {NoStop}%
\bibitem [{\citenamefont {Damski}(2014)}]{Damski2014Counterdiabatic}%
  \BibitemOpen
  \bibfield  {author} {\bibinfo {author} {\bibfnamefont {B.}~\bibnamefont
  {Damski}},\ }\bibfield  {title} {\bibinfo {title} {{Counterdiabatic driving
  of the quantum Ising model}},\ }\href
  {https://doi.org/10.1088/1742-5468/2014/12/P12019} {\bibfield  {journal}
  {\bibinfo  {journal} {J. Stat. Mech. Theory Exp.}\ }\textbf {\bibinfo
  {volume} {2014}},\ \bibinfo {pages} {P12019} (\bibinfo {year}
  {2014})}\BibitemShut {NoStop}%
\bibitem [{\citenamefont {Saberi}\ \emph {et~al.}(2014)\citenamefont {Saberi},
  \citenamefont {Opatrn\'y}, \citenamefont {M\o{}lmer},\ and\ \citenamefont
  {del Campo}}]{Saberi2014Adiabatic}%
  \BibitemOpen
  \bibfield  {author} {\bibinfo {author} {\bibfnamefont {H.}~\bibnamefont
  {Saberi}}, \bibinfo {author} {\bibfnamefont {T.}~\bibnamefont {Opatrn\'y}},
  \bibinfo {author} {\bibfnamefont {K.}~\bibnamefont {M\o{}lmer}},\ and\
  \bibinfo {author} {\bibfnamefont {A.}~\bibnamefont {del Campo}},\ }\bibfield
  {title} {\bibinfo {title} {{Adiabatic tracking of quantum many-body
  dynamics}},\ }\href {https://doi.org/10.1103/PhysRevA.90.060301} {\bibfield
  {journal} {\bibinfo  {journal} {Phys. Rev. A}\ }\textbf {\bibinfo {volume}
  {90}},\ \bibinfo {pages} {060301} (\bibinfo {year} {2014})}\BibitemShut
  {NoStop}%
\bibitem [{\citenamefont {Kolodrubetz}\ \emph {et~al.}(2017)\citenamefont
  {Kolodrubetz}, \citenamefont {Sels}, \citenamefont {Mehta},\ and\
  \citenamefont {Polkovnikov}}]{Kolodrubetz2017Geometry}%
  \BibitemOpen
  \bibfield  {author} {\bibinfo {author} {\bibfnamefont {M.}~\bibnamefont
  {Kolodrubetz}}, \bibinfo {author} {\bibfnamefont {D.}~\bibnamefont {Sels}},
  \bibinfo {author} {\bibfnamefont {P.}~\bibnamefont {Mehta}},\ and\ \bibinfo
  {author} {\bibfnamefont {A.}~\bibnamefont {Polkovnikov}},\ }\bibfield
  {title} {\bibinfo {title} {{Geometry and non-adiabatic response in quantum
  and classical systems}},\ }\href
  {https://doi.org/10.1016/j.physrep.2017.07.001} {\bibfield  {journal}
  {\bibinfo  {journal} {Phys. Rep.}\ }\textbf {\bibinfo {volume} {697}},\
  \bibinfo {pages} {1} (\bibinfo {year} {2017})}\BibitemShut {NoStop}%
\bibitem [{\citenamefont {Sels}\ and\ \citenamefont
  {Polkovnikov}(2017)}]{Sels2017Minimizing}%
  \BibitemOpen
  \bibfield  {author} {\bibinfo {author} {\bibfnamefont {D.}~\bibnamefont
  {Sels}}\ and\ \bibinfo {author} {\bibfnamefont {A.}~\bibnamefont
  {Polkovnikov}},\ }\bibfield  {title} {\bibinfo {title} {{Minimizing
  irreversible losses in quantum systems by local counterdiabatic driving}},\
  }\href {https://doi.org/10.1073/pnas.1619826114} {\bibfield  {journal}
  {\bibinfo  {journal} {Proc. Natl. Acad. Sci. USA}\ }\textbf {\bibinfo
  {volume} {114}},\ \bibinfo {pages} {E3909} (\bibinfo {year}
  {2017})}\BibitemShut {NoStop}%
\bibitem [{\citenamefont {Claeys}\ \emph {et~al.}(2019)\citenamefont {Claeys},
  \citenamefont {Pandey}, \citenamefont {Sels},\ and\ \citenamefont
  {Polkovnikov}}]{Claeys2019Floquet}%
  \BibitemOpen
  \bibfield  {author} {\bibinfo {author} {\bibfnamefont {P.~W.}\ \bibnamefont
  {Claeys}}, \bibinfo {author} {\bibfnamefont {M.}~\bibnamefont {Pandey}},
  \bibinfo {author} {\bibfnamefont {D.}~\bibnamefont {Sels}},\ and\ \bibinfo
  {author} {\bibfnamefont {A.}~\bibnamefont {Polkovnikov}},\ }\bibfield
  {title} {\bibinfo {title} {{Floquet-Engineering Counterdiabatic Protocols in
  Quantum Many-Body Systems}},\ }\href
  {https://doi.org/10.1103/PhysRevLett.123.090602} {\bibfield  {journal}
  {\bibinfo  {journal} {Phys. Rev. Lett.}\ }\textbf {\bibinfo {volume} {123}},\
  \bibinfo {pages} {090602} (\bibinfo {year} {2019})}\BibitemShut {NoStop}%
\bibitem [{\citenamefont {Takahashi}\ and\ \citenamefont {del
  Campo}(2024)}]{Takahashi2024Shortcuts}%
  \BibitemOpen
  \bibfield  {author} {\bibinfo {author} {\bibfnamefont {K.}~\bibnamefont
  {Takahashi}}\ and\ \bibinfo {author} {\bibfnamefont {A.}~\bibnamefont {del
  Campo}},\ }\bibfield  {title} {\bibinfo {title} {{Shortcuts to Adiabaticity
  in Krylov Space}},\ }\href {https://doi.org/10.1103/PhysRevX.14.011032}
  {\bibfield  {journal} {\bibinfo  {journal} {Phys. Rev. X}\ }\textbf {\bibinfo
  {volume} {14}},\ \bibinfo {pages} {011032} (\bibinfo {year}
  {2024})}\BibitemShut {NoStop}%
\bibitem [{\citenamefont {Bhattacharjee}(2023)}]{Bhattacharjee2023Lanczos}%
  \BibitemOpen
  \bibfield  {author} {\bibinfo {author} {\bibfnamefont {B.}~\bibnamefont
  {Bhattacharjee}},\ }\href@noop {} {\bibinfo {title} {{A Lanczos approach to
  the Adiabatic Gauge Potential}}} (\bibinfo {year} {2023}),\ \Eprint
  {https://arxiv.org/abs/2302.07228} {arXiv:2302.07228} \BibitemShut {NoStop}%
\bibitem [{\citenamefont {Chandarana}\ \emph {et~al.}(2022)\citenamefont
  {Chandarana}, \citenamefont {Hegade}, \citenamefont {Paul}, \citenamefont
  {Albarr\'an-Arriagada}, \citenamefont {Solano}, \citenamefont {del Campo},\
  and\ \citenamefont {Chen}}]{Chandarana2022Digitized}%
  \BibitemOpen
  \bibfield  {author} {\bibinfo {author} {\bibfnamefont {P.}~\bibnamefont
  {Chandarana}}, \bibinfo {author} {\bibfnamefont {N.~N.}\ \bibnamefont
  {Hegade}}, \bibinfo {author} {\bibfnamefont {K.}~\bibnamefont {Paul}},
  \bibinfo {author} {\bibfnamefont {F.}~\bibnamefont {Albarr\'an-Arriagada}},
  \bibinfo {author} {\bibfnamefont {E.}~\bibnamefont {Solano}}, \bibinfo
  {author} {\bibfnamefont {A.}~\bibnamefont {del Campo}},\ and\ \bibinfo
  {author} {\bibfnamefont {X.}~\bibnamefont {Chen}},\ }\bibfield  {title}
  {\bibinfo {title} {{Digitized-counterdiabatic quantum approximate
  optimization algorithm}},\ }\href
  {https://doi.org/10.1103/PhysRevResearch.4.013141} {\bibfield  {journal}
  {\bibinfo  {journal} {Phys. Rev. Res.}\ }\textbf {\bibinfo {volume} {4}},\
  \bibinfo {pages} {013141} (\bibinfo {year} {2022})}\BibitemShut {NoStop}%
\bibitem [{\citenamefont {Hegade}\ \emph {et~al.}(2022)\citenamefont {Hegade},
  \citenamefont {Chen},\ and\ \citenamefont {Solano}}]{Hegade2022Digitized}%
  \BibitemOpen
  \bibfield  {author} {\bibinfo {author} {\bibfnamefont {N.~N.}\ \bibnamefont
  {Hegade}}, \bibinfo {author} {\bibfnamefont {X.}~\bibnamefont {Chen}},\ and\
  \bibinfo {author} {\bibfnamefont {E.}~\bibnamefont {Solano}},\ }\bibfield
  {title} {\bibinfo {title} {{Digitized counterdiabatic quantum
  optimization}},\ }\href {https://doi.org/10.1103/PhysRevResearch.4.L042030}
  {\bibfield  {journal} {\bibinfo  {journal} {Phys. Rev. Res.}\ }\textbf
  {\bibinfo {volume} {4}},\ \bibinfo {pages} {L042030} (\bibinfo {year}
  {2022})}\BibitemShut {NoStop}%
\bibitem [{\citenamefont {Wurtz}\ and\ \citenamefont
  {Love}(2022)}]{Wurtz2022Counterdiabaticity}%
  \BibitemOpen
  \bibfield  {author} {\bibinfo {author} {\bibfnamefont {J.}~\bibnamefont
  {Wurtz}}\ and\ \bibinfo {author} {\bibfnamefont {P.~J.}\ \bibnamefont
  {Love}},\ }\bibfield  {title} {\bibinfo {title} {{Counterdiabaticity and the
  quantum approximate optimization algorithm}},\ }\href
  {https://doi.org/10.22331/q-2022-01-27-635} {\bibfield  {journal} {\bibinfo
  {journal} {{Quantum}}\ }\textbf {\bibinfo {volume} {6}},\ \bibinfo {pages}
  {635} (\bibinfo {year} {2022})}\BibitemShut {NoStop}%
\bibitem [{\citenamefont {Hegade}\ and\ \citenamefont
  {Solano}(2023)}]{hegade2023digitizedCDFact}%
  \BibitemOpen
  \bibfield  {author} {\bibinfo {author} {\bibfnamefont {N.~N.}\ \bibnamefont
  {Hegade}}\ and\ \bibinfo {author} {\bibfnamefont {E.}~\bibnamefont
  {Solano}},\ }\href@noop {} {\bibinfo {title} {{Digitized-counterdiabatic
  quantum factorization}}} (\bibinfo {year} {2023}),\ \Eprint
  {https://arxiv.org/abs/2301.11005} {arXiv:2301.11005 [quant-ph]} \BibitemShut
  {NoStop}%
\bibitem [{\citenamefont {Chandarana}\ \emph {et~al.}(2023)\citenamefont
  {Chandarana}, \citenamefont {Hegade}, \citenamefont {Montalban},
  \citenamefont {Solano},\ and\ \citenamefont
  {Chen}}]{Chandarana2023ProteinDigitizedCD}%
  \BibitemOpen
  \bibfield  {author} {\bibinfo {author} {\bibfnamefont {P.}~\bibnamefont
  {Chandarana}}, \bibinfo {author} {\bibfnamefont {N.~N.}\ \bibnamefont
  {Hegade}}, \bibinfo {author} {\bibfnamefont {I.}~\bibnamefont {Montalban}},
  \bibinfo {author} {\bibfnamefont {E.}~\bibnamefont {Solano}},\ and\ \bibinfo
  {author} {\bibfnamefont {X.}~\bibnamefont {Chen}},\ }\bibfield  {title}
  {\bibinfo {title} {{Digitized Counterdiabatic Quantum Algorithm for Protein
  Folding}},\ }\href {https://doi.org/10.1103/PhysRevApplied.20.014024}
  {\bibfield  {journal} {\bibinfo  {journal} {Phys. Rev. Appl.}\ }\textbf
  {\bibinfo {volume} {20}},\ \bibinfo {pages} {014024} (\bibinfo {year}
  {2023})}\BibitemShut {NoStop}%
\bibitem [{\citenamefont {Malla}\ \emph {et~al.}(2024)\citenamefont {Malla},
  \citenamefont {Sukeno}, \citenamefont {Yu}, \citenamefont {Wei},
  \citenamefont {Weichselbaum},\ and\ \citenamefont {Konik}}]{Malla2024}%
  \BibitemOpen
  \bibfield  {author} {\bibinfo {author} {\bibfnamefont {R.~K.}\ \bibnamefont
  {Malla}}, \bibinfo {author} {\bibfnamefont {H.}~\bibnamefont {Sukeno}},
  \bibinfo {author} {\bibfnamefont {H.}~\bibnamefont {Yu}}, \bibinfo {author}
  {\bibfnamefont {T.-C.}\ \bibnamefont {Wei}}, \bibinfo {author} {\bibfnamefont
  {A.}~\bibnamefont {Weichselbaum}},\ and\ \bibinfo {author} {\bibfnamefont
  {R.~M.}\ \bibnamefont {Konik}},\ }\href {https://arxiv.org/abs/2401.15303}
  {\bibinfo {title} {Feedback-based quantum algorithm inspired by
  counterdiabatic driving}} (\bibinfo {year} {2024}),\ \Eprint
  {https://arxiv.org/abs/2401.15303} {arXiv:2401.15303 [quant-ph]} \BibitemShut
  {NoStop}%
\bibitem [{\citenamefont {Chandarana}\ \emph {et~al.}(2024)\citenamefont
  {Chandarana}, \citenamefont {Paul}, \citenamefont {Swain}, \citenamefont
  {Chen},\ and\ \citenamefont {del Campo}}]{Chandarana2024}%
  \BibitemOpen
  \bibfield  {author} {\bibinfo {author} {\bibfnamefont {P.}~\bibnamefont
  {Chandarana}}, \bibinfo {author} {\bibfnamefont {K.}~\bibnamefont {Paul}},
  \bibinfo {author} {\bibfnamefont {K.~R.}\ \bibnamefont {Swain}}, \bibinfo
  {author} {\bibfnamefont {X.}~\bibnamefont {Chen}},\ and\ \bibinfo {author}
  {\bibfnamefont {A.}~\bibnamefont {del Campo}},\ }\href
  {https://arxiv.org/abs/2409.12525} {\bibinfo {title} {Lyapunov controlled
  counterdiabatic quantum optimization}} (\bibinfo {year} {2024}),\ \Eprint
  {https://arxiv.org/abs/2409.12525} {arXiv:2409.12525 [quant-ph]} \BibitemShut
  {NoStop}%
\bibitem [{\citenamefont {Hatomura}(2023)}]{Hatomura23}%
  \BibitemOpen
  \bibfield  {author} {\bibinfo {author} {\bibfnamefont {T.}~\bibnamefont
  {Hatomura}},\ }\bibfield  {title} {\bibinfo {title} {Scaling of errors in
  digitized counterdiabatic driving},\ }\href
  {https://doi.org/10.1088/1367-2630/acfd51} {\bibfield  {journal} {\bibinfo
  {journal} {New J. Phys.}\ }\textbf {\bibinfo {volume} {25}},\ \bibinfo
  {pages} {103025} (\bibinfo {year} {2023})}\BibitemShut {NoStop}%
\bibitem [{\citenamefont {Hatomura}(2024)}]{Hatomura24}%
  \BibitemOpen
  \bibfield  {author} {\bibinfo {author} {\bibfnamefont {T.}~\bibnamefont
  {Hatomura}},\ }\bibfield  {title} {\bibinfo {title} {Shortcuts to
  adiabaticity: theoretical framework, relations between different methods, and
  versatile approximations},\ }\href {https://doi.org/10.1088/1361-6455/ad38f1}
  {\bibfield  {journal} {\bibinfo  {journal} {J. Phys. B}\ }\textbf {\bibinfo
  {volume} {57}},\ \bibinfo {pages} {102001} (\bibinfo {year}
  {2024})}\BibitemShut {NoStop}%
\bibitem [{\citenamefont {Lucas~Lamata}\ and\ \citenamefont
  {Solano}(2018)}]{Lamata18}%
  \BibitemOpen
  \bibfield  {author} {\bibinfo {author} {\bibfnamefont {M.~S.}\ \bibnamefont
  {Lucas~Lamata}, \bibfnamefont {Adrian Parra-Rodriguez}}\ and\ \bibinfo
  {author} {\bibfnamefont {E.}~\bibnamefont {Solano}},\ }\bibfield  {title}
  {\bibinfo {title} {{Digital-analog quantum simulations with superconducting
  circuits}},\ }\href {https://doi.org/10.1080/23746149.2018.1457981}
  {\bibfield  {journal} {\bibinfo  {journal} {Adv. Phys. X}\ }\textbf {\bibinfo
  {volume} {3}},\ \bibinfo {pages} {1457981} (\bibinfo {year}
  {2018})}\BibitemShut {NoStop}%
\bibitem [{\citenamefont {Kumar}\ \emph {et~al.}(2024)\citenamefont {Kumar},
  \citenamefont {Hegade}, \citenamefont {Cadavid}, \citenamefont {de~Oliveira},
  \citenamefont {Solano},\ and\ \citenamefont {Albarrán-Arriagada}}]{Kumar24}%
  \BibitemOpen
  \bibfield  {author} {\bibinfo {author} {\bibfnamefont {S.}~\bibnamefont
  {Kumar}}, \bibinfo {author} {\bibfnamefont {N.~N.}\ \bibnamefont {Hegade}},
  \bibinfo {author} {\bibfnamefont {A.~G.}\ \bibnamefont {Cadavid}}, \bibinfo
  {author} {\bibfnamefont {M.~H.}\ \bibnamefont {de~Oliveira}}, \bibinfo
  {author} {\bibfnamefont {E.}~\bibnamefont {Solano}},\ and\ \bibinfo {author}
  {\bibfnamefont {F.}~\bibnamefont {Albarrán-Arriagada}},\ }\href@noop {}
  {\bibinfo {title} {{Digital-Analog Counterdiabatic Quantum Optimization with
  Trapped Ions}}} (\bibinfo {year} {2024}),\ \Eprint
  {https://arxiv.org/abs/2405.01447} {arXiv:2405.01447 [quant-ph]} \BibitemShut
  {NoStop}%
\bibitem [{\citenamefont {Campbell}\ \emph {et~al.}(2015)\citenamefont
  {Campbell}, \citenamefont {De~Chiara}, \citenamefont {Paternostro},
  \citenamefont {Palma},\ and\ \citenamefont {Fazio}}]{Campbell15}%
  \BibitemOpen
  \bibfield  {author} {\bibinfo {author} {\bibfnamefont {S.}~\bibnamefont
  {Campbell}}, \bibinfo {author} {\bibfnamefont {G.}~\bibnamefont {De~Chiara}},
  \bibinfo {author} {\bibfnamefont {M.}~\bibnamefont {Paternostro}}, \bibinfo
  {author} {\bibfnamefont {G.~M.}\ \bibnamefont {Palma}},\ and\ \bibinfo
  {author} {\bibfnamefont {R.}~\bibnamefont {Fazio}},\ }\bibfield  {title}
  {\bibinfo {title} {Shortcut to adiabaticity in the lipkin-meshkov-glick
  model},\ }\href {https://doi.org/10.1103/PhysRevLett.114.177206} {\bibfield
  {journal} {\bibinfo  {journal} {Phys. Rev. Lett.}\ }\textbf {\bibinfo
  {volume} {114}},\ \bibinfo {pages} {177206} (\bibinfo {year}
  {2015})}\BibitemShut {NoStop}%
\bibitem [{\citenamefont {Mukherjee}\ \emph {et~al.}(2016)\citenamefont
  {Mukherjee}, \citenamefont {Montangero},\ and\ \citenamefont
  {Fazio}}]{Mukherjee16}%
  \BibitemOpen
  \bibfield  {author} {\bibinfo {author} {\bibfnamefont {V.}~\bibnamefont
  {Mukherjee}}, \bibinfo {author} {\bibfnamefont {S.}~\bibnamefont
  {Montangero}},\ and\ \bibinfo {author} {\bibfnamefont {R.}~\bibnamefont
  {Fazio}},\ }\bibfield  {title} {\bibinfo {title} {Local shortcut to
  adiabaticity for quantum many-body systems},\ }\href
  {https://doi.org/10.1103/PhysRevA.93.062108} {\bibfield  {journal} {\bibinfo
  {journal} {Phys. Rev. A}\ }\textbf {\bibinfo {volume} {93}},\ \bibinfo
  {pages} {062108} (\bibinfo {year} {2016})}\BibitemShut {NoStop}%
\bibitem [{\citenamefont {Carolan}\ \emph {et~al.}(2022)\citenamefont
  {Carolan}, \citenamefont {Kiely},\ and\ \citenamefont
  {Campbell}}]{Carolan22}%
  \BibitemOpen
  \bibfield  {author} {\bibinfo {author} {\bibfnamefont {E.}~\bibnamefont
  {Carolan}}, \bibinfo {author} {\bibfnamefont {A.}~\bibnamefont {Kiely}},\
  and\ \bibinfo {author} {\bibfnamefont {S.}~\bibnamefont {Campbell}},\
  }\bibfield  {title} {\bibinfo {title} {Counterdiabatic control in the impulse
  regime},\ }\href {https://doi.org/10.1103/PhysRevA.105.012605} {\bibfield
  {journal} {\bibinfo  {journal} {Phys. Rev. A}\ }\textbf {\bibinfo {volume}
  {105}},\ \bibinfo {pages} {012605} (\bibinfo {year} {2022})}\BibitemShut
  {NoStop}%
\bibitem [{\citenamefont {Hartmann}\ \emph {et~al.}(2022)\citenamefont
  {Hartmann}, \citenamefont {Mbeng},\ and\ \citenamefont
  {Lechner}}]{Hartmann2022Polynomial}%
  \BibitemOpen
  \bibfield  {author} {\bibinfo {author} {\bibfnamefont {A.}~\bibnamefont
  {Hartmann}}, \bibinfo {author} {\bibfnamefont {G.~B.}\ \bibnamefont
  {Mbeng}},\ and\ \bibinfo {author} {\bibfnamefont {W.}~\bibnamefont
  {Lechner}},\ }\bibfield  {title} {\bibinfo {title} {{Polynomial scaling
  enhancement in the ground-state preparation of Ising spin models via
  counterdiabatic driving}},\ }\href
  {https://doi.org/10.1103/PhysRevA.105.022614} {\bibfield  {journal} {\bibinfo
   {journal} {Phys. Rev. A}\ }\textbf {\bibinfo {volume} {105}},\ \bibinfo
  {pages} {022614} (\bibinfo {year} {2022})}\BibitemShut {NoStop}%
\bibitem [{\citenamefont {\v{C}epait\.{e}}\ \emph {et~al.}(2023)\citenamefont
  {\v{C}epait\.{e}}, \citenamefont {Polkovnikov}, \citenamefont {Daley},\ and\
  \citenamefont {Duncan}}]{Cepaite2023Counterdiabatic}%
  \BibitemOpen
  \bibfield  {author} {\bibinfo {author} {\bibfnamefont {I.}~\bibnamefont
  {\v{C}epait\.{e}}}, \bibinfo {author} {\bibfnamefont {A.}~\bibnamefont
  {Polkovnikov}}, \bibinfo {author} {\bibfnamefont {A.~J.}\ \bibnamefont
  {Daley}},\ and\ \bibinfo {author} {\bibfnamefont {C.~W.}\ \bibnamefont
  {Duncan}},\ }\bibfield  {title} {\bibinfo {title} {{Counterdiabatic Optimized
  Local Driving}},\ }\href {https://doi.org/10.1103/PRXQuantum.4.010312}
  {\bibfield  {journal} {\bibinfo  {journal} {PRX Quantum}\ }\textbf {\bibinfo
  {volume} {4}},\ \bibinfo {pages} {010312} (\bibinfo {year}
  {2023})}\BibitemShut {NoStop}%
\bibitem [{\citenamefont {Mc~Keever}\ and\ \citenamefont
  {Lubasch}(2024)}]{McKeever24}%
  \BibitemOpen
  \bibfield  {author} {\bibinfo {author} {\bibfnamefont {C.}~\bibnamefont
  {Mc~Keever}}\ and\ \bibinfo {author} {\bibfnamefont {M.}~\bibnamefont
  {Lubasch}},\ }\bibfield  {title} {\bibinfo {title} {{Towards Adiabatic
  Quantum Computing Using Compressed Quantum Circuits}},\ }\href
  {https://doi.org/10.1103/PRXQuantum.5.020362} {\bibfield  {journal} {\bibinfo
   {journal} {PRX Quantum}\ }\textbf {\bibinfo {volume} {5}},\ \bibinfo {pages}
  {020362} (\bibinfo {year} {2024})}\BibitemShut {NoStop}%
\bibitem [{\citenamefont {Damski}(2005)}]{Damski05}%
  \BibitemOpen
  \bibfield  {author} {\bibinfo {author} {\bibfnamefont {B.}~\bibnamefont
  {Damski}},\ }\bibfield  {title} {\bibinfo {title} {{The Simplest Quantum
  Model Supporting the Kibble-Zurek Mechanism of Topological Defect Production:
  Landau-Zener Transitions from a New Perspective}},\ }\href
  {https://doi.org/10.1103/PhysRevLett.95.035701} {\bibfield  {journal}
  {\bibinfo  {journal} {Phys. Rev. Lett.}\ }\textbf {\bibinfo {volume} {95}},\
  \bibinfo {pages} {035701} (\bibinfo {year} {2005})}\BibitemShut {NoStop}%
\bibitem [{\citenamefont {Zurek}\ \emph {et~al.}(2005)\citenamefont {Zurek},
  \citenamefont {Dorner},\ and\ \citenamefont {Zoller}}]{Zurek2005Dynamics}%
  \BibitemOpen
  \bibfield  {author} {\bibinfo {author} {\bibfnamefont {W.~H.}\ \bibnamefont
  {Zurek}}, \bibinfo {author} {\bibfnamefont {U.}~\bibnamefont {Dorner}},\ and\
  \bibinfo {author} {\bibfnamefont {P.}~\bibnamefont {Zoller}},\ }\bibfield
  {title} {\bibinfo {title} {{Dynamics of a Quantum Phase Transition}},\ }\href
  {https://doi.org/10.1103/PhysRevLett.95.105701} {\bibfield  {journal}
  {\bibinfo  {journal} {Phys. Rev. Lett.}\ }\textbf {\bibinfo {volume} {95}},\
  \bibinfo {pages} {105701} (\bibinfo {year} {2005})}\BibitemShut {NoStop}%
\bibitem [{\citenamefont {Dziarmaga}(2005)}]{Dziarmaga05}%
  \BibitemOpen
  \bibfield  {author} {\bibinfo {author} {\bibfnamefont {J.}~\bibnamefont
  {Dziarmaga}},\ }\bibfield  {title} {\bibinfo {title} {{Dynamics of a Quantum
  Phase Transition: Exact Solution of the Quantum Ising Model}},\ }\href
  {https://doi.org/10.1103/PhysRevLett.95.245701} {\bibfield  {journal}
  {\bibinfo  {journal} {Phys. Rev. Lett.}\ }\textbf {\bibinfo {volume} {95}},\
  \bibinfo {pages} {245701} (\bibinfo {year} {2005})}\BibitemShut {NoStop}%
\bibitem [{\citenamefont {Polkovnikov}(2005)}]{Polkovnikov2005Universal}%
  \BibitemOpen
  \bibfield  {author} {\bibinfo {author} {\bibfnamefont {A.}~\bibnamefont
  {Polkovnikov}},\ }\bibfield  {title} {\bibinfo {title} {{Universal adiabatic
  dynamics in the vicinity of a quantum critical point}},\ }\href
  {https://doi.org/10.1103/PhysRevB.72.161201} {\bibfield  {journal} {\bibinfo
  {journal} {Phys. Rev. B}\ }\textbf {\bibinfo {volume} {72}},\ \bibinfo
  {pages} {161201(R)} (\bibinfo {year} {2005})}\BibitemShut {NoStop}%
\bibitem [{\citenamefont {Damski}\ and\ \citenamefont
  {Zurek}(2006)}]{Damski06}%
  \BibitemOpen
  \bibfield  {author} {\bibinfo {author} {\bibfnamefont {B.}~\bibnamefont
  {Damski}}\ and\ \bibinfo {author} {\bibfnamefont {W.~H.}\ \bibnamefont
  {Zurek}},\ }\bibfield  {title} {\bibinfo {title} {{Adiabatic-impulse
  approximation for avoided level crossings: From phase-transition dynamics to
  Landau-Zener evolutions and back again}},\ }\href
  {https://doi.org/10.1103/PhysRevA.73.063405} {\bibfield  {journal} {\bibinfo
  {journal} {Phys. Rev. A}\ }\textbf {\bibinfo {volume} {73}},\ \bibinfo
  {pages} {063405} (\bibinfo {year} {2006})}\BibitemShut {NoStop}%
\bibitem [{\citenamefont {del Campo}\ and\ \citenamefont
  {Zurek}(2014)}]{delCampo2014Universality}%
  \BibitemOpen
  \bibfield  {author} {\bibinfo {author} {\bibfnamefont {A.}~\bibnamefont {del
  Campo}}\ and\ \bibinfo {author} {\bibfnamefont {W.~H.}\ \bibnamefont
  {Zurek}},\ }\bibfield  {title} {\bibinfo {title} {{Universality of phase
  transition dynamics: Topological defects from symmetry breaking}},\ }\href
  {https://doi.org/10.1142/S0217751X1430018X} {\bibfield  {journal} {\bibinfo
  {journal} {Intl. J. Mod. Phys. A}\ }\textbf {\bibinfo {volume} {29}},\
  \bibinfo {pages} {1430018} (\bibinfo {year} {2014})}\BibitemShut {NoStop}%
\bibitem [{\citenamefont {Qiu}\ \emph {et~al.}(2020)\citenamefont {Qiu},
  \citenamefont {Liang}, \citenamefont {Yang}, \citenamefont {Yang},
  \citenamefont {Tian}, \citenamefont {Xu},\ and\ \citenamefont
  {Duan}}]{Qiu20}%
  \BibitemOpen
  \bibfield  {author} {\bibinfo {author} {\bibfnamefont {L.-Y.}\ \bibnamefont
  {Qiu}}, \bibinfo {author} {\bibfnamefont {H.-Y.}\ \bibnamefont {Liang}},
  \bibinfo {author} {\bibfnamefont {Y.-B.}\ \bibnamefont {Yang}}, \bibinfo
  {author} {\bibfnamefont {H.-X.}\ \bibnamefont {Yang}}, \bibinfo {author}
  {\bibfnamefont {T.}~\bibnamefont {Tian}}, \bibinfo {author} {\bibfnamefont
  {Y.}~\bibnamefont {Xu}},\ and\ \bibinfo {author} {\bibfnamefont {L.-M.}\
  \bibnamefont {Duan}},\ }\bibfield  {title} {\bibinfo {title} {{Observation of
  generalized Kibble-Zurek mechanism across a first-order quantum phase
  transition in a spinor condensate}},\ }\href
  {https://doi.org/10.1126/sciadv.aba7292} {\bibfield  {journal} {\bibinfo
  {journal} {Sci. Adv.}\ }\textbf {\bibinfo {volume} {6}},\ \bibinfo {pages}
  {eaba7292} (\bibinfo {year} {2020})}\BibitemShut {NoStop}%
\bibitem [{\citenamefont {Suzuki}\ and\ \citenamefont
  {Zurek}(2024)}]{Suzuki24}%
  \BibitemOpen
  \bibfield  {author} {\bibinfo {author} {\bibfnamefont {F.}~\bibnamefont
  {Suzuki}}\ and\ \bibinfo {author} {\bibfnamefont {W.~H.}\ \bibnamefont
  {Zurek}},\ }\bibfield  {title} {\bibinfo {title} {{Topological Defect
  Formation in a Phase Transition with Tunable Order}},\ }\href
  {https://doi.org/10.1103/PhysRevLett.132.241601} {\bibfield  {journal}
  {\bibinfo  {journal} {Phys. Rev. Lett.}\ }\textbf {\bibinfo {volume} {132}},\
  \bibinfo {pages} {241601} (\bibinfo {year} {2024})}\BibitemShut {NoStop}%
\bibitem [{\citenamefont {Cadavid}\ \emph {et~al.}(2024)\citenamefont
  {Cadavid}, \citenamefont {Dalal}, \citenamefont {Simen}, \citenamefont
  {Solano},\ and\ \citenamefont {Hegade}}]{Cadavid2024bias}%
  \BibitemOpen
  \bibfield  {author} {\bibinfo {author} {\bibfnamefont {A.~G.}\ \bibnamefont
  {Cadavid}}, \bibinfo {author} {\bibfnamefont {A.}~\bibnamefont {Dalal}},
  \bibinfo {author} {\bibfnamefont {A.}~\bibnamefont {Simen}}, \bibinfo
  {author} {\bibfnamefont {E.}~\bibnamefont {Solano}},\ and\ \bibinfo {author}
  {\bibfnamefont {N.~N.}\ \bibnamefont {Hegade}},\ }\href@noop {} {\bibinfo
  {title} {{Bias-field digitized counterdiabatic quantum optimization}}}
  (\bibinfo {year} {2024}),\ \Eprint {https://arxiv.org/abs/2405.13898}
  {arXiv:2405.13898 [quant-ph]} \BibitemShut {NoStop}%
\bibitem [{\citenamefont {Passarelli}\ \emph {et~al.}(2020)\citenamefont
  {Passarelli}, \citenamefont {Cataudella}, \citenamefont {Fazio},\ and\
  \citenamefont {Lucignano}}]{Passarelli2020Counterdiabatic}%
  \BibitemOpen
  \bibfield  {author} {\bibinfo {author} {\bibfnamefont {G.}~\bibnamefont
  {Passarelli}}, \bibinfo {author} {\bibfnamefont {V.}~\bibnamefont
  {Cataudella}}, \bibinfo {author} {\bibfnamefont {R.}~\bibnamefont {Fazio}},\
  and\ \bibinfo {author} {\bibfnamefont {P.}~\bibnamefont {Lucignano}},\
  }\bibfield  {title} {\bibinfo {title} {{Counterdiabatic driving in the
  quantum annealing of the $p$-spin model: A variational approach}},\ }\href
  {https://doi.org/10.1103/PhysRevResearch.2.013283} {\bibfield  {journal}
  {\bibinfo  {journal} {Phys. Rev. Res.}\ }\textbf {\bibinfo {volume} {2}},\
  \bibinfo {pages} {013283} (\bibinfo {year} {2020})}\BibitemShut {NoStop}%
\bibitem [{\citenamefont {Prielinger}\ \emph {et~al.}(2021)\citenamefont
  {Prielinger}, \citenamefont {Hartmann}, \citenamefont {Yamashiro},
  \citenamefont {Nishimura}, \citenamefont {Lechner},\ and\ \citenamefont
  {Nishimori}}]{Prielinger2021Two}%
  \BibitemOpen
  \bibfield  {author} {\bibinfo {author} {\bibfnamefont {L.}~\bibnamefont
  {Prielinger}}, \bibinfo {author} {\bibfnamefont {A.}~\bibnamefont
  {Hartmann}}, \bibinfo {author} {\bibfnamefont {Y.}~\bibnamefont {Yamashiro}},
  \bibinfo {author} {\bibfnamefont {K.}~\bibnamefont {Nishimura}}, \bibinfo
  {author} {\bibfnamefont {W.}~\bibnamefont {Lechner}},\ and\ \bibinfo {author}
  {\bibfnamefont {H.}~\bibnamefont {Nishimori}},\ }\bibfield  {title} {\bibinfo
  {title} {{Two-parameter counter-diabatic driving in quantum annealing}},\
  }\href {https://doi.org/10.1103/PhysRevResearch.3.013227} {\bibfield
  {journal} {\bibinfo  {journal} {Phys. Rev. Res.}\ }\textbf {\bibinfo {volume}
  {3}},\ \bibinfo {pages} {013227} (\bibinfo {year} {2021})}\BibitemShut
  {NoStop}%
\bibitem [{\citenamefont {Gong}\ \emph {et~al.}(2016)\citenamefont {Gong},
  \citenamefont {Wen}, \citenamefont {Sun} \emph {et~al.}}]{Gong2016}%
  \BibitemOpen
  \bibfield  {author} {\bibinfo {author} {\bibfnamefont {M.}~\bibnamefont
  {Gong}}, \bibinfo {author} {\bibfnamefont {X.}~\bibnamefont {Wen}}, \bibinfo
  {author} {\bibfnamefont {G.}~\bibnamefont {Sun}}, \emph {et~al.},\ }\bibfield
   {title} {\bibinfo {title} {{Simulating the Kibble-Zurek mechanism of the
  Ising model with a superconducting qubit system}},\ }\href
  {https://doi.org/10.1038/srep22667} {\bibfield  {journal} {\bibinfo
  {journal} {Sci. Rep.}\ }\textbf {\bibinfo {volume} {6}},\ \bibinfo {pages}
  {22667} (\bibinfo {year} {2016})}\BibitemShut {NoStop}%
\bibitem [{\citenamefont {Cui}\ \emph {et~al.}(2016)\citenamefont {Cui},
  \citenamefont {Huang}, \citenamefont {Wang} \emph {et~al.}}]{Cui16}%
  \BibitemOpen
  \bibfield  {author} {\bibinfo {author} {\bibfnamefont {J.-M.}\ \bibnamefont
  {Cui}}, \bibinfo {author} {\bibfnamefont {Y.-F.}\ \bibnamefont {Huang}},
  \bibinfo {author} {\bibfnamefont {Z.}~\bibnamefont {Wang}}, \emph {et~al.},\
  }\bibfield  {title} {\bibinfo {title} {{Experimental Trapped-ion Quantum
  Simulation of the {K}ibble-{Z}urek dynamics in momentum space}},\ }\href
  {https://doi.org/10.1038/srep33381} {\bibfield  {journal} {\bibinfo
  {journal} {Sci. Rep.}\ }\textbf {\bibinfo {volume} {6}},\ \bibinfo {pages}
  {33381} (\bibinfo {year} {2016})}\BibitemShut {NoStop}%
\bibitem [{\citenamefont {Cui}\ \emph {et~al.}(2020)\citenamefont {Cui},
  \citenamefont {G{\'o}mez-Ruiz}, \citenamefont {Huang}, \citenamefont {Li},
  \citenamefont {Guo},\ and\ \citenamefont {del Campo}}]{Cui20}%
  \BibitemOpen
  \bibfield  {author} {\bibinfo {author} {\bibfnamefont {J.-M.}\ \bibnamefont
  {Cui}}, \bibinfo {author} {\bibfnamefont {F.~J.}\ \bibnamefont
  {G{\'o}mez-Ruiz}}, \bibinfo {author} {\bibfnamefont {Y.-F.}\ \bibnamefont
  {Huang}}, \bibinfo {author} {\bibfnamefont {C.-F.}\ \bibnamefont {Li}},
  \bibinfo {author} {\bibfnamefont {G.-C.}\ \bibnamefont {Guo}},\ and\ \bibinfo
  {author} {\bibfnamefont {A.}~\bibnamefont {del Campo}},\ }\bibfield  {title}
  {\bibinfo {title} {{Experimentally testing quantum critical dynamics beyond
  the Kibble--Zurek mechanism}},\ }\href
  {https://doi.org/10.1038/s42005-020-0306-6} {\bibfield  {journal} {\bibinfo
  {journal} {Commun. Phys.}\ }\textbf {\bibinfo {volume} {3}},\ \bibinfo
  {pages} {44} (\bibinfo {year} {2020})}\BibitemShut {NoStop}%
\bibitem [{\citenamefont {Keesling}\ \emph {et~al.}(2019)\citenamefont
  {Keesling}, \citenamefont {Omran}, \citenamefont {Levine} \emph
  {et~al.}}]{Keesling2019}%
  \BibitemOpen
  \bibfield  {author} {\bibinfo {author} {\bibfnamefont {A.}~\bibnamefont
  {Keesling}}, \bibinfo {author} {\bibfnamefont {A.}~\bibnamefont {Omran}},
  \bibinfo {author} {\bibfnamefont {H.}~\bibnamefont {Levine}}, \emph
  {et~al.},\ }\bibfield  {title} {\bibinfo {title} {{Quantum Kibble--Zurek
  mechanism and critical dynamics on a programmable Rydberg simulator}},\
  }\href {https://doi.org/10.1038/s41586-019-1070-1} {\bibfield  {journal}
  {\bibinfo  {journal} {Nature}\ }\textbf {\bibinfo {volume} {568}},\ \bibinfo
  {pages} {207} (\bibinfo {year} {2019})}\BibitemShut {NoStop}%
\bibitem [{\citenamefont {Gardas}\ \emph {et~al.}(2018)\citenamefont {Gardas},
  \citenamefont {Dziarmaga}, \citenamefont {Zurek},\ and\ \citenamefont
  {Zwolak}}]{Gardas2018Defects}%
  \BibitemOpen
  \bibfield  {author} {\bibinfo {author} {\bibfnamefont {B.}~\bibnamefont
  {Gardas}}, \bibinfo {author} {\bibfnamefont {J.}~\bibnamefont {Dziarmaga}},
  \bibinfo {author} {\bibfnamefont {W.~H.}\ \bibnamefont {Zurek}},\ and\
  \bibinfo {author} {\bibfnamefont {M.}~\bibnamefont {Zwolak}},\ }\bibfield
  {title} {\bibinfo {title} {{Defects in Quantum Computers}},\ }\href
  {https://doi.org/10.1038/s41598-018-22763-2} {\bibfield  {journal} {\bibinfo
  {journal} {Sci. Rep.}\ }\textbf {\bibinfo {volume} {8}},\ \bibinfo {pages}
  {4539} (\bibinfo {year} {2018})}\BibitemShut {NoStop}%
\bibitem [{\citenamefont {Weinberg}\ \emph {et~al.}(2020)\citenamefont
  {Weinberg}, \citenamefont {Tylutki}, \citenamefont {R\"onkk\"o},
  \citenamefont {Westerholm}, \citenamefont {\AA{}str\"om}, \citenamefont
  {Manninen}, \citenamefont {T\"orm\"a},\ and\ \citenamefont
  {Sandvik}}]{Weinberg2020Scaling}%
  \BibitemOpen
  \bibfield  {author} {\bibinfo {author} {\bibfnamefont {P.}~\bibnamefont
  {Weinberg}}, \bibinfo {author} {\bibfnamefont {M.}~\bibnamefont {Tylutki}},
  \bibinfo {author} {\bibfnamefont {J.~M.}\ \bibnamefont {R\"onkk\"o}},
  \bibinfo {author} {\bibfnamefont {J.}~\bibnamefont {Westerholm}}, \bibinfo
  {author} {\bibfnamefont {J.~A.}\ \bibnamefont {\AA{}str\"om}}, \bibinfo
  {author} {\bibfnamefont {P.}~\bibnamefont {Manninen}}, \bibinfo {author}
  {\bibfnamefont {P.}~\bibnamefont {T\"orm\"a}},\ and\ \bibinfo {author}
  {\bibfnamefont {A.~W.}\ \bibnamefont {Sandvik}},\ }\bibfield  {title}
  {\bibinfo {title} {{Scaling and Diabatic Effects in Quantum Annealing with a
  D-Wave Device}},\ }\href {https://doi.org/10.1103/PhysRevLett.124.090502}
  {\bibfield  {journal} {\bibinfo  {journal} {Phys. Rev. Lett.}\ }\textbf
  {\bibinfo {volume} {124}},\ \bibinfo {pages} {090502} (\bibinfo {year}
  {2020})}\BibitemShut {NoStop}%
\bibitem [{\citenamefont {Bando}\ \emph {et~al.}(2020)\citenamefont {Bando},
  \citenamefont {Susa}, \citenamefont {Oshiyama}, \citenamefont {Shibata},
  \citenamefont {Ohzeki}, \citenamefont {G\'omez-Ruiz}, \citenamefont {Lidar},
  \citenamefont {Suzuki}, \citenamefont {del Campo},\ and\ \citenamefont
  {Nishimori}}]{Bando2020Probing}%
  \BibitemOpen
  \bibfield  {author} {\bibinfo {author} {\bibfnamefont {Y.}~\bibnamefont
  {Bando}}, \bibinfo {author} {\bibfnamefont {Y.}~\bibnamefont {Susa}},
  \bibinfo {author} {\bibfnamefont {H.}~\bibnamefont {Oshiyama}}, \bibinfo
  {author} {\bibfnamefont {N.}~\bibnamefont {Shibata}}, \bibinfo {author}
  {\bibfnamefont {M.}~\bibnamefont {Ohzeki}}, \bibinfo {author} {\bibfnamefont
  {F.~J.}\ \bibnamefont {G\'omez-Ruiz}}, \bibinfo {author} {\bibfnamefont
  {D.~A.}\ \bibnamefont {Lidar}}, \bibinfo {author} {\bibfnamefont
  {S.}~\bibnamefont {Suzuki}}, \bibinfo {author} {\bibfnamefont
  {A.}~\bibnamefont {del Campo}},\ and\ \bibinfo {author} {\bibfnamefont
  {H.}~\bibnamefont {Nishimori}},\ }\bibfield  {title} {\bibinfo {title}
  {{Probing the universality of topological defect formation in a quantum
  annealer: Kibble-Zurek mechanism and beyond}},\ }\href
  {https://doi.org/10.1103/PhysRevResearch.2.033369} {\bibfield  {journal}
  {\bibinfo  {journal} {Phys. Rev. Res.}\ }\textbf {\bibinfo {volume} {2}},\
  \bibinfo {pages} {033369} (\bibinfo {year} {2020})}\BibitemShut {NoStop}%
\bibitem [{\citenamefont {King}\ \emph {et~al.}(2022)\citenamefont {King},
  \citenamefont {Suzuki}, \citenamefont {Raymond} \emph
  {et~al.}}]{King2022Coherent}%
  \BibitemOpen
  \bibfield  {author} {\bibinfo {author} {\bibfnamefont {A.~D.}\ \bibnamefont
  {King}}, \bibinfo {author} {\bibfnamefont {S.}~\bibnamefont {Suzuki}},
  \bibinfo {author} {\bibfnamefont {J.}~\bibnamefont {Raymond}}, \emph
  {et~al.},\ }\bibfield  {title} {\bibinfo {title} {{Coherent quantum annealing
  in a programmable 2,000-qubit Ising chain}},\ }\href
  {https://doi.org/10.1038/s41567-022-01741-6} {\bibfield  {journal} {\bibinfo
  {journal} {Nat. Phys.}\ }\textbf {\bibinfo {volume} {18}},\ \bibinfo {pages}
  {1324} (\bibinfo {year} {2022})}\BibitemShut {NoStop}%
\bibitem [{\citenamefont {King}\ \emph {et~al.}(2024)\citenamefont {King},
  \citenamefont {Nocera}, \citenamefont {Rams} \emph {et~al.}}]{King2024}%
  \BibitemOpen
  \bibfield  {author} {\bibinfo {author} {\bibfnamefont {A.~D.}\ \bibnamefont
  {King}}, \bibinfo {author} {\bibfnamefont {A.}~\bibnamefont {Nocera}},
  \bibinfo {author} {\bibfnamefont {M.~M.}\ \bibnamefont {Rams}}, \emph
  {et~al.},\ }\href@noop {} {\bibinfo {title} {{Computational supremacy in
  quantum simulation}}} (\bibinfo {year} {2024}),\ \Eprint
  {https://arxiv.org/abs/2403.00910} {arXiv:2403.00910 [quant-ph]} \BibitemShut
  {NoStop}%
\bibitem [{\citenamefont {Roberts}\ \emph {et~al.}(2020)\citenamefont
  {Roberts}, \citenamefont {Cincio}, \citenamefont {Saxena}, \citenamefont
  {Petukhov},\ and\ \citenamefont {Knysh}}]{Roberts2020Noise}%
  \BibitemOpen
  \bibfield  {author} {\bibinfo {author} {\bibfnamefont {D.}~\bibnamefont
  {Roberts}}, \bibinfo {author} {\bibfnamefont {L.}~\bibnamefont {Cincio}},
  \bibinfo {author} {\bibfnamefont {A.}~\bibnamefont {Saxena}}, \bibinfo
  {author} {\bibfnamefont {A.}~\bibnamefont {Petukhov}},\ and\ \bibinfo
  {author} {\bibfnamefont {S.}~\bibnamefont {Knysh}},\ }\bibfield  {title}
  {\bibinfo {title} {{Noise amplification at spin-glass bottlenecks of quantum
  annealing: A solvable model}},\ }\href
  {https://doi.org/10.1103/PhysRevA.101.042317} {\bibfield  {journal} {\bibinfo
   {journal} {Phys. Rev. A}\ }\textbf {\bibinfo {volume} {101}},\ \bibinfo
  {pages} {042317} (\bibinfo {year} {2020})}\BibitemShut {NoStop}%
\bibitem [{\citenamefont {Berry}(1984)}]{Berry1984Quantal}%
  \BibitemOpen
  \bibfield  {author} {\bibinfo {author} {\bibfnamefont {M.~V.}\ \bibnamefont
  {Berry}},\ }\bibfield  {title} {\bibinfo {title} {{Quantal phase factors
  accompanying adiabatic changes}},\ }\href
  {https://doi.org/10.1098/rspa.1984.0023} {\bibfield  {journal} {\bibinfo
  {journal} {Proc. R. Soc. London A}\ }\textbf {\bibinfo {volume} {392}},\
  \bibinfo {pages} {45} (\bibinfo {year} {1984})}\BibitemShut {NoStop}%
\bibitem [{\citenamefont {Yao}\ \emph {et~al.}(2021)\citenamefont {Yao},
  \citenamefont {Lin},\ and\ \citenamefont {Bukov}}]{Yao2021}%
  \BibitemOpen
  \bibfield  {author} {\bibinfo {author} {\bibfnamefont {J.}~\bibnamefont
  {Yao}}, \bibinfo {author} {\bibfnamefont {L.}~\bibnamefont {Lin}},\ and\
  \bibinfo {author} {\bibfnamefont {M.}~\bibnamefont {Bukov}},\ }\bibfield
  {title} {\bibinfo {title} {{Reinforcement Learning for Many-Body Ground-State
  Preparation Inspired by Counterdiabatic Driving}},\ }\href
  {https://doi.org/10.1103/PhysRevX.11.031070} {\bibfield  {journal} {\bibinfo
  {journal} {Phys. Rev. X}\ }\textbf {\bibinfo {volume} {11}},\ \bibinfo
  {pages} {031070} (\bibinfo {year} {2021})}\BibitemShut {NoStop}%
\bibitem [{\citenamefont {Ohkuwa}\ \emph {et~al.}(2018)\citenamefont {Ohkuwa},
  \citenamefont {Nishimori},\ and\ \citenamefont {Lidar}}]{Ohkuwa2018Reverse}%
  \BibitemOpen
  \bibfield  {author} {\bibinfo {author} {\bibfnamefont {M.}~\bibnamefont
  {Ohkuwa}}, \bibinfo {author} {\bibfnamefont {H.}~\bibnamefont {Nishimori}},\
  and\ \bibinfo {author} {\bibfnamefont {D.~A.}\ \bibnamefont {Lidar}},\
  }\bibfield  {title} {\bibinfo {title} {{Reverse annealing for the fully
  connected $p$-spin model}},\ }\href
  {https://doi.org/10.1103/PhysRevA.98.022314} {\bibfield  {journal} {\bibinfo
  {journal} {Phys. Rev. A}\ }\textbf {\bibinfo {volume} {98}},\ \bibinfo
  {pages} {022314} (\bibinfo {year} {2018})}\BibitemShut {NoStop}%
\bibitem [{\citenamefont {King}\ \emph {et~al.}(2018)\citenamefont {King},
  \citenamefont {Carrasquilla}, \citenamefont {Raymond} \emph
  {et~al.}}]{King2018Observation}%
  \BibitemOpen
  \bibfield  {author} {\bibinfo {author} {\bibfnamefont {A.~D.}\ \bibnamefont
  {King}}, \bibinfo {author} {\bibfnamefont {J.}~\bibnamefont {Carrasquilla}},
  \bibinfo {author} {\bibfnamefont {J.}~\bibnamefont {Raymond}}, \emph
  {et~al.},\ }\bibfield  {title} {\bibinfo {title} {{Observation of topological
  phenomena in a programmable lattice of 1,800 qubits}},\ }\href
  {https://doi.org/10.1038/s41586-018-0410-x} {\bibfield  {journal} {\bibinfo
  {journal} {Nature}\ }\textbf {\bibinfo {volume} {560}},\ \bibinfo {pages}
  {456} (\bibinfo {year} {2018})}\BibitemShut {NoStop}%
\bibitem [{\citenamefont {Yamashiro}\ \emph {et~al.}(2019)\citenamefont
  {Yamashiro}, \citenamefont {Ohkuwa}, \citenamefont {Nishimori},\ and\
  \citenamefont {Lidar}}]{Yamashiro2019Dynamics}%
  \BibitemOpen
  \bibfield  {author} {\bibinfo {author} {\bibfnamefont {Y.}~\bibnamefont
  {Yamashiro}}, \bibinfo {author} {\bibfnamefont {M.}~\bibnamefont {Ohkuwa}},
  \bibinfo {author} {\bibfnamefont {H.}~\bibnamefont {Nishimori}},\ and\
  \bibinfo {author} {\bibfnamefont {D.~A.}\ \bibnamefont {Lidar}},\ }\bibfield
  {title} {\bibinfo {title} {{Dynamics of reverse annealing for the fully
  connected $p$-spin model}},\ }\href
  {https://doi.org/10.1103/PhysRevA.100.052321} {\bibfield  {journal} {\bibinfo
   {journal} {Phys. Rev. A}\ }\textbf {\bibinfo {volume} {100}},\ \bibinfo
  {pages} {052321} (\bibinfo {year} {2019})}\BibitemShut {NoStop}%
\bibitem [{\citenamefont {Boixo}\ \emph {et~al.}(2014)\citenamefont {Boixo},
  \citenamefont {R{\o}nnow}, \citenamefont {Isakov}, \citenamefont {Wang},
  \citenamefont {Wecker}, \citenamefont {Lidar}, \citenamefont {Martinis},\
  and\ \citenamefont {Troyer}}]{Boixo2014Evidence}%
  \BibitemOpen
  \bibfield  {author} {\bibinfo {author} {\bibfnamefont {S.}~\bibnamefont
  {Boixo}}, \bibinfo {author} {\bibfnamefont {T.~F.}\ \bibnamefont
  {R{\o}nnow}}, \bibinfo {author} {\bibfnamefont {S.~V.}\ \bibnamefont
  {Isakov}}, \bibinfo {author} {\bibfnamefont {Z.}~\bibnamefont {Wang}},
  \bibinfo {author} {\bibfnamefont {D.}~\bibnamefont {Wecker}}, \bibinfo
  {author} {\bibfnamefont {D.~A.}\ \bibnamefont {Lidar}}, \bibinfo {author}
  {\bibfnamefont {J.~M.}\ \bibnamefont {Martinis}},\ and\ \bibinfo {author}
  {\bibfnamefont {M.}~\bibnamefont {Troyer}},\ }\bibfield  {title} {\bibinfo
  {title} {{Evidence for quantum annealing with more than one hundred
  qubits}},\ }\href {https://doi.org/10.1038/nphys2900} {\bibfield  {journal}
  {\bibinfo  {journal} {Nat. Phys.}\ }\textbf {\bibinfo {volume} {10}},\
  \bibinfo {pages} {218} (\bibinfo {year} {2014})}\BibitemShut {NoStop}%
\bibitem [{\citenamefont {Carlini}\ \emph {et~al.}(2006)\citenamefont
  {Carlini}, \citenamefont {Hosoya}, \citenamefont {Koike},\ and\ \citenamefont
  {Okudaira}}]{Carlini06}%
  \BibitemOpen
  \bibfield  {author} {\bibinfo {author} {\bibfnamefont {A.}~\bibnamefont
  {Carlini}}, \bibinfo {author} {\bibfnamefont {A.}~\bibnamefont {Hosoya}},
  \bibinfo {author} {\bibfnamefont {T.}~\bibnamefont {Koike}},\ and\ \bibinfo
  {author} {\bibfnamefont {Y.}~\bibnamefont {Okudaira}},\ }\bibfield  {title}
  {\bibinfo {title} {{Time-Optimal Quantum Evolution}},\ }\href
  {https://doi.org/10.1103/PhysRevLett.96.060503} {\bibfield  {journal}
  {\bibinfo  {journal} {Phys. Rev. Lett.}\ }\textbf {\bibinfo {volume} {96}},\
  \bibinfo {pages} {060503} (\bibinfo {year} {2006})}\BibitemShut {NoStop}%
\bibitem [{\citenamefont {Bender}\ and\ \citenamefont
  {Brody}(2009)}]{Bender09}%
  \BibitemOpen
  \bibfield  {author} {\bibinfo {author} {\bibfnamefont {C.~M.}\ \bibnamefont
  {Bender}}\ and\ \bibinfo {author} {\bibfnamefont {D.~C.}\ \bibnamefont
  {Brody}},\ }\bibinfo {title} {{Optimal Time Evolution for Hermitian and
  Non-Hermitian Hamiltonians}},\ in\ \href
  {https://doi.org/10.1007/978-3-642-03174-8_12} {\emph {\bibinfo {booktitle}
  {Time in Quantum Mechanics - Vol. 2}}},\ \bibinfo {editor} {edited by\
  \bibinfo {editor} {\bibfnamefont {G.}~\bibnamefont {Muga}}, \bibinfo {editor}
  {\bibfnamefont {A.}~\bibnamefont {Ruschhaupt}},\ and\ \bibinfo {editor}
  {\bibfnamefont {A.}~\bibnamefont {del Campo}}}\ (\bibinfo  {publisher}
  {Springer Berlin Heidelberg},\ \bibinfo {address} {Berlin, Heidelberg},\
  \bibinfo {year} {2009})\ pp.\ \bibinfo {pages} {341--361}\BibitemShut
  {NoStop}%
\bibitem [{\citenamefont {Takahashi}(2013{\natexlab{b}})}]{Takahashi13}%
  \BibitemOpen
  \bibfield  {author} {\bibinfo {author} {\bibfnamefont {K.}~\bibnamefont
  {Takahashi}},\ }\bibfield  {title} {\bibinfo {title} {{How fast and robust is
  the quantum adiabatic passage?}},\ }\href
  {https://doi.org/10.1088/1751-8113/46/31/315304} {\bibfield  {journal}
  {\bibinfo  {journal} {J. Phys. A}\ }\textbf {\bibinfo {volume} {46}},\
  \bibinfo {pages} {315304} (\bibinfo {year} {2013}{\natexlab{b}})}\BibitemShut
  {NoStop}%
\bibitem [{\citenamefont {Romero}\ \emph
  {et~al.}(2025{\natexlab{a}})\citenamefont {Romero}, \citenamefont {Visuri},
  \citenamefont {Cadavid}, \citenamefont {Simen}, \citenamefont {Solano},\ and\
  \citenamefont {Hegade}}]{Romero2025}%
  \BibitemOpen
  \bibfield  {author} {\bibinfo {author} {\bibfnamefont {S.~V.}\ \bibnamefont
  {Romero}}, \bibinfo {author} {\bibfnamefont {A.-M.}\ \bibnamefont {Visuri}},
  \bibinfo {author} {\bibfnamefont {A.~G.}\ \bibnamefont {Cadavid}}, \bibinfo
  {author} {\bibfnamefont {A.}~\bibnamefont {Simen}}, \bibinfo {author}
  {\bibfnamefont {E.}~\bibnamefont {Solano}},\ and\ \bibinfo {author}
  {\bibfnamefont {N.~N.}\ \bibnamefont {Hegade}},\ }\bibfield  {title}
  {\bibinfo {title} {Bias-field digitized counterdiabatic quantum algorithm for
  higher-order binary optimization},\ }\href
  {https://doi.org/10.1038/s42005-025-02270-3} {\bibfield  {journal} {\bibinfo
  {journal} {Communications Physics}\ }\textbf {\bibinfo {volume} {8}},\
  \bibinfo {pages} {348} (\bibinfo {year} {2025}{\natexlab{a}})}\BibitemShut
  {NoStop}%
\bibitem [{\citenamefont {Romero}\ \emph
  {et~al.}(2025{\natexlab{b}})\citenamefont {Romero}, \citenamefont {Cadavid},
  \citenamefont {Nikačević}, \citenamefont {Solano}, \citenamefont {Hegade},
  \citenamefont {Lopez-Ruiz}, \citenamefont {Girotto}, \citenamefont {Yamada},
  \citenamefont {Barkoutsos}, \citenamefont {Kaushik},\ and\ \citenamefont
  {Roetteler}}]{romero2025proteinfolding}%
  \BibitemOpen
  \bibfield  {author} {\bibinfo {author} {\bibfnamefont {S.~V.}\ \bibnamefont
  {Romero}}, \bibinfo {author} {\bibfnamefont {A.~G.}\ \bibnamefont {Cadavid}},
  \bibinfo {author} {\bibfnamefont {P.}~\bibnamefont {Nikačević}}, \bibinfo
  {author} {\bibfnamefont {E.}~\bibnamefont {Solano}}, \bibinfo {author}
  {\bibfnamefont {N.~N.}\ \bibnamefont {Hegade}}, \bibinfo {author}
  {\bibfnamefont {M.~A.}\ \bibnamefont {Lopez-Ruiz}}, \bibinfo {author}
  {\bibfnamefont {C.}~\bibnamefont {Girotto}}, \bibinfo {author} {\bibfnamefont
  {M.}~\bibnamefont {Yamada}}, \bibinfo {author} {\bibfnamefont {P.~K.}\
  \bibnamefont {Barkoutsos}}, \bibinfo {author} {\bibfnamefont
  {A.}~\bibnamefont {Kaushik}},\ and\ \bibinfo {author} {\bibfnamefont
  {M.}~\bibnamefont {Roetteler}},\ }\href {https://arxiv.org/abs/2506.07866}
  {\bibinfo {title} {Protein folding with an all-to-all trapped-ion quantum
  computer}} (\bibinfo {year} {2025}{\natexlab{b}}),\ \Eprint
  {https://arxiv.org/abs/2506.07866} {arXiv:2506.07866 [quant-ph]} \BibitemShut
  {NoStop}%
\bibitem [{\citenamefont {Abah}\ \emph {et~al.}(2019)\citenamefont {Abah},
  \citenamefont {Puebla}, \citenamefont {Kiely}, \citenamefont {Chiara},
  \citenamefont {Paternostro},\ and\ \citenamefont
  {Campbell}}]{Abah_2019_Energetic_costs}%
  \BibitemOpen
  \bibfield  {author} {\bibinfo {author} {\bibfnamefont {O.}~\bibnamefont
  {Abah}}, \bibinfo {author} {\bibfnamefont {R.}~\bibnamefont {Puebla}},
  \bibinfo {author} {\bibfnamefont {A.}~\bibnamefont {Kiely}}, \bibinfo
  {author} {\bibfnamefont {G.~D.}\ \bibnamefont {Chiara}}, \bibinfo {author}
  {\bibfnamefont {M.}~\bibnamefont {Paternostro}},\ and\ \bibinfo {author}
  {\bibfnamefont {S.}~\bibnamefont {Campbell}},\ }\bibfield  {title} {\bibinfo
  {title} {{Energetic cost of quantum control protocols}},\ }\href
  {https://doi.org/10.1088/1367-2630/ab4c8c} {\bibfield  {journal} {\bibinfo
  {journal} {New J. Phys.}\ }\textbf {\bibinfo {volume} {21}},\ \bibinfo
  {pages} {103048} (\bibinfo {year} {2019})}\BibitemShut {NoStop}%
\bibitem [{\citenamefont {Torrontegui}\ \emph {et~al.}(2017)\citenamefont
  {Torrontegui}, \citenamefont {Lizuain}, \citenamefont {Gonz\'alez-Resines},
  \citenamefont {Tobalina}, \citenamefont {Ruschhaupt}, \citenamefont
  {Kosloff},\ and\ \citenamefont {Muga}}]{Torrontegui_2017_EnergyConsumeSTA}%
  \BibitemOpen
  \bibfield  {author} {\bibinfo {author} {\bibfnamefont {E.}~\bibnamefont
  {Torrontegui}}, \bibinfo {author} {\bibfnamefont {I.}~\bibnamefont
  {Lizuain}}, \bibinfo {author} {\bibfnamefont {S.}~\bibnamefont
  {Gonz\'alez-Resines}}, \bibinfo {author} {\bibfnamefont {A.}~\bibnamefont
  {Tobalina}}, \bibinfo {author} {\bibfnamefont {A.}~\bibnamefont
  {Ruschhaupt}}, \bibinfo {author} {\bibfnamefont {R.}~\bibnamefont
  {Kosloff}},\ and\ \bibinfo {author} {\bibfnamefont {J.~G.}\ \bibnamefont
  {Muga}},\ }\bibfield  {title} {\bibinfo {title} {{Energy consumption for
  shortcuts to adiabaticity}},\ }\href
  {https://doi.org/10.1103/PhysRevA.96.022133} {\bibfield  {journal} {\bibinfo
  {journal} {Phys. Rev. A}\ }\textbf {\bibinfo {volume} {96}},\ \bibinfo
  {pages} {022133} (\bibinfo {year} {2017})}\BibitemShut {NoStop}%
\bibitem [{\citenamefont {Campbell}\ and\ \citenamefont
  {Deffner}(2017)}]{Campbell2017}%
  \BibitemOpen
  \bibfield  {author} {\bibinfo {author} {\bibfnamefont {S.}~\bibnamefont
  {Campbell}}\ and\ \bibinfo {author} {\bibfnamefont {S.}~\bibnamefont
  {Deffner}},\ }\bibfield  {title} {\bibinfo {title} {{Trade-Off Between Speed
  and Cost in Shortcuts to Adiabaticity}},\ }\href
  {https://doi.org/10.1103/PhysRevLett.118.100601} {\bibfield  {journal}
  {\bibinfo  {journal} {Phys. Rev. Lett.}\ }\textbf {\bibinfo {volume} {118}},\
  \bibinfo {pages} {100601} (\bibinfo {year} {2017})}\BibitemShut {NoStop}%
\bibitem [{\citenamefont {Funo}\ \emph {et~al.}(2017)\citenamefont {Funo},
  \citenamefont {Zhang}, \citenamefont {Chatou}, \citenamefont {Kim},
  \citenamefont {Ueda},\ and\ \citenamefont {del Campo}}]{Funo2017}%
  \BibitemOpen
  \bibfield  {author} {\bibinfo {author} {\bibfnamefont {K.}~\bibnamefont
  {Funo}}, \bibinfo {author} {\bibfnamefont {J.-N.}\ \bibnamefont {Zhang}},
  \bibinfo {author} {\bibfnamefont {C.}~\bibnamefont {Chatou}}, \bibinfo
  {author} {\bibfnamefont {K.}~\bibnamefont {Kim}}, \bibinfo {author}
  {\bibfnamefont {M.}~\bibnamefont {Ueda}},\ and\ \bibinfo {author}
  {\bibfnamefont {A.}~\bibnamefont {del Campo}},\ }\bibfield  {title} {\bibinfo
  {title} {Universal work fluctuations during shortcuts to adiabaticity by
  counterdiabatic driving},\ }\href
  {https://doi.org/10.1103/PhysRevLett.118.100602} {\bibfield  {journal}
  {\bibinfo  {journal} {Phys. Rev. Lett.}\ }\textbf {\bibinfo {volume} {118}},\
  \bibinfo {pages} {100602} (\bibinfo {year} {2017})}\BibitemShut {NoStop}%
\bibitem [{\citenamefont {Farhi}\ \emph {et~al.}(2012)\citenamefont {Farhi},
  \citenamefont {Gosset}, \citenamefont {Hen}, \citenamefont {Sandvik},
  \citenamefont {Shor}, \citenamefont {Young},\ and\ \citenamefont
  {Zamponi}}]{FarhiMAXCUT_XORSAT2012}%
  \BibitemOpen
  \bibfield  {author} {\bibinfo {author} {\bibfnamefont {E.}~\bibnamefont
  {Farhi}}, \bibinfo {author} {\bibfnamefont {D.}~\bibnamefont {Gosset}},
  \bibinfo {author} {\bibfnamefont {I.}~\bibnamefont {Hen}}, \bibinfo {author}
  {\bibfnamefont {A.~W.}\ \bibnamefont {Sandvik}}, \bibinfo {author}
  {\bibfnamefont {P.}~\bibnamefont {Shor}}, \bibinfo {author} {\bibfnamefont
  {A.~P.}\ \bibnamefont {Young}},\ and\ \bibinfo {author} {\bibfnamefont
  {F.}~\bibnamefont {Zamponi}},\ }\bibfield  {title} {\bibinfo {title}
  {Performance of the quantum adiabatic algorithm on random instances of two
  optimization problems on regular hypergraphs},\ }\href
  {https://doi.org/10.1103/PhysRevA.86.052334} {\bibfield  {journal} {\bibinfo
  {journal} {Phys. Rev. A}\ }\textbf {\bibinfo {volume} {86}},\ \bibinfo
  {pages} {052334} (\bibinfo {year} {2012})}\BibitemShut {NoStop}%
\bibitem [{\citenamefont {Albash}\ and\ \citenamefont
  {Lidar}(2018)}]{Albash2018Adiabatic}%
  \BibitemOpen
  \bibfield  {author} {\bibinfo {author} {\bibfnamefont {T.}~\bibnamefont
  {Albash}}\ and\ \bibinfo {author} {\bibfnamefont {D.~A.}\ \bibnamefont
  {Lidar}},\ }\bibfield  {title} {\bibinfo {title} {{Adiabatic quantum
  computation}},\ }\href {https://doi.org/10.1103/RevModPhys.90.015002}
  {\bibfield  {journal} {\bibinfo  {journal} {Rev. Mod. Phys.}\ }\textbf
  {\bibinfo {volume} {90}},\ \bibinfo {pages} {015002} (\bibinfo {year}
  {2018})}\BibitemShut {NoStop}%
\bibitem [{\citenamefont {Tindall}\ \emph {et~al.}(2025)\citenamefont
  {Tindall}, \citenamefont {Mello}, \citenamefont {Fishman}, \citenamefont
  {Stoudenmire},\ and\ \citenamefont {Sels}}]{tindall2025}%
  \BibitemOpen
  \bibfield  {author} {\bibinfo {author} {\bibfnamefont {J.}~\bibnamefont
  {Tindall}}, \bibinfo {author} {\bibfnamefont {A.}~\bibnamefont {Mello}},
  \bibinfo {author} {\bibfnamefont {M.}~\bibnamefont {Fishman}}, \bibinfo
  {author} {\bibfnamefont {M.}~\bibnamefont {Stoudenmire}},\ and\ \bibinfo
  {author} {\bibfnamefont {D.}~\bibnamefont {Sels}},\ }\href
  {https://arxiv.org/abs/2503.05693} {\bibinfo {title} {Dynamics of disordered
  quantum systems with two- and three-dimensional tensor networks}} (\bibinfo
  {year} {2025}),\ \Eprint {https://arxiv.org/abs/2503.05693} {arXiv:2503.05693
  [quant-ph]} \BibitemShut {NoStop}%
\bibitem [{\citenamefont {Mauron}\ and\ \citenamefont
  {Carleo}(2025)}]{mauron2025}%
  \BibitemOpen
  \bibfield  {author} {\bibinfo {author} {\bibfnamefont {L.}~\bibnamefont
  {Mauron}}\ and\ \bibinfo {author} {\bibfnamefont {G.}~\bibnamefont
  {Carleo}},\ }\href {https://arxiv.org/abs/2503.08247} {\bibinfo {title}
  {Challenging the quantum advantage frontier with large-scale classical
  simulations of annealing dynamics}} (\bibinfo {year} {2025}),\ \Eprint
  {https://arxiv.org/abs/2503.08247} {arXiv:2503.08247 [quant-ph]} \BibitemShut
  {NoStop}%
\bibitem [{\citenamefont {Morawetz}\ and\ \citenamefont
  {Polkovnikov}(2024)}]{LCD_Morawetz_2024}%
  \BibitemOpen
  \bibfield  {author} {\bibinfo {author} {\bibfnamefont {S.}~\bibnamefont
  {Morawetz}}\ and\ \bibinfo {author} {\bibfnamefont {A.}~\bibnamefont
  {Polkovnikov}},\ }\bibfield  {title} {\bibinfo {title} {Efficient paths for
  local counterdiabatic driving},\ }\href
  {https://doi.org/10.1103/PhysRevB.110.024304} {\bibfield  {journal} {\bibinfo
   {journal} {Phys. Rev. B}\ }\textbf {\bibinfo {volume} {110}},\ \bibinfo
  {pages} {024304} (\bibinfo {year} {2024})}\BibitemShut {NoStop}%
\bibitem [{\citenamefont {Visuri}\ \emph {et~al.}(2025)\citenamefont {Visuri},
  \citenamefont {Cadavid}, \citenamefont {Bhargava}, \citenamefont {Romero},
  \citenamefont {Grabarits}, \citenamefont {Chandarana}, \citenamefont
  {Solano}, \citenamefont {del Campo},\ and\ \citenamefont
  {Hegade}}]{visuri2025digitized}%
  \BibitemOpen
  \bibfield  {author} {\bibinfo {author} {\bibfnamefont {A.-M.}\ \bibnamefont
  {Visuri}}, \bibinfo {author} {\bibfnamefont {A.~G.}\ \bibnamefont {Cadavid}},
  \bibinfo {author} {\bibfnamefont {B.~A.}\ \bibnamefont {Bhargava}}, \bibinfo
  {author} {\bibfnamefont {S.~V.}\ \bibnamefont {Romero}}, \bibinfo {author}
  {\bibfnamefont {A.}~\bibnamefont {Grabarits}}, \bibinfo {author}
  {\bibfnamefont {P.}~\bibnamefont {Chandarana}}, \bibinfo {author}
  {\bibfnamefont {E.}~\bibnamefont {Solano}}, \bibinfo {author} {\bibfnamefont
  {A.}~\bibnamefont {del Campo}},\ and\ \bibinfo {author} {\bibfnamefont
  {N.~N.}\ \bibnamefont {Hegade}},\ }\href {https://arxiv.org/abs/2502.15100}
  {\bibinfo {title} {Digitized counterdiabatic quantum critical dynamics}}
  (\bibinfo {year} {2025}),\ \Eprint {https://arxiv.org/abs/2502.15100}
  {arXiv:2502.15100 [quant-ph]} \BibitemShut {NoStop}%
\bibitem [{\citenamefont {Bellitti}\ \emph {et~al.}(2021)\citenamefont
  {Bellitti}, \citenamefont {Ricci-Tersenghi},\ and\ \citenamefont
  {Scardicchio}}]{Bellitti2021Entropic}%
  \BibitemOpen
  \bibfield  {author} {\bibinfo {author} {\bibfnamefont {M.}~\bibnamefont
  {Bellitti}}, \bibinfo {author} {\bibfnamefont {F.}~\bibnamefont
  {Ricci-Tersenghi}},\ and\ \bibinfo {author} {\bibfnamefont {A.}~\bibnamefont
  {Scardicchio}},\ }\bibfield  {title} {\bibinfo {title} {{Entropic barriers as
  a reason for hardness in both classical and quantum algorithms}},\ }\href
  {https://doi.org/10.1103/PhysRevResearch.3.043015} {\bibfield  {journal}
  {\bibinfo  {journal} {Phys. Rev. Res.}\ }\textbf {\bibinfo {volume} {3}},\
  \bibinfo {pages} {043015} (\bibinfo {year} {2021})}\BibitemShut {NoStop}%
\bibitem [{\citenamefont {McClean}\ \emph {et~al.}(2018)\citenamefont
  {McClean}, \citenamefont {Boixo}, \citenamefont {Smelyanskiy}, \citenamefont
  {Babbush},\ and\ \citenamefont {Neven}}]{McClean2018Barren}%
  \BibitemOpen
  \bibfield  {author} {\bibinfo {author} {\bibfnamefont {J.~R.}\ \bibnamefont
  {McClean}}, \bibinfo {author} {\bibfnamefont {S.}~\bibnamefont {Boixo}},
  \bibinfo {author} {\bibfnamefont {V.~N.}\ \bibnamefont {Smelyanskiy}},
  \bibinfo {author} {\bibfnamefont {R.}~\bibnamefont {Babbush}},\ and\ \bibinfo
  {author} {\bibfnamefont {H.}~\bibnamefont {Neven}},\ }\bibfield  {title}
  {\bibinfo {title} {{Barren plateaus in quantum neural network training
  landscapes}},\ }\href {https://doi.org/10.1038/s41467-018-07090-4} {\bibfield
   {journal} {\bibinfo  {journal} {Nat. Commun.}\ }\textbf {\bibinfo {volume}
  {9}},\ \bibinfo {pages} {4812} (\bibinfo {year} {2018})}\BibitemShut
  {NoStop}%
\bibitem [{\citenamefont {Day}\ \emph {et~al.}(2019)\citenamefont {Day},
  \citenamefont {Bukov}, \citenamefont {Weinberg}, \citenamefont {Mehta},\ and\
  \citenamefont {Sels}}]{Day2019Glassy}%
  \BibitemOpen
  \bibfield  {author} {\bibinfo {author} {\bibfnamefont {A.~G.~R.}\
  \bibnamefont {Day}}, \bibinfo {author} {\bibfnamefont {M.}~\bibnamefont
  {Bukov}}, \bibinfo {author} {\bibfnamefont {P.}~\bibnamefont {Weinberg}},
  \bibinfo {author} {\bibfnamefont {P.}~\bibnamefont {Mehta}},\ and\ \bibinfo
  {author} {\bibfnamefont {D.}~\bibnamefont {Sels}},\ }\bibfield  {title}
  {\bibinfo {title} {{Glassy Phase of Optimal Quantum Control}},\ }\href
  {https://doi.org/10.1103/PhysRevLett.122.020601} {\bibfield  {journal}
  {\bibinfo  {journal} {Phys. Rev. Lett.}\ }\textbf {\bibinfo {volume} {122}},\
  \bibinfo {pages} {020601} (\bibinfo {year} {2019})}\BibitemShut {NoStop}%
\bibitem [{\citenamefont {Larocca}\ \emph {et~al.}(2024)\citenamefont
  {Larocca}, \citenamefont {Thanasilp}, \citenamefont {Wang} \emph
  {et~al.}}]{Larocca2024Review}%
  \BibitemOpen
  \bibfield  {author} {\bibinfo {author} {\bibfnamefont {M.}~\bibnamefont
  {Larocca}}, \bibinfo {author} {\bibfnamefont {S.}~\bibnamefont {Thanasilp}},
  \bibinfo {author} {\bibfnamefont {S.}~\bibnamefont {Wang}}, \emph {et~al.},\
  }\href@noop {} {\bibinfo {title} {{A Review of Barren Plateaus in Variational
  Quantum Computing}}} (\bibinfo {year} {2024}),\ \Eprint
  {https://arxiv.org/abs/2405.00781} {arXiv:2405.00781} \BibitemShut {NoStop}%
\bibitem [{\citenamefont {Beato}\ \emph {et~al.}(2024)\citenamefont {Beato},
  \citenamefont {Patil},\ and\ \citenamefont {Bukov}}]{Beato2024Theory}%
  \BibitemOpen
  \bibfield  {author} {\bibinfo {author} {\bibfnamefont {N.}~\bibnamefont
  {Beato}}, \bibinfo {author} {\bibfnamefont {P.}~\bibnamefont {Patil}},\ and\
  \bibinfo {author} {\bibfnamefont {M.}~\bibnamefont {Bukov}},\ }\href@noop {}
  {\bibinfo {title} {{Towards a theory of phase transitions in quantum control
  landscapes}}} (\bibinfo {year} {2024}),\ \Eprint
  {https://arxiv.org/abs/2408.11110} {arXiv:2408.11110} \BibitemShut {NoStop}%
\bibitem [{\citenamefont {Rams}\ \emph {et~al.}(2019)\citenamefont {Rams},
  \citenamefont {Dziarmaga},\ and\ \citenamefont {Zurek}}]{Rams2019}%
  \BibitemOpen
  \bibfield  {author} {\bibinfo {author} {\bibfnamefont {M.~M.}\ \bibnamefont
  {Rams}}, \bibinfo {author} {\bibfnamefont {J.}~\bibnamefont {Dziarmaga}},\
  and\ \bibinfo {author} {\bibfnamefont {W.~H.}\ \bibnamefont {Zurek}},\
  }\bibfield  {title} {\bibinfo {title} {{Symmetry Breaking Bias and the
  Dynamics of a Quantum Phase Transition}},\ }\href
  {https://doi.org/10.1103/PhysRevLett.123.130603} {\bibfield  {journal}
  {\bibinfo  {journal} {Phys. Rev. Lett.}\ }\textbf {\bibinfo {volume} {123}},\
  \bibinfo {pages} {130603} (\bibinfo {year} {2019})}\BibitemShut {NoStop}%
\bibitem [{\citenamefont {Asb{\'o}th}\ \emph {et~al.}(2016)\citenamefont
  {Asb{\'o}th}, \citenamefont {Oroszl{\'a}ny},\ and\ \citenamefont
  {P{\'a}lyi}}]{Asboth2016Short}%
  \BibitemOpen
  \bibfield  {author} {\bibinfo {author} {\bibfnamefont {J.~K.}\ \bibnamefont
  {Asb{\'o}th}}, \bibinfo {author} {\bibfnamefont {L.}~\bibnamefont
  {Oroszl{\'a}ny}},\ and\ \bibinfo {author} {\bibfnamefont {A.}~\bibnamefont
  {P{\'a}lyi}},\ }\href {https://doi.org/10.1007/978-3-319-25607-8} {\emph
  {\bibinfo {title} {{A Short Course on Topological Insulators: Band Structure
  and Edge States in One and Two Dimensions}}}}\ (\bibinfo  {publisher}
  {Springer International},\ \bibinfo {address} {Cham},\ \bibinfo {year}
  {2016})\BibitemShut {NoStop}%
\bibitem [{\citenamefont {Seiberg}\ and\ \citenamefont
  {Shao}(2024)}]{Seiberg2024Majorana}%
  \BibitemOpen
  \bibfield  {author} {\bibinfo {author} {\bibfnamefont {N.}~\bibnamefont
  {Seiberg}}\ and\ \bibinfo {author} {\bibfnamefont {S.-H.}\ \bibnamefont
  {Shao}},\ }\bibfield  {title} {\bibinfo {title} {{Majorana chain and Ising
  model - (non-invertible) translations, anomalies, and emanant symmetries}},\
  }\href {https://doi.org/10.21468/SciPostPhys.16.3.064} {\bibfield  {journal}
  {\bibinfo  {journal} {SciPost Phys.}\ }\textbf {\bibinfo {volume} {16}},\
  \bibinfo {pages} {064} (\bibinfo {year} {2024})}\BibitemShut {NoStop}%
\bibitem [{\citenamefont {Lieb}\ \emph {et~al.}(1961)\citenamefont {Lieb},
  \citenamefont {Schultz},\ and\ \citenamefont {Mattis}}]{Lieb1967Two}%
  \BibitemOpen
  \bibfield  {author} {\bibinfo {author} {\bibfnamefont {E.}~\bibnamefont
  {Lieb}}, \bibinfo {author} {\bibfnamefont {T.}~\bibnamefont {Schultz}},\ and\
  \bibinfo {author} {\bibfnamefont {D.}~\bibnamefont {Mattis}},\ }\bibfield
  {title} {\bibinfo {title} {{Two soluble models of an antiferromagnetic
  chain}},\ }\href {https://doi.org/10.1016/0003-4916(61)90115-4} {\bibfield
  {journal} {\bibinfo  {journal} {Ann. Phys.}\ }\textbf {\bibinfo {volume}
  {16}},\ \bibinfo {pages} {407} (\bibinfo {year} {1961})}\BibitemShut
  {NoStop}%
\bibitem [{\citenamefont {Pro\'{s}niak}(2019)}]{Prosniak2019Size}%
  \BibitemOpen
  \bibfield  {author} {\bibinfo {author} {\bibfnamefont {O.~A.}\ \bibnamefont
  {Pro\'{s}niak}},\ }\bibfield  {title} {\bibinfo {title} {{On the size of
  boundary effects in the Ising chain}},\ }\href
  {https://doi.org/10.1088/1402-4896/ab1492} {\bibfield  {journal} {\bibinfo
  {journal} {Phys. Scripta}\ }\textbf {\bibinfo {volume} {94}},\ \bibinfo
  {pages} {085201} (\bibinfo {year} {2019})}\BibitemShut {NoStop}%
\bibitem [{\citenamefont {Mbeng}\ \emph {et~al.}(2024)\citenamefont {Mbeng},
  \citenamefont {Russomanno},\ and\ \citenamefont
  {Santoro}}]{mbeng2020quantum}%
  \BibitemOpen
  \bibfield  {author} {\bibinfo {author} {\bibfnamefont {G.~B.}\ \bibnamefont
  {Mbeng}}, \bibinfo {author} {\bibfnamefont {A.}~\bibnamefont {Russomanno}},\
  and\ \bibinfo {author} {\bibfnamefont {G.~E.}\ \bibnamefont {Santoro}},\
  }\bibfield  {title} {\bibinfo {title} {{The quantum Ising chain for
  beginners}},\ }\href {https://doi.org/10.21468/SciPostPhysLectNotes.82}
  {\bibfield  {journal} {\bibinfo  {journal} {SciPost Phys. Lect. Notes}\ ,\
  \bibinfo {pages} {82}} (\bibinfo {year} {2024})}\BibitemShut {NoStop}%
\bibitem [{\citenamefont {Surace}\ and\ \citenamefont
  {Tagliacozzo}(2022)}]{Surace2022Fermionic}%
  \BibitemOpen
  \bibfield  {author} {\bibinfo {author} {\bibfnamefont {J.}~\bibnamefont
  {Surace}}\ and\ \bibinfo {author} {\bibfnamefont {L.}~\bibnamefont
  {Tagliacozzo}},\ }\bibfield  {title} {\bibinfo {title} {{Fermionic Gaussian
  states: an introduction to numerical approaches}},\ }\href
  {https://doi.org/10.21468/SciPostPhysLectNotes.54} {\bibfield  {journal}
  {\bibinfo  {journal} {SciPost Phys. Lect. Notes}\ ,\ \bibinfo {pages} {54}}
  (\bibinfo {year} {2022})}\BibitemShut {NoStop}%
\end{thebibliography}%

\end{document}